\documentclass[
  aps, prb,
  reprint,
  amsmath
  ]{revtex4-2}

\usepackage[T1]{fontenc}
\usepackage{libertinus}
\usepackage{libertinust1math}
\usepackage[cal=stixtwoplain]{mathalpha}
\usepackage{booktabs}
\usepackage{csquotes}

\usepackage{enumitem}

\usepackage{microtype}
\usepackage{bm}

\usepackage{xcolor}
\usepackage{graphicx}
\graphicspath{{./img/}}

\usepackage{siunitx}
\DeclareSIUnit{\belmilliwatt}{Bm}
\DeclareSIUnit{\dBm}{\deci\belmilliwatt}
\usepackage[version=4]{mhchem}

\usepackage{hyperref}
\colorlet{hypercolor}{blue!50!black}
\hypersetup{
  colorlinks,
  linkcolor=hypercolor,
  filecolor=hypercolor,
  citecolor=hypercolor,
  urlcolor=hypercolor,
}

\makeatletter
  \patchcmd\frontmatter@setup{\normalfont}{\normalfont\sffamily}{}{}
  \patchcmd\frontmatter@title@format{\bfseries}{\sffamily\bfseries}{}{}
  \patchcmd\frontmatter@affiliationfont{\it}{\slshape}
  \patchcmd\frontmatter@title@format{\large}{\Large}{}{}
  \patchcmd\section{\bfseries}{\sffamily\bfseries}{}{}
  \patchcmd\subsection{\bfseries}{\sffamily\bfseries}{}{}
  \patchcmd\subsubsection{\itshape}{\sffamily}{}{}
  \patchcmd\@makecaption{\rmfamily}{\sffamily}{}{}
\makeatother

\makeatletter
  \def\bibsection{%
    \expandafter\section\expandafter*\expandafter{\refname}%
    \@nobreaktrue
  }%
\makeatother

\def\e{\mathrm{e}}
\def\i{\mathrm{i}}

\NewDocumentCommand\titlecaption{ m m }{%
  \caption[#1]{\textbf{#1.}\hskip 1.5\fontdimen2\font plus 1em minus 1.5\fontdimen4\font\relax#2}%
}
\NewDocumentCommand\sublabel{ m }{\textbf{#1}}

\NewDocumentCommand\lisaplus{}{LISA\!\textsuperscript{\bfseries +}}

\usepackage{ninecolors}
\usepackage{marginnote}
\makeatletter
  \appto\@floatboxreset{}
\makeatother
\usepackage[
  textsize=scriptsize,
  linecolor=red4,
  bordercolor=red4,
  backgroundcolor=red9,
  ]{todonotes}
\usepackage{changes}
\DeclareHookRule{enddocument}{loc}{before}{changes}
\DeclareHookRule{enddocument}{loc}{before}{revtex4-2}

\newcommand*\sn{\mathrm{Si}_3\mathrm{N}_4}

\DeclareMathOperator\atanii{atan2}

\begin{document}

\title{Tunable and nonlinearity-enhanced \\ dispersive-plus-dissipative coupling in photon-pressure circuits}

\def\pitaffil{%
  \affiliation{Physikalisches Institut, Center for Quantum Science~(CQ) and \lisaplus, Universität Tübingen, 72076~Tübingen, Germany}
}
\author{Mohamad~Kazouini}
\email{mohamad.adnan-el-kazouini@uni-tuebingen.de}
\pitaffil
\author{Janis~Peter}
\pitaffil
\author{Zisu~Emily~Guo}
\pitaffil
\author{Benedikt~Wilde}
\pitaffil
\author{Kevin~Uhl}
\pitaffil
\author{Dieter~Koelle}
\pitaffil
\author{Reinhold~Kleiner}
\pitaffil
\author{Daniel~Bothner}
\email{daniel.bothner@uni-tuebingen.de}
\pitaffil

\begin{abstract}
Photon-pressure circuits are the circuit implementation of the cavity optomechanical Hamiltonian and discussed for qubit readout, low-frequency quantum photonics and dark matter axion detection.
Due to the enormous design flexibility of superconducting circuits, photon-pressure systems provide fascinating possibilities to explore unusual parameter regimes of the optomechanical Hamiltonian.
Here, we report the realization of a photon-pressure platform, in which a GHz circuit interacts with a MHz circuit via a magnetic-flux-tunable combination of dispersive and dissipative photon-pressure.
In addition, both coupling rates are considerably enhanced by nonlinearities of the GHz-mode, which leads to the multi-photon coupling rates scaling stronger with the pump photon number $n_\mathrm{c}$ than the usual $\sqrt{n_\mathrm{c}}$ dependence.
We demonstrate that interference of the two interaction paths leads to a Fano-like response in photon-pressure induced transparency, and that the dynamical backaction is considerably modified compared to the dispersive case, including a parametric instability caused by a red-detuned pump tone.
\end{abstract}

\makeatletter
  \appto\titleblock@produce{\addvspace{-4\p@}\vspace{10pt}}
  \skip\footins=50pt plus 4pt minus 12pt
\makeatother

\maketitle

\let\oldaddcontentsline\addcontentsline
\renewcommand{\addcontentsline}[3]{}

\section{Introduction}
\vspace{-0mm}
Superconducting microwave circuits have developed into one of the most exciting and versatile platforms for emerging technologies in both the quantum and the classical domain \cite{Krantz2019, Blais2021}.
Quantum processors \cite{Arute2019, Kim2023}, quantum simulators \cite{Houck2012, Carusotto2020}, frequency transducers \cite{Andrews2014, Forsch2020}, electron-spin-resonance detectors \cite{Bienfait2016, Wang2023}, or quantum-limited signal amplifiers \cite{CastellanosBeltran2008, Bergeal2010, Macklin2015} are just a few of the many groundbreaking results of the recent decade.
Being highly flexible in frequency, linewidth and nonlinearity, it seems one can engineer a superconducting circuit implementation for any imaginable application on demand.
One of the more recent developments in superconducting microwave platforms are photon-pressure circuits, i.e., the circuit quantum electrodynamics implementation of the optomechanical Hamiltonian, in which two oscillators are parametrically interacting through a radiation-pressure-like nonlinear coupling \cite{Johansson2014, Eichler2018, Bothner2021}.
Compared to cavity optomechanics \cite{Aspelmeyer2014}, photon-pressure circuits provide a large degree of designability and possess orders of magnitude larger single-photon coupling rates \cite{Rodrigues2021, Potts2025}, which makes them not only interesting for applications, but also to study new regimes of radiation-pressure physics.
Beyond a high-quality replication of optomechanical gold-standard experiments such as photon-pressure induced transparency \cite{Eichler2018, Bothner2021}, parametric strong coupling \cite{Bothner2021} or sideband-cooling into the quantum regime \cite{Rodrigues2021}, photon-pressure circuits have allowed the realization of more sophisticated experiments with parametrically amplified coupling rates, nonreciprocal bath dynamics or effective negative mass modes based on their intrinsic Kerr nonlinearity~\cite{Rodrigues2022, Rodrigues2024}.
All the existing experiments to date are based on dispersive photon-pressure, which is realized when the amplitude of one LC oscillator modulates the resonance frequency of a second.
However, this is not the only possible interaction type, the amplitude of one oscillator might also modulate the decay rate of the other.
The latter then corresponds to so-called dissipative photon-pressure, a type of radiation-pressure that has attracted considerable attention in optomechanical systems \cite{Elste2009, Xuereb2011, Weiss2013, Kilda2015, Yanay2016, Tagantsev2019, Baraillon2020, Tagantsev2021, Monsel2024, Li2009, Wu2014, Sawadsky2015, Primo2023}, especially from the theoretical viewpoint \cite{Elste2009, Xuereb2011, Weiss2013, Kilda2015, Yanay2016, Tagantsev2019, Baraillon2020, Tagantsev2021, Monsel2024} since it is not as straightforward to realize experimentally as its dispersive counterpart \cite{Li2009, Wu2014, Sawadsky2015, Primo2023}.
Dissipative coupling would for instance allow groundstate-cooling in the sideband-unresolved limit \cite{Elste2009}, i.e., to extend the quantum regime to lower photon frequencies, or to enable alternative protocols for squeezing and quantum-limited sensing \cite{Kilda2015, Yanay2016, Tagantsev2019}.
Therefore, the implementation of dissipative coupling would greatly enrich the capabilities of photon-pressure circuits, constitute a flexible testbed for dissipative optomechanics and enable kHz to MHz quantum photonics as e.g. required for dark matter axion detection with radio-frequency upconverters \cite{Chaudhuri2019, Brouwer2022, Kuenstner2025}.
\begin{figure*}
	\includegraphics[trim={0 0mm 0 0},clip = true]{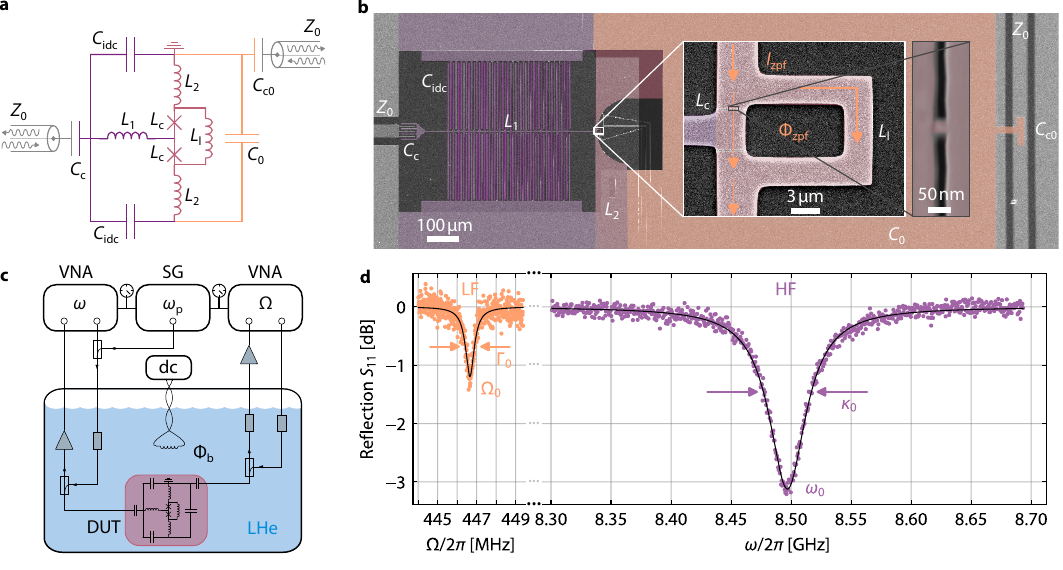}
	\titlecaption{Niobium photon-pressure circuits based on a monolithic nano-constriction SQUID and operated in liquid helium}{%
		\sublabel{a}~Simplified circuit equivalent. The HF mode (left half, purple and pink) combines two interdigitated capacitors $C_\mathrm{idc}$ with the inductors $L_1$, $L_2$ and $L_\mathrm{c}$. The LF mode (right half, orange and pink) comprises a parallel-plate capacitor $C_0$ with an inductance composed of $L_2$, $L_\mathrm{l}$ and $L_\mathrm{c}$. Both modes share the SQUID and parts of the linear inductance (in pink), and each mode is coupled to an individual feedline by coupling capacitors $C_\mathrm{c}$ and $C_\mathrm{c0}$, respectively.
		\sublabel{b}~False-color scanning electron microscopy image of the device. Colored and light gray parts are Nb, dark gray and transparent parts are Si substrate and Si$_3$N$_4$ dielectric of the LF capacitor, respectively. Color-code equivalent to \textbf{a}.
		White box inset shows a magnification of the interaction part. The rectangle is the SQUID loop, and the LF zero-point-fluctuation current $I_\mathrm{zpf}$ flows through the SQUID from top to bottom, hereby threading the loop with a zero-point-fluctuation flux $\Phi_\mathrm{zpf}$. Black box inset shows one of the monolithic 3D nano-constrictions with inductance $L_\mathrm{c}$ (image rotated by $90\degree$).
		\sublabel{c}~Simplified experimental setup. The device under test (DUT) is immersed in liquid helium (LHe) and connected to two coaxial lines for readout of the individual modes by a vector network analyzer (VNA). A HF sideband-pump field with frequency $\omega_\mathrm{p}$ from a signal generator (SG) is combined with the HF VNA signal in a directional coupler. Input and output of each readout line are combined by means of further directional couplers and the HF return signal is amplified by a cryogenic amplifier depicted as triangle; the LF mode return signal is amplified by a room-temperature amplifier. A small magnetic-field coil is driven with a direct current (dc) source to induce a bias flux $\Phi_\mathrm{b}$ into the SQUID. Gray rectangles represent attenuators. For more details cf.~Supplementary Note~I.
		\sublabel{d}~Reflection $S_{11}$ from both circuits ($\Phi_\mathrm{b} = 0$) measured through their individual feedlines. Colored symbols are experimental data, black lines are fit curves. From the fits, we extract the resonance frequencies $\omega_0 = 2\pi\times\qty{8.497}{\giga\hertz}$ and $\Omega_0 = 2\pi\times \qty{446.7}{\mega\hertz}$, as well as the total decay rates $\kappa_0 = 2\pi\times\qty{44.9}{\mega\hertz}$ and $\Gamma_0 = 2\pi\times\qty{525}{\kilo\hertz}$.
	}
	\label{fig:figure1}
  \vspace{2pt}
\end{figure*}
Here, we report the implementation of niobium-based photon-pressure circuits with both a dispersive and a dissipative interaction.
The two participating LC circuits are coupled to each other through an asymmetrically shared nano-constriction-based superconducting quantum interference device (SQUID).
The dissipative single-photon coupling rate of the system is flux-tunable with large values up to $g_{0\kappa}/2\pi \sim \qty{150}{kHz}$, and -- in contrast to earlier experiments with dissipative radiation-pressure in optomechanical systems -- it is solely attributed to a modulation of the internal decay rate.
As a consequence of the interference between the dispersive and the dissipative interactions, we observe a Fano-like modification of photon-pressure induced transparency, that provides an independent and simple measure to quantify the ratio between the two coupling rates.
Since the GHz SQUID circuit possesses both a flux-dependent Kerr nonlinearity and nonlinear damping, we furthermore observe that the multi-photon coupling rates are not scaling ${\propto} \sqrt{n_\mathrm{c}}$ with the pump photon number $n_\mathrm{c}$, but have contributions ${\propto}\sqrt{n_\mathrm{c}^3}$ and ${\propto} \sqrt{n_\mathrm{c}^7}$ as (partially) predicted by theoretical work \cite{Mikkelsen2017, Djorwe2019} but not yet observed in an experiment.
Finally, we characterize the generalized dynamical backaction of the high-frequency (HF) intracavity fields to the low-frequency (LF) circuit in the presence of a red-sideband pump field.
We find characteristic signatures of dissipative coupling contributions as well as a pump-power scaling in agreement with the nonlinearity-enhanced cooperativities.
Our results lay the foundation for the experimental investigation of dissipative and nonlinearity-enhanced interactions in photon-pressure systems and related platforms like SQUID optomechanics \cite{Rodrigues2019, Zoepfl2020, Schmidt2020, Bera2021, Bothner2022, Luschmann2022, Zoepfl2023, Schmidt2024}, and constitute a stepping stone towards photon-photon backaction-cooling in the unresolved sideband regime and kHz quantum photonics.
Facilitated by the use of niobium, all experiments presented here could be conducted at liquid helium temperatures, also opening the door for investigating radiation-pressure physics in the thermal regime and for photon-pressure experiments without the need for dilution refrigerators.
\vspace{-5mm}
\section{Results}
\vspace{-2mm}
\subsection{Device and setup}
\vspace{-2mm}
Our photon-pressure device, discussed in Fig.~\ref{fig:figure1} and Supplementary Notes~II and III, comprises a low-frequency LC circuit with a sweetspot resonance frequency $\Omega_0 = 2\pi\times\qty{446.7}{\mega\hertz}$ and a high-frequency quantum interference circuit with a sweetspot resonance frequency $\omega_0 = 2\pi\times\qty{8.497}{\giga\hertz}$.
Both circuits are composed of linear capacitors and multiple linear inductors (cf.~Supplementary Note~II) and share as nonlinear inductance and central coupling element a rectangular SQUID.
The Josephson elements in the SQUID are monolithic 3D nano-constriction junctions, fabricated by focused neon-ion-beam milling \cite{Uhl2024, Uhl2024a}.
For maximized photon-pressure coupling rates, the SQUID is galvanically shared by the two circuits, and the LF zero-point-fluctuation currents induce a zero-point-fluctuation flux $\Phi_\mathrm{zpf} \approx \qty{198.3}\,$\textmu$\Phi_0$ in the SQUID loop (containing both magnetic and kinetic contributions) with the flux quantum $\Phi_0 \approx \qty{2.068e-15}{\tesla\meter\squared}$, cf.~Supplementary Note~III.
Each of the two circuits is capacitively coupled to an individual coplanar waveguide (CPW) feedline with characteristic impedance $Z_0\approx \qty{50}{\ohm}$ for driving and readout.
Due to the galvanic coupling, the circuits can also be considered a single circuit with a LF and a HF mode, and we will use the words circuit and mode interchangeably in this manuscript.
As superconducting material for all the metallic parts we chose dc-magnetron sputtered niobium with a critical temperature $T_\mathrm{c} \approx \qty{9}{\kelvin}$, and as substrate we use high-resistivity intrinsic silicon with a substrate thickness $t_\mathrm{Si} = \qty{525}{\micro\meter}$.
The bottom niobium layer -- defining the HF circuit, the HF CPW feedline center conductor, the SQUID and the bottom electrode of the LF circuit capacitance -- has a thickness $t_\mathrm{Nb1} = \qty{120}{\nano\meter}$ and the top layer -- defining the top electrode of the LF capacitance, the LF CPW feedline center conductor, and all ground planes -- has $t_\mathrm{Nb2} = \qty{300}{\nano\meter}$.
As dielectric for the LF parallel-plate capacitors $C_0$ and $C_\mathrm{c0}$ we deposited $\qty{200}{\nano\meter}$ of silicon-nitride by means of plasma-enhanced chemical vapour deposition.
All layers have been patterned by maskless optical lithography and reactive-ion-etching (bottom Nb layer) or liftoff (Si$_3$N$_4$ and top Nb layer).
Further details regarding device fabrication can be found in Methods Sec.~\ref{sec:meth_fab}.
For the experiments, the $10\times \qty{10}{\milli\meter\squared}$-large chip is wire-bonded to a microwave printed circuit board (PCB) and enclosed in a radiation-tight copper housing.
A small electromagnetic coil is mechanically attached to the copper housing in order to apply a dc magnetic field perpendicular to the chip surface and to flux-bias the SQUID.
For microwave control and response measurements, both on-chip feedlines are connected via CPWs on the PCB and PCB-SMP connectors to two individual coaxial cables, which combine both input and output (I/O) for each of the circuits by means of a directional coupler.
The HF coaxial input line is highly attenuated to equilibrate the feedline thermal noise to the experiment temperature $T_\mathrm{s} = \qty{4.2}{\kelvin}$, and the HF output line is equipped with a cryogenic high-electron-mobility transistor (HEMT) amplifier to maximize the signal-to-noise ratio.
On the LF side, both input and output line are slightly attenuated and the returning signal is amplified by a room-temperature HEMT.
The two I/O line pairs are connected to individual vector network analyzers (VNAs), and the HF input line is additionally connected to a microwave signal generator, which provides the sideband-pump tone.
All experiments reported here were conducted with the sample directly immersed in liquid helium.
As first experiment, we characterize the reflection scattering matrix element $S_{11}$ of the two circuits via their individual feedlines around their respective resonance frequencies.
From fits to the absorption dips, cf.~Fig.~\ref{fig:figure1}\textbf{d} and Methods~Sec.~\ref{sec:meth_fitting}, we extract the resonance frequencies of both modes (given above) as well as their internal and external decay rates $\kappa_\mathrm{int}= 2\pi\times \qty{38.2}{\mega\hertz}$, $\kappa_\mathrm{ext} = 2\pi\times\qty{6.7}{\mega\hertz}$ for the HF circuit and $\Gamma_\mathrm{int} = 2\pi \times \qty{489}{\kilo\hertz}$, $\Gamma_\mathrm{ext} = 2\pi\times\qty{36}{\kilo\hertz}$ for the LF circuit; data were taken at the flux sweetspot.
Both circuits are undercoupled with $\kappa_\mathrm{int}/\kappa_\mathrm{ext} > 1$ and $\Gamma_\mathrm{int}/\Gamma_\mathrm{ext} > 1$.
The high value for $\kappa_\mathrm{int}$ can be attributed to the nano-constrictions, which have been shown to have a suppressed critical temperature and therefore considerable quasiparticle losses at $\qty{4.2}{\kelvin}$ \cite{Uhl2024, Uhl2024a}.
However, this is not detrimental for the current experiment, but rather desired, since the high $\kappa_\mathrm{int}$ is directly related to the dissipative coupling contribution.
The large $\Gamma_\mathrm{int}/\Gamma_\mathrm{ext}$ has the advantage that the LF circuit does not need much attenuation in addition to the cable-attenuation to reach an effective mode temperature $T_\mathrm{LF} \lesssim \qty{10}{\kelvin}$, which is a big advantage if a cryogenic LF amplifier is not available (as was the case for our experiment).
\subsection{Flux-tunable photon-pressure interactions}
\label{sec:FTPP}
\begin{figure*}
	\includegraphics[trim={0 2mm 0 2mm},clip = true]{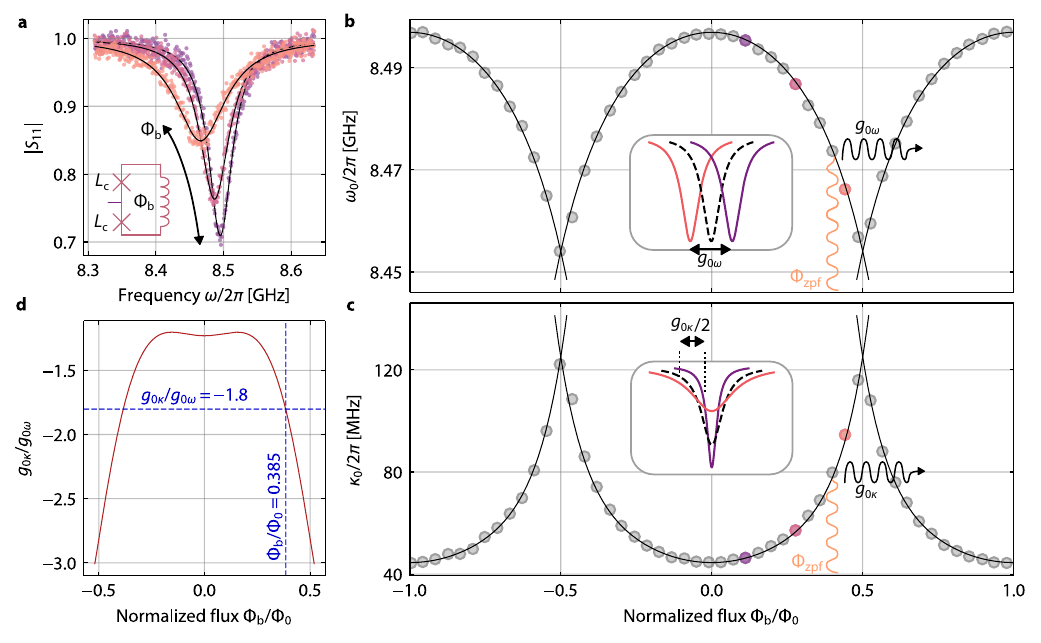}
	\titlecaption{Dispersive and dissipative photon-pressure coupling rates and their tuning with bias flux through the SQUID}{%
		\sublabel{a}~Reflection $|S_{11}|$ of the HF mode for three different bias-flux values $0.1 \lesssim\Phi_\mathrm{b}/\Phi_0 \lesssim 0.45$. With increasing bias flux, the resonance frequency shifts to lower values and the total linewidth increases. Symbols are data, lines are fits. The frequency-shift can be understood as an increase of the nonlinear constriction inductance $L_\mathrm{c}$ with increasing flux through the SQUID loop (cf.~inset), the increase in linewidth by current-induced quasiparticles.
		\sublabel{b}~Resonance frequencies $\omega_0(\Phi_\mathrm{b})$, extracted from fits to reflection data $S_{11}$ (cf.~\textbf{a}), which show a periodic modulation with period $\Phi_0$. Circles are data, line is a fit (cf.~Supplementary Note~III). If the SQUID is biased to a constant value $\Phi_\mathrm{b}/\Phi_0 \approx 0.385$ and additional oscillating zero-point-fluctuation flux from the LF mode $\Phi_\mathrm{zpf} \approx 198.3\,$\textmu$\Phi_0$ threads the SQUID, this leads to a modulation of the HF resonance frequency. The total frequency-modulation induced by $\Phi_\mathrm{zpf}$ is equivalent to the dispersive single-photon coupling rate $g_{0\omega} = 2\pi\times\qty{27.1}{\kilo\hertz}$. Inset schematic visualizes $g_{0\omega}$.
		\sublabel{c}~Analogous to panel \textbf{b}, but for the total HF linewidth $\kappa_0$, which periodically modulates between $\qty{\sim 44.5}{\mega\hertz}$ and $\qty{\sim 125}{\mega\hertz}$, revealing that despite the large decay rates the device is in the sideband-resolved limit with $\Omega_0/\kappa_0\gtrsim 3.5$ for all $\Phi_\mathrm{b}$. Circles are data, line is a polynomial fit (cf.~Supplementary Note~III). At the operation point $\Phi_\mathrm{b}/\Phi_0 = 0.385$, the LF flux $\Phi_\mathrm{zpf}$ induces a dissipative photon-pressure interaction with single-photon coupling rate $g_{0\kappa} = -2\pi\times\qty{48.8}{\kilo\hertz}$.  Inset schematic visualizes $g_{0\kappa}$.
		\sublabel{d}~Ratio of dissipative to dispersive single-photon coupling rate $g_{0\kappa}/g_{0\omega}$ vs.~$\Phi_\mathrm{b}$, which is calculated from the fit curves in \textbf{b} and \textbf{c}. At the operation point for this work (marked by the crossing point of the blue dashed lines), we find $g_{0\kappa}/g_{0\omega} \approx -1.8$.
	}
	\vspace{-2pt}
	\label{fig:figure2}
\end{figure*}
The dispersive and dissipative single-photon coupling rates of a photon-pressure system are given by
\begin{eqnarray}
	g_{0\omega} & = & - \frac{\partial\omega_0}{\partial\Phi_\mathrm{b}}\Phi_\mathrm{zpf} \\
	g_{0\kappa} & = & - \frac{\partial\kappa_0}{\partial\Phi_\mathrm{b}}\Phi_\mathrm{zpf},
\end{eqnarray}
respectively, with the bias flux though the SQUID $\Phi_\mathrm{b}$.
Hence, to characterize the system and to select a good working point for the further experiments, as a next step we investigate the response of the HF resonance frequency and linewidth with respect to applied flux $\Phi_\mathrm{b}$.
We sweep the current through the attached coil, and for each current value we take a trace of both the HF and the LF reflection with the VNAs.
Although the LF circuit is also weakly flux dependent (cf.~Supplementary Note~III), we focus our discussion on the HF mode here, since its properties determine the interaction between the circuits.
Our findings are summarized in Fig.~\ref{fig:figure2}.
Upon increasing $\Phi_\mathrm{b}$ starting at the sweetspot $\Phi_\mathrm{b} = 0$, we observe a redshift of the HF resonance frequency and an increase of the linewidth.
Over a larger flux range, we observe that both quantities periodically modulate with period $\Phi_0$ as expected for a SQUID due to fluxoid quantization.
Note however, that $\omega_0$ and $\kappa_0 = \kappa_\mathrm{int} + \kappa_\mathrm{ext}$ modulate with opposite trend, i.e., when $\omega_0$ decreases $\kappa_0$ increases and vice versa.
The maximum resonance frequency is found at the sweetspots $\Phi_\mathrm{b}/\Phi_0 = n$ with $n\in\mathbb{Z}$, while this is the flux of the lowest decay rate.
The modulation range for the resonance frequency is $\omega_0^\mathrm{max} - \omega_0^\mathrm{min}\approx 2\pi \times \qty{43}{\mega\hertz}$ and for the linewidth $\kappa_0^\mathrm{max} - \kappa_0^\mathrm{min} \approx 2\pi\times\qty{81}{\mega\hertz}$, i.e., $\kappa_0$ changes by about twice the amount $\omega_0$ does.
From fits to the data points (for details see Supplementary Note~III), we determine the derivatives for both parameters to calculate the single-photon coupling rates $g_{0\omega}$ and $g_{0\kappa}$.
Additionally, we obtain the ratio of the derivatives as a function of $\Phi_\mathrm{b}$, which is equal to the flux-dependent ratio of the coupling rates $g_{0\kappa}/g_{0\omega}$.
As a good compromise between low circuit nonlinearity, medium total decay rate and maximum slope of both the tuning arcs, we choose the operation point $\Phi_\mathrm{b}/\Phi_0 \approx 0.385$.
Combining all parameters, we determine the dispersive and dissipative single-photon coupling rates at this flux point as $g_\mathrm{0\omega} = 2\pi\times \qty{27.1}{\kilo\hertz}$ and $g_{0\kappa} = 2\pi\times\qty{-48.8}{\kilo\hertz}$, respectively, which correspond to a large ratio $g_{0\kappa}/g_{0\omega} \approx -1.8$.
Note that depending on the exact flux-bias point, this ratio can be tuned between $-1.2$ and $-3$, cf.~Fig.~\ref{fig:figure2}\sublabel{d}.
The HF resonance frequency and linewidth at the working point are $\omega_0 \approx 2\pi \times 8.476\,$GHz and $\kappa_0 \approx 2\pi\times\qty{73.9}{\mega\hertz}$; the LF parameters are $\Omega_0 \approx 2\pi\times\qty{446.3}{\mega\hertz}$ and $\Gamma_0 \approx 2\pi\times\qty{601}{\kilo\hertz}$.
The external decay rates are nearly constant as a function of flux, and so all the variation of $\kappa_0$ with $\Phi_\mathrm{b}$ and $\Phi_\mathrm{zpf}$ can be attributed to $\kappa_\mathrm{int}$, cf.~Supplementary Note~III.
Note though, that in general it is important to discriminate between internal and external dissipative photon-pressure, since they can lead to qualitatively and quantitatively different consequences; here $g_{0\kappa_\mathrm{int}} = g_{0\kappa}$ and $g_{0\kappa_\mathrm{ext}} \approx 0$.
Combined with the self-Kerr nonlinearity of the HF mode $\mathcal{K} \approx 2\pi\times\qty{-5.4}{\kilo\hertz}$ (the LF self-Kerr is negligibly small), cf.~Supplementary Note~VII, the device is well in the sideband-resolved regime with $\Omega_0/\kappa_0 \approx 6$ while in principle allowing for strong sideband pump tones due to $\mathcal{K}/\kappa_0 \ll 1$.
\subsection{Photon-pressure induced Fano-transparency}
\vspace{-3mm}
\begin{figure*}
	\includegraphics[trim={0 2mm 0 2mm},clip = true]{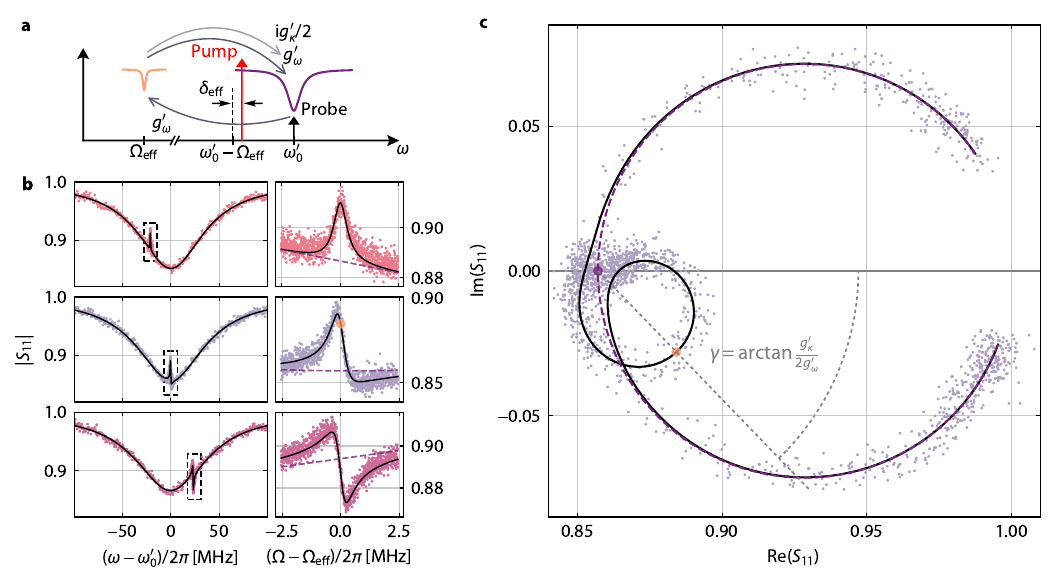}
	\titlecaption{Fano-transparency indicates interference of dispersive and dissipative photon-pressure}{%
		\sublabel{a}~Schematic of the experimental setting for the observation of photon-pressure induced transparency (PPIT). A strong microwave pump tone (red vertical arrow) with power $P_\mathrm{p}$ and frequency $\omega_\mathrm{p} = \omega_0' - \Omega_\mathrm{eff} + \delta_\mathrm{eff}$ is sent to the device around the red sideband of the HF cavity. A small VNA probe tone (black vertical arrow) with relative frequency $\Omega = \omega - \omega_\mathrm{p}$ scans the HF reflection around $\omega_0'$. The beating of the pump and probe tones coherently drives the LF mode with a strength proportional to the dispersive coupling rate $g_\omega'$. The resulting LF amplitude in turn induces both a dispersively and a dissipatively generated sideband to the pump tone at $\omega_\mathrm{p} + \Omega$, which are ${\propto}\,g_\omega'$ and ${\propto}\,\i g_\kappa'/2$, respectively, and which both interfere with the original probe tone.
		\sublabel{b}~HF cavity reflection $|S_{11}|$ in the presence of a pump tone for three different pump detunings $\delta_\mathrm{eff} = \left(-\kappa_0'/4, 0, +\kappa_0'/4\right)$ (from top to bottom). Data in left graphs are vs.~detuning from HF cavity resonance $\omega - \omega_0'$, and clear PPIT signatures can be identified when $\omega - \omega_0' \approx \delta_\mathrm{eff}$. The smaller graphs on the right show zooms to the transparency resonances (dashed boxes in left graphs) plotted vs.~detuning from $\omega_\mathrm{p} + \Omega_\mathrm{eff}$. Symbols are data, lines are fits. The point of PPIT resonance in the zoom for $\delta_\mathrm{eff} \approx 0$ is marked with an orange disk, and the cavity reflection for $g_\omega' = g_\kappa' = 0$ is added as dashed purple line in all zooms. \textbf{c} The dataset for $\delta_\mathrm{eff} \approx 0$ in complex representation. Points are data, black line is a theory curve with all parameters from the fit and $\delta_\mathrm{eff} = 0$. The cavity resonance traces a large circle anchored at $(1, 0)$, the PPIT signature traces a smaller circle anchored on the resonance point of the bare cavity (purple disk); the resonance point of the PPIT is marked with the orange disk. For $g_\kappa' = 0$, the PPIT circle would point towards $(1, 0)$ and the dashed line connecting the two resonance points would be the real axis. However, the presence of dissipative coupling leads to a rotation of the PPIT circle around its anchor point by an angle $\gamma = \arctan{\left(g_{\kappa}'/2g_\omega'\right)}$. Here, we obtain $\gamma \approx -46\,\degree$, which corresponds to $g_\kappa'/g_\omega' \approx -2.1$.
	}
	\label{fig:figure3}
\end{figure*}
To experimentally investigate the effects of the dissipative coupling contribution to the overall dynamics of the circuits, we begin with the protocol of photon-pressure induced transparency (PPIT).
Here, a strong, fixed-frequency sideband-pump field is sent to the HF cavity around its red sideband $\omega_\mathrm{p} = \omega_0' - \Omega_\mathrm{eff} + \delta_\mathrm{eff}$ with $|\delta_\mathrm{eff}| \leq \kappa_0'/4$, and a much weaker VNA probe tone with frequency $\omega \sim \omega_0'$ is used to characterize the modified HF reflection response in a frequency span of few $\kappa_0'$ around $\omega_0'$, cf.~Fig.~\ref{fig:figure3}\textbf{a}.
We prime all HF mode quantities here to indicate that they shift with the intracavity pump photon number $n_\mathrm{c}$ due to the Kerr nonlinearity and a nonlinear damping.
The effective LF mode frequency $\Omega_\mathrm{eff} = \Omega_0' + \delta\Omega_\mathrm{pp}$ contains both, the power-shifted LF frequency due to a potential cross-Kerr effect $\Omega_0'$ and possible dynamical backaction contributions $\delta\Omega_\mathrm{pp}$, which we do not know at this point and therefore choose $\Omega_\mathrm{eff}$ as reference for the cavity red sideband.
However, for all our data $\Omega_\mathrm{eff} - \Omega_0 < \Gamma_0$, hence the difference is very small.
The beating of the pump and probe tones in the PPIT protocol coherently drives the LF circuit into oscillation, which in turn generates a sideband field to the pump tone, that interferes with the original probe tone with a phase-relation determined by the interaction and the detuning.
As a consequence, a narrow transparency window with the shape of the effective LF susceptibility appears within the HF cavity resonance, a phenomenon closely related to EIT in atomic systems \cite{Fleischhauer2005} and OMIT in optomechanical devices \cite{Weis2010}.
An interesting question is now whether and how the presence of dissipative coupling alters the usual, purely dispersive PPIT signature.
To model and understand the experiment, we derive the linearized equations of motion for the HF cavity field $\hat{c}$ and the LF mode field $\hat{d}$, which are given by
\begin{eqnarray}
	\dot{\hat{c}} & = & \left[-\i\Delta_\mathrm{p}' - \frac{\kappa_0'}{2}\right]\hat{c} - \i\left[g_\omega' + \i\frac{g_\kappa'}{2}\right]\left(\hat{d} + \hat{d}^\dagger\right) + \i\sqrt{\kappa_\mathrm{ext}'}\hat{c}_\mathrm{in} \\
	\dot{\hat{d}} & = & \left[\i\Omega_0' - \frac{\Gamma_0'}{2}\right]\hat{d} - \i g_\omega'\left(\hat{c} + \hat{c}^\dagger \right)
\end{eqnarray}
where $\Delta_\mathrm{p}' = \omega_\mathrm{p} - \omega_0'$, $\hat{c}_\mathrm{in}$ represents the input probe tone, and $g_\omega'$ and $g_\kappa'$ are the dispersive and dissipative multiphoton coupling rates.
We omit any input noise here, since it can be neglected to first order in a PPIT experiment, that is only considering the response to a coherent input tone.
A complete derivation of the equations of motion can be found in Supplementary Notes~IV and V.
One very interesting detail in the equations of motion is an asymmetry in the coupling terms.
While the HF mode is coupled to the LF mode with a term ${\propto} g' = g_\omega' +  \i g_\kappa'/2$, only the dispersive interaction $g_\omega'$ contributes explicitly to changes of the LF mode field.
In other words, the LF mode is driven proportional to $g_\omega'$ only, while the HF sideband is generated via $g_\omega' + \i g_\kappa'/2$.
By solving the equations of motion using Fourier transformation, combining the results, and applying the high-$Q$ limit for the LF mode (cf.~Supplementary Note~V), we arrive at an expression for the HF reflection, when probed around resonance
\begin{equation}
	S_{11} = 1 - \kappa_\mathrm{ext}'\chi_\mathrm{c}'\left[1 - g_\omega'\left(g_\omega' + \i\frac{g_\kappa'}{2}\right)\chi_\mathrm{c}'\chi_0^\mathrm{eff}\right]
	\label{eqn:S11_PPIT_main}
\end{equation}
with the HF susceptibility $\chi_\mathrm{c}' = \left[\kappa_0'/2 + \i\left(\Delta_\mathrm{p}' + \Omega\right) \right]^{-1}$, the effective LF susceptibility $\chi_0^\mathrm{eff} = \left[\Gamma_0'/2 + \i\left(\Omega - \Omega_0'\right) + \Sigma\right]^{-1}$, the dynamical backaction $\Sigma = g_\omega'\left(g'\chi_\mathrm{c0}' - g'^*\overline{\chi}_\mathrm{c0}' \right)$, $\chi_\mathrm{c0}' = \chi_\mathrm{c}'(\Omega_0')$, $\overline{\chi}_\mathrm{c0}' = \chi_\mathrm{c}'^*(-\Omega_0')$ and the probe frequency relative to the pump $\Omega = \omega - \omega_\mathrm{p}$.
Note that without dispersive coupling $g_\omega' = 0$ there would be no PPIT at all and no dynamical backaction, but both effects are nevertheless considerably modified by the presence of $g_\kappa'$ in $g'$.
If we had an external-dissipative interaction instead of (or in addition to) the internal-dissipative one, the equations of motion and the reflection expression would contain additional terms and neither the dynamical backaction nor the PPIT signature would vanish for $g_\omega' = 0$~\cite{Primo2023}.
As a consequence of $g_\kappa' \neq 0$, experiment and theory both reveal an interference-based and Fano-like modification of the PPIT resonance within the HF cavity resonance, cf.~Fig.~\ref{fig:figure3}\textbf{b}.
Most striking are two features.
First, we have an asymmetric PPIT lineshape when the pump is directly on the effective red sideband $\delta_\mathrm{eff} = \omega_\mathrm{p} - \left(\omega_0' - \Omega_\mathrm{eff} \right) \approx 0$ in contrast to dispersive coupling only.
And secondly, the usual mirror symmetry for $+\delta_\mathrm{eff}$ and $-\delta_\mathrm{eff}$ is not preserved anymore.
In the complex representation of $S_{11}$, cf.~Fig.~\ref{fig:figure3}\textbf{c}, the origin of this additional tilt becomes apparent.
Usually (i.e.~for $g_\kappa' = 0$), the PPIT describes a small circle within the much larger HF cavity resonance circle, but both have their anchor points and their centers on the real axis for the pump on the red sideband.
Due to the additional phase shift of $\pi/2$ of the dissipative coupling term ($\i = \e^{\i\pi/2}$), however, the PPIT circle gets rotated around its anchor point.
Evaluating Eq.~(\ref{eqn:S11_PPIT_main}) at the corresponding frequency points (cf.~Methods Sec.~\ref{sec:fano_angle} and Supplementary Note~VI) reveals that in fact the angle $\gamma$ between the two circle axes is given by
\begin{equation}
	\tan{\gamma} = \frac{g_\kappa'}{2g_\omega'}.
\end{equation}
For the dataset discussed in Fig.~\ref{fig:figure3}, we find $\gamma =- 46\degree$, which is equivalent to $g_\kappa'/g_\omega' \approx -2.1$.
Observing this tilt is therefore not only a clear confirmation for the presence of internal-dissipative photon-pressure coupling, but also might be a very useful and extraordinarily fast and simple method to quantify internal-dissipative coupling rates in the first place.
It only requires a single complex resonance measurement and especially in devices, which do not provide the possibility to tune a system parameter $p$ to extract the derivatives of $\partial\omega_0/\partial p$ and $\partial \kappa_0/\partial p$ (e.g.~in most optomechanical systems), the transparency rotation angle can be of high relevance.
In this context it is also noteworthy, that external-dissipative photon-pressure does not result in the same Fano-interference in the sideband-resolved limit and with a pump tone around one of the sidebands, since its main consequence is a detuning-dependent re-scaling of $g_\omega'$ instead of adding an imaginary component to the total coupling rate~\cite{Primo2023}.
The value of $g_\kappa'/g_\omega' \approx -2.1$ we find here from the PPIT data is only close to the single-photon equivalent $g_{0\kappa}/g_{0\omega} \approx -1.8$ obtained in the context of Fig.~\ref{fig:figure2}, but not exactly the same.
The latter, i.e., $g_{0\kappa}/g_{0\omega} = g_\kappa'/g_\omega'$, would be expected if the multiphoton coupling rates scaled with pump photon number $n_\mathrm{c}$ as usual as $g_\omega' = \sqrt{n_\mathrm{c}}g_{0\omega}$ and $g_\kappa' = \sqrt{n_\mathrm{c}}g_{0\kappa}$.
The difference found here is not caused by inaccuracies, however, it rather points towards another highly interesting effect, which will be addressed in the next section.
\begin{figure*}
	\includegraphics[trim={0 2mm 0 2mm},clip = true]{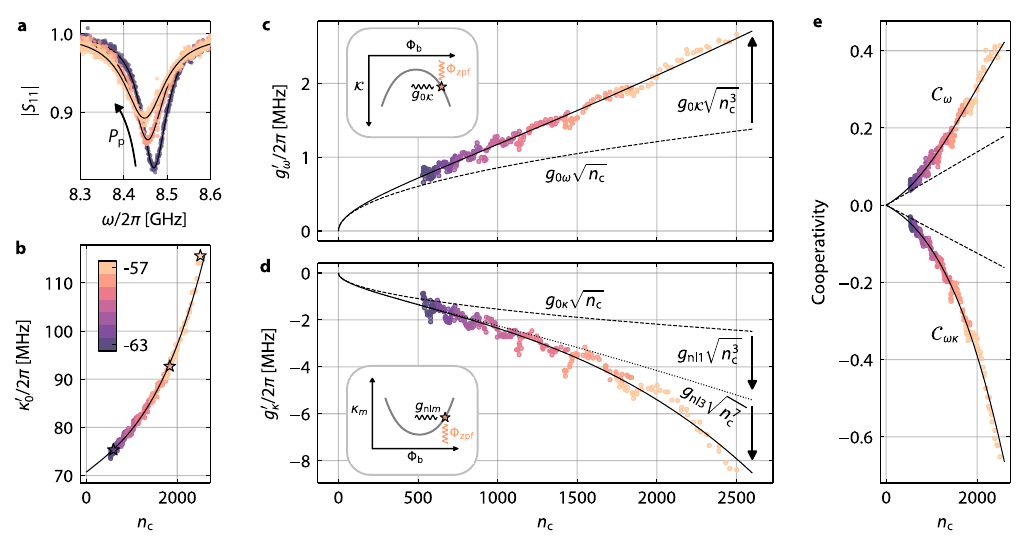}
	\vspace{-4pt}
	\titlecaption{Nonlinear damping and nonlinearity-enhanced photon-pressure coupling}{%
		\sublabel{a} HF cavity reflection $|S_{11}|$ for fixed $\omega_\mathrm{p} = 2\pi\times \qty{8.031}{\giga\hertz}$ and increasing sideband pump-power $P_\mathrm{p}$, measured by a small VNA probe tone. With increasing pump power, the cavity shifts to lower frequencies due to its Kerr anharmonicity, and its linewidth increases due to nonlinear damping. Symbols are experimental data, lines are fits.
		\sublabel{b} Total HF mode linewidth $\kappa_0'$ for seven different pump powers (color-coded as given by the color bar in units of \qty{}{\dBm}) and multiple detunings between $ -\kappa_0'/4$ and $ +\kappa_0'/4$ in pump-frequency steps of $\qty{2}{\mega\hertz}$, all plotted vs.~their corresponding intracavity pump photon-number $n_\mathrm{c}$. Symbols are data, star symbols correspond to the datasets in \sublabel{a}. Line is a fit with $\kappa_0' = \kappa_0 + 2\kappa_1 n_\mathrm{c} + 3\kappa_2 n_\mathrm{c}^2 + 4\kappa_3 n_\mathrm{c}^3$.
		\sublabel{c}~Dispersive multiphoton coupling rate $g_\omega'$ vs.~$n_\mathrm{c}$ as extracted from PPIT. Pump powers and detunings are identical to \sublabel{b}. Symbols are data, solid line is a fit with $g_\omega' = g_{0\omega}\sqrt{n_\mathrm{c}} + g_{0\mathcal{K}}\sqrt{n_\mathrm{c}^3}$; dashed line shows $g_\omega = g_{0\omega}\sqrt{n_\mathrm{c}}$ without the Kerr enhancement. Inset: Schematic of the HF self-Kerr nonlinearity $\mathcal{K}$ as a function of bias flux and how the LF zero-point-fluctuation flux modulates it by $g_{0\mathcal{K}} = -\Phi_\mathrm{zpf}\partial\mathcal{K}/\partial\Phi_\mathrm{b}$.
		\sublabel{d}~Dissipative multiphoton coupling rate $g_\kappa'$ vs.~$n_\mathrm{c}$ as extracted from PPIT. Pump powers and detunings are identical to \sublabel{b}. Symbols are data, solid line is a fit with $g_\kappa' = g_{0\kappa}\sqrt{n_\mathrm{c}} + g_\mathrm{nl1}\sqrt{n_\mathrm{c}^3} + g_\mathrm{nl2}\sqrt{n_\mathrm{c}^5} + g_\mathrm{nl3}\sqrt{n_\mathrm{c}^7}$; dashed line shows $g_\kappa = g_{0\kappa}\sqrt{n_\mathrm{c}}$ without nonlinearity-enhancement, dotted line shows $g_{0\kappa}\sqrt{n_\mathrm{c}} + g_\mathrm{nl1}\sqrt{n_\mathrm{c}^3}$. From the fit, we obtain $g_\mathrm{nl2} \approx 0$. Inset: Schematic of a generic nonlinear damping coefficient $\kappa_m$, $m \in \mathbb{N}$, as a function of bias flux and how the LF zero-point-fluctuation flux modulates it by $g_{\mathrm{nl}m} = -\Phi_\mathrm{zpf}\partial\kappa_m/\partial\Phi_\mathrm{b}$.
		\sublabel{e}~Dispersive cooperativity $\mathcal{C}_\omega = 4g_\omega'^2/\kappa_0'\Gamma_0$ and cross-cooperativity $\mathcal{C}_{\omega\kappa} = 2g_\omega'g_\kappa'/\kappa_0'\Gamma_0$ vs.~$n_\mathrm{c}$. Symbols are derived from data, solid lines follow from the fits in \sublabel{b}--\sublabel{d} with $\Gamma_0 = 2\pi \times \qty{601}{\kilo\hertz}$, dashed line is the reference without the nonlinearity-enhancement and with constant $\kappa_0' = \kappa_0$. The effects from increasing $\kappa_0'$ and nonlinearity-enhanced $g_\omega', g_\kappa'$ compete, but overall still lead to a significant enhancement of both cooperativities by factors up to ${\sim}2.4$ and ${\sim}4.1$, respectively.
	}
	\vspace{-4pt}
	\label{fig:figure4}
\end{figure*}
\vspace{-4mm}
\subsection{Nonlinearity-enhanced coupling rates}
\vspace{-2mm}
When performing the PPIT experiment with varying pump power and detuning, we observe that resonance frequencies, linewidths and coupling rates, and surprisingly even $g_\kappa'/g_\omega'$, depend on pump power.
The shift of $\omega_0'$ is explained by the Kerr anharmonicity of the HF mode, which stems from the nonlinear superconducting inductances.
The origin of the nonlinear linewidth broadening, which is clearly faster than linear in pump photon number, cf.~Fig.~\ref{fig:figure4}\sublabel{b}, are ac-current-induced quasiparticles and an ac-current-induced suppression of the superconducting gap in the constrictions.
We observe, consistent with that interpretation, that $\kappa_\mathrm{ext}'$ is not significantly modified by the pump.
The pump-broadened linewidth is phenomenologically modeled using $\kappa_0' = \kappa_0 + 2\kappa_1 n_\mathrm{c} + 3\kappa_2 n_\mathrm{c}^2 + 4\kappa_3 n_\mathrm{c}^3$ with $\kappa_0 = 2\pi\times\qty{70.7}{\mega\hertz}$ and the nonlinear-damping coefficients $\kappa_1 \approx 2\pi\times\qty{3.7}{\kilo\hertz}$, $\kappa_2 \approx 0$, and $\kappa_3 \approx 2\pi\times\qty{0.37}{\milli\hertz}$; all values are obtained from a fit to the data.
For the origin of the prefactors $2, 3$ and $4$ in the nonlinear contributions, cf.~Supplementary Note~IV.
Finally, we investigate $g_\omega'$ and $g_\kappa'$ as a function of pump photon number.
A description of how we obtained the values from PPIT data is given in Methods Sec.~\ref{sec:coupl_rates}.
In Fig.~\ref{fig:figure4}, we show all the multiphoton coupling rates obtained for multiple pump powers and detunings vs.~$n_\mathrm{c}$, and what we observe is surprising and exciting.
Both coupling rates do not scale ${\propto}\sqrt{n_\mathrm{c}}$ as expected from a simple first-order theory, but they increase considerably faster.
While $g_\omega'$ seems to grow almost linearly, but with a finite value at $n_\mathrm{c} = 0$, $g_\kappa'$ grows even faster, in particular at the highest photon numbers, similar to the increase of $\kappa_0'$ with $n_\mathrm{c}$.
For the lowest pump powers, the ratio $g_\kappa'/g_\omega'$ is very close to $g_{0\kappa}/g_\mathrm{0\omega} = -1.8$, for the highest photon numbers though $g_\kappa'/g_\omega' \rightarrow -3.1$.
The PPIT data in Fig.~\ref{fig:figure3} are in the intermediate $g_\kappa'/g_\omega'$-regime (taken at $P_\mathrm{p} = \qty{-58}{\dBm}$, i.e., $n_\mathrm{c} \approx 1570$ for $\delta_\mathrm{eff} \approx 0$), which explains the discrepancy between $g_{0\kappa}/g_{0\omega} = -1.8$ and $g_\kappa'/g_\omega' = -2.1$ found from $\gamma$.
Nonlinearity-enhanced multiphoton coupling rates are an intriguing and potentially very useful observation, which can be explained as follows.
Each of the nonlinearity parameters $\mathcal{K}$, $\kappa_1$, $\kappa_2$ and $\kappa_3$ is flux-dependent \cite{Uhl2024}, i.e., modulated by $\Phi_\mathrm{zpf}$, and hence each can be the origin of an additional type of photon-pressure coupling.
To first order, these additional coupling rates on the single-photon level are
\begin{equation}
	g_{0\mathcal{K}} = -\frac{\partial\mathcal{K}}{\partial \Phi_\mathrm{b}}\Phi_\mathrm{zpf}, ~~~ g_{\mathrm{nl}m} = -\frac{\partial\kappa_{m}}{\partial \Phi_\mathrm{b}}\Phi_\mathrm{zpf}, ~~~ m\in \mathbb{N}
\end{equation}
and intrinsically they are several orders of magnitude smaller than $g_{0\omega}$ and $g_{0\kappa}$.
In the framework of the linearized equations of motion, however, they get integrated into the multiphoton coupling rates as
\begin{eqnarray}
	g_\omega' & = & g_{0\omega}\sqrt{n_\mathrm{c}} + g_{0\mathcal{K}}\sqrt{n_\mathrm{c}^3} \\
	g_\kappa' & = & g_{0\kappa}\sqrt{n_\mathrm{c}} + g_\mathrm{nl1}\sqrt{n_\mathrm{c}^3} + g_\mathrm{nl2}\sqrt{n_\mathrm{c}^5} + g_\mathrm{nl3}\sqrt{n_\mathrm{c}^7},
\end{eqnarray}
which means that despite their smallness on the single-photon level they can be very significant in the high-pump-power regime.
Interestingly, also here we find a vanishing second order $g_\mathrm{nl2} \approx 0$ from a corresponding fit, and for the finite coupling rates we get $g_{0\mathcal{K}} \approx 2\pi\times\qty{10.0}{\hertz}$, $g_{\mathrm{nl1}} \approx 2\pi\times\qty{-22.1}{\hertz}$ and $g_{\mathrm{nl3}} \approx 2\pi\times\qty{-3.5}{\micro\hertz}$.
Table~\ref{tab:table1_main} summarizes all the relevant nonlinearity coefficients for $\omega_0'$, $\kappa_0', g_\omega'$ and $g_\kappa'$.
\begin{table}[b]
	\vspace{-4mm}
	\titlecaption{Nonlinearity coefficients for $\omega_0', \kappa_0', g_\omega'$ and $g_\kappa'$}{Values extracted from fits, numbers given in \qty{}{\kilo\hertz} for $\mathcal{K}, \kappa_1, \kappa_2$ and $\kappa_3$, and in \qty{}{\hertz} for $g_{0\mathcal{K}}, g_{\mathrm{nl}1}, g_{\mathrm{nl}2}$ and $g_{\mathrm{nl}3}$.\\}
	\begin{tabular}{ c   c   c   c   c   c   c   c}
		\toprule
		$\mathcal{K}/2\pi$ & $\kappa_1/2\pi$ & $\kappa_2/2\pi$ & $\kappa_3/2\pi$ & $g_{0\mathcal{K}}/2\pi$ & $g_{\mathrm{nl}1}/2\pi$ & $g_{\mathrm{nl}2}/2\pi$ & $g_{\mathrm{nl}3}/2\pi$ \\ \midrule
		$-5.4$ & $3.7$ & $0$ & $3.7\times 10^{-7}$ & $10.0$ & $-22.1$ & $0$ & $-3.5\times 10^{-6}$  \\ \bottomrule
	\end{tabular}
	\label{tab:table1_main}
\end{table}
For the Kerr constant, this effect has been predicted by theoretical work \cite{Mikkelsen2017}, and a related observation of much smaller magnitude has been reported in a multi-tone driven photon-pressure experiment~\cite{Rodrigues2022}.
The dissipative case, however, is even more impactful in our work than the dispersive one.
While $g_\omega'$ is roughly doubled in the high-power regime compared to $g_{0\omega}\sqrt{n_\mathrm{c}}$, which is already impressive, the dissipative multiphoton coupling $g_\kappa'$ gets boosted by up to a factor ${\sim} 3.4$ compared to $g_\kappa = g_{\kappa 0}\sqrt{n_\mathrm{c}}$, cf.~Fig.~\ref{fig:figure4}.
Often, it is insightful to consider the cooperativity instead of the coupling rates, especially, when also the mode linewidth is a function of $n_\mathrm{c}$ as is the case here.
Unfortunately, the purely dissipative cooperativity $\mathcal{C}_\kappa = g_\kappa'^2/\kappa_0'\Gamma_0$ plays no role in this experiment due to the nonreciprocity of the interaction, but it would be enhanced by almost one order of magnitude.
The dispersive cooperativity $\mathcal{C}_\omega = 4g_\omega'^2/\kappa_0'\Gamma_0$ and the cross-cooperativity $\mathcal{C}_{\omega\kappa} = 2g_\omega'g_\kappa'/\kappa_0'\Gamma_0$, on the other hand, are important for e.g.~dynamical backaction (see below) and perspectively for sideband-cooling or parametric amplification.
Those two cooperativities are still enhanced by a factor ${\sim} 2.4$ and ${\sim} 4.1$, respectively, which shows that the enhancement of the coupling rates by far overcompensates the increase of $\kappa_0'$.
\begin{figure*}
	\includegraphics[trim={2.5mm 2mm 0 2mm},clip = true]{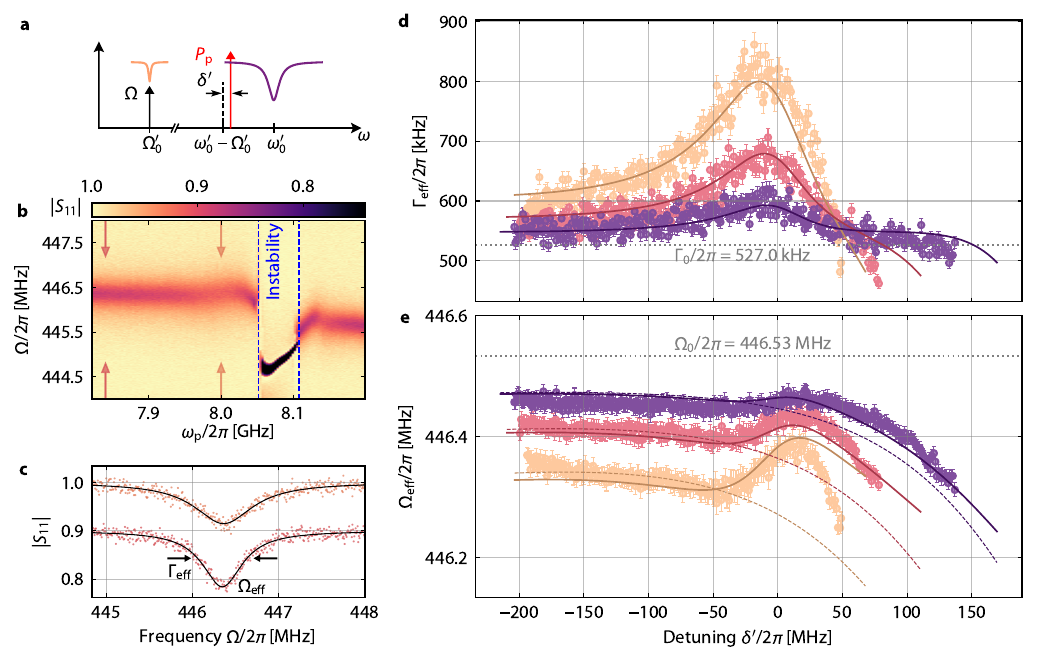}
	\titlecaption{Nonlinearity-enhanced dynamical backaction with dissipative coupling}{%
		\sublabel{a}~Schematic of the experiment. A pump tone (red arrow) with frequency $\omega_\mathrm{p}$ and power $P_\mathrm{p}$ is applied around the red sideband of the HF mode. Detuning of pump from the red sideband is $\delta' = \omega_\mathrm{p} -\left(\omega_0' - \Omega_0'\right)$. For each pair $(\omega_\mathrm{p}, P_\mathrm{p})$, the LF mode reflection is probed directly around $\Omega_0'$ using the LF feedline and a VNA (black arrow).
		\sublabel{b}~Color-map of the LF reflection $|S_{11}|$ vs.~pump frequency $\omega_\mathrm{p}$. When the pump is far red-detuned from the HF mode red-sideband, the resonance is nearly unmodified compared to the pump-free case. Just below $\delta' = 0$ or $\omega_\mathrm{p}/2\pi \approx \qty{8.0}{\giga\hertz}$, the absorption gets wider and shallower and slightly shifts in frequency; both effects indicate dynamical backaction at work. Strikingly, for $\omega_\mathrm{p}/2\pi \approx \qty{8.05}{\giga\hertz}$, which is still around the cavity red-sideband, a parametric instability region appears, which is experimentally identified by a strongly deformed resonance feature in $S_{11}$, the simultaneous disappearance of the HF mode in the HF reflection, and by a large amplitude spectral feature in the HF output spectrum even without the LF probe tone (not shown). The two pairs of arrows indicate the linescans shown in \sublabel{c}. Top curve in \sublabel{c} shows $|S_{11}|$ at $\omega_\mathrm{p}/2\pi \approx \qty{8.0}{\giga\hertz}$, the point of maximum photon-pressure damping, bottom curve is manually offset by $-0.1$ for clarity and is for $\omega_\mathrm{p}/2\pi \approx \qty{7.84}{\giga\hertz}$. Symbols are data, lines are fits, from which we extract $\Omega_\mathrm{eff}$ and $\Gamma_\mathrm{eff}$.
		\sublabel{d}, \sublabel{e} Effective LF mode linewidth $\Gamma_\mathrm{eff}$ and effective resonance frequency $\Omega_\mathrm{eff}$ vs.~detuning $\delta'$ for three different $P_\mathrm{p}$, color code for $P_\mathrm{p}$ identical to Fig.~\ref{fig:figure4}, cf.~Supplementary Fig.~15 for the corresponding $n_\mathrm{c}$. Symbols are data and error bars are standard errors obtained from the fit routine, lines are theory curves. Note the two clearest signatures for dissipative coupling: first, the maximum for the photon-pressure damping is at negative detunings from the red sideband. And second, for positive detunings from the red sideband the total damping rate is smaller than the intrinsic damping rate $\Gamma_\mathrm{eff} < \Gamma_0$, where $\Gamma_0$ is shown as dashed line. This indicates negative backaction damping with a red-detuned pump, and is the precursor for the approaching parametric instability. To fully model the data, we included a cross-Kerr frequency shift to the LF frequency $\Omega_0' = \Omega_0 + \mathcal{K}_\mathrm{c}n_\mathrm{c}$ with $\mathcal{K}_\mathrm{c} = 2\pi\times \qty{-130.8}{\hertz}$, and a small nonlinear cross-damping $\Gamma_0' = \Gamma_0 + \kappa_\mathrm{c}n_\mathrm{c}$ with $\kappa_\mathrm{c} = 2\pi\times\qty{38.4}{\hertz}$. Dashed lines in \textbf{e} show $\Omega_0'$.
	}
	\vspace{-0pt}
	\label{fig:figure5}
\end{figure*}
Potentially, such cooperativity enhancements could lead to low-power groundstate-cooling or increased flux/displacement sensitivity, with a particular usefulness in optomechanics, where single-photon coupling rates are often very small compared to photon-pressure circuits and where device or setup heating by large pump powers is sometimes a limiting factor.
Especially in superconducting circuits, it is furthermore possible to design $\mathcal{K}$ and correspondingly $\partial\mathcal{K}/\partial\Phi_\mathrm{b}$ by carefully considered junction arrays \cite{Frattini2018}, which might facilitate $g_{0\mathcal{K}}$ to be increased further by orders of magnitude in future systems.
The data in Fig.~\ref{fig:figure4}\sublabel{c} and~\sublabel{d} also confirm once more, that the external-dissipative coupling rate plays no role here.
If a significant contribution from $g_{\kappa_\mathrm{ext}}$ was present in the device, this would lead to a dependence of the effective $g_\omega'$ on the detuning between pump and HF resonance $\Delta_\mathrm{p}'$~\cite{Primo2023}, which would have two notable consequences.
First, not all the data for $g_\omega'$ in \sublabel{c} would fall onto a single line, but within each color the values for high photon numbers would be pushed down (up), if $g_{0\kappa_\mathrm{ext}}<0$ ($g_{0\kappa_\mathrm{ext}}>0$), and the values for low photon numbers would be pushed up (down).
And secondly, the ratio $g_\kappa'/g_\omega'$ would deviate considerably from the flux arc-based values even for very small $n_\mathrm{c}$, cf. also Supplementary Note~VI.
Even though some scattering of both coupling rates is visible, none of the two effects is observed within the experimental accuracy.
\vspace{-5mm}
\subsection{Dynamical backaction}
\vspace{-2mm}
Finally, we analyze the dynamical backaction exerted from the HF intracavity fields to the LF circuit as a function of power and detuning.
This is not only an essential quantity for future experiments and applications like sideband-cooling with dissipative interactions, but an agreement with theoretical predictions will also serve as independent confirmation of our above conclusions, including the existence and strength of the dissipative coupling, the nonlinear enhancement of $\mathcal{C}_\omega$ and $\mathcal{C}_{\omega\kappa}$, and the photon-number dependent ratio of $g_\omega'/g_\kappa'$.
Dynamical backaction arises from the time-delayed adjustments of the pump fields in the HF mode to changes of $\omega_0$ \cite{Aspelmeyer2014, Teufel2008}, here in our system caused by a change of flux through the SQUID.
Hence, it depends on detuning between pump and cavity and on the cavity decay rate, which determines the time scale, on which this adjustment will happen.
The effect can be separated into an in-phase and an out-of-phase component relative to the change of flux in the SQUID, which are equivalent to a change of effective LF frequency and a change of the LF decay rate.
Since $g_\kappa'$ comes with an additional phase-lag compared to $g_\omega'$, the dynamical backaction will likely be modified considerably in the presence of dissipative photon-pressure.
And indeed, on a formal level we can directly see that the photon-pressure damping $\Gamma_\mathrm{pp}$ and the photon-pressure frequency shift $\delta\Omega_\mathrm{pp}$, which are given by
\begin{eqnarray}
	\Gamma_\mathrm{pp} & = & 2 \mathrm{Re}\left[ g_\omega'\left(g'\chi_\mathrm{c0}' - g'^*\overline{\chi}_\mathrm{c0}'\right) \right] \label{eqn:PPdamping_main} \\
	\delta\Omega_\mathrm{pp} & = & -\mathrm{Im}\left[ g_\omega'\left(g'\chi_\mathrm{c0}' - g'^*\overline{\chi}_\mathrm{c0}'\right) \right] \label{eqn:PPshift_main}
\end{eqnarray}
deviate from the purely dispersive case, where the PP damping and PP frequency shift just encode the real and imaginary part of the cavity susceptibility, i.e., they are described by a complex Lorentzian.
Strictly speaking, this is only completely true in the sideband-resolved regime, where $\chi_\mathrm{c0}' - \overline{\chi}_\mathrm{c0}' \approx \chi_\mathrm{c0}'$, but our device falls well within that approximation.
With the phase-shifted dissipative coupling and the complex valued $g'$, the two backaction effects mix, i.e., they are rotated and scaled in the complex plane, which is no different to saying they acquire a Fano-like modification, very similar to what occurred in PPIT.
In Figure~\ref{fig:figure5} we present the results of the experimental characterization of $\Gamma_\mathrm{eff} = \Gamma_0' + \Gamma_\mathrm{pp}$ and $\Omega_\mathrm{eff} = \Omega_0' + \delta\Omega_\mathrm{pp}$ and indeed find a significant deviation from the Lorentzian-like shape in the dispersive-only case.
The data were acquired differently to the photon-pressure experiments discussed above.
Here, we directly probe the LF mode via its own feedline, while stepping the HF red-sideband pump tone through various powers and detunings.
The LF on-chip probe power was $\qty{-97}{\dBm}$, which -- except for few data points very close to instability -- is below the power required to observe LF nonlinearities, more details can be found in Supplementary Notes~VII and VIII.
One of the main advantages of this approach is the ability to investigate far more detunings, since, in contrast to the PPIT resonance, the LF mode visibility does not depend on the pump detuning and can be probed for small pump photons numbers $n_\mathrm{c}$, i.e., for large detunings between HF cavity and pump.
A non-intuitive result of these experiments is that we observe a parametric instability, when the pump has a slightly higher frequency than $\omega_0' - \Omega_0'$, but is still far red-detuned.
This is also predicted by theory, cf.~Supplementary Note~VI, and it is an intrinsic signature for the presence of dissipative coupling, which has also been studied in related optomechanical systems \cite{Weiss2013, Sawadsky2015}.
We can understand it from Eq.~(\ref{eqn:PPdamping_main}), which includes (for simplicity again in the sideband-resolved limit) that the photon-pressure damping $\Gamma_\mathrm{pp} = g_\omega'^2 \kappa_0'|\chi_\mathrm{c0}'|^2 + g_\omega'g_\kappa'\delta' |\chi_\mathrm{c0}'|^2$ has two contributions, the second of which changes sign with the pump detuning from the red sideband $\delta' = \omega_\mathrm{p} - \left(\omega_0' - \Omega_0'\right)$ and is $<0$ for $\delta' > 0$.
So it enhances the dispersive PP damping for $\delta' < 0$, and counteracts it for $\delta' > 0$.
In our case it even overcompensates it to $\Gamma_\mathrm{eff} \leq 0$ (theoretical instability criterion) around $\delta' \approx +\kappa_0'$ due to the large $|g_\kappa'/g_\omega'|$ ratio for large $n_\mathrm{c}$.
Similar considerations are valid for $\delta\Omega_\mathrm{pp}$.
An additional signature for backaction interference is that the maximum of $\Gamma_\mathrm{eff}$ occurs at $\delta' < 0$, cf.~Fig.~\ref{fig:figure5}\textbf{d}, instead of at $\delta' \approx 0$ as expected for $g_\kappa' = 0$.
The slight shift of the maximum  $\Gamma_\mathrm{eff}$ to lower frequencies with incressing $P_\mathrm{p}$ on the other hand reflects the increasing ratio $|g_\kappa'/g_\omega'|$.
A more detailed comparison between purely dispersive and mixed photon-pressure can be found in Supplementary Note~VI.
To quantitatively model $\Gamma_\mathrm{eff}$ and $\Omega_\mathrm{eff}$, we include also a cross-nonlinear damping and a cross-Kerr frequency shift, i.e.,
\begin{eqnarray}
	\Gamma_\mathrm{eff} & = & \Gamma_0 + \kappa_\mathrm{c}n_\mathrm{c} + \Gamma_\mathrm{pp} \label{eqn:Gamma_eff_main} \\
	\Omega_\mathrm{eff} & = & \Omega_0 + \mathcal{K}_\mathrm{c}n_\mathrm{c} + \delta\Omega_\mathrm{pp}, \label{eqn:Omega_eff_main}
\end{eqnarray}
where both nonlinear effects are occurring naturally due to the constrictions being part of the LF mode in the galvanic coupling scheme.
We fit the backaction data using all the HF mode and photon-pressure parameters obtained independently, and with the free parameters $\Gamma_0$, $\Omega_0$, $\mathcal{K}_\mathrm{c}$ and $\kappa_\mathrm{c}$.
The cross-parameters we obtain from this are rather small with $\mathcal{K}_\mathrm{c} = 2\pi\times\qty{-130.8}{\hertz}$ and $\kappa_\mathrm{c} = 2\pi\times\qty{38.4}{Hz}$, but the shift due to $\mathcal{K}_\mathrm{c}$ is still dominating the total pump-induced LF frequency shift.
The cross-nonlinear damping on the other hand only contributes less than $10\%$ to the deviations of $\Gamma_\mathrm{eff}$ from $\Gamma_0$ at the maximum of $\Gamma_\mathrm{eff}$, and the biggest part is due to $\Gamma_\mathrm{pp}$.
With the cross-nonlinearities included, we find good agreement between theory and data.
The remaining deviations we attribute to uncertainties in the frequency-dependent pump-photon number, higher-order nonlinearities for the largest photon numbers and the onset of the parametric instability for the points close to the instability threshold.
The recovery of the LF mode from the instability regime at $\omega_\mathrm{p}/2\pi \gtrsim \qty{8.1}{\giga\hertz}$, which is visible in Fig.~\ref{fig:figure5}\sublabel{b} but not predicted by theory, is accompanied by the absence of the HF mode in the HF reflection, which we believe is due to the pump photon number being so large that the cJJ SQUID is driven into the voltage state by overcritical peak HF currents.
Overall though, the agreement between backaction data and model clearly confirms our above conclusions regarding Fano-like effects by dissipative coupling contributions and the nonlinearity-enhanced coupling rates, without which the datasets would lie much closer together for the pump powers used in Fig.~\ref{fig:figure5}\textbf{d}, \textbf{e}.
\vspace{-5mm}
\section{Discussion}
\vspace{-2mm}
In this work, we have reported the realization of niobium-based superconducting photon-pressure circuits, which are operated at liquid helium temperature in the highly dissipative circuit regime.
Due to the elevated temperature and dissipation compared to earlier implementations, we obtained a device with a considerable flux-tunable dissipative interaction $g_{0\kappa}$ in addition to the usual dispersive coupling $g_{0\omega}$, and with $-1.2 \gtrsim g_{0\kappa}/g_{0\omega} \gtrsim -3$.
Furthermore, we observed at least three additional photon-pressure coupling terms due to the flux-modulation of inherent device nonlinearities, namely a Kerr-anharmonicity interaction and both a first- and a third-order nonlinear-damping-based interaction.
The additional coupling terms were shown to lead to a strong multiphoton enhancement of the original interaction terms by up to a factor $3.4$ and of the cooperativities by up to a factor $4.1$.
We have revealed with both our experiments and our theoretical description that the presence of dissipative photon-pressure leads to a nonreciprocal interaction between the modes, and to a Fano-like distortion of photon-pressure induced transparency due to the phase-shifted interference of the two coupling types, which can be used to determine $g_\kappa/g_\omega$ from a single response trace in frequency.
A related Fano-like distortion is present in the frequency-dependence of dynamical backaction, when $g_\kappa \neq 0$, which leads to enhanced photon-pressure damping when the sideband-pump is red-detuned from the cavity red-sideband and to a counter-intuitive instability when the pump tone is placed between the red cavity sideband and the cavity resonance.
In summary, our results reveal and describe several unexplored types of interaction between two superconducting circuits and demonstrate a series of experimental consequences of the presence of a purely-internal-dissipative radiation-pressure coupling.
This work opens the door for the investigation and application of dissipative photon-pressure in circuit QED, for unexplored low-frequency photon control protocols, and for photon-pressure experiments in the thermal and dissipative regime due to its compatibility with liquid helium.
Such tools are relevant for research fields like dark matter axion detection, since they provide new possibilities for LF photon-sensing and control, potentially down to the kHz frequency range.
Our report also confirms that photon-pressure circuits are an ideal testbed for general radiation-pressure systems.
It is directly relevant for and applicable to optomechanical systems, in particular to SQUID optomechanics, where similar effects can be expected to be observed and utilized in the future.
Finally, the experiments presented here suggest interesting questions and directions for future developments.
The most immediate question is how the dissipative contribution modifies sideband-cooling.
Generally, it will furthermore be very interesting to lower the LF mode resonance frequency to investigate the unresolved-sideband regime and corresponding effects like cooling on resonance or novel squeezing protocols, since in this regime dissipative coupling is predicted to have its biggest strengths.
Engineering an external-dissipative coupling by a modulation of $\kappa_\mathrm{ext}$ remains an open challenge, that would provide an additional degree of freedom and would bring new dynamics into play, since in such a system the LF mode directly couples to the pump tone on the HF feedline, while simultaneously not necessarily requiring a low $Q$ of the HF mode.
Once such an external-dissipative coupling can be realized, i.e., a flux-dependent coupling between the HF mode and a CPW feedline, the same approach could be used to implement effective internal-dissipative couplings even in the case of flux-independent constriction losses, e.g.~at \qty{}{\milli\kelvin} temperatures, by using a second \enquote{internal} feedline or on-chip resistor which is coupled to the SQUID circuit in a flux-dependent fashion.
Lastly, the observed nonlinearity-enhancements suggest that it might be worth to target stronger or even different types of nonlinearities both in photon-pressure circuits and in optomechanics.
\vspace{-4mm}
\begin{acknowledgments}
	\vspace{-2mm}
	The authors thank Markus Turad, Ronny Löffler (instrument scientists of the core facility \lisaplus), and Christoph Back for technical support.
	Furthermore, we thank Monika Fleischer and Ralf Stiefel for providing access to the PECVD system.
	This research received funding from the Deutsche Forschungsgemeinschaft (DFG) via grant numbers 490939971 (BO 6068/1-1) and 511315638 (BO 6068/2-1).
	M.K.~gratefully acknowledges financial support by the Studienstiftung des deutschen Volkes, J.P.~acknowledges support from the Cusanuswerk, Bischöfliche Studienförderung.
	We also gratefully acknowledge support by the COST actions NANOCOHYBRI (CA16218) and SUPERQUMAP (CA21144).
\end{acknowledgments}

\vspace{-4mm}
\subsection*{Data availability}
\vspace{-2mm}
	All data presented in this paper and the Supplementary Material (including raw data) and the corresponding processing scripts used during the analysis will be made publicly available on the repository Zenodo upon peer-reviewed publication of this work.

\vspace{-4mm}
\subsection*{Supplementary Material}
\vspace{-2mm}
	The Supplementary Material contains detailed descriptions of the experimental setups, a characterization of the device before insertion of the nano-constrictions, the complete theoretical framework, and a discussion of the differences between purely dispersive and dispersive-plus-dissipative photon-pressure.
	It furthermore presents details on data processing, data analysis and additional data, that support the findings and conclusions discussed in this work, as well as the additional references \cite{Igreja2004, Wenner2011, Rieger2023, Khapaev2001, Uhl2023, Gely2024, Gardiner1985}.
\vspace{-4mm}
\subsection*{Author contributions}
\vspace{-2mm}
M.K.~designed and fabricated the device, carried out the experiments, performed data analysis, prepared the figures and contributed to the first draft of the manuscript.
J.P.~performed data analysis, prepared the figures, and contributed to the circuit design and to the first draft of the manuscript.
Z.E.G.~contributed to fabrication recipe, data acquisition, data analysis, figure preparation, and to the first draft of the manuscript.
B.W.~developed the measurement code and contributed to theory.
K.U.~contributed to the device fabrication.
D.K.~and R.K.~contributed to the project funding and participated in scientific discussions.
D.B.~conceived the experiment, supervised all parts of the project, acquired the project funding, developed the theoretical framework and wrote the first draft of the manuscript.
All authors discussed the results and conclusions and contributed to manuscript revisions.
\vspace{-4mm}
\subsection*{Competing interests}
\vspace{-2mm}
	The authors declare no competing interests.

\vspace{-2mm}
\section{Methods}
\vspace{-2mm}
\subsection{Device fabrication}
\label{sec:meth_fab}
\vspace{-2mm}
\begin{itemize}[leftmargin=*, label = {}]
	\item \textit{Step 1: Bottom niobium layer}
	\\
	The fabrication starts with dc-magnetron-sputtering of $\qty{120}{\nano\meter}$ thick niobium (Nb) on top of a high-resistivity ($\rho>\qty{10}{\kilo\ohm\centi\meter}$ at room temperature) intrinsic two inch silicon wafer.
	The wafer has a thickness of $\qty{525}{\micro\meter}$, and after Nb deposition it is covered with a positive photoresist (ma-P~1205) by means of spin coating (resist thickness $\qty{\sim 0.5}{\micro\meter}$).
	The bottom niobium layer design is transferred to the photoresist using maskless photolithography ($\lambda_\mathrm{litho} = \qty{365}{\nano\meter}$).
	The irradiated resist is developed with ma-D 331/S for $\qty{25}{\second}$, and then the excess Nb film is removed by $\mathrm{SF}_6$ reactive ion etching.
	Finally, the wafer is cleaned with multiple subsequent baths of acetone and isopropanol.
	\item \textit{Step 2: Dielectric layer}
	\\
	The second step is to deposit a dielectric layer on the patterned first layer of Nb, where the parallel-plate-capacitors (PPCs) will be formed.
	Before deposition, the wafer is covered with the same photoresist as in step~1, and the areas where the PPCs are formed are defined again by maskless optical lithography.
	After the resist is developed, the wafer is completely covered by plasma-enhanced chemical vapor deposition (PECVD) with $\qty{200}{\nano\meter}$ of silicon nitride ($\sn$) in a low-temperature process ($T \sim \qty{100}{\celsius}$).
	Then, a lift-off procedure is performed in acetone for $\qty{15}{\minute}$ to remove the resist and all the excess $\sn$.
	Finally, the wafer is rinsed in multiple baths of acetone and isopropanol.
	\item \textit{Step 3: Top niobium layer}
	\\
	The third step in the fabrication procedure is similar to step~2, but instead of $\sn$, $\qty{300}{\nano\meter}$ of Nb is deposited onto the resist-covered and patterned wafer, again by dc-magnetron sputtering.
	After lift-off of the excess Nb in acetone, the second Nb layer forms the PPC top plate, the transmission line of the low-frequency (LF) resonator, and the ground planes of both high-frequency (HF) and LF resonators.
	\item \textit{Step 4: Mounting and pre-characterization}
	\\
	After fabrication, the wafer is diced into individual $10\times\qty{10}{\square\milli\meter}$ large chips.
	Then, a single chip is mounted into a matching cutout of a microwave printed circuit board (PCB), and is wirebonded to the microwave feedlines and ground plane of the PCB.
	Both chip and PCB are packed in a radiation-tight copper housing.
	Finally, the chip is mounted into the measurement setup, and the device is pre-characterized in liquid helium by means of microwave reflectometry.
	Here, each of the modes is characterized via its own CPW feedline.
	\item \textit{Step 5: Constriction fabrication}
	\\
	To cut the constriction Josephson junctions (cJJs) into the pre-characterized device, the sample is removed from the PCB and mounted onto an aluminum stub; it is wirebonded to the stub to prevent any charging of the sample during the ion irradiation.
	The mounted sample is placed in the neon ion microscope, which allows high-precision milling with a focused neon ion beam (Ne-FIB).
	The cJJs are of the monolithic 3D-type~\cite{Uhl2024, Uhl2024a}.
	They are formed by cutting two $\qty{\sim 40}{\nano\meter}$ narrow slot-shaped rectangles from both sides into the $\qty{3}{\micro\meter}$ wide bridges with a dose of $\qty{20000}{\mathrm{ions}\per\nano\meter\squared}$ and an accelerating voltage of $\qty{20}{\kilo\volt}$.
	In addition, the constrictions are milled from the top with a third rectangle, but with a lower dose of $\qty{3000}{\mathrm{ions}\per\nano\meter\squared}$.
	\item \textit{Step 6: Mounting and experiments}
	\\
	After the Ne-FIB cutting process, the sample is mounted again into the measurement setup, similar to step~4, but this time a small coil for the application of a magnetic field perpendicular to the chip surface is added.
	Then, the photon-pressure experiments begin.
\end{itemize}
\vspace{-4mm}
\subsection{Resonance fitting}
\label{sec:meth_fitting}
\vspace{-2mm}
In this Methods part, we describe the fitting routine, which we used to fit all pumped and unpumped HF and LF resonances as well as the PPIT zoom measurements.
We will generically use the unprimed HF quantities here $\omega_0, \kappa_\mathrm{ext}, \kappa_\mathrm{int}$, and $\kappa_0$ to do so, but the same formalism is valid for the primed quantities and the capitalized LF quantities.
Some more details and a corresponding figure can be found in Supplementary Note~VII.
The ideal reflection-response function of a high-$Q$ ($Q \gg 10$) parallel RLC circuit, which is capacitively coupled to a feedline with characteristic impedance $Z_0$, is given by
\begin{equation}
	S_{11}^\mathrm{ideal} = 1 - \frac{2\kappa_\mathrm{ext}}{\kappa_0 + 2\i\left(\omega - \omega_0 \right)}
	\label{eqn:s11_ideal_main}
\end{equation}
with the angular excitation frequency $\omega$, the corresponding resonance frequency $\omega_0$, the external linewidth $\kappa_\mathrm{ext}$ and the total linewidth $\kappa_0 = \kappa_\mathrm{int} + \kappa_\mathrm{ext}$.
Due to the cabling and all the microwave components in between the vector network analyzer and the circuit, the ideal transmission is not what we measure though.
To take frequency-dependent attenuation, the electrical cable length and possible interferences (e.g.~parasitic transmission through the directional coupler or parasitic reflections) into account, we model the actual reflection as
\begin{equation}
	S_{11}^\mathrm{real} = \left(a_0 + a_1\omega + a_2\omega^2\right)\left(1 - \frac{2\kappa_\mathrm{ext}\e^{\i\theta}}{\kappa_0 + 2\i\left(\omega - \omega_0 \right)}\right)\e^{\i\left(\phi_0 + \phi_1\omega \right)}.
	\label{eqn:s11_real_main}
\end{equation}
The factors $a_0, a_1, a_2, \phi_0, \phi_1$ and $\theta$ are real-valued fit parameters.
Before we apply this equation to the large-linewidth HF data, however, we divide the experimental $S_{11}^\mathrm{exp}$ data by a corresponding high-power background dataset $S_{11}^\mathrm{bg, exp}$, that has been taken in a VNA power-regime, in which the resonance is completely suppressed due to its nonlinearity.
As a result, we obtain $S_{11}^\mathrm{cor} = S_{11}^\mathrm{exp}/S_{11}^\mathrm{bg, exp}$, details are described in Supplementary Note~VII.
Afterwards, the background reflection is sufficiently smooth to apply Eq.~(\ref{eqn:s11_real_main}).
During our automated data fitting routine we first remove the absorption resonance from the dataset (leaving a gapped $S_{11}$-dataset) and fit the remaining $S_{11}$-response in an appropriate frequency window (typically five to ten times $\kappa_0$) with the background function
\begin{equation}
	S_{11}^\mathrm{bg} = \left(a_0 + a_1\omega + a_2\omega^2\right)\e^{\i\left(\phi_0 + \phi_1\omega \right)}.
\end{equation}
We obtain preliminary values for $a_0, a_1, a_2, \phi_0$ and $\phi_1$.
Then, we calculate $S_{11}^\mathrm{cor}/S_{11}^\mathrm{bg}$ for the complete dataset and fit the resulting data with
\begin{equation}
	S_{11}^\mathrm{\theta} = 1 - \frac{2\kappa_\mathrm{ext}\e^{\i\theta}}{\kappa_0 + 2\i\left(\omega - \omega_0 \right)},
\end{equation}
from which we obtain a preliminary set of values for $\omega_0, \kappa_0, \kappa_\mathrm{ext}$ and $\theta$.
Finally, we use all the preliminary values for $a_0, a_1, a_2, \phi_1, \phi_2, \omega_0, \kappa, \kappa_\mathrm{ext}$ and $\theta$ as starting parameters to re-fit the original dataset with the complete Eq.~(\ref{eqn:s11_real_main}).
Here, we find that $\theta$ is small and nearly constant and during the last step, we keep $\theta = \theta_0 = 0.08$ constant to improve the quality of the fits and the values of the remaining parameters.
All the HF $S_{11}$-datasets after constriction cutting, which are shown in the manuscript and Supplementary Material figures, as well as their corresponding fit curves have been background-corrected using this method.
Additionally, we have rotated off the interference angle $\theta_0$.
\vspace{-3mm}
\subsection{Fano-angle in transparency experiment}
\label{sec:fano_angle}
\vspace{-2mm}
The reflection response of the PPIT experiment is modeled using Eq.~(\ref{eqn:S11_PPIT_main}) (for the derivation see Supplementary Note~V)
\begin{equation}
	S_{11} = 1 - \kappa_\mathrm{ext}'\chi_\mathrm{c}'\left[1 - g_\omega' g' \chi_\mathrm{c}'\chi_0^\mathrm{eff}\right]
\end{equation}
with the multi-photon coupling rates $g_\omega', g_\kappa'$ and $g' = g_\omega' + \i\frac{g_\kappa'}{2}$, the HF cavity susceptibility
\begin{equation}
	\chi_\mathrm{c}' = \frac{1}{\frac{\kappa_0'}{2} + \i(\Delta_\mathrm{p}' + \Omega)}
\end{equation}
and the effective LF susceptibility
\begin{equation}
	\chi_0^\mathrm{eff} = \frac{1}{\frac{\Gamma_\mathrm{eff}}{2} + \i\left(\Omega - \Omega_\mathrm{eff}\right)}.
\end{equation}
Exactly on resonance of both HF cavity and PPIT, i.e., when $\Delta_\mathrm{p}' + \Omega = \Omega - \Omega_\mathrm{eff} = 0$ (imaginary parts of both susceptibilities vanish) or alternatively $\Delta_\mathrm{p}' = -\Omega_\mathrm{eff}$, we get for the reflection at $\Omega = +\Omega_\mathrm{eff}$
\begin{equation}
	S_{11}^\mathrm{res} = 1 - 2\frac{\kappa_\mathrm{ext}'}{\kappa_0'}\left[1 - 4\frac{g_\omega' g'}{\kappa_0'\Gamma_\mathrm{eff}}\right].
\end{equation}
For vanishing coupling $g' = g_\omega' = 0$, the resulting cavity resonance point is the usual one
\begin{equation}
	S_{11}^\mathrm{HF} = 1 - 2\frac{\kappa_\mathrm{ext}'}{\kappa_0'},
\end{equation}
i.e., exactly on the real axis.
To find the angle between the real axis and the connecting line between that point and the PPIT resonance, we shift the PPIT reflection by this value and get
\begin{equation}
	S_{11}^\mathrm{res, shift} = 8\frac{\kappa_\mathrm{ext}'}{\kappa_0'}\frac{g_\omega' g'}{\kappa_0'\Gamma_\mathrm{eff}}.
\end{equation}
The angle is then obtained via
\begin{eqnarray}
	\gamma & = & \arctan\frac{\mathrm{Im}\left[S_{11}^\mathrm{res, shift}\right]}{\mathrm{Re}\left[S_{11}^\mathrm{res, shift}\right]} \\
	& = & \arctan{\frac{g_\kappa'}{2g_\omega'}}.
\end{eqnarray}
\vspace{-3mm}
\subsection{Extracting the multiphoton coupling rates}
\label{sec:coupl_rates}
\vspace{-2mm}
To study the scaling of the multiphoton coupling rates with intracavity pump photon number, we perform the experiment described in Fig.~\ref{fig:figure3} for several different pump powers and detunings in the range $\delta_\mathrm{eff} \in \left[-\kappa_0'/4, \kappa_0'/4\right]$.
The PPIT data and fits for the resonant cases $\delta_\mathrm{eff} \approx 0$ are presented for all $P_\mathrm{p}$ in Supplementary Note~VIII.
The range of detunings within the $\kappa_0'/2$-wide interval around the red sideband is chosen to ensure a clear PPIT signature for all $\delta_\mathrm{eff}$ and $P_\mathrm{p}$, while simultaneously avoiding parametric instabilities.
For the analysis, we first fit the cavity resonance only of each dataset by removing the narrow PPIT window from it and obtain as fit parameters $\omega_0'$ and $\kappa_0'$, which fully determine $\chi_\mathrm{c}'$; additionally, we obtain $\kappa_\mathrm{ext}'$.
Secondly, we fit the isolated high-resolution PPIT resonance in its narrow frequency span and obtain as fit parameters $\Gamma_\mathrm{eff}$ and $\Omega_\mathrm{eff}$, i.e., the effective LF linewidth and resonance frequency including dynamical backaction effects, which together determine $\chi_0^\mathrm{eff}$.
In a last step, we fit the combined cavity-plus-PPIT data using Eq.~(\ref{eqn:S11_PPIT_main}) with all parameters fixed from the individual fits except for $g_\omega'$ and $g_\kappa'$.
The deviations of $g_\omega'$ and $g_\kappa'$ from the fit lines in Fig.~\ref{fig:figure4} seem to increase with $|\delta_\mathrm{eff}|$, and we attribute them to a frequency-dependence of the cavity background parameters such as $\theta$, which are not completely accounted for.
The intracavity pump photon number $n_\mathrm{c}$, which is required to analyze $g_\omega'(n_\mathrm{c})$ and $g_\kappa'(n_\mathrm{c})$, we obtain from the experimental ac Stark shift of the HF mode $\delta\omega_0 = \omega_0' - \omega_0$ in combination with the theoretical value for $g_{0\omega}$.
A detailed description is presented in Supplementary Notes~IV and VII.
\vspace{-3mm}
\subsection{Fit approach for dynamical backaction}
\label{sec:fit_DB}
\vspace{-2mm}
To model the dynamical backaction curves in Fig.~\ref{fig:figure5}, we use Eqs.~(\ref{eqn:PPdamping_main}) and (\ref{eqn:PPshift_main}) to calculate $\Gamma_\mathrm{pp}$ and $\delta\Omega_\mathrm{pp}$ without any free parameters.
The scaling of $g_\omega'$, $g_\kappa'$ and $\kappa_0'$ with $n_\mathrm{c}$ we obtain from the corresponding fits shown in Fig.~\ref{fig:figure4}, while the HF mode frequency $\omega_0'$ as a function of $n_\mathrm{c}$ we calculate as
\begin{equation}
	\omega_0' = \omega_\mathrm{p} + \sqrt{\left(\Delta_\mathrm{p} - \mathcal{K}n_\mathrm{c} \right)\left(\Delta_\mathrm{p} - 3\mathcal{K}n_\mathrm{c} \right) - \frac{\kappa_\mathrm{aux}^2}{4}}
\end{equation}
and $\kappa_\mathrm{aux} = \kappa_1 n_\mathrm{c} + 2\kappa_2n_\mathrm{c}^2 + 3\kappa_3n_\mathrm{c}^3$, cf.~Supplementary Notes~IV and~VII.
The only missing ingredient then is $n_\mathrm{c}(\omega_\mathrm{p})$, which we calculate by numerically solving the characteristic polynomial of a superconducting circuit with a dispersive Kerr nonlinearity and up to third order nonlinear damping
\begin{eqnarray}
	& &\frac{\kappa_3^2}{4}n_\mathrm{c}^7 + \frac{\kappa_2\kappa_3}{2}n_\mathrm{c}^6 + \frac{\kappa_2^2 + 2\kappa_1\kappa_3}{4}n_\mathrm{c}^5 \nonumber\\
	& & + \frac{\kappa_0\kappa_3 + \kappa_1\kappa_2}{2}n_\mathrm{c}^4 + \left[ \mathcal{K}^2 + \frac{\kappa_1^2 + 2\kappa_0\kappa_2}{4}\right]n_\mathrm{c}^3 \nonumber \\
	& & + \left[\frac{\kappa_0\kappa_1}{2} - 2\mathcal{K}\Delta_\mathrm{p}\right]n_\mathrm{c}^2  +\left[\Delta_\mathrm{p}^2 + \frac{\kappa_0^2}{4}\right]n_\mathrm{c} - \kappa_\mathrm{ext}n_\mathrm{in} = 0
\end{eqnarray}
cf.~Supplementary Note~IV.
The input photon flux
\begin{equation}
	n_\mathrm{in} = \mathcal{G}_\mathrm{att}(\omega_\mathrm{p})\frac{P_\mathrm{sg}}{\hbar\omega_\mathrm{p}}
\end{equation}
contains the experimentally determined pump-attenuation $\mathcal{G}_\mathrm{att}(\omega_\mathrm{p})$ and the output power of the signal generator $P_\mathrm{sg}$, cf.~Supplementary Note~VII.
Finally, we fit the total effective linewidth and resonance frequency for all pump powers simultaneously, using Eqs.~(\ref{eqn:Gamma_eff_main}) and (\ref{eqn:Omega_eff_main}) with $\Omega_0$, $\Gamma_0$, $\kappa_\mathrm{c}$ and $\mathcal{K}_\mathrm{c}$ as fit parameters.
Notably, the fit values for $\Gamma_0$ and $\Omega_0$ obtained from this fit deviate a bit from the ones obtained from the flux arc fits given in Sec.~\ref{sec:FTPP}, cf.~also Supplementary Note~IV.
To be more precise, they differ by $\qty{\sim 70}{\kilo\hertz}$ and $ \qty{\sim 190}{\kilo\hertz}$, respectively.
A good part of the deviation in $\Omega_0$ can be attributed to the LF flux arc fit (and the data extracted from it) slightly deviating from the data points, which accounts for $\qty{\sim 100}{\kilo\hertz}$ in $\Omega_0$.
We believe the remaining mismatches are mainly due to a $\qty{5}{\decibel}$ difference in probe powers used for the LF reflection measurements during the flux arc sweep (higher probe power) and the dynamical backaction experiment (lower probe power).
The latter would also be consistent with probe-power dependence measurements of the LF circuit during another cooldown, where we observed slight shifts of $\Omega_0$ and $\Gamma_0$ as function of probe power even below the onset of significant self-Kerr nonlinearities.

\let\addcontentsline\oldaddcontentsline
\onecolumngrid
\clearpage

\makeatletter
\AddToHook{__hyp/target/setname}{%
	\ifHy@localanchorname
	\expandafter\preto
	\else
	\expandafter\gpreto
	\fi
	\@currentHref{suppl.}%
}%
\makeatother

\begin{center}
	\noindent\textbf{\large Supplementary Material for: \vspace{5mm}\\ Tunable and nonlinearity-enhanced \\ dispersive-plus-dissipative coupling in photon-pressure circuits}

	\normalsize
	\vspace{.8cm}

	\noindent{M.~Kazouini, J.~Peter, Z.~E.~Guo, B.~Wilde, K.~Uhl, D.~Koelle, R.~Kleiner, and D.~Bothner}
	\\
\end{center}
\vspace{0.2cm}

\tableofcontents

\setcounter{section}{0}
\setcounter{figure}{0}
\setcounter{equation}{0}
\setcounter{table}{0}
\renewcommand{\figurename}{Supplementary Figure}
\renewcommand{\tablename}{Supplementary Table}
\renewcommand{\theequation}{S\arabic{equation}}
\renewcommand{\bibnumfmt}[1]{[S#1]}
\renewcommand{\citenumfont}[1]{S#1}

\clearpage
\section*{List of most important parameters and variables}
\vspace{-8mm}
\renewcommand{\arraystretch}{1.3}

\begin{table}[h!]
	\titlecaption{List of the most important basic variables used in the manuscript and supplementary material}  \\
	\begin{tabular}{ l   l   l }
		\toprule
		Symbol & Meaning & Important relation to other quantities \\
		\midrule
		$\Omega_0$ & Bare resonance frequency of the LF mode & -- \\
		$\Gamma_\mathrm{int}$ & Bare internal decay rate of the LF mode & -- \\
		$\Gamma_\mathrm{ext}$ & Bare external decay rate of the LF mode & -- \\
		$\Gamma_0$ & Bare total decay rate of the LF mode & $\Gamma_0 = \Gamma_\mathrm{int} + \Gamma_\mathrm{ext}$ \\
		$\omega_0$ & Bare resonance frequency of the HF mode & -- \\
		$\kappa_\mathrm{int}$ & Bare internal decay rate of the HF mode & -- \\
		$\kappa_\mathrm{ext}$ & Bare external decay rate of the HF mode & -- \\
		$\kappa_0$ & Bare total decay rate of the HF mode & $\kappa_0 = \kappa_\mathrm{int} + \kappa_\mathrm{ext}$ \\
		$g_{0\omega}$ & Dispersive single-photon coupling rate & -- \\
		$g_{0\kappa}$ & Dissipative single-photon coupling rate & -- \\
		$\omega_\mathrm{p}$ & Frequency of the HF sideband pump tone & -- \\
		$\Delta_\mathrm{p}$ & Detuning between pump tone and bare HF resonance & $\Delta_\mathrm{p} = \omega_\mathrm{p} - \omega_0$ \\
		$P_\mathrm{p}$ & On-chip power of the HF sideband pump tone & -- \\
		$n_\mathrm{c}$ & Intracavity pump-photon number in the HF mode & -- \\
		$\Omega_0'$ & Pump-shifted angular resonance frequency of the LF mode & $\Omega_0' = \Omega_0 + \mathcal{K}_\mathrm{c}n_\mathrm{c}$ \\
		$\mathcal{K}_\mathrm{c}$ & Cross-Kerr constant; LF dispersive shift per HF photon & -- \\
		$\Gamma_0'$ & Pump-broadened total decay rate of the LF mode & $\Gamma_0' = \Gamma_0 + \kappa_\mathrm{c}n_\mathrm{c}$ \\
		$\kappa_\mathrm{c}$ & Cross-damping constant; LF linewidth-broadening per HF photon  & -- \\
		$\omega_0'$ & Pump-shifted resonance frequency of the HF mode for a probe tone & $\omega_0' = \omega_\mathrm{p} + \sqrt{\left(\Delta_\mathrm{p} - \mathcal{K}n_\mathrm{c} \right)\left(\Delta_\mathrm{p} - 3\mathcal{K}n_\mathrm{c} \right) - \frac{\kappa_\mathrm{aux}^2}{4}}$ \\
		$\mathcal{K}$ & Self-Kerr constant of the HF mode; frequency shift per photon  & -- \\
		$\kappa_0'$ & Pump-broadened linewidth of the HF mode for a probe tone & $\kappa_0' = \kappa_0 + 2\kappa_1 n_\mathrm{c} + 3\kappa_2 n_\mathrm{c}^2 + 4\kappa_3 n_\mathrm{c}^3$ \\
		$\kappa_m$ & $m$-th order nonlinear damping coefficient, $m \in \left\{1, 2, 3\right\}$ & -- \\
		$ \kappa_\mathrm{eff} $ & Pump-broadened linewidth of the HF mode for the pump tone &  $\kappa_\mathrm{eff} = \kappa_0 + \kappa_1 n_\mathrm{c} + \kappa_2 n_\mathrm{c}^2 + \kappa_3 n_\mathrm{c}^3$ \\
		$\kappa_\mathrm{aux}$ & Auxiliary linewidth variable introduced for brevity & $\kappa_\mathrm{aux} = \kappa_0' - \kappa_\mathrm{eff}$ \\
		$g_\omega'$ & Pump- and nonlinearity-enhanced total dispersive coupling rate & $g_\omega' = g_{0\omega}\sqrt{n_\mathrm{c}} + g_{0\mathcal{K}}\sqrt{n_\mathrm{c}^3}$ \\
		$g_{0\mathcal{K}}$ & Single-photon coupling rate due to Kerr nonlinearity  & -- \\
		$g_\kappa'$ & Pump- and nonlinearity-enhanced total dissipative coupling rate & $g_\kappa' = g_{0\kappa}\sqrt{n_\mathrm{c}} + g_{\mathrm{nl}1}\sqrt{n_\mathrm{c}^3} + g_{\mathrm{nl}2}\sqrt{n_\mathrm{c}^5} + g_{\mathrm{nl}3}\sqrt{n_\mathrm{c}^7}$ \\
		$g_{\mathrm{nl}m}$ & Single-photon coupling rate due to $m$-th order nonlinear damping, $m \in \left\{1, 2, 3\right\}$ & -- \\
		$\Delta_\mathrm{p}'$ & Detuning between pump tone and pump-shifted HF probe resonance & $\Delta_\mathrm{p}' = \omega_\mathrm{p} - \omega_0'$ \\
		$\delta$ & Detuning between pump tone and bare HF mode red sideband & $\delta = \omega_\mathrm{p} - \left(\omega_0 - \Omega_0\right)$ \\
		$\delta'$ & Detuning between pump tone and pump-shifted HF mode red sideband & $\delta' = \omega_\mathrm{p} - \left(\omega_0' - \Omega_0'\right)$ \\
		$\delta\Omega_\mathrm{pp}$ & Photon-pressure backaction frequency shift & -- \\
		$\Omega_\mathrm{eff}$ & LF resonance frequency including backaction and cross-Kerr shift & $\Omega_\mathrm{eff} = \Omega_0' + \delta\Omega_\mathrm{pp}$ \\
		$\Gamma_\mathrm{pp}$ & Photon-pressure backaction damping rate & -- \\
		$\Gamma_\mathrm{eff}$ & LF damping rate including backaction and cross-damping & $\Gamma_\mathrm{eff} = \Gamma_0' + \Gamma_\mathrm{pp}$ \\
		$\delta_\mathrm{eff}$ & Detuning between pump tone and effective HF mode red sideband & $\delta_\mathrm{eff} = \omega_\mathrm{p} - \left(\omega_0' - \Omega_\mathrm{eff} \right)$ \\
		$\chi_0$ & Bare LF mode susceptibility & $\chi_0 = \left[\Gamma_0/2 + \i\left(\Omega - \Omega_0 \right)\right]^{-1}$  \\
		$\chi_0^\mathrm{eff}$ & Effective LF mode susceptibility including backaction and cross-effects & $\chi_0^\mathrm{eff} = \left[\Gamma_\mathrm{eff}/2 + \i\left(\Omega - \Omega_\mathrm{eff} \right)\right]^{-1}$ \\
		$\chi_\mathrm{c}$ & Bare HF mode susceptibility & $\chi_\mathrm{c} = \left[\kappa_0/2 + \i\left(\Delta_\mathrm{p} + \Omega \right)\right]^{-1}$  \\
		$\chi_\mathrm{c}'$ & Pump-modified HF mode susceptibility for a probe tone & $\chi_\mathrm{c}' = \left[\kappa_0'/2 + \i\left(\Delta_\mathrm{p}' + \Omega\right)\right]^{-1}$ \\
		\bottomrule
	\end{tabular}
	\label{tab:Table_0}
\end{table}
\renewcommand{\arraystretch}{1}
\clearpage

\section{Supplementary Note I: Measurement setup}

\label{sec:Note1}
During our experiments, we have used two different experimental setups, a simple one for pre-characterization of the device before the constriction Josephson junctions (cJJs) were milled into the circuit, and a more sophisticated one for after cJJ nano-patterning and for all the photon-pressure (PP) experiments.
Both variants are schematically shown in Supplementary~Fig.~\ref{fig:FigS1}.

\begin{figure*}[!b]
	\includegraphics{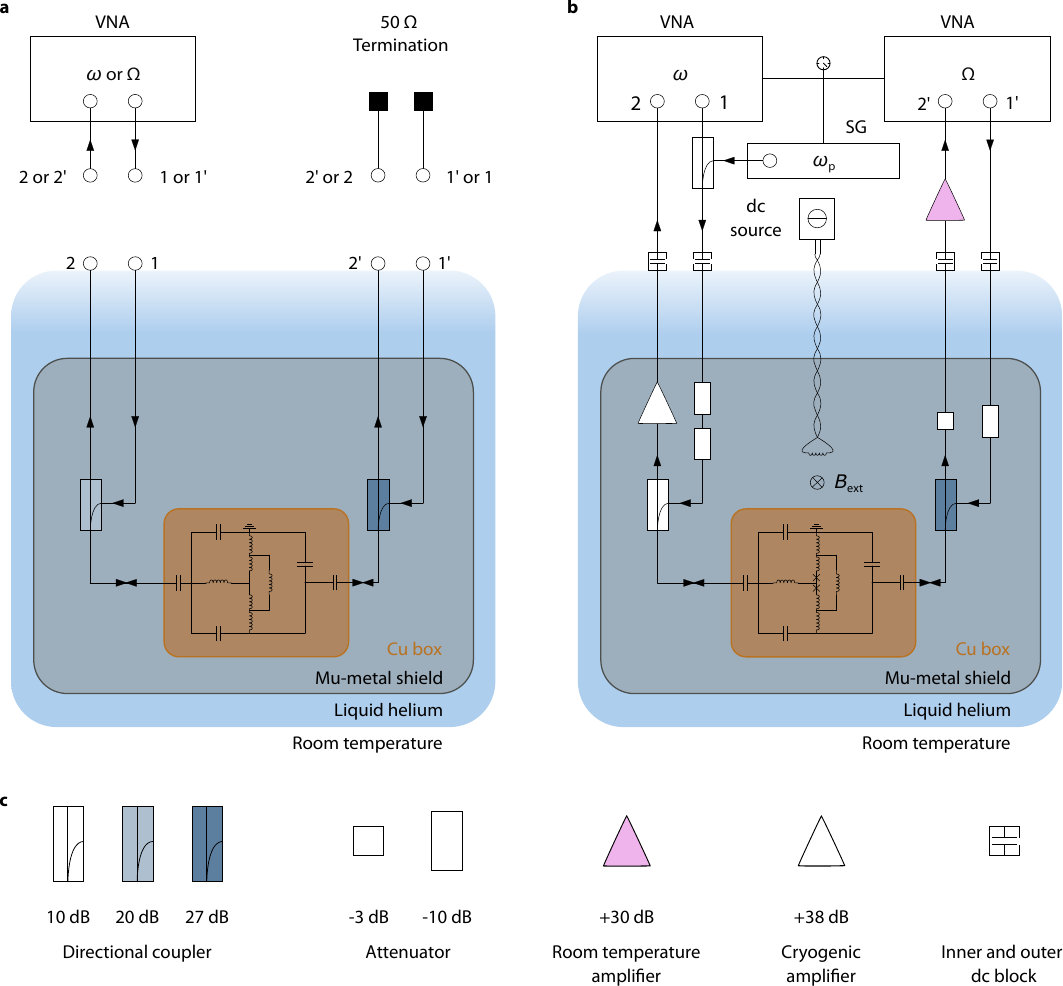}
	\titlecaption{Schematics of the measurement setups}{
		\sublabel{a}~Setup for measurements without cJJs.
		\sublabel{b}~Setup for measurements with cJJs and for photon-pressure experiments. \sublabel{c}~Symbol legend for setup elements not directly labeled in \sublabel{a}, \sublabel{b}. Details are given in the text.
	}
	\label{fig:FigS1}
\end{figure*}

\begin{figure*}
	\includegraphics{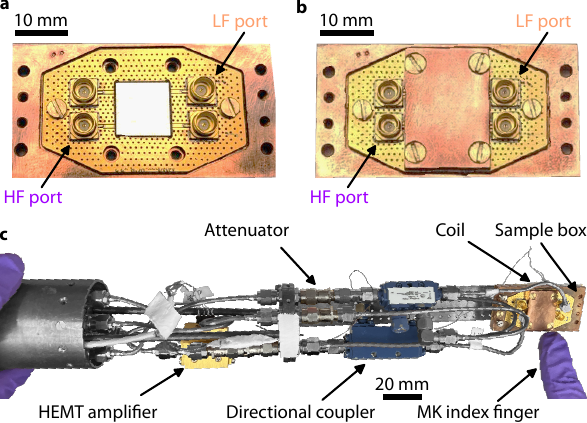}
	\titlecaption{Photographs of sample box and dipstick tip}{
		\sublabel{a}~The square-shaped $\qty{10}{\milli\meter}\times\qty{10}{\milli\meter}$ large microchip with the PP device (image center, light gray) is mounted into the square-shaped cutout of a microwave PCB (golden) with four coplanar waveguide feedlines and four SMP connectors. Both chip and PCB are mounted onto the bottom part of the copper sample box; the chip is fixed by carbon tape and the PCB is fixed by screws. PCB and chip are thoroughly contacted by wirebonds all around the chip. The top right SMP connector leads to the on-chip feedline for the LF mode, the bottom left SMP connector to the feedline for the HF mode.
		\sublabel{b}~For the experiments, the chip is covered with a copper lid, which is fixed by four additional screws to the copper sample holder with chip and PCB as it is depicted in~\sublabel{a}.
		\sublabel{c}~Tip of the microwave dipstick for measurements in a liquid helium transport dewar. At the right end, the sample box shown in \sublabel{b} is mounted with a small electromagnetic coil below it. Both HF and LF SMP ports are connected to copper coaxial cables. Also visible are the two directional couplers, multiple attenuators and the cryogenic HEMT amplifier. For clearer emphasis on the most important components, all microwave cables, dc wires and mounting parts are grayed out. Visible in purple are the gloved fingers of M.~Kazouini.
	}
	\label{fig:FigS2}
\end{figure*}
All experiments were conducted directly in liquid helium at the sample temperature $T_\mathrm{s} \approx \qty{4.2}{\kelvin}$.
The copper housing of the sample was always attached to a copper or brass mounting bracket at the end of a dipstick for microwave experiments in liquid helium transport dewars,~cf.~Supplementary Fig.~\ref{fig:FigS2} for pictures of the sample box and the dipstick tip of the more sophisticated setup.
Both of the dipsticks we used have four coaxial lines and the whole sample space is enclosed in a $ \qty{\sim 30}{\centi\meter}$ long cryoperm magnetic shielding tube with a diameter of $ \qty{\sim 5}{\centi\meter}$.
The two circuits of the device, high-frequency (HF) and low-frequency (LF) circuit, were in both cases simultaneously connected to the coaxial lines by means of a directional coupler, which allowed us to measure the reflection $S_{11}$ as transmission $S_{21}$ using a vector network analyzer (VNA) and separate input and output lines, thus minimizing the accumulation of parasitic reflection contributions in the measurement signal.
In the simpler measurements before cJJ-milling, cf.~Supplementary Fig.~\ref{fig:FigS1}\sublabel{a}, we connected a VNA to either of the two circuits and terminated the two coaxial lines of the other circuit at the top of the dipstick at room temperature with a $\qty{50}{\ohm}$ termination.
Then, we either characterized the device response around the resonance frequency of the HF mode, i.e., at high frequencies $\omega$, or around the resonance frequency of the LF mode, i.e., at low frequencies $\Omega$.
Once the cJJs are introduced and we characterize the photon-pressure interaction, the setup gets much more sophisticated, cf.~Supplementary Fig.~\ref{fig:FigS1}\sublabel{b} and Supplementary Fig.~\ref{fig:FigS2}\sublabel{c}.
Firstly, we need a variable magnetic field $B_\mathrm{ext}$ perpendicular to the chip surface to change the magnetic flux in the superconducting quantum interference device (SQUID) and to pick the ideal operation point.
To this end, a homemade electromagnetic coil is attached to the copper housing of the sample, which is connected by a pair of twisted copper wires to a battery-powered direct current (dc) source at room-temperature.
Secondly, we need to minimize incoming thermal noise on the coaxial lines and to equilibrate the equivalent temperature of that noise to $\qty{\sim 4.2}{\kelvin}$, at least on the HF side of the device.
The LF mode is much less nonlinear than the HF mode due to the circuit design, and since at the time of this experiment we did not have a cryogenic amplifier for the frequencies of the LF mode, we could not completely attenuate the corresponding coaxial lines.
Hence, the HF input line is attenuated at $\qty{4.2}{\kelvin}$ with two $\qty{10}{\decibel}$ attenuators and additionally with a $\qty{\sim 10}{\decibel}$ directional coupler.
After reflection at the HF circuit, the signal is routed straight through the directional coupler to a cryogenic high-electron-mobility-transistor (HEMT) amplifier with a gain of $\qty{\sim 38}{\decibel}$.
The LF input line has a single $\qty{10}{\decibel}$ attenuator in series with additional $\qty{27}{\decibel}$ attenuation at its directional coupler.
The output line on the other hand has a small attenuator with $\qty{3}{\decibel}$ to provide at least some more attenuation of input noise than just the coaxial cable damping, while preserving the detectability of the reflected signal.
At room temperature, the reflected LF signal is also amplified by a HEMT amplifier with a gain of $\qty{\sim 30}{\decibel}$ to improve the signal-to-noise ratio.
Note that all four coaxial lines in the photon-pressure setup are made of stainless steel cables (and short copper cables in the cryogenic region), which provide additional attenuation of $\qty{5}{\decibel}$ to $\qty{10}{\decibel}$ per line.
Furthermore, directly at the top of the dipstick on the coaxial feedthrough flange, four dc blocks were inserted into the four coaxial lines.
Regarding the microwave electronics, we conducted all experiments after cJJ-cutting with a single setup.
For the spectroscopic characterization of the HF and LF reflection, we permanently had two independent VNAs connected to the two device modes.
In addition, we connected a microwave signal generator (SG), that provided the sideband-pump tone for the photon-pressure experiments.
The SG pump tone with frequency $\omega_\mathrm{p}$ is combined with the VNA input signal of the HF circuit by means of a directional coupler at room temperature.
The three microwave electronics components were referenced to the $\qty{10}{\mega\hertz}$ oscillator of the signal generator.
All active electronics except for the statically operated HEMT amplifiers are controlled and read out during the measurements via a single computer, using Python-based measurement scripts.
\section{Supplementary Note II: The individual circuits without constrictions}
\label{sec:Note2}
We describe both circuits by lumped element parallel RLC circuits, that are capacitively coupled to a single transmission line with characteristic impedance $Z_0$.
\subsection{The high-frequency circuit without constrictions}
The high-frequency circuit without the constrictions consists of two interdigitated capacitors $C_\mathrm{idc}$ that we formally combine into a single capacitance $C = 2C_\mathrm{idc}$, and of several inductors, that we combine into a single inductance $L$.
The two small parts of the loop, into which the constrictions are milled, have each the inductance $L_\mathrm{arm}$ and therefore the high-frequency inductance without these bridges would be given by $L' = L - L_\mathrm{arm}/2$.
The total loop inductance we denote as $L_\mathrm{loop}$ and we estimate $L_\mathrm{arm} \approx L_\mathrm{loop}/10$.
By means of a coupling capacitor $C_\mathrm{c}$, the circuit is capacitively coupled to its coplanar waveguide feedline with characteristic impedance $Z_0$.
All internal losses we summarize in a resistor $R$.
Using these parameters and assuming a low-loss oscillator, the resonance frequency of the junction-less circuit is given by
\begin{equation}
	\omega_\mathrm{b} = \frac{1}{\sqrt{LC_\mathrm{tot}}}, ~~~~~ C_\mathrm{tot} = C + C_\mathrm{c}
\end{equation}
and the internal and external linewidths by
\begin{equation}
	\kappa_\mathrm{int, b} = \frac{1}{R C_\mathrm{tot}}, ~~~~~ \kappa_\mathrm{ext, b} = \frac{Z_0 C_\mathrm{c}^2}{L C_\mathrm{tot}^2}. \label{eqn:kappa_intext}
\end{equation}
\subsection{The low-frequency circuit without constrictions}
The low-frequency circuit consists of a parallel plate capacitance $C_0$ and an effective inductance $L_0$, and it is coupled by means of a coupling capacitor $C_\mathrm{c0}$ to its coplanar waveguide feedline, also with characteristic impedance $Z_0$.
Analogous to the high-frequency case, we can write the relevant relations as
\begin{equation}
	\Omega_\mathrm{b} = \frac{1}{\sqrt{L_0 C_\Sigma }}, ~~~~~ C_\Sigma = C_0 + C_\mathrm{c0}
\end{equation}
and
\begin{equation}
	\Gamma_\mathrm{int, b} = \frac{1}{R_0 C_\Sigma}, ~~~~~ \Gamma_\mathrm{ext, b} = \frac{Z_0 C_\mathrm{c0}^2}{L_0 C_\Sigma^2}. \label{eqn:Gamma_intext}
\end{equation}
\subsection{Circuit parameters}
We determine the circuit parameters starting from the capacitances $C_0$ and $C$.
The low-frequency parallel plate capacitance is calculated by
\begin{equation}
	C_0 = \epsilon_0\epsilon_\mathrm{r}\frac{A}{d_\mathrm{SiN}}
\end{equation}
and with the silicon-nitride dielectric constant $\epsilon_\mathrm{r} \approx 7$, the silicon-nitride thickness $d_\mathrm{SiN} \approx \qty{200}{\nano\meter}$, and the capacitor area $A = \qty{1.583}{\milli\meter\squared}$ we find $C_0 = \qty{490.5}{\pico\farad}$.
The high-frequency capacitance we obtain using the analytical equations of Ref.~\cite{Igreja2004x} for interdigitated capacitors~(IDCs), given by
\begin{eqnarray}
	C_\mathrm{idc} & = & \left(N_\mathrm{idc} - 3\right)\frac{C_1}{2} + 2\frac{C_1 C_2}{C_1 + C_2} \\
	C_i & = & 2\epsilon_0\epsilon_\mathrm{eff}l_\mathrm{idc}\frac{K(k_i)}{K(k_i')}, ~~~~~ i = 1, 2
\end{eqnarray}
where $K(k_i)$ are complete elliptic integrals of the first kind, $l_\mathrm{idc} = \qty{210}{\micro\meter}$ is the finger length, $\epsilon_\mathrm{eff} = (\epsilon_\mathrm{Si} + 1)/2$ is the effective permittivity with the silicon permittivity $\epsilon_\mathrm{Si} = 11.8$, and $N_\mathrm{idc} = 48$ is the total number of fingers in the capacitor.
The remaining parameters $k_i$ are given by
\begin{eqnarray}
	k_1 & = & \sin{\left( \frac{\pi}{2}\frac{a}{a + b} \right)} \\
	k_2 & = & 2\frac{\sqrt{a(a+b)}}{2a + b} \\
	k_i' & = & \sqrt{1 - k_i^2}
\end{eqnarray}
with $a = \qty{3}{\micro\meter}$ the finger width and $b = \qty{5}{\micro\meter}$ the gap width in between two fingers.
Finally, we obtain $C = 2C_\mathrm{idc} = \qty{941.5}{\femto\farad}$.
Similarly, we calculate the coupling capacitors as $C_\mathrm{c} = \qty{15.3}{\femto\farad}$ by the IDC model and $C_\mathrm{c0} = \qty{559}{\femto\farad}$ using the parallel plate expression.
\begin{figure*}
	\includegraphics{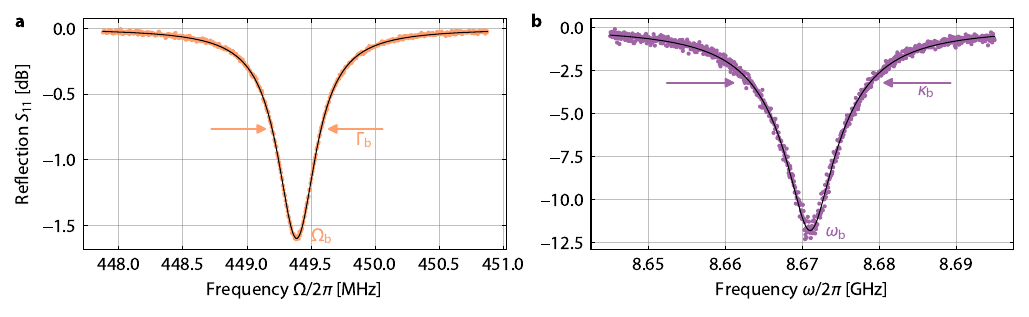}
	\titlecaption{Device resonances before cutting the nano-constrictions}{
		\sublabel{a} Reflection $S_{11}$ of the low-frequency (LF) circuit around its resonance frequency.
		Orange symbols are data, black line is a fit.
		From the fit curve, we obtain the resonance frequency before constriction cutting $\Omega_\mathrm{b} = 2\pi\times \qty{449.4}{\mega\hertz}$ and the total linewidth $\Gamma_\mathrm{b} = 2\pi\times\qty{401}{\kilo\hertz}$.
		\sublabel{b} Reflection $S_{11}$ of the high-frequency (HF) circuit around its resonance frequency.
		Purple symbols are data, black line is a fit.
		From the fit curve, we obtain the resonance frequency before constriction cutting $\omega_\mathrm{b} = 2\pi\times \qty{8.671}{\giga\hertz}$ and the total linewidth $\kappa_\mathrm{b} = 2\pi\times\qty{14.8}{\mega\hertz}$.
		Internal and external contributions to $\Gamma_\mathrm{b}$ and $\kappa_\mathrm{b}$ can be found in the text and in Supplementary Tables~\ref{tab:Table1} and~\ref{tab:Table2}.
	}
	\label{fig:FigS3}
\end{figure*}
Next, we measure the resonance frequencies $\omega_\mathrm{b} = 2\pi \times \qty{8.671}{\giga\hertz}$ and $\Omega_\mathrm{b} = 2\pi \times \qty{449.4}{\mega\hertz}$, cf.~Supplementary Fig.~\ref{fig:FigS3} for the two resonance datsets.
In combination with $C_\mathrm{tot} = \qty{956.8}{\femto\farad}$ and $C_\Sigma = \qty{491}{\pico\farad}$ we calculate $L = \qty{352}{\pico\henry}$ and $L_0 = \qty{255}{\pico\henry}$.
From the fits to the reflection data, we furthermore obtain the linewidths $\Gamma_\mathrm{int, b} = 2\pi\times \qty{367}{\kilo\hertz}$, $\Gamma_\mathrm{ext, b} = 2\pi\times \qty{34}{\kilo\hertz}$, $\kappa_\mathrm{int, b} = 2\pi\times \qty{5.1}{\mega\hertz}$ and $\kappa_\mathrm{ext, b} = 2\pi\times \qty{9.7}{\mega\hertz}$.
With the characteristic impedance $Z_0 = \qty{50}{\ohm}$, we can also calculate theoretical numbers for $\Gamma_\mathrm{ext, b}^\mathrm{th} = 2\pi\times \qty{40}{\kilo\hertz}$ and $\kappa_\mathrm{ext, b}^\mathrm{th} = 2\pi\times \qty{5.76}{\mega\hertz}$ using Eqs.~(\ref{eqn:kappa_intext}) and (\ref{eqn:Gamma_intext}).
The agreement with the experimental result is not extremely high, but acceptable.
Deviations between theory and experiment in the external coupling rates are very common and can be explained by stray capacitances, parasitic reflections (e.g.~at the crossover from cable to PCB or from PCB to chip), impedance mismatches in the microwave cabling leading to the sample or interferences by the imperfect isolation e.g.~in the directional couplers, cf.~Refs.~\cite{Wenner2011x, Rieger2023x}.
Finally, we use the software package \textit{3D-MLSI}~\cite{Khapaev2001x} with the London penetration depth $\lambda_\mathrm{L} = \qty{130}{\nano\meter}$ (typical value for our Nb films at $\qty{4.2}{\kelvin}$) to calculate the SQUID loop self-inductance $L_\mathrm{loop}^\mathrm{sim} = \qty{19.3}{\pico\henry}$.
The loop has a rectangular shape, the inner hole is $11\times\qty{6}{\micro\meter\squared}$ large, the outer dimensions are $17\times \qty{12}{\micro\meter\squared}$.
Supplementary Tables~\ref{tab:Table1} and \ref{tab:Table2} summarize all the relevant parameters of our device before constriction cutting and as determined from theoretical calculations, simulations and experiments.
Note that these values are good estimates, but not exact, since there are various sources of uncertainty ($\epsilon_\mathrm{r}$, $d_\mathrm{SiN}$, $\epsilon_\mathrm{eff}$ or $\lambda_\mathrm{L}$) and we also observe that the resonance frequencies of our devices can change from cooldown to cooldown on the order $0.1\% - 0.5\%$.
Hence, we will not fix the values of the circuit elements rigidly in our further analyses except for the capacitances, and allow the inductances to slightly vary in the description of different cooldowns/experiments.
We label the preliminary inductances $ L = L^\mathrm{pre}$ and $L_0 = L_0^\mathrm{pre}$ in the tables.\clearpage
\begin{table}
	\titlecaption{Preliminary circuit parameters of the LF mode estimated before cutting the nanobridge junctions}  \\
	\begin{tabular}{ c   c   c   c   c   c   c   c }
		\toprule
		$A\,$[$\qty{}{\milli\meter\squared}$] & $d_\mathrm{SiN}\,$[$\qty{}{\nano\meter}$] & $C_0\,$[$\qty{}{\pico\farad}$] & $C_\mathrm{c0}\,$[$\qty{}{\femto\farad}$] & $L_0^\mathrm{pre}\,$[$\qty{}{\pico\henry}$] & $\Omega_\mathrm{b}/2\pi\,$[$\qty{}{\mega\hertz}$] & $\Gamma_\mathrm{int, b}/2\pi\,$[$\qty{}{\kilo\hertz}$] & $\Gamma_\mathrm{ext, b}/2\pi\,$[$\qty{}{\kilo\hertz}$] \\
		\midrule
		1.583 & 200 & 490.5 & 559 & 255.4 & 449.4 & 367 & 34 \\ \bottomrule
	\end{tabular}
	\label{tab:Table1}
\end{table}
\begin{table}
	\titlecaption{Preliminary circuit parameters of the HF mode estimated before cutting the nanobridge junctions}  \\
	\begin{tabular}{ c   c   c   c   c   c   c   c   c   c   c }
		\toprule
		$N_\mathrm{idc}$ & $l_\mathrm{idc}\,$[$\qty{}{\micro\meter}$] & $a\,$[$\qty{}{\micro\meter}$] & $b\,$[$\qty{}{\micro\meter}$] & $C\,$[$\qty{}{\femto\farad}$] & $C_{c}\,$[$\qty{}{\femto\farad}$] & $L^\mathrm{pre}\,$[$\qty{}{\pico\henry}$] & $\omega_\mathrm{b}/2\pi\,$[$\qty{}{\giga\hertz}$] & $\kappa_\mathrm{int, b}/2\pi\,$[$\qty{}{\mega\hertz}$] & $\kappa_\mathrm{ext, b}/2\pi\,$[$\qty{}{\mega\hertz}$] & $L_\mathrm{loop}^\mathrm{sim}\,$[$\qty{}{\pico\henry}$] \\ \midrule
		48 & 210 & 3 & 5 & 941.5 & 15.3 & 352.1 & 8.671 & 5.1 & 9.7 & 19.3 \\ \bottomrule
		\vspace{2mm}
	\end{tabular}
	\label{tab:Table2}
\end{table}
\section{Supplementary Note III: The individual circuits with constrictions}
\label{sec:Note3}
\subsection{Including the constrictions}
Our treatment is analogous to the descriptions given in Refs.~\cite{Rodrigues2021x, Uhl2023x, Uhl2024x}.
For low microwave powers we treat each constriction as a bias-flux-dependent but linear inductance $L_\mathrm{c}$, for high powers we add a Kerr nonlinearity and nonlinear damping.
It is a very simple but quite accurate approach to describe the constriction inductance as a series combination of a linear inductance $L_\mathrm{lin}$ and a Josephson inductance $L_\mathrm{J}$, i.e.,
\begin{equation}
	L_\mathrm{c} = L_\mathrm{lin} + L_\mathrm{J}
\end{equation}
where the linear contribution considers deviations from the sinusoidal current-phase-relation \cite{Uhl2024x, Uhl2024ax} and the Josephson part is given by
\begin{equation}
	L_\mathrm{J} = \frac{\Phi_0}{2\pi I_0 \cos{\delta_0}}
\end{equation}
with $\delta_0$ the bias-flux-induced equilibrium phase difference, $I_0$ the critical current of the cJJ, and $\Phi_0 \approx \qty{2.068e-15}{\tesla\meter\squared}$ the flux quantum.
The equilibrium phase difference is related to the total flux $\Phi$ in the SQUID by
\begin{equation}
	\delta_0 = \pi\frac{\Phi}{\Phi_0},
\end{equation}
and so the Josephson inductance of a single constriction is
\begin{equation}
	L_\mathrm{J} = \frac{\Phi_0}{2\pi I_0 \cos{\left(\pi\frac{\Phi}{\Phi_0}\right)}}.
\end{equation}
In general, the constrictions also change the internal decay rate of the two modes, which could be considered by a (flux-dependent) constriction resistance $R_\mathrm{c}$ in parallel to $L_\mathrm{c}$.
Since the impact of this resistance on the resonance frequency is typically very small in our parameter regime \cite{Uhl2024x}, we omit it for all inductive calculations and phenomenologically model the linewidth as a function of flux by a polynomial.
\subsection{Describing the SQUID}
To describe the flux-biased SQUID, we start with the flux equation for a symmetric SQUID without bias current, which is given by
\begin{equation}
	\frac{\Phi}{\Phi_0} = \frac{\Phi_\mathrm{b}}{\Phi_0} + \frac{\tilde{L}_\mathrm{loop}J}{\Phi_0}
	\label{eqn:flux_basic}
\end{equation}
where $\Phi_\mathrm{b}$ is the external flux threading the SQUID loop, $J$ is the induced circulating current and $\tilde{L}_\mathrm{loop} = L_\mathrm{loop} + 2L_\mathrm{lin}$ is the effective linear loop inductance, which contains both geometric and kinetic contributions.
Without a bias current and for two identical junctions, the circulating current is described by
\begin{equation}
	J = -I_0\sin{\left(\pi\frac{\Phi}{\Phi_0}\right)}.
	\label{eqn:J_flux}
\end{equation}
The minus sign in this relation can be intuitively understood as follows: for small but finite external flux $0 < \Phi_\mathrm{b}/\Phi_0 < 0.5$ (and zero flux trapped in the SQUID loop), the screening-current flux $\tilde L_\mathrm{loop}J$ reduces the total flux in the SQUID compared to the external flux, i.e., it must be $J < 0$ to satisfy the condition $\Phi < \Phi_\mathrm{b}$.
Since also $0 < \Phi/\Phi_0 < 0.5$ in this situation we find $\sin{\left(\pi\Phi/\Phi_0 \right)} > 0$ and the condition $J < 0$ can only be fulfilled with the minus sign in Eq.~(\ref{eqn:J_flux}).
With the screening parameter $\beta_L = 2\tilde{L}_\mathrm{loop}I_0/\Phi_0 = \tilde{L}_\mathrm{loop}/\pi L_\mathrm{J0}$ and $L_\mathrm{J0} = \Phi_0/2\pi I_0$ we can finally re-write the flux relation Eq.~(\ref{eqn:flux_basic}) as
\begin{equation}
	\frac{\Phi}{\Phi_0} = \frac{\Phi_\mathrm{b}}{\Phi_0} - \frac{\beta_L}{2}\sin{\left(\pi\frac{\Phi}{\Phi_0} \right)}.
\end{equation}
For each external flux $\Phi_\mathrm{b}$, we numerically solve this equation to find the total flux $\Phi$, and subsequently the corresponding Josephson inductance $L_\mathrm{J}(\Phi)$.
\subsection{The total circuit including constrictions -- Dispersive part}
\begin{figure*}[b]
	\includegraphics{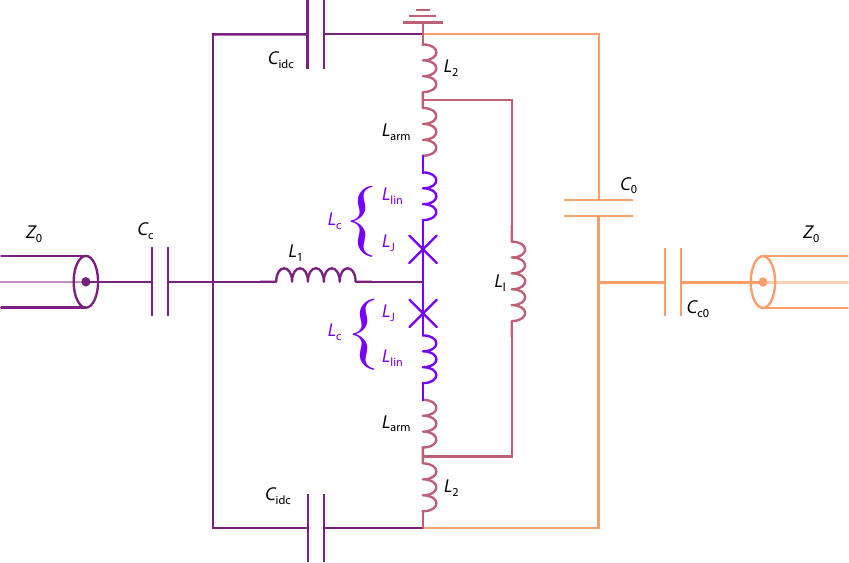}
	\titlecaption{The complete equivalent circuit of the device}{
		The low-frequency mode (right half) comprises a capacitor $C_0$ and the inductive branch, consisting of all inductors except for $L_1$. The mode is capacitively coupled to a feedline with characteristic impedance $Z_0$ by means of a coupling capacitance $C_\mathrm{c0}$.
		The high-frequency mode (left half) has two interdigitated capacitors $C_\mathrm{idc}$, the linear inductance $L_1$ and a parallel combination of two composite inductive branches. The mode is coupled to a $Z_0$-feedline by a coupling capacitance $C_\mathrm{c}$.
		The SQUID loop has the inductance $L_\mathrm{loop} = L_\mathrm{l} + 2L_\mathrm{arm}$, each constriction contributes $L_\mathrm{c} = L_\mathrm{J} + L_\mathrm{lin}$ and the two modes share the inductors $L_2$, which connect from top and bottom to the SQUID loop.
		More details and the values for all inductors can be found in the text and in Supplementary Table~\ref{tab:Table3}.
	}
	\label{fig:FigS4}
\end{figure*}
\begin{figure*}
	\includegraphics{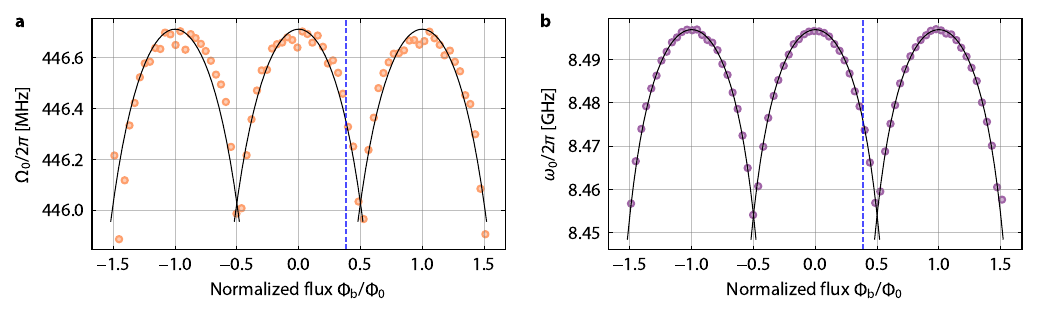}
	\titlecaption{Flux-tuning of LF and HF resonance frequencies}{
		\sublabel{a}~Resonance frequency of the LF mode $\Omega_0$ vs.~normalized bias flux $\Phi_\mathrm{b}/\Phi_0$. Total tuning range is $\qty{\sim 700}{\kilo\hertz}$. \sublabel{b}~Resonance frequency of the HF mode $\omega_0$ vs.~normalized bias flux $\Phi_\mathrm{b}/\Phi_0$. Total tuning range is $\qty{\sim 43}{\mega\hertz}$. In both panels symbols are data and lines are a simultaneous fit of both modes with a single parameter set. From the fit, we obtain the parameters $L = \qty{355.2}{\pico\henry}$, $L_0 = \qty{249.4}{\pico\henry}$, $L_\mathrm{loop} = \qty{21.5}{\pico\henry}$, $L_\mathrm{lin} = \qty{6.5}{\pico\henry}$ and $L_\mathrm{J0} = \qty{16.5}{\pico\henry}$. Blue dashed vertical lines indicate the operation point $\Phi_\mathrm{b}/\Phi_0 = 0.385$ for all photon-pressure experiments in this work.
	}
	\label{fig:FigS5}
\end{figure*}

To model the complete circuit including the constrictions, we separate the inductances into multiple parts, cf.~Supplementary Fig.~\ref{fig:FigS4}.
First, we take a part of the HF inductance as not contributing to the LF mode and label this part with~$L_1$.
Then we add two symmetric inductors~$L_2$, which contribute to both the HF and the LF circuit.
Note, however, that they contribute in series to the LF mode and in parallel to the HF mode.
The most complicated part is the SQUID loop including the cJJs.
In combination with the linear part of the constrictions, each loop part with a constriction has the linear inductance $L_\mathrm{lin} + L_\mathrm{arm} \approx L_\mathrm{lin} + 0.1 L_\mathrm{loop}$, and the total inductance
\begin{equation}
	L_\mathrm{p} = L_\mathrm{J} + L_\mathrm{lin} + 0.1 L_\mathrm{loop}.
\end{equation}
The large part of the loop without the constrictions has then the inductance $L_\mathrm{l} \approx 0.8 L_\mathrm{loop}$, and we will assume that this part does not contribute significantly to the HF mode inductance, since the HF currents will not significantly pass through that loop section.
Subsequently, we get for the total inductances of LF and HF mode, respectively,
\begin{eqnarray}
	L_\mathrm{LF} & = & 2L_2 + \frac{1.6 L_\mathrm{p}L_\mathrm{loop}}{2L_\mathrm{p} + 0.8 L_\mathrm{loop}} \\
	L_\mathrm{HF} & = & L_1 + \frac{L_\mathrm{p} + L_2}{2}.
\end{eqnarray}
Plugging these relations into the equations for the resonance frequencies
\begin{eqnarray}
	\omega_0 & = & \frac{1}{\sqrt{C_\mathrm{tot}L_\mathrm{HF}}} \\
	\Omega_0 & = & \frac{1}{\sqrt{C_\Sigma L_\mathrm{LF}}}
\end{eqnarray}
allows us to apply a combined, complex-valued fit function
\begin{equation}
	\tilde{\omega}_0 = \omega_0(\Phi)+ \i\Omega_0(\Phi)
\end{equation}
to both datasets of flux-dependence.
Supplementary Fig.~\ref{fig:FigS5} shows the resonance frequency of both modes as a function of bias flux $\Phi_\mathrm{b}$ and the corresponding simultaneous fit curves.
In practice we multiply the LF resonance $\Omega_0$ by a factor $19$ (both dataset and theory expression) for the fitting, in order to bring both resonances to the same frequency scale and to get identical weights for the fits from both modes.
\begin{table}[b]
	\titlecaption{Best-fit inductance parameters extracted from a simultaneous fit of the frequency flux-tuning curves} \\
	\begin{tabular}{ c   c   c   c   c   c   c   c   c }
		\toprule
		$L\,$[$\qty{}{\pico\henry}$] & $L_0\,$[$\qty{}{\pico\henry}$] & $L_1\,$[$\qty{}{\pico\henry}$] & $L_2\,$[$\qty{}{\pico\henry}$] & $L_\mathrm{lin}\,$[$\qty{}{\pico\henry}$] & $L_\mathrm{J0}\,$[$\qty{}{\pico\henry}$] & $I_0\,$[$\qty{}{\micro\ampere}$] & $L_\mathrm{loop}\,$[$\qty{}{\pico\henry}$] & $\beta_L\,$ \\ \midrule
		355.2 & 249.4 & 292.7 & 123 & 6.5 & 16.5 & 20 & 21.5 & 0.67 \\ \bottomrule
	\end{tabular}
	\label{tab:Table3}
\end{table}

For the fit, we keep $C_\mathrm{tot}$ and $C_\Sigma$ fixed from the calculations and we fit for $L, L_0, L_\mathrm{loop}, L_\mathrm{lin}$ and $L_\mathrm{J0}$ (in practice $I_0$), which together determine $\beta_L$ and the flux-dependence of $\omega_0$ and $\Omega_0$.
From the fit, we get $L = \qty{355.2}{\pico\henry}$, $L_0 = \qty{249.4}{\pico\henry}$, $L_\mathrm{loop} = \qty{21.5}{\pico\henry}$, $L_\mathrm{lin} = \qty{6.5}{\pico\henry}$ and $L_\mathrm{J0} = \qty{16.5}{\pico\henry}$, which corresponds to a critical constriction current $I_0 = \qty{20}{\micro\ampere}$.
The corresponding screening parameter is $\beta_L = 0.67$.
By inverting above relations, we can also get the values for $L_1 = \qty{292.7}{\pico\henry}$ and $L_2 = \qty{123}{\pico\henry}$.
Supplementary Table~\ref{tab:Table3} summarizes all the relevant parameters of this section, i.e., all inductances of the device as extracted from the flux-tuning fits.
The values for $L$ and $L_0$ differ from the original ones obtained from the measurements before constriction-cutting by $0.6\%$ and $-1.9\%$, respectively, which is an acceptable and non-significant deviation considering our typical fluctuations between different cooldowns.
Additionally the device has been re-wirebonded after the cJJ cutting, which can also modify the effective feedline impedance and contribute to small frequency shifts.
\subsection{The total circuit including constrictions -- Dissipative part}
The introduction of the constrictions not only changes the resonance frequencies of the circuits and makes them flux-tunable, also the circuit linewidths are impacted similarly.
The origin of the additional dissipation is likely thermal quasiparticles in the constrictions, since the cJJs have a reduced transition temperature $T_\mathrm{cc} < T_\mathrm{c}$ compared to the critical temperature of the bare niobium film $T_\mathrm{c} \approx \SI{9}{\kelvin}$ \cite{Uhl2024x, Uhl2024ax}.
All our measurements have been taken at a sample temperature $T_\mathrm{s} = \SI{4.2}{\kelvin}$ in a liquid helium bath.
At the flux sweetspots, the fit-curve linewidths we find with the constrictions are $\Gamma_\mathrm{ext} = 2\pi\times\qty{37}{\kilo\hertz}$, $\Gamma_\mathrm{int} = 2\pi\times\qty{503}{\kilo\hertz}$, $\kappa_\mathrm{ext} = 2\pi\times\qty{6.7}{\mega\hertz}$ and $\kappa_\mathrm{int} = 2\pi\times\qty{38}{\mega\hertz}$, for the corresponding $S_{11}$ data of one specific dataset cf.~main paper Fig.~1.
As expected, mainly the internal linewidths are affected by the constriction cutting.
The external linewidths stay roughly the same as before the cutting, and the differences once again might be related to interferences and impedance mismatches in the cabling; the change from before to after cutting could either be due to the circuits having different resonance frequencies after the cutting (interferences e.g.~are frequency-dependent), or because the chip has been taken out of the sample holder for the constriction fabrication and afterwards was re-wirebonded to the PCB, a process that is never completely identical when repeated.
Interestingly, the HF external linewidth $\kappa_\mathrm{ext}$ is now much closer to its theoretical value than before the cutting.
\begin{figure*}[b]
	\includegraphics{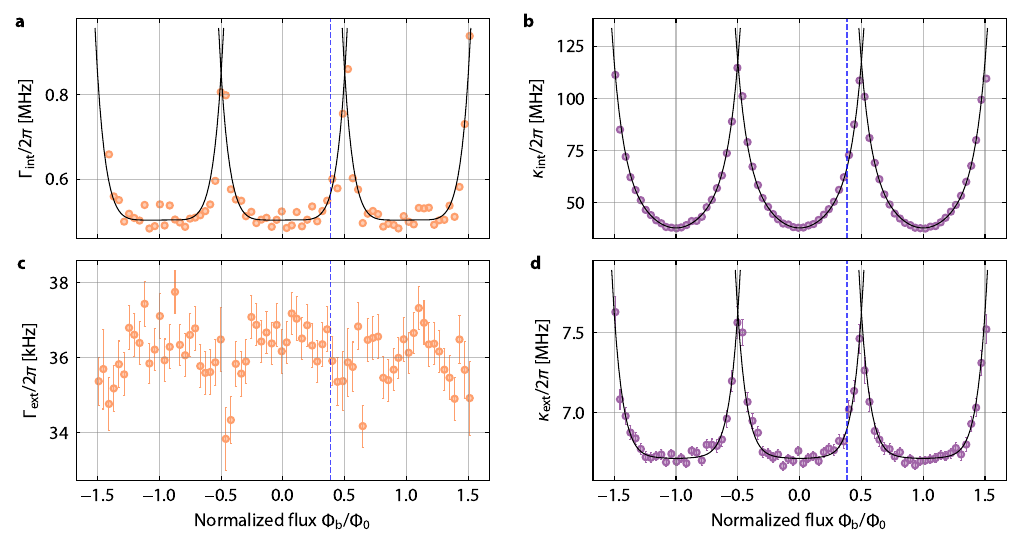}
	\titlecaption{Change of linewidths with flux through the SQUID}{
		The internal and external linewidths of HF and LF circuit as a function of normalized external flux $\Phi_\mathrm{b}/\Phi_0$ through the SQUID. \sublabel{a} and \sublabel{c} show internal and external linewidth of the LF circuit $\Gamma_\mathrm{int}$ and $\Gamma_\mathrm{ext}$, respectively. \sublabel{b} and \sublabel{d} show the internal and external linewidths of the HF mode $\kappa_\mathrm{int}$ and $\kappa_\mathrm{ext}$, respectively. Symbols are data extracted from fits of the corresponding resonances, lines are fits with Eqs.~(\ref{eqn:kappai_Phi}, \ref{eqn:kappae_Phi}, \ref{eqn:Gamma_Phi}). Error bars are the standard errors obtained from the fit. Blue dashed vertical line in each panel indicates the operation point $\Phi_\mathrm{b}/\Phi_0 = 0.385$ for all photon-pressure experiments in this work.
	}
	\label{fig:FigS6}
\end{figure*}
When a magnetic flux is applied to the SQUID, we observe that the linewidths modulate periodically with flux, similar to earlier observations in niobium Ne-FIB constriction circuits \cite{Uhl2024x, Uhl2024ax}.
The extracted flux-dependent linewidths $\Gamma_\mathrm{int}, \Gamma_\mathrm{ext}$ and $\kappa_\mathrm{int}, \kappa_\mathrm{ext}$ are shown as a function of $\Phi_\mathrm{b}/\Phi_0$ in Supplementary Fig.~\ref{fig:FigS6}.
Mainly the internal circuit linewidths are modulating, which is consistent with the initial change of predominantly the internal linewidth by the cJJ patterning and its interpretation based on thermal quasiparticles, which are now additionally increased by the screening current when the SQUID is flux biased.
The internal linewidth of the HF circuit increases from $\kappa_\mathrm{int}/2\pi \approx \qty{38}{\mega\hertz}$ at the flux-sweetspot to $\qty{\sim 118}{\mega\hertz}$ at half-integer flux values.
The external HF linewidth increases in the same flux range from $\kappa_\mathrm{ext}/2\pi = \qty{6.7}{\mega\hertz}$ at the sweetspot to $\qty{\sim 7.6}{\mega\hertz}$.
However, the seeming flux-tuning of $\kappa_\mathrm{ext}$ is most likely an artifact of an imperfect fit and not an actual flux-dependence of the external decay rate.
In Ref.~\cite{Rieger2023x} it was shown, that the smaller one of the two linewidths $\kappa_\mathrm{int}$ and $\kappa_\mathrm{ext}$, which here is $\kappa_\mathrm{ext}$, will have a considerable uncertainty in the case that $\kappa_\mathrm{int}/\kappa_\mathrm{ext}$ deviates significantly from 1.
This would mean, that easily $10\% - 20\%$ of what we obtain as $\kappa_\mathrm{ext}$ could actually be a part of $\kappa_\mathrm{int}$, which is misattributed to $\kappa_\mathrm{ext}$ in the fitting routine by the presence of e.g.~a small leakage signal in the directional coupler.
If this small part follows the same trend as $\kappa_\mathrm{int}$, i.e.~triples by flux, it can explain the seeming increase of $\kappa_\mathrm{ext}$ by $\qty{0.9}{\mega\hertz}$ with flux without the external decay rate actually being flux-dependent.
Alternatively, the interference between the actual reflection and the leakage signal itself and the small fraction of $\kappa_\mathrm{int}$, that is falsely attributed to $\kappa_\mathrm{ext}$, can be frequency-dependent, leading to similar consequences.
If $\kappa_\mathrm{ext}$ really was changing with flux in the SQUID, i.e., with the resonance frequency of the HF mode, we would expect to see a similar trend of increasing $\kappa_\mathrm{ext}$ with decreasing $\omega_0$, when the HF mode is frequency-shifted by a pump tone instead of by flux.
However, when we do so at the operation point during the PPIT (photon-pressure induced transparency) experiments, we find a nearly constant $\kappa_\mathrm{ext}$ with a gentle trend to even decrease instead of increase.
The fact that the fitting procedure for the PPIT data slightly deviates from the one used for the flux arc response data (e.g.~by cutting out the PPIT window, cf.~Supplementary Note~\ref{sec:Note7}) leads us to the conclusion that it is indeed an artifact, that depends on the exact fit protocol.
Note, that just based on the theoretical expression for the external decay rate $\kappa_\mathrm{ext} = \omega_0^2 Z_0 C_\mathrm{c}^2/C_\mathrm{tot}$, one would actually expect a decrease of $\kappa_\mathrm{ext}$ by ${\sim}1\%$ over the flux range from $0$ to $0.5\Phi_0$ instead of the seeming increase by ${\sim}12\%$, but such a small shift would be below our analysis accuracy.
Since in the LF circuit the cJJs are in parallel to a second junction-free inductance, the impact of their presence and their flux-tuning is smaller.
The internal LF linewidth increases from $\Gamma_\mathrm{int}/2\pi = \qty{503}{\kilo\hertz}$ at the sweetspot to $\Gamma_\mathrm{int}/2\pi \approx \qty{851}{\kilo\hertz}$ at the anti-sweetspot.
The external LF linewidth is almost constant and only slightly modulates around $\Gamma_\mathrm{ext}/2\pi \approx \qty{36}{\kilo\hertz} \pm \qty{2}{\kilo\hertz}$, which again is likely impacted by signal interferences and the fitting routine for the strongly undercoupled circuit with $\Gamma_\mathrm{ext}/\Gamma_\mathrm{int} \ll 1$.
We fit the internal linewidths and the external HF linewidth for a single flux arc with sixth-order even polynomials
\begin{eqnarray}
	\kappa_\mathrm{int}(\Phi_\mathrm{b}) & = & \kappa_\mathrm{int, ss} + \kappa_\mathrm{int, 2}\Phi_\mathrm{b}^2 + \kappa_\mathrm{int, 4}\Phi_\mathrm{b}^4 + \kappa_\mathrm{int, 6}\Phi_\mathrm{b}^6 \label{eqn:kappai_Phi}\\
	\kappa_\mathrm{ext}(\Phi_\mathrm{b}) & = & \kappa_\mathrm{ext, ss} + \kappa_\mathrm{ext, 2}\Phi_\mathrm{b}^2 + \kappa_\mathrm{ext, 4}\Phi_\mathrm{b}^4 + \kappa_\mathrm{ext, 6}\Phi_\mathrm{b}^6 \label{eqn:kappae_Phi}\\
	\Gamma_\mathrm{int}(\Phi_\mathrm{b}) & = & \Gamma_\mathrm{int, ss} + \Gamma_\mathrm{int, 2}\Phi_\mathrm{b}^2 + \Gamma_\mathrm{int, 4}\Phi_\mathrm{b}^4 + \Gamma_\mathrm{int, 6}\Phi_\mathrm{b}^6
	\label{eqn:Gamma_Phi}
\end{eqnarray}
where $\kappa_\mathrm{int, ss}, \kappa_\mathrm{ext, ss}, \Gamma_\mathrm{int, ss}$ are the sweetspot values and all $\kappa_{\mathrm{int}, j}, \kappa_{\mathrm{ext}, j}, \Gamma_{\mathrm{int}, j}$ for even $j > 1$ are fit parameters.
The resulting fit lines (and their replicas at $\pm 1\Phi_0$) are shown in Supplementary Fig.~\ref{fig:FigS6}.
The level of agreement with the data, in particular with that of the HF mode, is satisfactory.
For $\Gamma_\mathrm{ext}$ we did not add a fit curve, since there is too little actual modulation and since we nowhere need it.
At the photon-pressure operation point of this work $\Phi_\mathrm{b}/\Phi_0 \approx 0.385$, marked in Supplementary Fig.~\ref{fig:FigS6} with dashed vertical lines, we find for the four linewidths $\kappa_\mathrm{int}/2\pi = \qty{67}{\mega\hertz}$, $\kappa_\mathrm{ext}/2\pi = \qty{7}{\mega\hertz}$, $\Gamma_\mathrm{int}/2\pi = \qty{565}{\kilo\hertz}$ and $\Gamma_\mathrm{ext}/2\pi = \qty{36}{\kilo\hertz}$.
\vspace{-2mm}
\subsection{Single-photon coupling rates}
\vspace{-2mm}
As a last step in the estimation of the device parameters, we calculate the two single-photon coupling rates
\begin{eqnarray}
	g_{0\omega} & = & -\frac{\partial\omega_0}{\partial\Phi_\mathrm{b}}\Phi_\mathrm{zpf} \\
	g_{0\kappa} & = & -\frac{\partial\kappa_0}{\partial\Phi_\mathrm{b}}\Phi_\mathrm{zpf}
\end{eqnarray}
where $\Phi_\mathrm{zpf}$ is the zero-point flux coupled into the SQUID loop by the LF resonator.
The two derivatives (responsivities) we obtain numerically from the fits of the $\omega_0(\Phi_\mathrm{b})$ and $\kappa_0(\Phi_\mathrm{b})$ flux arcs, cf.~main paper Fig.~2.
The zero-point flux on the other hand we calculate through the zero-point current of the LF mode
\begin{equation}
	I_\mathrm{zpf} = \sqrt{\frac{\hbar\Omega_0}{2L_\mathrm{LF}}} \approx \qty{23.9}{\nano\ampere}.
\end{equation}
Next, we use the flux equation for an asymmetric SQUID, biased with a current $I_\mathrm{zpf}$, which flows along the LF inductance,
\begin{equation}
	\Phi = \Phi_\mathrm{b} + \tilde{L}_\mathrm{loop}J - \alpha_L \tilde{L}_\mathrm{loop}\frac{I_\mathrm{zpf}}{2}
\end{equation}
where the inductive asymmetry parameter is
\begin{eqnarray}
	\alpha_L & = & \frac{2L_\mathrm{lin} + 2L_\mathrm{arm} - \left(L_\mathrm{loop} - 2L_\mathrm{arm} \right)}{\tilde{L}_\mathrm{loop}} \\
	& = & \frac{ 4(L_\mathrm{lin} + L_\mathrm{arm}) - \tilde{L}_\mathrm{loop} }{\tilde{L}_\mathrm{loop}},
\end{eqnarray}
i.e., in our case $\alpha_L \approx -0.058$.
Note, that the bias current $I_\mathrm{zpf}$ flows unusually through the three-terminal SQUID considered here: one SQUID arm for this current contains no cJJ, while the second SQUID arm contains both cJJs.
\begin{figure*}
	\includegraphics{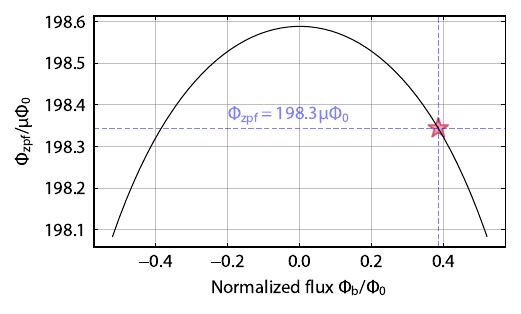}
	\titlecaption{Zero-point-fluctuation amplitude of effective LF flux coupling into the SQUID}{
		Line shows the zero-point-flux fluctuation amplitude of the LF mode calculated via Eq.~(\ref{eqn:Phizpf}). It is only weakly dependent on $\Phi_\mathrm{b}$ and at the operation point of this work $\Phi_\mathrm{b}/\Phi_0 = 0.385$ (marked with dashed lines and a star symbol), it is $\Phi_\mathrm{zpf} \approx 198.3\,$\textmu$\Phi_0$.
	}
	\label{fig:FigS7}
\end{figure*}
\begin{figure*}
	\includegraphics{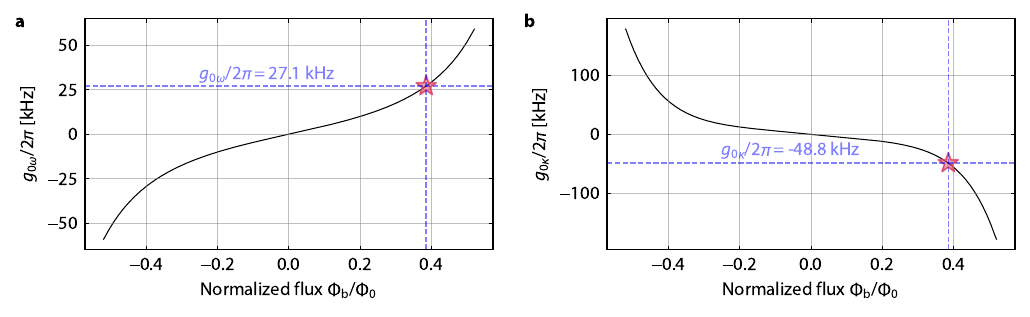}
	\titlecaption{Dispersive and dissipative single-photon coupling rates}{
		From the flux-tuning fits of $\omega_0$ and $\kappa_0$, we numerically calculate their derivatives and by multiplication with $\Phi_\mathrm{zpf}$ we obtain the photon-pressure single-photon coupling rates $g_{0\omega}$ and $g_{0\kappa}$, respectively. The former is the dispersive coupling rate and shown in panel \sublabel{a}, the latter is the dissipative coupling rate and displayed in panel \sublabel{b}. At the operation point for this work $\Phi_\mathrm{b}/\Phi_0 = 0.385$ (marked by dashed lines and star symbols), the rates we obtain are $g_{0\omega}/2\pi = \qty{27.1}{\kilo\hertz}$ and $g_{0\kappa}/2\pi = \qty{-48.8}{\kilo\hertz}$.
	}
	\label{fig:FigS8}
\end{figure*}

Using the expression for the circulating current with a bias current $I_\mathrm{zpf}$
\begin{equation}
	J = -I_0\sin{\left(\pi\frac{\Phi}{\Phi_0} \right)} + \frac{I_\mathrm{zpf}}{2}
\end{equation}
we get in total
\begin{equation}
	\Phi = \Phi_\mathrm{b} + \left(1 - \alpha_L\right) \tilde{L}_\mathrm{loop}\frac{I_\mathrm{zpf}}{2} - \tilde{L}_\mathrm{loop}I_0\sin{\left(\pi\frac{\Phi}{\Phi_0} \right)}.
\end{equation}
Hence, the zero-point current is equivalent to an additional external flux
\begin{eqnarray}
	\Phi_\mathrm{zpf} & = & \tilde{L}_\mathrm{loop}\left(1 - \alpha_L\right)\frac{I_\mathrm{zpf}}{2} \\
	& = & L_\mathrm{l}I_\mathrm{zpf} \label{eqn:Phizpf}\\
	& \approx & 198.3\,\textrm{\textmu} \Phi_0.
\end{eqnarray}
In general $\Phi_\mathrm{zpf}$ is also flux-dependent due to $I_\mathrm{zpf}$ being dependent on $\Omega_0$ and $L_\mathrm{LF}$, cf.~Supplementary Fig.~\ref{fig:FigS7}, but in our case it is nearly constant since the LF mode is so weakly dependent on flux.
Finally, we plug everything into the equations for $g_{0\omega}$ and $g_{0\kappa}$ and show both parameters as a function of $\Phi_\mathrm{b}$ in Supplementary Fig.~\ref{fig:FigS8}.
Again, the relevant operation point $\Phi_\mathrm{b}/\Phi_0 = 0.385$ is marked with vertical dashed lines and star symbols, and the corresponding single-photon coupling rates are $g_{0\omega} = 2\pi\times\qty{27.1}{\kilo\hertz}$ and $g_{0\kappa} = 2\pi\times\qty{-48.8}{\kilo\hertz}$.
In general it is important to distinguish between internal-dissipative and external-dissipative coupling, since they can lead to qualitatively and quantitatively different effects.
From the fits in Supplementary~Fig.~\ref{fig:FigS6} one would expect now that $g_{0\kappa_\mathrm{ext}}/g_{0\kappa_\mathrm{int}} \approx 10^{-2}$ or $g_{0\kappa_\mathrm{ext}} \approx 2\pi\times\qty{-0.55}{\kilo\hertz}$, which despite its smallness would not be negligible for large pump-cavity detunings, cf.~Ref.~\cite{Primo2023x} and Supplementary Note~\ref{sec:Note6D}.
However, based on our above conclusion, that the seeming flux-tuning of $\kappa_\mathrm{ext}$ is an artifact, we estimate the actual external-dissipative coupling rate to be only $g_{\kappa_\mathrm{ext}} = 2\pi\times\qty{35.3}{\hertz}$, i.e., $g_{0\kappa_\mathrm{ext}}/g_{0\kappa_\mathrm{int}} = -7.2\times 10^{-4} $, using the equation $\kappa_\mathrm{ext} = \omega_0^2 Z_0 C_\mathrm{c}^2/C_\mathrm{tot}$, from which we find
\begin{align}
	g_{0\kappa_\mathrm{ext}} & = \frac{\partial\kappa_\mathrm{ext}}{\partial\omega_0}g_{0\omega} \\
	& = \frac{2\omega_0 Z_0 C_\mathrm{c}^2}{C_\mathrm{tot}}g_{0\omega}. \\
	& = 1.3\times 10^{-3}g_{0\omega}.
\end{align}
Such a small $g_{0\kappa_\mathrm{ext}}$ will not have measurable consequences within our experimental accuracy and can be safely neglected.
The approach to completely neglect $g_{0\kappa_\mathrm{ext}}$ is backed up by our results being in good agreement with $g_{0\kappa} = g_{0\kappa_\mathrm{int}}$ and $g_{0\kappa_\mathrm{ext}} = 0$, in particular for large detunings between pump and HF mode and low pump photon numbers, where the impact of an external-dissipative coupling would be largest.
For an additional discussion on the potential impact of an external-dissipative contribution to the results of this manuscript, see Supplementary Note~\ref{sec:Note6D}.
\section{Supplementary Note IV: Theoretical response of the individual circuits}
\label{sec:Note4}
\vspace{-2mm}
\subsection{The low-frequency circuit -- Complete equation of motion}
\vspace{-2mm}
We model the dynamics of the low-frequency resonator in the low-power regime with the equation of motion (EOM)
\begin{equation}
	\dot{\hat{b}} = \i\left(\Omega_0 + \mathcal{K}_\mathrm{LF}\hat{b}^\dagger\hat{b}\right)\hat{b} - \frac{\Gamma_0}{2}\hat{b} + \i\sqrt{\Gamma_\mathrm{ext}}\hat{b}_\mathrm{in}
\end{equation}
where $\hat{b}$ is the annihilation operator for quanta in the LF mode, $\Omega_0$ is the LF resonance frequency, $\mathcal{K}_\mathrm{LF}$ is the LF mode Kerr anharmonicity, $\Gamma_0$ is the total LF damping rate, $\Gamma_\mathrm{ext}$ is the external damping rate and $\hat{b}_\mathrm{in}$ is the input field.
Note that in all equations of this work, which are relevant for the analysis and interpretation of the results, we drop the stochastic Langevin input term describing the coupling of the circuits to their thermal baths, since those are not relevant to first order for the experiments we perform.

\vspace{-2mm}
\subsection{The low-frequency circuit -- Single-tone response in the linear regime}
\vspace{-2mm}
For the vast majority of cases in this work, except for a few data points in the dynamical backaction analysis (cf.~Supplementary Notes~\ref{sec:Note7C} and \ref{sec:Note8B}), the LF Kerr anharmonicity is negligible and we will usually work with $\mathcal{K}_\mathrm{LF} = 0$.
We can solve the remaining EOM by the method of Fourier transform and obtain
\begin{equation}
	\hat{b}(\Omega) = \frac{\i\sqrt{\Gamma_\mathrm{ext}}}{\frac{\Gamma_0}{2} + \i(\Omega - \Omega_0)}\hat{b}_\mathrm{in}(\Omega).
\end{equation}
We note that in most instances, we will omit the explicit frequency-dependence and use for simplicity the same operators/variables for frequency-domain and time-domain.
Which one it is in each instance, should be clear from the context.
For the output field we have the input-output relation
\begin{equation}
	\hat{b}_\mathrm{out} = \hat{b}_\mathrm{in} + \i\sqrt{\Gamma_\mathrm{ext}}\hat{b}
\end{equation}
which leads to the reflection coefficient
\begin{eqnarray}
	S_{11} & = & \frac{\langle\hat{b}_\mathrm{out}\rangle}{\langle\hat{b}_\mathrm{in}\rangle} \\
	& = & 1 - \frac{\Gamma_\mathrm{ext}}{\frac{\Gamma_0}{2} + \i(\Omega - \Omega_0)}.
	\label{eqn:S11_LF}
\end{eqnarray}
\vspace{-2mm}
\subsection{The low-frequency circuit -- Single-tone response in the nonlinear regime}
\vspace{-2mm}
As mentioned before, this case is only relevant for few LF $S_{11}$ traces during the experiment on dynamical backaction and close to the instability, cf.~Supplementary Note~\ref{sec:Note8B}.
Since $\mathcal{K}_\mathrm{LF}$ is very small, we can treat the situation completely classically and set $\hat{b}_\mathrm{in} = \beta_\mathrm{in}\e^{\i\phi_\beta}$ and $\hat{b} = \beta$ to get to
\begin{equation}
	\dot{\beta} = \i\left(\Omega_0 + \mathcal{K}_\mathrm{LF}|\beta|^2\right)\beta - \frac{\Gamma_0}{2}\beta + \i\sqrt{\Gamma_\mathrm{ext}}\beta_\mathrm{in}\e^{\i\phi_\beta}
\end{equation}
which is normalized such that $|\beta|^2 = n_\beta$ is the probe photon number in the LF mode and $|\beta_\mathrm{in}|^2 = P_\mathrm{\beta}/\hbar\Omega = n_{\mathrm{in}, \beta}$ is the LF input photon flux with on-chip LF probe power $P_\beta$.
Both $\beta$ and $\beta_\mathrm{in}$ shall be real-valued and the phase lag between them is encoded in $\phi_\beta$.
Using again Fourier transformation and the detuning between LF probe and LF resonance $\Delta_\beta = \Omega - \Omega_0$, we find
\begin{equation}
	\i\left[\Delta_\beta - \mathcal{K}_\mathrm{LF}n_\beta\right]\beta + \frac{\Gamma_0}{2}\beta = \i\sqrt{\Gamma_\mathrm{ext}}\beta_\mathrm{in}e^{\i\phi_\beta}
\end{equation}
or magnitude-squared
\begin{equation}
	\left|\Delta_\beta - \mathcal{K}_\mathrm{LF}n_\beta\right|^2 n_\beta + \frac{\Gamma_0^2}{4}n_\beta = \Gamma_\mathrm{ext}n_{\mathrm{in}, \beta}.
	\label{eqn:charac_poly_beta}
\end{equation}
The latter is a third-order polynomial and can be solved numerically to find $n_\beta$ or $\beta = \sqrt{n_\beta}$, and all we need in addition to find the reflection
\begin{equation}
	S_{11} = 1 + \i\sqrt{\Gamma_\mathrm{ext}}\frac{\beta}{\beta_\mathrm{in}}\e^{-\i\phi_\beta}
\end{equation}
is the phase $\phi_\beta$, which can be obtained via
\begin{equation}
	\phi_\beta = \atanii{\left( - \frac{\Gamma_0}{2}, \Delta_\beta - \mathcal{K}_\mathrm{LF}n_\beta \right)}.
	\label{eqn:phi_beta}
\end{equation}
\vspace{-2mm}
\subsection{The high-frequency circuit -- Complete equation of motion}
\vspace{-2mm}
The EOM for the high-frequency circuit is similar to the one of the LF circuit, but in addition to the HF Kerr nonlinearity $\mathcal{K}$ induced by the nano-constrictions we include a nonlinear damping up to third order using $\kappa_1$, $\kappa_2$ amd $\kappa_3$ and get \cite{Gely2023x, Uhl2024x}
\begin{equation}
	\dot{\hat{a}} = \i\left(\omega_0 + \mathcal{K}\hat{a}^\dagger \hat{a}\right)\hat{a} - \frac{\kappa_0 + \kappa_1\hat{a}^\dagger \hat{a} + \kappa_2\left(\hat{a}^\dagger \hat{a}\right)^2 + \kappa_3\left(\hat{a}^\dagger \hat{a}\right)^3}{2}\hat{a} + \i\sqrt{\kappa_\mathrm{ext}}\hat{a}_\mathrm{in}
\end{equation}
where $\omega_0$ is the low-excitation resonance frequency, $\kappa_0$ is the total low-excitation damping rate, $\kappa_\mathrm{ext}$ is the external damping rate and $\hat{a}_\mathrm{in}$ is the input field.
\vspace{-2mm}
\subsection{The high-frequency circuit -- Single-tone response in the linear regime}
\vspace{-2mm}
In the low-power single-tone regime, we set $\mathcal{K} = \kappa_1 = \kappa_2 = \kappa_3 = 0$ and get
\begin{equation}
	\dot{\hat{a}} = \i\omega_0 \hat{a} - \frac{\kappa_0}{2}\hat{a} + \i\sqrt{\kappa_\mathrm{ext}}\hat{a}_\mathrm{in}
\end{equation}
which -- in analogy to the low-frequency circuit -- can be solved by Fourier transform to get
\begin{equation}
	\hat{a} = \frac{\i\sqrt{\kappa_\mathrm{ext}}}{\frac{\kappa_0}{2} + \i\left(\omega - \omega_0\right)}\hat{a}_\mathrm{in}
\end{equation}
and to finally obtain the reflection
\begin{equation}
	S_{11} = 1 - \frac{\kappa_\mathrm{ext}}{\frac{\kappa_0}{2} + \i\left(\omega - \omega_0 \right)}.
\end{equation}
\vspace{-2mm}
\subsection{The high-frequency circuit -- The linearized two-tone response}
\vspace{-2mm}
In the PPIT experiments we apply a strong pump tone to the HF cavity and additionally a weak, near-resonant probe tone.
For this case, we linearize the system by assuming input and intracavity field to consist of a classical, coherent amplitude and a fluctuation field (e.g.~a weak probe tone), i.e.,
\begin{eqnarray}
	\hat{a}_\mathrm{in} & = & \left[\alpha_\mathrm{in}\e^{\i\phi_\mathrm{p}} + \hat{c}_\mathrm{in}\right]\e^{\i\omega_\mathrm{p}t} \\
	\hat{a} & = & \left[\alpha_0 + \hat{c}\right]\e^{\i\omega_\mathrm{p}t}
\end{eqnarray}
with real-valued and time-independent $\alpha_\mathrm{in}$ and $\alpha_0$, complex-valued and time-dependent $\hat{c}_\mathrm{in}$ and $\hat{c}$, and with the pump frequency $\omega_\mathrm{p}$.
A possible (frequency-dependent) phase lag between $\alpha_\mathrm{in}$ and $\alpha_0$ is included via $\phi_\mathrm{p}$.
Plugging the ansatz into the EOM and omitting afterwards the $\e^{\i\omega_\mathrm{p}t}$ factor (rotating frame) as well as nonlinear terms in $\hat{c}$, we find
\begin{eqnarray}
	\i\omega_\mathrm{p}\alpha_0 + \i\omega_\mathrm{p}\hat{c} + \dot{\hat{c}} & = & \i\left(\omega_0 + \mathcal{K}\alpha_0^2\right)\alpha_0 - \frac{\kappa_0 + \kappa_1\alpha_0^2 + \kappa_2\alpha_0^4 + \kappa_3\alpha_0^6}{2}\alpha_0 + \i\sqrt{\kappa_\mathrm{ext}}\alpha_\mathrm{in}\e^{\i\phi_\mathrm{p}} \\
	& & +~\i\left( \omega_0 + 2\mathcal{K}\alpha_0^2\right)\hat{c} - \frac{\kappa_0 + 2\kappa_1\alpha_0^2 + 3\kappa_3\alpha_0^4 + 4\kappa_2\alpha_0^6}{2}\hat{c} + \i\sqrt{\kappa_\mathrm{ext}}\hat{c}_\mathrm{in} \\
	& & +~\i\mathcal{K}\alpha_0^2\hat{c}^\dagger - \frac{\kappa_1\alpha_0^2 + 2\kappa_2\alpha_0^4 + 3\kappa_3\alpha_0^6}{2}\hat{c}^\dagger.
\end{eqnarray}
The first line on the right hand side (RHS) describes the steady-state terms of the pumped circuit with respect to the pump, the second line are the dynamic terms for $\hat{c}$ and the third line corresponds to amplification, squeezing and four-wave-mixing.
We split the equation into steady-state constant parts
\begin{equation}
	\i\left[\Delta_\mathrm{p} -\mathcal{K}n_\mathrm{c}\right]\alpha_0 + \frac{\kappa_\mathrm{eff}}{2}\alpha_0 = \i\sqrt{\kappa_\mathrm{ext}}\alpha_\mathrm{in}\e^{\i\phi_\mathrm{p}}
	\label{eqn:SteadyState_alpha0}
\end{equation}
and dynamical parts
\begin{equation}
	\dot{\hat{c}} = -\i\left[\Delta_\mathrm{p} - 2\mathcal{K}n_\mathrm{c}\right]\hat{c} - \frac{\kappa_0'}{2}\hat{c} + \left[\i\mathcal{K} - \frac{\kappa_\mathrm{nl}}{2}\right]n_\mathrm{c}\hat{c}^\dagger + \i\sqrt{\kappa_\mathrm{ext}}\hat{c}_\mathrm{in}
	\label{eqn:EOM_TT_c}
\end{equation}
using the detuning between the unpumped HF cavity and the pump $\Delta_\mathrm{p} = \omega_\mathrm{p} - \omega_0$, the intracavity pump photon number $n_\mathrm{c} = \alpha_0^2$, and the linewidths
\begin{eqnarray}
	\kappa_0' & = & \kappa_0 + 2\kappa_1n_\mathrm{c} + 3\kappa_2n_\mathrm{c}^2 + 4\kappa_3n_\mathrm{c}^3 \\
	\kappa_\mathrm{eff} & = & \kappa_0 + \kappa_1n_\mathrm{c} + \kappa_2n_\mathrm{c}^2 + \kappa_3n_\mathrm{c}^3 \\
	\kappa_\mathrm{nl} & = & \kappa_1 + 2\kappa_2n_\mathrm{c} + 3\kappa_3n_\mathrm{c}^2.
\end{eqnarray}
Although not relevant for the further theory, note that $\kappa_0' = \kappa_\mathrm{eff} + \kappa_\mathrm{nl}n_\mathrm{c}$.
We multiply Eq.~(\ref{eqn:SteadyState_alpha0}) by its complex conjugate and find the characteristic polynomial of a strongly driven, weakly-nonlinear Kerr oscillator with up to third order nonlinear damping
\begin{equation}
	\left[\left(\Delta_\mathrm{p} - \mathcal{K}n_\mathrm{c} \right)^2 + \frac{\kappa_\mathrm{eff}^2}{4}\right]n_\mathrm{c} - \kappa_\mathrm{ext}n_\mathrm{in} = 0
	\label{eqn:charac_poly}
\end{equation}
where $n_\mathrm{in} = \alpha_\mathrm{in}^2$ is the input flux of pump photons.
This seventh-order polynomial, which in principle can be solved numerically to find the stable solution(s) for the photon number $n_\mathrm{c}$, is also given in its expanded form in the main manuscript Methods section as~Eq.~(27).
To get the complete solution $\alpha_0\e^{\i\phi_\mathrm{p}}$, we can also calculate the phase between pump and intracircuit field as
\begin{equation}
	\phi_\mathrm{p} = \mathrm{atan2}{\left( -\frac{\kappa_\mathrm{eff}}{2}, \Delta_\mathrm{p} - \mathcal{K}n_\mathrm{c}\right)},
\end{equation}
although this phase is not important for the probe response as long as pump and probe are at considerably different frequencies.
To calculate the probe response, we additionally need the EOM for $\hat{c}^\dagger$, which is
\begin{equation}
	\dot{\hat{c}}^\dagger = \i\left[\Delta_\mathrm{p} - 2\mathcal{K}n_\mathrm{c}\right]\hat{c}^\dagger - \frac{\kappa_0'}{2}\hat{c}^\dagger - \left[\i\mathcal{K} + \frac{\kappa_\mathrm{nl}}{2}\right]n_\mathrm{c}\hat{c} - \i\sqrt{\kappa_\mathrm{ext}}\hat{c}_\mathrm{in}^\dagger.
	\label{eqn:EOM_TT_cdag}
\end{equation}
We Fourier transform both Eq.~(\ref{eqn:EOM_TT_c}) and Eq.~(\ref{eqn:EOM_TT_cdag}) and get
\begin{eqnarray}
	\frac{\hat{c}(\Omega)}{\chi_\mathrm{p}(\Omega)} & = & \left[\i\mathcal{K} - \frac{\kappa_\mathrm{nl}}{2} \right]n_\mathrm{c}\hat{c}^\dagger(-\Omega) + \i\sqrt{\kappa_\mathrm{ext}}\hat{c}_\mathrm{in}(\Omega) \\
	\frac{\hat{c}^\dagger(-\Omega)}{\chi_\mathrm{p}^*(-\Omega)} & = & -\left[\i\mathcal{K} + \frac{\kappa_\mathrm{nl}}{2} \right]n_\mathrm{c}\hat{c}(\Omega) - \i\sqrt{\kappa_\mathrm{ext}}\hat{c}_\mathrm{in}^\dagger(-\Omega)
\end{eqnarray}
where
\begin{equation}
	\chi_\mathrm{p}(\Omega) = \frac{1}{\frac{\kappa_0'}{2} + \i\left(\Delta_\mathrm{p} - 2\mathcal{K}n_\mathrm{c} + \Omega\right)},
\end{equation}
and $\Omega = \omega - \omega_\mathrm{p}$ the frequency relative to the pump.
We drop the input term in the equation for $\hat{c}^\dagger$ at this point, since we only detect the Fourier component of $\hat{c}$ at $\Omega$ with the VNA.
Next, we define $\overline{\chi}_\mathrm{p} = \chi_\mathrm{p}^*(-\Omega)$, we omit the explicit statement of $\left(\pm\Omega\right)$ and combine the two equations to
\begin{equation}
	\frac{\hat{c}}{\chi_\mathrm{g}} = \i\sqrt{\kappa_\mathrm{ext}}\hat{c}_\mathrm{in}
\end{equation}
with
\begin{equation}
	\chi_\mathrm{g} = \frac{\chi_\mathrm{p}}{1 - \left[\mathcal{K}^2 + \frac{ \kappa_\mathrm{nl}^2}{4} \right]n_\mathrm{c}^2\chi_\mathrm{p}\overline{\chi}_\mathrm{p}}.
\end{equation}
The probe reflection follows as
\begin{equation}
	S_{11} = 1 - \kappa_\mathrm{ext}\chi_\mathrm{g}.
	\label{eqn:S11_HF}
\end{equation}

\vspace{-2mm}
\subsection{The pumped Kerr eigenmodes}
\vspace{-2mm}
\label{sec:Note4_E}
We can find the solutions to the pumped susceptibility $\chi_\mathrm{g}$ by the condition $\chi_\mathrm{g}^{-1} = 0$, which is equivalent to
\begin{equation}
	1 - \left[\mathcal{K}^2 + \frac{\kappa_\mathrm{nl}^2}{4}\right]n_\mathrm{c}^2\chi_\mathrm{p}\overline{\chi}_\mathrm{p} = 0
\end{equation}
and which is solved by the complex frequencies
\begin{equation}
	\tilde{\omega}_{1/2} = \omega_\mathrm{p} + \i\frac{\kappa_0'}{2} \pm \sqrt{\left(\Delta_\mathrm{p} - \mathcal{K}n_\mathrm{c} \right)\left(\Delta_\mathrm{p} - 3\mathcal{K}n_\mathrm{c} \right) - \frac{\kappa_\mathrm{nl}^2 n_\mathrm{c}^2}{4}}.
\end{equation}
The real parts of this solution correspond to the resonance frequencies and the imaginary parts to half the linewidths.
We always work in a regime where only one of the two modes plays a significant role (moderate pumping with a far-red-detuned pump tone), whose solution is
\begin{equation}
	\tilde{\omega}_{0} = \omega_\mathrm{p} + \i\frac{\kappa_0'}{2} + \sqrt{\left(\Delta_\mathrm{p} - \mathcal{K}n_\mathrm{c} \right)\left(\Delta_\mathrm{p} - 3\mathcal{K}n_\mathrm{c} \right) - \frac{\kappa_\mathrm{nl}^2 n_\mathrm{c}^2}{4}}.
\end{equation}
This means the new probe-mode is broadened to a linewidth $\kappa_0' = \kappa_0 + 2\kappa_1n_\mathrm{c} + 3\kappa_2 n_\mathrm{c}^2 + 4\kappa_3 n_\mathrm{c}^3$ and ac-Stark-shifted by
\begin{equation}
	\delta\omega_0 = \Delta_\mathrm{p} + \sqrt{\left(\Delta_\mathrm{p} - \mathcal{K}n_\mathrm{c} \right)\left(\Delta_\mathrm{p} - 3\mathcal{K}n_\mathrm{c} \right) - \frac{\kappa_\mathrm{nl}^2 n_\mathrm{c}^2}{4}}.
	\label{eqn:ac_Stark}
\end{equation}
We call the pumped resonance frequency $\omega_0' = \omega_0 + \delta\omega_0$ and get in good approximation for all our experiments
\begin{equation}
	\chi_\mathrm{g} \approx \frac{1}{\frac{\kappa_0'}{2} + \i(\omega - \omega_0')}.
\end{equation}
Note that in general one would also have to consider the probe gain $\mathcal{G}$ due to parametric amplification \cite{Rodrigues2022x, Rodrigues2024x}, but we always work in a regime where $\mathcal{G} \approx 1$.
To see the latter relation and $\mathcal{G} \approx 1$, we write $\chi_\mathrm{g}$ as
\begin{equation}
	\chi_\mathrm{g} = \frac{\chi_1\chi_2}{\overline{\chi}_\mathrm{p}}
\end{equation}
with
\begin{equation}
	\chi_1 = \frac{1}{\frac{\kappa_0'}{2} + \i(\omega - \omega_0')}, ~~~~~ \chi_2 = \frac{1}{\frac{\kappa_0'}{2} + \i(\omega - \omega_1')}
\end{equation}
where
\begin{eqnarray}
	\omega_0' & = & \omega_\mathrm{p} + \sqrt{\left(\Delta_\mathrm{p} - \mathcal{K}n_\mathrm{c} \right)\left(\Delta_\mathrm{p} - 3\mathcal{K}n_\mathrm{c} \right) - \frac{\kappa_\mathrm{nl}^2 n_\mathrm{c}^2}{4}} \label{eqn:pumped_freq}\\
	\omega_1' & = & \omega_\mathrm{p} - \sqrt{\left(\Delta_\mathrm{p} - \mathcal{K}n_\mathrm{c} \right)\left(\Delta_\mathrm{p} - 3\mathcal{K}n_\mathrm{c} \right) - \frac{\kappa_\mathrm{nl}^2 n_\mathrm{c}^2}{4}}.
\end{eqnarray}
It can also easily be confirmed, that
\begin{equation}
	\chi_1\chi_2 = \frac{\chi_1 + \chi_2}{\kappa_0' + 2\i\Omega}.
\end{equation}
With all this, we can write
\begin{equation}
	\chi_\mathrm{g} = \mathcal{G}\left(\chi_1 + \chi_2\right)
\end{equation}
with the frequency-dependent gain
\begin{eqnarray}
	\mathcal{G} & = & \frac{\frac{\kappa_0'}{2} - \i\left(\Delta_\mathrm{p} - 2\mathcal{K}n_\mathrm{c} - \Omega \right)}{\kappa_0' + 2 \i\Omega}.
\end{eqnarray}
For the parameter range in the experiment presented in this paper (i.e.~a red-sideband pump with $\Delta_\mathrm{p} \approx -\Omega_0 + 2\mathcal{K}n_\mathrm{c}$ and probe frequencies around $\omega_0'$, equivalent to $\Omega \approx \Omega_0$), we find
\begin{eqnarray}
	\mathcal{G} & \approx & \frac{\frac{\kappa_0'}{2} +2\i\Omega_0}{\kappa_0' + 2 \i\Omega_0} \nonumber \\
	& \approx & 1 \\
	\chi_2 & \approx & 0
\end{eqnarray}
and hence
\begin{equation}
	\mathcal{G}\left(\chi_1 + \chi_2\right) \approx \chi_1.
\end{equation}
\vspace{-2mm}
\subsection{Intracircuit pump photon number}
\label{sec:Note4F}
\vspace{-2mm}
Due to the unknown and strongly frequency-dependent attenuation in our setup, originating from a pump-span covering more than $\qty{300}{\mega\hertz}$, we extract $n_\mathrm{c}$ from the HF cavity frequency shift due to its Kerr anharmonicity in combination with the photon-pressure coupling strength.
In the preceding Sec.~\ref{sec:Note4_E}, we derived the expression Eq.~(\ref{eqn:ac_Stark}) for the frequency shift
\begin{equation}
	\delta\omega_0 = \Delta_\mathrm{p} + \sqrt{\left(\Delta_\mathrm{p} - \mathcal{K}n_\mathrm{c} \right)\left(\Delta_\mathrm{p} - 3\mathcal{K}n_\mathrm{c} \right) - \frac{\kappa_\mathrm{nl}^2 n_\mathrm{c}^2}{4}}.
\end{equation}
However, in contrast to the case of only first order nonlinear damping \cite{Uhl2024x, Uhl2024ax} the quantity $\kappa_\mathrm{nl}n_\mathrm{c} = \kappa_1 n_\mathrm{c} + 2\kappa_2 n_\mathrm{c}^2 + 3\kappa_3 n_\mathrm{c}^3$ is not experimentally accessible with ease.
What we do have access to though is
\begin{equation}
	2\kappa_1 n_\mathrm{c} + 3\kappa_2 n_\mathrm{c}^2 + 4\kappa_3 n_\mathrm{c}^3 = \kappa_0' - \kappa_0 \geq \kappa_\mathrm{nl}n_\mathrm{c} ~~~~~ \mathrm{for}~~~\kappa_1, \kappa_2, \kappa_3 \geq 0
\end{equation}
and so if we write the resonance frequency shift as
\begin{equation}
	\delta\omega_0 = \Delta_\mathrm{p} + |\Delta_\mathrm{p}|\sqrt{1 - \frac{4\mathcal{K}n_\mathrm{c}}{\Delta_\mathrm{p}} + \frac{3\mathcal{K}^2 n_\mathrm{c}^2}{\Delta_\mathrm{p}^2} - \frac{\kappa_\mathrm{nl}^2 n_\mathrm{c}^2}{4\Delta_\mathrm{p}^2}}
\end{equation}
and observe experimentally
\begin{equation}
	\kappa_0' - \kappa_0 \ll 2|\Delta_\mathrm{p}|
\end{equation}
we also know immediately that
\begin{equation}
	\frac{\kappa_\mathrm{nl}^2 n_\mathrm{c}^2}{4\Delta_\mathrm{p}^2} \leq \frac{\left( \kappa_0' - \kappa_0\right)^2}{4\Delta_\mathrm{p}^2} \ll 1
\end{equation}
and so we can omit it.
Also the second-last term is small and we can write
\begin{equation}
	\delta\omega_0 \approx \Delta_\mathrm{p} + |\Delta_\mathrm{p}|\sqrt{1 - \frac{4\mathcal{K}n_\mathrm{c}}{\Delta_\mathrm{p}}},
	\label{eqn:shift}
\end{equation}
which for small shifts compared to the detuning $|\mathcal{K}n_\mathrm{c}/\Delta_\mathrm{p}| \ll 1$ and $\Delta_\mathrm{p} < 0$ could be Taylor-approximated as
\begin{equation}
	\delta\omega_0 \approx 2\mathcal{K}n_\mathrm{c},
\end{equation}
i.e., as the well-known relation.
Nevertheless we work with Eq.~(\ref{eqn:shift}) here and rearrange it to
\begin{equation}
	n_\mathrm{c} = \frac{ \delta\omega_0\left(2 - \delta\omega_0/\Delta_\mathrm{p} \right)}{4\mathcal{K}}.
	\label{eqn:nc_Kerr}
\end{equation}
There is another useful relation, which connects the on-chip pump power $P_\mathrm{p} = n_\mathrm{in}\hbar\omega_\mathrm{p}$ to the intracircuit pump photon number $n_\mathrm{c}$.
To get there, we just rearrange the characteristic polynomial Eq.~(\ref{eqn:charac_poly}) and find
\begin{equation}
	n_\mathrm{c} = \frac{4 P_\mathrm{p}}{\hbar\omega_\mathrm{p}} \frac{\kappa_\mathrm{ext}}{\kappa_\mathrm{eff}^2 + 4\tilde{\Delta}_\mathrm{p}^2}
	\label{eqn:nc_new},
\end{equation}
where $\tilde\Delta_\mathrm{p} = \Delta_\mathrm{p} - \mathcal{K}n_\mathrm{c}$.
\section{Supplementary Note V: Theoretical model of the coupled circuits}
\label{sec:Note5}
\vspace{-2mm}
\subsection{Hamiltonian and equations of motion}
\vspace{-2mm}
We start our considerations with the Hamiltonian of the coupled circuits
\begin{equation}
	\hat{H}/\hbar = \omega_0\hat{a}^\dagger\hat{a} + \Omega_0\hat{b}^\dagger \hat{b} - g_{0\omega}\hat{a}^\dagger\hat{a}\left(\hat{b} + \hat{b}^\dagger \right) + \frac{\mathcal{K}}{2}\left( \hat{a}^\dagger\hat{a} \right)^2 - \frac{g_{0\mathcal{K}}}{2}\left( \hat{a}^\dagger\hat{a} \right)^2\left(\hat{b} + \hat{b}^\dagger \right) + \mathcal{K_\mathrm{c}}\hat{a}^\dagger \hat{a}\hat{b}^\dagger \hat{b} + \hat{H}_\mathrm{bath}/\hbar
\end{equation}
which contains the self-Kerr term $\frac{\mathcal{K}}{2}\left( \hat{a}^\dagger\hat{a} \right)^2$ and a possible cross-Kerr term $\mathcal{K_\mathrm{c}}\hat{a}^\dagger \hat{a}\hat{b}^\dagger \hat{b}$.
The two single-photon coupling rates $g_{0\omega}$ and $g_{0\mathcal{K}}$ originate from the flux-dependence of $\omega_0$ and $\mathcal{K}$ and are given by
\begin{eqnarray}
	g_{0\omega} & = & -\frac{\partial \omega_0}{\partial \Phi_\mathrm{b}}\Phi_\mathrm{zpf} \\
	g_{0\mathcal{K}} & = & -\frac{\partial\mathcal{K}}{\partial \Phi_\mathrm{b}}\Phi_\mathrm{zpf}.
\end{eqnarray}
Furthermore, we will assume for the bath Hamiltonian that the internal HF linewidth is given by
\begin{eqnarray}
	\kappa_0 - g_{0\kappa}\left(\hat{b} + \hat{b}^\dagger \right) ~~~~~ \mathrm{with} ~~~~~ g_{0\kappa} = -\frac{\partial\kappa_0}{\partial \Phi_\mathrm{b}}\Phi_\mathrm{zpf}
\end{eqnarray}
the dissipative coupling rate, while $\kappa_\mathrm{ext}, \Gamma_\mathrm{ext}, \Gamma_\mathrm{int}$ are constant.
Later, we will introduce a phenomenological pump-power dependence of both $\kappa_\mathrm{int}$ and $\Gamma_\mathrm{int}$, and for the former also the corresponding PP coupling terms.
The bath Hamiltonian in RWA according to Ref.~\cite{Gardiner1985x, Elste2009x, Xuereb2011x, Weiss2013x} is given by (we choose the sum notation for the bath modes here, but it is equivalent to the integral notation)
\begin{equation}
	\hat{H}_\mathrm{bath}/\hbar = \sum_n\omega_n \hat{e}_n^\dagger\hat{e}_n + \sum_m\omega_m \hat{i}_m^\dagger\hat{i}_m - \i\sqrt{\frac{\kappa_\mathrm{ext}}{2\pi\rho_\mathrm{ext}}}\sum_n\left( \hat{a}^\dagger \hat{e}_n - \hat{e}_n^\dagger \hat{a} \right) - \i\sqrt{\frac{\kappa_\mathrm{int}}{2\pi\rho_\mathrm{int}}}\sum_m\left( \hat{a}^\dagger \hat{i}_m - \hat{i}_m^\dagger \hat{a} \right) + \hat{H}_\Gamma/\hbar
\end{equation}
where the $\hat{e}$ and $\hat{i}$ operators are describing the external and internal HF cavity bath modes, $\rho_\mathrm{int}$ and $\rho_\mathrm{ext}$ are the frequency-independent bath density of states, and
\begin{equation}
	\hat{H}_\Gamma/\hbar = \sum_l\Omega_l \hat{E}_l^\dagger\hat{E}_l + \sum_j\Omega_j \hat{I}_j^\dagger\hat{I}_j - \i\sqrt{\frac{\Gamma_\mathrm{ext}}{2\pi\tilde{\rho}_\mathrm{ext}}}\sum_l\left( \hat{b}^\dagger \hat{E}_l - \hat{E}_l^\dagger \hat{b} \right) - \i\sqrt{\frac{\Gamma_\mathrm{int}}{2\pi\tilde{\rho}_\mathrm{int}}}\sum_j\left( \hat{b}^\dagger \hat{I}_j - \hat{I}_j^\dagger \hat{b} \right)
\end{equation}
takes into account the bath and dissipation of the low-frequency circuit.
To get a useful expression with the LF-flux-dependent $\kappa$-terms, we calculate
\begin{eqnarray}
	\hat{H}_\mathrm{bath} & \approx & \hat{H}_\mathrm{bath, 0} + \frac{\partial \hat{H}_\mathrm{bath}}{\partial \left( \hat{b} + \hat{b}^\dagger\right)}\left( \hat{b} + \hat{b}^\dagger\right) \\
	& = & \hat{H}_\mathrm{bath, 0} + \frac{\partial \hat{H}_\mathrm{bath}}{\partial \kappa_\mathrm{int}}\frac{\partial\kappa_\mathrm{int}}{\partial \left( \hat{b} + \hat{b}^\dagger\right)}\left( \hat{b} + \hat{b}^\dagger\right) \\
	& = & \hat{H}_\mathrm{bath, 0} - \frac{\partial \hat{H}_\mathrm{bath}}{\partial \kappa_\mathrm{int}}g_\mathrm{0\kappa}\left( \hat{b} + \hat{b}^\dagger\right) \\
	& = & \hat{H}_\mathrm{bath, 0} + \i\frac{\hbar g_\mathrm{0\kappa}}{2}\sqrt{\frac{1}{2\pi\rho_\mathrm{int}\kappa_\mathrm{int}}}\left( \hat{b} + \hat{b}^\dagger\right)\sum_m\left( \hat{a}^\dagger \hat{i}_m - \hat{i}_m^\dagger \hat{a} \right)
\end{eqnarray}
since for our device $\partial\kappa_0/\partial\Phi_\mathrm{b} \approx \partial\kappa_\mathrm{int}/\partial\Phi_\mathrm{b}$.
From here, we get the coupled EOMs as
\begin{eqnarray}
	\dot{\hat{e}}_n & = & \i\omega_n\hat{e}_n - \sqrt{\frac{\kappa_\mathrm{ext}}{2\pi\rho_\mathrm{ext}}}\hat{a}  \\
	\dot{\hat{i}}_m & = & \i\omega_m\hat{i}_m - \sqrt{\frac{\kappa_\mathrm{int}}{2\pi\rho_\mathrm{int}}}\hat{a} + \frac{g_\mathrm{0\kappa}}{2}\sqrt{\frac{1}{2\pi\rho_\mathrm{int}\kappa_\mathrm{int}}}\left( \hat{b} + \hat{b}^\dagger\right)\hat{a} \\
	\dot{\hat{E}}_l & = & \i\Omega_l\hat{E}_l - \sqrt{\frac{\Gamma_\mathrm{ext}}{2\pi\tilde{\rho}_\mathrm{ext}}}\hat{b}    \\
	\dot{\hat{I}}_j & = & \i\Omega_j\hat{I}_j - \sqrt{\frac{\Gamma_\mathrm{int}}{2\pi\tilde{\rho}_\mathrm{int}}}\hat{b}    \\
	\dot{\hat{a}} & = & \i\left[\omega_0 - g_{0\omega}\left( \hat{b} + \hat{b}^\dagger \right) + \mathcal{K}\left( \hat{a}^\dagger\hat{a} + \frac{1}{2} \right) - g_{0\mathcal{K}}\left( \hat{a}^\dagger\hat{a} + \frac{1}{2} \right)\left( \hat{b} + \hat{b}^\dagger \right) + \mathcal{K}_\mathrm{c}\hat{b}^\dagger\hat{b}\right]\hat{a} \nonumber \\
	& & + \sqrt{\frac{\kappa_\mathrm{ext}}{2\pi\rho_\mathrm{ext}}}\sum_n\hat{e}_n + \sqrt{\frac{\kappa_\mathrm{int}}{2\pi\rho_\mathrm{int}}}\sum_m\hat{i}_m - \frac{g_\mathrm{0\kappa}}{2}\sqrt{\frac{1}{2\pi\rho_\mathrm{int}\kappa_\mathrm{int}}}\left( \hat{b} + \hat{b}^\dagger\right)\sum_m\hat{i}_m \\
	\dot{\hat{b}} & = & \i\left[\Omega_0 + \mathcal{K}_\mathrm{c}\hat{a}^\dagger \hat{a} \right]\hat{b} - \i g_{0\omega}\hat{a}^\dagger \hat{a} - \i\frac{g_{0\mathcal{K}}}{2}\left(\hat{a}^\dagger \hat{a}\right)^2 + \sqrt{\frac{\Gamma_\mathrm{ext}}{2\pi\tilde{\rho}_\mathrm{ext}}}\sum_l\hat{E}_l + \sqrt{\frac{\Gamma_\mathrm{int}}{2\pi\tilde{\rho}_\mathrm{int}}}\sum_j\hat{I}_j -  \frac{g_\mathrm{0\kappa}}{2}\sqrt{\frac{1}{2\pi\rho_\mathrm{int}\kappa_\mathrm{int}}}\sum_m\left( \hat{a}^\dagger \hat{i}_m - \hat{i}_m^\dagger \hat{a} \right) ~~~~~~~~~~~
\end{eqnarray}
and we solve the equations for the baths as
\begin{eqnarray}
	\hat{E}_l & = & \hat{E}_{l, 0}\e^{\i\Omega_l(t-t_0)} - \sqrt{\frac{\Gamma_\mathrm{ext}}{2\pi\tilde{\rho}_\mathrm{ext}}}\int_{t_0}^t \e^{\i\Omega_l(t-t')}\hat{b}(t')dt' \\
	\hat{I}_j & = & \hat{I}_{j, 0}\e^{\i\Omega_j(t-t_0)} - \sqrt{\frac{\Gamma_\mathrm{int}}{2\pi\tilde{\rho}_\mathrm{int}}}\int_{t_0}^t \e^{\i\Omega_j(t-t')}\hat{b}(t')dt' \\
	\hat{e}_n & = & \hat{e}_{n, 0}\e^{\i\omega_n(t-t_0)} - \sqrt{\frac{\kappa_\mathrm{ext}}{2\pi\rho_\mathrm{ext}}}\int_{t_0}^t \e^{\i\omega_n(t-t')}\hat{a}(t')dt' \\
	\hat{i}_m & = & \hat{i}_{m, 0}\e^{\i\omega_m(t-t_0)} - \sqrt{\frac{\kappa_\mathrm{int}}{2\pi\rho_\mathrm{int}}}\int_{t_0}^t \e^{\i\omega_m(t-t')}\hat{a}(t')\left[1 - \frac{g_{0\kappa}}{2\kappa_\mathrm{int}}\left(\hat{b}(t') + \hat{b}^\dagger(t')\right)\right]dt' \\
	\hat{i}_m^\dagger & = & \hat{i}_{m, 0}^\dagger \e^{-\i\omega_m(t-t_0)} - \sqrt{\frac{\kappa_\mathrm{int}}{2\pi\rho_\mathrm{int}}}\int_{t_0}^t \e^{-\i\omega_m(t-t')}\hat{a}^\dagger(t')\left[1 - \frac{g_{0\kappa}}{2\kappa_\mathrm{int}}\left(\hat{b}(t') + \hat{b}^\dagger(t')\right)\right]dt'.
\end{eqnarray}
Then we can write the input terms in the EOMs for $\hat{a}$ as
\begin{eqnarray}
	\sqrt{\frac{\kappa_\mathrm{ext}}{2\pi\rho_\mathrm{ext}}}\sum_n\hat{e}_n & = & \sqrt{\frac{\kappa_\mathrm{ext}}{2\pi\rho_\mathrm{ext}}}\sum_n \left[\hat{e}_{n, 0}\e^{\i\omega_n(t-t_0)} - \sqrt{\frac{\kappa_\mathrm{ext}}{2\pi\rho_\mathrm{ext}}}\int_{t_0}^t \e^{\i\omega_n(t-t')}\hat{a}(t')dt'\right] \\
	& = & \sqrt{\frac{\kappa_\mathrm{ext}}{2\pi\rho_\mathrm{ext}}}\sum_n \hat{e}_{n, 0}\e^{\i\omega_n(t-t_0)} - \frac{\kappa_\mathrm{ext}}{2\pi\rho_\mathrm{ext}}\sum_n\int_{t_0}^t \e^{\i\omega_n(t-t')}\hat{a}(t')dt' \\
	& = & \sqrt{\kappa_\mathrm{ext}}\hat{a}_\mathrm{ext, in} - \frac{\kappa_\mathrm{ext}}{2}\hat{a}
\end{eqnarray}
\begin{eqnarray}
	\sqrt{\frac{\kappa_\mathrm{int}}{2\pi\rho_\mathrm{int}}}\sum_m\hat{i}_m & = & \sqrt{\kappa_\mathrm{int}}\hat{a}_\mathrm{int, in} - \frac{\kappa_\mathrm{int}}{2}\hat{a} + \frac{g_{0\kappa}}{4}\hat{a}\left(\hat{b} + \hat{b}^\dagger \right) \\
	\frac{g_\mathrm{0\kappa}}{2}\sqrt{\frac{1}{2\pi\rho_\mathrm{int}\kappa_\mathrm{int}}}\left( \hat{b} + \hat{b}^\dagger\right)\sum_m\hat{i}_m & = &  \frac{g_{0\kappa}}{2\sqrt{\kappa_\mathrm{int}}}\hat{a}_\mathrm{int, in}\left( \hat{b} + \hat{b}^\dagger\right) - \frac{g_{0\kappa}}{4}\hat{a}\left( \hat{b} + \hat{b}^\dagger\right) + \frac{g_{0\kappa}^2}{8\kappa_\mathrm{int}}\hat{a}\left(\hat{b} + \hat{b}^\dagger\right)^2.
\end{eqnarray}
On the other hand we get for the input terms of the LF resonator
\begin{eqnarray}
	\sqrt{\frac{\Gamma_\mathrm{ext}}{2\pi\tilde{\rho}_\mathrm{ext}}}\sum_l\hat{E}_l & = & \sqrt{\Gamma_\mathrm{ext}}\hat{b}_\mathrm{ext, in} - \frac{\Gamma_\mathrm{ext}}{2}\hat{b} \\
	\sqrt{\frac{\Gamma_\mathrm{int}}{2\pi\tilde{\rho}_\mathrm{int}}}\sum_j\hat{I}_j & = & \sqrt{\Gamma_\mathrm{int}}\hat{b}_\mathrm{int, in} - \frac{\Gamma_\mathrm{int}}{2}\hat{b} \\
	\frac{g_\mathrm{0\kappa}}{2}\sqrt{\frac{1}{2\pi\rho_\mathrm{int}\kappa_\mathrm{int}}}\sum_m\left( \hat{a}^\dagger \hat{i}_m - \hat{i}_m^\dagger \hat{a} \right) & = & \frac{g_\mathrm{0\kappa}}{2\kappa_\mathrm{int}}\hat{a}^\dagger\sqrt{\frac{\kappa_\mathrm{int}}{2\pi\rho_\mathrm{int}}}\sum_m \hat{i}_m - \frac{g_\mathrm{0\kappa}}{2\kappa_\mathrm{int}}\hat{a}\sqrt{\frac{\kappa_\mathrm{int}}{2\pi\rho_\mathrm{int}}}\sum_m\hat{i}_m^\dagger \\
	& = & \frac{g_\mathrm{0\kappa}}{2\kappa_\mathrm{int}}\hat{a}^\dagger\left[\sqrt{\kappa_\mathrm{int}}\hat{a}_\mathrm{int, in} - \frac{\kappa_\mathrm{int}}{2}\hat{a} + \frac{g_{0\kappa}}{4}\hat{a}\left(\hat{b} + \hat{b}^\dagger \right)\right] \\
	& & - \frac{g_\mathrm{0\kappa}}{2\kappa_\mathrm{int}}\hat{a}\left[\sqrt{\kappa_\mathrm{int}}\hat{a}^\dagger_\mathrm{int, in} - \frac{\kappa_\mathrm{int}}{2}\hat{a}^\dagger + \frac{g_{0\kappa}}{4}\hat{a}^\dagger\left(\hat{b} + \hat{b}^\dagger \right)\right] \\
	& = & \frac{g_{0\kappa}}{2\sqrt{\kappa_\mathrm{int}}}\left(\hat{a}^\dagger \hat{a}_\mathrm{int, in} - \hat{a}\hat{a}^\dagger_\mathrm{int, in}\right) + \frac{g_{0\kappa}}{4} - \frac{g_{0\kappa}^2}{8\kappa_\mathrm{int}}\left(\hat{b} + \hat{b}^\dagger\right).
\end{eqnarray}
Combining all results leads to the set of EOMs
\begin{eqnarray}
	\dot{\hat{a}} & = & \i\left[\omega_0 - g_{0\omega}\left( \hat{b} + \hat{b}^\dagger \right) + \mathcal{K}\left( \hat{a}^\dagger\hat{a} + \frac{1}{2} \right) - g_{0\mathcal{K}}\left( \hat{a}^\dagger\hat{a} + \frac{1}{2} \right)\left( \hat{b} + \hat{b}^\dagger \right) + \mathcal{K}_\mathrm{c}\hat{b}^\dagger\hat{b}\right]\hat{a} \nonumber \\
	& & - \frac{\kappa_0}{2}\hat{a} + \frac{g_{0\kappa}}{2}\hat{a}\left(\hat{b} + \hat{b}^\dagger\right) - \frac{g_{0\kappa}^2}{8\kappa_\mathrm{int}}\hat{a}\left(\hat{b} + \hat{b}^\dagger\right)^2 \nonumber \\
	& &  + \sqrt{\kappa_\mathrm{ext}}\hat{a}_\mathrm{ext, in} + \sqrt{\kappa_\mathrm{int}}\hat{a}_\mathrm{int, in} - \frac{g_{0\kappa}}{2\sqrt{\kappa_\mathrm{int}}}\hat{a}_\mathrm{int, in}\left(\hat{b} + \hat{b}^\dagger\right) \label{eqn:EOM_NL_a}\\
	\dot{\hat{b}} & = & \i\left[\Omega_0 + \mathcal{K}_\mathrm{c}\hat{a}^\dagger \hat{a} \right]\hat{b} - \i g_{0\omega}\hat{a}^\dagger \hat{a} - \i\frac{g_{0\mathcal{K}}}{2}\left(\hat{a}^\dagger \hat{a}\right)^2 - \frac{\Gamma_0}{2}\hat{b} - \frac{g_{0\kappa}}{4} + \frac{g_{0\kappa}^2}{8\kappa_\mathrm{int}}\left(\hat{b} + \hat{b}^\dagger\right) \nonumber \\
	& & + \sqrt{\Gamma_\mathrm{ext}}\hat{b}_\mathrm{ext, in} + \sqrt{\Gamma_\mathrm{int}}\hat{b}_\mathrm{int, in} - \frac{g_{0\kappa}}{2\sqrt{\kappa_\mathrm{int}}}\left(\hat{a}^\dagger \hat{a}_\mathrm{int, in} - \hat{a}\hat{a}^\dagger_\mathrm{int, in}\right).
	\label{eqn:EOM_NL_b}
\end{eqnarray}

\subsection{Dropping constants, small shifts, the quadratic term and input noise}
\vspace{-2mm}
For the coherent pump-probe experiments we are performing here, we will drop all noise input terms first since they are not relevant to first order, so the only input terms we keep are $\hat{a}_\mathrm{ext, in} = \i\hat{a}_\mathrm{in}$ and $\hat{b}_\mathrm{ext, in} = \i\hat{b}_\mathrm{in}$, which represent pump and probe signals.
At the same time, we drop HF cavity quantum noise contributions (the $+1/2$ contributions in the parenthesis expressions, which are negligibly small in the thermal regime), higher order contributions in $\hat{b}, \hat{b}^\dagger$, the cross-Kerr term from the LF to the HF mode (the resulting frequency shift is negligible for the high-frequency, large-linewidth HF circuit of this work), and the small term ${\propto}g_{0\kappa}^2/8\kappa_\mathrm{int}$ in Eq.~(\ref{eqn:EOM_NL_b}).
After all this we get
\begin{eqnarray}
	\dot{\hat{a}} & = & \i\left[\omega_0 - g_{0\omega}\left( \hat{b} + \hat{b}^\dagger \right) + \mathcal{K}\hat{a}^\dagger\hat{a} - g_{0\mathcal{K}}\hat{a}^\dagger\hat{a}\left( \hat{b} + \hat{b}^\dagger \right)\right]\hat{a} - \frac{\kappa_0}{2}\hat{a} + \frac{g_{0\kappa}}{2}\hat{a}\left(\hat{b} + \hat{b}^\dagger\right) + \i\sqrt{\kappa_\mathrm{ext}}\hat{a}_\mathrm{in} \\
	\dot{\hat{b}} & = & \i\left[\Omega_0 + \mathcal{K}_\mathrm{c}\hat{a}^\dagger \hat{a} \right]\hat{b} - \i g_{0\omega}\hat{a}^\dagger \hat{a} - \i\frac{g_{0\mathcal{K}}}{2}\left(\hat{a}^\dagger \hat{a}\right)^2 - \frac{\Gamma_0}{2}\hat{b} + \i\sqrt{\Gamma_\mathrm{ext}}\hat{b}_\mathrm{in}.
\end{eqnarray}
\subsection{Adding nonlinear damping}
\vspace{-2mm}
Finally we add up to third order nonlinear damping and cross-damping phenomenologically, cf.~also Sec.~\ref{sec:Note4}, and get
\begin{eqnarray}
	\dot{\hat{a}} & = & \i\left[ \omega_0 - g_{0\omega} \left(\hat{b} + \hat{b}^\dagger \right) + \mathcal{K}\hat{a}^\dagger\hat{a} - g_{0\mathcal{K}}\hat{a}^\dagger\hat{a}\left( \hat{b} + \hat{b}^\dagger \right) \right]\hat{a} + \i\sqrt{\kappa_\mathrm{ext}}\hat{a}_\mathrm{in}  \\
	& & - \frac{1}{2}\left[\kappa_0 - g_{0\kappa}\left( \hat{b} + \hat{b}^\dagger \right) + \kappa_1\hat{a}^\dagger\hat{a} - g_\mathrm{nl1}\hat{a}^\dagger\hat{a}\left(\hat{b} + \hat{b}^\dagger \right) + \kappa_2\left(\hat{a}^\dagger\hat{a}\right)^2 - g_\mathrm{nl2}\left(\hat{a}^\dagger\hat{a}\right)^2\left(\hat{b} + \hat{b}^\dagger \right) + \kappa_3\left(\hat{a}^\dagger\hat{a}\right)^3 - g_\mathrm{nl3}\left(\hat{a}^\dagger\hat{a}\right)^3\left(\hat{b} + \hat{b}^\dagger \right)\right]\hat{a} \nonumber \\
	\dot{\hat{b}} & = & \i\left[\Omega_0 + \mathcal{K}_\mathrm{c}\hat{a}^\dagger\hat{a}\right]\hat{b} - \i g_{0\omega} \hat{a}^\dagger\hat{a} - \i\frac{g_{0\mathcal{K}}}{2}\left(\hat{a}^\dagger \hat{a}\right)^2
	- \frac{\Gamma_0 + \kappa_\mathrm{c}\hat{a}^\dagger \hat{a}}{2}\hat{b} + \i\sqrt{\Gamma_\mathrm{ext}}\hat{b}_\mathrm{in}.
\end{eqnarray}
As a reminder, the photon-pressure single-photon coupling rates here are defined via
\begin{eqnarray}
	g_{0\omega} & = & -\frac{\partial\omega_0}{\partial\Phi_\mathrm{b}}\Phi_\mathrm{zpf} \\
	g_{0\mathcal{K}} & = & -\frac{\partial \mathcal{K}}{\partial\Phi_\mathrm{b}}\Phi_\mathrm{zpf} \\
	g_{0\kappa} & = & -\frac{\partial\kappa_0}{\partial\Phi_\mathrm{b}}\Phi_\mathrm{zpf} \\
	g_\mathrm{nl1} & = & -\frac{\partial\kappa_1}{\partial\Phi_\mathrm{b}}\Phi_\mathrm{zpf} \\
	g_\mathrm{nl2} & = & -\frac{\partial\kappa_2}{\partial\Phi_\mathrm{b}}\Phi_\mathrm{zpf} \\
	g_\mathrm{nl3} & = & -\frac{\partial\kappa_3}{\partial\Phi_\mathrm{b}}\Phi_\mathrm{zpf}
\end{eqnarray}
with the SQUID bias flux $\Phi_\mathrm{b}$ and the zero-point-fluctuation flux of the LF mode coupling into the SQUID $\Phi_\mathrm{zpf}$.
In principle also the cross-Kerr and cross-damping constants will depend on magnetic flux and lead to a coupling term, but since the terms themselves are very small here, we neglect their coupling contributions.
\vspace{-2mm}
\subsection{Low-pump-power regime -- Equations of motion}
\vspace{-2mm}
For low pump powers, we can set all nonlinearities to zero, i.e., $\mathcal{K}, \mathcal{K}_\mathrm{c}, \kappa_1, \kappa_2, \kappa_3, \kappa_\mathrm{c} = 0$; we also drop the corresponding coupling terms.
Then we get
\begin{eqnarray}
	\dot{\hat{a}} & = & \left[ \i\omega_0 - \frac{\kappa_0}{2} \right]\hat{a} -\i\left[g_{0\omega} + \i\frac{g_{0\kappa}}{2} \right]\left( \hat{b} + \hat{b}^\dagger \right)\hat{a} + \i\sqrt{\kappa_\mathrm{ext}}\hat{a}_\mathrm{in} \\
	\dot{\hat{b}} & = & \left[\i\Omega_0 - \frac{\Gamma_0}{2}\right]\hat{b} - \i g_{0\omega} \hat{a}^\dagger\hat{a} + \i\sqrt{\Gamma_\mathrm{ext}}\hat{b}_\mathrm{in}.
\end{eqnarray}
Now, we simplify and linearize the system by setting
\begin{eqnarray}
	\hat{a} & = & \left[\alpha_0 + \hat{c}\right]\e^{\i\omega_\mathrm{p}t} \\
	\hat{b} & = & \beta_0 + \hat{d} \\
	\hat{a}_\mathrm{in} & = & \left[\alpha_\mathrm{in}\e^{\i\phi_\mathrm{p}} + \hat{c}_\mathrm{in}\right]\e^{\i\omega_\mathrm{p}t} \\
	\hat{b}_\mathrm{in} & = & \hat{d}_\mathrm{in}
\end{eqnarray}
with the pump frequency $\omega_\mathrm{p}$.
After dropping all terms not linear in $\hat{c}$ and $\hat{d}$ and neglecting the small contributions due to the terms containing the LF steady-state term $\left(\beta_0 + \beta_0^*\right) \approx 2g_{0\omega}n_\mathrm{c}/\Omega_0$, we find for the dynamical quantities
\begin{eqnarray}
	\dot{\hat{c}} & = & \left[-\i\Delta_\mathrm{p} - \frac{\kappa_0}{2} \right]\hat{c} - \i\left[g_\omega + \i\frac{g_\kappa}{2} \right]\left( \hat{d} + \hat{d}^\dagger \right) + \i\sqrt{\kappa_\mathrm{ext}}\hat{c}_\mathrm{in} \label{eqn:EOM_LP_c} \\
	\dot{\hat{d}} & = & \left[ \i\Omega_0 - \frac{\Gamma_0}{2} \right]\hat{d} - \i g_\omega\left( \hat{c} + \hat{c}^\dagger\right) + \i\sqrt{\Gamma_\mathrm{ext}}\hat{d}_\mathrm{in} \label{eqn:EOM_LP_d}
\end{eqnarray}
with $\Delta_\mathrm{p} = \omega_\mathrm{p} - \omega_0$, $g_\omega = \alpha_0 g_{0\omega}$ and $g_\kappa = \alpha_0 g_{0\kappa}$.
Note that $\alpha_0 = \sqrt{n_\mathrm{c}}$ with the pump-induced intracavity photon number $n_\mathrm{c}$.
We solve these by Fourier transform and get
\begin{eqnarray}
	\hat{c} & = & - \i\chi_\mathrm{c}\left[g_\omega + \i\frac{g_\kappa}{2} \right]\left( \hat{d} + \hat{d}^\dagger \right) + \i\chi_\mathrm{c}\sqrt{\kappa_\mathrm{ext}}\hat{c}_\mathrm{in} \label{eqn:EOM_LP_c_FT}\\
	\hat{d} & = & - \i\chi_0g_\omega\left( \hat{c} + \hat{c}^\dagger\right) + \i\chi_0\sqrt{\Gamma_\mathrm{ext}}\hat{d}_\mathrm{in} \label{eqn:EOM_LP_d_FT}
\end{eqnarray}
with the susceptibilities
\begin{equation}
	\chi_\mathrm{c} = \frac{1}{\frac{\kappa_0}{2} + \i\left(\Delta_\mathrm{p} + \Omega \right)}, ~~~ \chi_0 = \frac{1}{\frac{\Gamma_0}{2} + \i\left( \Omega - \Omega_0 \right)}
\end{equation}
and the frequency with respect to the pump tone $\Omega = \omega - \omega_\mathrm{p}$.
Note again, that we do not discriminate by notation between time-domain quantities $\hat{c}(t), \hat{d}(t)$ and their Fourier transforms $\hat{c}(\Omega), \hat{d}(\Omega)$ for simplicity of the equations.
In all equations discussed here, this implies that $\hat{c}^\dagger, \hat{d^\dagger}$ in the frequency domain have an inverted frequency argument, i.e., that the Fourier transforms $\hat{c}^\dagger, \hat{d}^\dagger$ shall be read as $\hat{c}^\dagger(-\Omega)$ and $\hat{d}^\dagger(-\Omega)$, respectively.
\vspace{-2mm}
\subsection{Dynamical backaction in the low-pump-power regime}
\vspace{-2mm}
To calculate the dynamical backaction exerted from the HF intracavity pump fields to the LF mode, we drop in Eq.~(\ref{eqn:EOM_LP_c_FT}) the HF input term and write $g = g_\omega + \i g_\kappa/2$.
We get
\begin{eqnarray}
	\hat{c} & = & -\i g\chi_\mathrm{c}\left( \hat{d} + \hat{d}^\dagger\right) \\
	\hat{c}^\dagger & = & \i g^*\overline{\chi}_\mathrm{c}\left( \hat{d} + \hat{d}^\dagger\right),
\end{eqnarray}
or combined
\begin{equation}
	\hat{c} + \hat{c}^\dagger = -\i\left[g\chi_\mathrm{c} - g^*\overline{\chi}_\mathrm{c}\right]\left( \hat{d} + \hat{d}^\dagger\right).
\end{equation}
Hence, we find for the LF mode from Eq.~(\ref{eqn:EOM_LP_d_FT})
\begin{eqnarray}
	\hat{d} & = & -g_\omega\chi_0 \left[g\chi_\mathrm{c} - g^*\overline{\chi}_\mathrm{c}\right]\left( \hat{d} + \hat{d}^\dagger\right) + \i\chi_0\sqrt{\Gamma_\mathrm{ext}}\hat{d}_\mathrm{in} \\
	\hat{d}^\dagger & = & g_\omega\overline{\chi}_0 \left[g\chi_\mathrm{c} - g^*\overline{\chi}_\mathrm{c}\right]\left( \hat{d} + \hat{d}^\dagger\right) - \i\overline{\chi}_0\sqrt{\Gamma_\mathrm{ext}}\hat{d}_\mathrm{in}^\dagger,
\end{eqnarray}
or again combined
\begin{equation}
	\hat{d} + \hat{d}^\dagger = -g_\omega \left[\chi_0 - \overline{\chi}_0\right] \left[g\chi_\mathrm{c} - g^*\overline{\chi}_\mathrm{c}\right] \left(\hat{d} + \hat{d}^\dagger  \right) + \i\sqrt{\Gamma_\mathrm{ext}} \left( \chi_0 \hat{d}_\mathrm{in} - \overline{\chi}_0\hat{d}_\mathrm{in}^\dagger \right).
\end{equation}
The latter can also be written as
\begin{equation}
	\frac{\hat{d} + \hat{d}^\dagger}{\chi_0 - \overline{\chi}_0} = -g_\omega \left[g\chi_\mathrm{c} - g^*\overline{\chi}_\mathrm{c}\right] \left(\hat{d} + \hat{d}^\dagger  \right) + \i\sqrt{\Gamma_\mathrm{ext}}\frac{\chi_0 \hat{d}_\mathrm{in} - \overline{\chi}_0\hat{d}_\mathrm{in}^\dagger}{\chi_0 - \overline{\chi}_0}.
\end{equation}
Using the high-$Q$ results
\begin{eqnarray}
	\chi_\Phi
	& = & \frac{1}{\Omega_0^2 - \Omega^2 + \i\Omega \Gamma_0} \\
	& \approx & 2i\Omega_0\left(\chi_0 - \overline{\chi}_0\right) \\
	\frac{\chi_0}{\chi_0 - \overline{\chi}_0} & \approx & ~1 ~~~~~\textrm{for} ~~~~~ \Omega \approx +\Omega_0 \\
	-\frac{\overline{\chi}_0}{\chi_0 - \overline{\chi}_0} & \approx & ~1 ~~~~~ \textrm{for} ~~~~~ \Omega \approx -\Omega_0
\end{eqnarray}
as well as the flux $\hat{\Phi} = \Phi_\mathrm{zpf}\left( \hat{d} + \hat{d}^\dagger \right)$, we arrive at
\begin{equation}
	\frac{\hat{\Phi}}{\chi_\Phi} = -2\i\Omega_0 g_\omega \left[g\chi_\mathrm{c} - g^*\overline{\chi}_\mathrm{c} \right]\hat{\Phi} -2\Omega_0\sqrt{\Gamma_\mathrm{ext}}\hat{\Phi}_\mathrm{in}
\end{equation}
with $\hat{\Phi}_\mathrm{in} = \Phi_\mathrm{zpf}\left(\hat{d}_\mathrm{in} + \hat{d}_\mathrm{in}^\dagger \right)$.
Finally, we combine all terms and get
\begin{equation}
	\hat{\Phi} = -2\Omega_0\chi_\Phi^\mathrm{eff} \sqrt{\Gamma_\mathrm{ext}}\hat{\Phi}_\mathrm{in}
\end{equation}
with the effective LF susceptibility
\begin{equation}
	\chi_\Phi^\mathrm{eff} = \frac{1}{\Omega_0^2 - \Omega^2 + \i\Omega \Gamma_0 + 2\i\Omega_0 g_\omega \left[g\chi_\mathrm{c} - g^*\overline{\chi}_\mathrm{c}\right]}.
\end{equation}
The dynamical backaction is easier to identify in high-$Q$ approximation for $\Omega \approx \Omega_0$ though, for which the susceptibility becomes
\begin{eqnarray}
	\chi_\Phi^\mathrm{eff} & \approx & \frac{1}{2\i\Omega_0} \chi_0^\mathrm{eff} \\
	& = &  \frac{1}{2\i\Omega_0} \frac{1}{\frac{\Gamma_0}{2} + \i\left(\Omega - \Omega_0\right) + g_\omega\left[g\chi_\mathrm{c0} - g^*\overline{\chi}_\mathrm{c0}\right]},
\end{eqnarray}
from where we can see that photon-pressure damping $\Gamma_\mathrm{pp}$ and photon-pressure frequency shift $\delta\Omega_\mathrm{pp}$ are given by
\begin{eqnarray}
	\Gamma_\mathrm{pp} & = & 2\textrm{Re}\left[ g_\omega\left(g\chi_\mathrm{c0} - g^*\overline{\chi}_\mathrm{c0}\right) \right] \label{eqn:Gpp_lin}\\
	\delta\Omega_\mathrm{pp} & = & -\textrm{Im}\left[ g_\omega\left(g\chi_\mathrm{c0} - g^*\overline{\chi}_\mathrm{c0}\right) \right] \label{eqn:dOmega_lin}
\end{eqnarray}
and where $\chi_\mathrm{c0}$ and $\overline{\chi}_\mathrm{c0}$ are the HF susceptibilities evaluated at $\Omega = \Omega_0$, i.e.,
\begin{equation}
	\chi_\mathrm{c0} = \frac{1}{\frac{\kappa_0}{2} + \i\left( \Delta_\mathrm{p} + \Omega_0 \right)}, ~~~~~ \overline{\chi}_\mathrm{c0} = \frac{1}{\frac{\kappa_0}{2} - \i\left( \Delta_\mathrm{p} - \Omega_0 \right)}.
	\label{eqn:chi_c0_chi_c0_b}
\end{equation}
Note that these relations do not contain any sideband-resolution approximation, they are fully valid for arbitrary pump detunings and contain the information from both Stokes and anti-Stokes scattering (cooling and amplification).
In the sideband-resolved limit $\Omega_0^2 \gg \kappa_0^2$, which here for $\Delta_\mathrm{p} \approx -\Omega_0$ is equivalent to $|\chi_\mathrm{c0}| \gg |\overline{\chi}_\mathrm{c0}|$, we would find for a red-sideband pump
\begin{eqnarray}
	\Gamma_\mathrm{pp}^\mathrm{sbr} & = & 2\textrm{Re}\left[ g_\omega g\chi_\mathrm{c0} \right] \\
	\delta\Omega_\mathrm{pp}^\mathrm{sbr} & = & -\textrm{Im}\left[ g_\omega g\chi_\mathrm{c0}\right].
\end{eqnarray}
\vspace{-2mm}
\subsection{Photon-pressure induced transparency in the low-pump-power regime}
\vspace{-2mm}
To describe the transparency experiment, we need to calculate the reflection of the HF cavity in the presence of a sideband pump.
For this, we go back to Eq.~(\ref{eqn:EOM_LP_c_FT})
\begin{equation}
	\frac{\hat{c}}{\chi_\mathrm{c}} = -\i g\left(\hat{d} + \hat{d}^\dagger\right) + \i\sqrt{\kappa_\mathrm{ext}}\hat{c}_\mathrm{in}
\end{equation}
and combine it with the input-less relation
\begin{equation}
	\hat{d} + \hat{d}^\dagger = -\i g_\omega\left[ \chi_0 - \overline{\chi}_0\right]\left( \hat{c} + \hat{c}^\dagger \right)
\end{equation}
and with
\begin{equation}
	\hat{c} + \hat{c}^\dagger = -\i\left[g\chi_\mathrm{c} - g^*\overline{\chi}_\mathrm{c}\right]\left(\hat{d} + \hat{d}^\dagger\right) + \i\sqrt{\kappa_\mathrm{ext}}\left(\chi_\mathrm{c}\hat{c}_\mathrm{in} - \overline{\chi}_\mathrm{c}\hat{c}_\mathrm{in}^\dagger\right).
\end{equation}
This leads to
\begin{equation}
	\hat{d} + \hat{d}^\dagger = \frac{g_\omega \left[\chi_0 - \overline{\chi}_0\right]}{1 + g_\omega\left[\chi_0 - \overline{\chi}_0\right]\left[g\chi_\mathrm{c} - g^*\overline{\chi}_\mathrm{c}\right]}\sqrt{\kappa_\mathrm{ext}}\left(\chi_\mathrm{c}\hat{c}_\mathrm{in} - \overline{\chi}_\mathrm{c}\hat{c}_\mathrm{in}^\dagger \right)
\end{equation}
which we use to finally find for the relevant Fourier component
\begin{equation}
	\frac{\hat{c}}{\chi_\mathrm{c}} = \i\sqrt{\kappa_\mathrm{ext}}\left[1 - \frac{g_\omega g \left(\chi_0 - \overline{\chi}_0\right) \chi_\mathrm{c}}{1 + g_\omega\left(\chi_0 - \overline{\chi}_0 \right)\left( g\chi_\mathrm{c} - g^*\overline{\chi}_\mathrm{c}\right)}\right]\hat{c}_\mathrm{in}.
\end{equation}
Now we use again the high-$Q$ approximation $2\i\Omega_0\chi_\Phi = \chi_0 - \overline{\chi}_0 $ to get
\begin{eqnarray}
	\hat{c} & = & \i\sqrt{\kappa_\mathrm{ext}}\chi_\mathrm{c}\left[1 - \frac{2\i\Omega_0g_\omega g \chi_\Phi \chi_\mathrm{c}}{1 + 2\i\Omega_0g_\omega\chi_\Phi\left( g\chi_\mathrm{c} - g^*\overline{\chi}_\mathrm{c}\right)}\right]\hat{c}_\mathrm{in} \\
	& = & \i\sqrt{\kappa_\mathrm{ext}}\chi_\mathrm{c}\left[1 - 2\i\Omega_0 g_\omega g \chi_\mathrm{c}\chi_\Phi^\mathrm{eff}\right]\hat{c}_\mathrm{in}
\end{eqnarray}
and ultimately
\begin{equation}
	S_{11} = 1 - \kappa_\mathrm{ext}\chi_\mathrm{c}\left[1 - 2\i\Omega_0 g_\omega g \chi_\mathrm{c}\chi_\Phi^\mathrm{eff}\right].
\end{equation}
Using again $2\i\Omega_0\chi_\Phi^\mathrm{eff} = \chi_0^\mathrm{eff}$ for $\Omega \sim \Omega_0$, we find
\begin{equation}
	S_{11} = 1 - \kappa_\mathrm{ext}\chi_\mathrm{c}\left[1 - g_\omega g \chi_\mathrm{c}\chi_0^\mathrm{eff}\right].
	\label{eqn:S11_PPIT}
\end{equation}
Note that we can write the effective LF suceptibility also just as
\begin{equation}
	\chi_0^\mathrm{eff} = \frac{1}{\frac{\Gamma_\mathrm{eff}}{2} + \i\left(\Omega - \Omega_\mathrm{eff} \right)}
\end{equation}
where $\Gamma_\mathrm{eff}$ and $\Omega_\mathrm{eff}$ contain the dynamical backaction already.
\vspace{-2mm}
\subsection{High-pump-power regime -- Equations of motion}
\vspace{-2mm}
With higher pump powers, we still use the same ansatz for solving the EOMs and arrive (after dropping constants, negligibly small terms and higher order terms) at the linearized version
\begin{eqnarray}
	\dot{\hat{c}} & = & -\i\left[\Delta_\mathrm{p} - 2\mathcal{K}n_\mathrm{c}\right]\hat{c} - \frac{\kappa_0 + 2\kappa_1n_\mathrm{c} + 3\kappa_2 n_\mathrm{c}^2 + 4\kappa_3 n_\mathrm{c}^3}{2}\hat{c} - \i\sqrt{n_\mathrm{c}}\left[g_{0\omega} + g_{0\mathcal{K}}n_\mathrm{c} + \i\frac{g_{0\kappa} + g_\mathrm{nl1}n_\mathrm{c} + g_\mathrm{nl2}n_\mathrm{c}^2 + g_\mathrm{nl3}n_\mathrm{c}^3}{2}\right]\left(\hat{d} + \hat{d}^\dagger \right) \nonumber \\
	& & + \left[\i\mathcal{K} - \frac{\kappa_1 + 2\kappa_2 n_\mathrm{c} + 3\kappa_3 n_\mathrm{c}^2}{2}\right]n_\mathrm{c}\hat{c}^\dagger + \i\sqrt{\kappa_\mathrm{ext}}\hat{c}_\mathrm{in} \\
	\dot{\hat{d}} & = & \i\left[\Omega_0 + \mathcal{K}_\mathrm{c}n_\mathrm{c} \right]\hat{d} - \frac{\Gamma_0 + \kappa_\mathrm{c}n_\mathrm{c}}{2}\hat{d} - \i\sqrt{n_\mathrm{c}}\left[g_{0\omega} + g_{0\mathcal{K}}n_\mathrm{c}\right]\left(\hat{c} + \hat{c}^\dagger \right) + \i\sqrt{\Gamma_\mathrm{ext}}\hat{d}_\mathrm{in}.
\end{eqnarray}
Now, we can define the effective multi-photon coupling rates
\begin{eqnarray}
	g_\omega' & = & g_{0\omega}\sqrt{n_\mathrm{c}} + g_{0\mathcal{K}}\sqrt{n_\mathrm{c}^3} \\
	g_\kappa' & = & g_{0\kappa}\sqrt{n_\mathrm{c}} + g_\mathrm{nl1}\sqrt{n_\mathrm{c}^3} + g_\mathrm{nl2}\sqrt{n_\mathrm{c}^5} + g_\mathrm{nl3}\sqrt{n_\mathrm{c}^7}
\end{eqnarray}
which for large photon numbers scale much stronger with power than in usual optomechanics or photon-pressure.
Since with our pump tone, we work in a regime of very small intracavity gain and very small four-wave mixing due to the large detuning and the high nonlinear damping (for a more strict treatment see Sec.~\ref{sec:Note4} and Refs.~\cite{Bothner2022x, Rodrigues2022x, Rodrigues2024x}), the Kerr frequency shifts and nonlinear dampings can just be included by using modified frequencies and linewidths, labeled as $\omega_0', \kappa_0', \Omega_0'$ and $\Gamma_0'$, just as described above in Sec.~\ref{sec:Note4} and we get
\begin{eqnarray}
	\dot{\hat{c}} & = & -\left[\i\Delta_\mathrm{p}' + \frac{\kappa_0'}{2}\right]\hat{c} - \i\left[g_\omega' + \i\frac{g_\kappa'}{2}\right]\left(\hat{d} + \hat{d}^\dagger \right) + \i\sqrt{\kappa_\mathrm{ext}}\hat{c}_\mathrm{in} \label{eqn:NL_lin_EOM_HF}\\
	\dot{\hat{d}} & = & \left[\i\Omega_0' - \frac{\Gamma_0'}{2} \right]\hat{d} - \i g_\omega'\left(\hat{c} + \hat{c}^\dagger \right) + \i\sqrt{\Gamma_\mathrm{ext}}\hat{d}_\mathrm{in}
	\label{eqn:NL_lin_EOM_LF}
\end{eqnarray}
where for our experimental conditions (red-sideband pumping)
\begin{eqnarray}
	\omega_0' & = & \omega_0 + \Delta_\mathrm{p} + \sqrt{\left(\Delta_\mathrm{p} - \mathcal{K}n_\mathrm{c} \right)\left( \Delta_\mathrm{p} - 3\mathcal{K}n_\mathrm{c} \right) - \frac{\tilde{\kappa}_\mathrm{nl}^2 n_\mathrm{c}^2}{4}} \\
	\kappa_0' & = & \kappa_0 + 2\kappa_1n_\mathrm{c} + 3\kappa_2 n_\mathrm{c}^2 + 4\kappa_3 n_\mathrm{c}^3 \\
	\Omega_0' & = & \Omega_0 + \mathcal{K}_\mathrm{c} n_\mathrm{c} \\
	\Gamma_0' & = & \Gamma_0 + \kappa_\mathrm{c} n_\mathrm{c}.
\end{eqnarray}
Beautifully, however, Eqs.~(\ref{eqn:NL_lin_EOM_HF}, \ref{eqn:NL_lin_EOM_LF}) are completely identical to their corresponding equations in the low-power regime, just with primed (i.e.~pump-shifted but constant during each VNA scan) quantities.
Therefore, we can just apply all our former analyses and results.
\vspace{-2mm}
\subsection{Dynamical backaction in the high-pump-power regime}
\vspace{-2mm}
We find, analogous to Eqs.~(\ref{eqn:Gpp_lin}, \ref{eqn:dOmega_lin}),
\begin{eqnarray}
	\Gamma_\mathrm{pp} & = & 2\textrm{Re}\left[ g_\omega'\left(g'\chi_\mathrm{c0}' - g'^*\overline{\chi}_\mathrm{c0}'\right) \right] \label{eqn:Gpp_nonlin}\\\\
	\delta\Omega_\mathrm{pp} & = & -\textrm{Im}\left[ g_\omega'\left(g'\chi_\mathrm{c0}' - g'^*\overline{\chi}_\mathrm{c0}'\right) \right] \label{eqn:dOmega_nonlin}
\end{eqnarray}
where $\chi_\mathrm{c0}$ and $\overline{\chi}_\mathrm{c0}$ are the HF susceptibilities evaluated at $\Omega = \Omega_0'$, cf.~Eq.~(\ref{eqn:chi_c0_chi_c0_b}), i.e.,
\begin{equation}
	\chi_\mathrm{c0}' = \frac{1}{\frac{\kappa_0'}{2} + \i\left( \Delta_\mathrm{p}' + \Omega_0' \right)}, ~~~~~ \overline{\chi}_\mathrm{c0}' = \frac{1}{\frac{\kappa_0'}{2} - \i\left( \Delta_\mathrm{p}' - \Omega_0' \right)}.
\end{equation}
\subsection{Photon-pressure induced transparency in the high-pump-power regime}
For the reflection we obtain (high-$Q$ limit, cf.~Eq.~(\ref{eqn:S11_PPIT}))
\begin{equation}
	S_{11} = 1 - \kappa_\mathrm{ext}\chi_\mathrm{c}' \left[1 - g_\omega' g' \chi_\mathrm{c}' \chi_\mathrm{0}^\mathrm{eff} \right].
\end{equation}
where
\begin{eqnarray}
	\chi_\mathrm{c}' & = & \frac{1}{\frac{\kappa_0'}{2} + \i\left(\Delta_\mathrm{p}' + \Omega \right)} \\
	\overline\chi_\mathrm{c}' & = & \frac{1}{\frac{\kappa_0'}{2} - \i\left(\Delta_\mathrm{p}' - \Omega \right)} \\
	\chi_\mathrm{0}^\mathrm{eff} & = & \frac{1}{\frac{\Gamma_0'}{2} + \i\left( \Omega - \Omega_0'\right) + g_\omega'\left(g'\chi_\mathrm{c}' - g'^*\overline{\chi}_\mathrm{c}' \right)}.
\end{eqnarray}
It seems we somewhat inconsistently did not prime $\chi_0^\mathrm{eff}$, but it is on purpose to illustrate that we still write it as
\begin{eqnarray}
	\chi_0^\mathrm{eff} & = & \frac{1}{\frac{\Gamma_\mathrm{eff}}{2} + \i\left(\Omega - \Omega_\mathrm{eff} \right)} \\
	\Gamma_\mathrm{eff} & = & \Gamma_0 + \kappa_\mathrm{c} n_\mathrm{c} + \Gamma_\mathrm{pp} \\
	\Omega_\mathrm{eff} & = & \Omega_0 + \mathcal{K}_\mathrm{c} n_\mathrm{c} + \delta\Omega_\mathrm{pp},
\end{eqnarray}
since effective quantities stay effective quantities, although in detail they have one more term now.
\section{Supplementary Note VI: Effects of dissipative coupling}
\label{sec:Note6}
This section is meant to visualize the qualitative and quantitative changes for both PPIT and dynamical backaction, when a dissipative contribution is added to the dispersive photon-pressure interaction, and to introduce in this context the cross-cooperativity discussed in the main manuscript.
Additionally, we show how the normal-mode-splitting in the strong-coupling regime would reveal itself in the presence of $g_\kappa \neq 0$, with an exciting and surprising prospect.
\vspace{-2mm}
\subsection{Dynamical backaction and cross-cooperativity}
\label{sec:Note6A}
\vspace{-2mm}
\begin{figure*}
	\centerline{\includegraphics{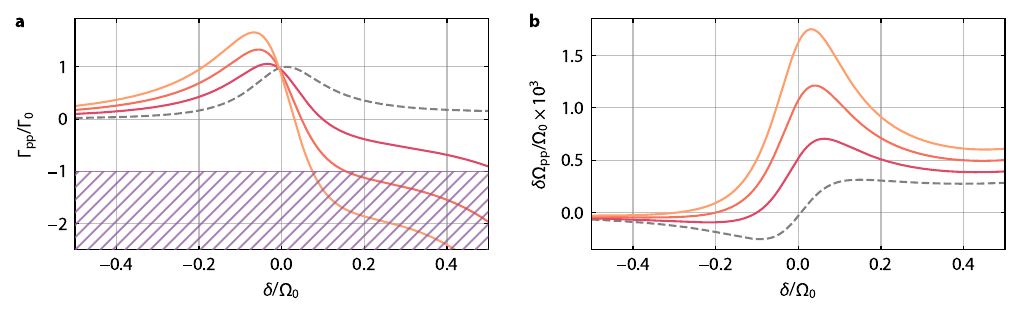}}
	\titlecaption{Red-sideband dynamical backaction for increasing contribution of internal-dissipative photon-pressure}{
		\sublabel{a} Calculated photon-pressure damping $\Gamma_\mathrm{pp}$ and \sublabel{b} calculated photon-pressure frequency shift $\delta\Omega_\mathrm{pp}$ vs.~pump detuning $\delta = \omega_\mathrm{p} - \left(\omega_0 - \Omega_0 \right)$ from the red sideband for four different ratios of $g_\kappa/g_\omega = (0, -2, -4, -6)$ and $g_\omega(\delta = 0) = 2\pi\times \qty{3.52}{\mega\hertz}$. Dashed gray line is the purely dispersive case $g_\kappa = 0$, and lines in red, dark orange, and bright orange correspond to $g_\kappa/g_\omega = -2$, $g_\kappa/g_\omega = -4$, and $g_\kappa/g_\omega = -6$, respectively. Quantities are normalized to the bare LF mode characteristics $\Gamma_0$ and $\Omega_0$, parameters of the calculation are $\omega_0 = 2\pi \times\qty{8.5}{\giga\hertz}$, $\kappa_0 = 2\pi\times\qty{100}{\mega\hertz}$, $\Omega_0 = 2\pi \times\qty{450}{\mega\hertz}$, $\Gamma_0 = 2\pi \times\qty{500}{\kilo\hertz}$, $g_{0\omega} = 2\pi\times\qty{30}{\kilo\hertz}$. Pump power $P_\mathrm{p} = \qty{-48.7}{\dBm}$ is constant, which means $n_\mathrm{c}$ increases with the pump moving to higher frequencies, cf.~Eq.~(\ref{eqn:nc_DBA}), which leads to e.g.~the curves for $g_\kappa = 0$ being clearly asymmetric with respect to $\delta = 0$. With increasing $|g_\kappa/g_\omega|$, both the absolute values of the dynamical backaction and the Fano-like distortion grow. Note that $\Gamma_\mathrm{pp}$ is enhanced for $\delta \lesssim 0$ by $g_\kappa$ and pushed to smaller and eventually even negative values for $\delta \gtrsim 0$. For $\Gamma_\mathrm{pp}/\Gamma_0 \le -1$, the system will experience parametric instability (indicated by the hatched area in \sublabel{a}).
	}
	\label{fig:FigS9}
\end{figure*}
In Supplementary Fig.~\ref{fig:FigS9} we plot the two components of dynamical backaction (neglecting any nonlinearities, so no primed quantities here)
\begin{eqnarray}
	\Gamma_\mathrm{pp} & = & 2\mathrm{Re}\left[g_\omega\left(g\chi_\mathrm{c0} - g^*\overline{\chi}_\mathrm{c0} \right)\right] \label{eqn:DBA_noNL_Gpp}\\
	\delta\Omega_\mathrm{pp} & = & -\mathrm{Im}\left[g_\omega\left(g\chi_\mathrm{c0} - g^*\overline{\chi}_\mathrm{c0} \right)\right] \label{eqn:DBA_noNL_dWpp}
\end{eqnarray}
for four different ratios of $g_\kappa/g_\omega = (0, -2, -4, -6)$ and for parameters close to the ones of our experiment.
In detail, we chose $\omega_0 = 2\pi\times\qty{8.5}{\giga\hertz}$, $\kappa_0 = 2\pi\times\qty{100}{\mega\hertz}$, $\Omega_0 = 2\pi\times\qty{450}{\mega\hertz}$, $\Gamma_0 = 2\pi\times\qty{500}{\kilo\hertz}$, $g_{0\omega} = 2\pi \times\qty{30}{\kilo\hertz}$, and calculate the detuning-dependent pump photon number for $g_\omega = \sqrt{n_\mathrm{c}}g_{0\omega}$ via the linear-cavity relation
\begin{equation}
	n_\mathrm{c} = \frac{4P_\mathrm{p}}{\hbar\omega_\mathrm{p}} \frac{\kappa_\mathrm{ext}}{\kappa_0^2 + 4\Delta_\mathrm{p}^2}
	\label{eqn:nc_DBA}
\end{equation}
where $\kappa_\mathrm{ext} = 2\pi\times\qty{7}{\mega\hertz}$ and $\Delta_\mathrm{p} = \omega_\mathrm{p} - \omega_0$.
The theoretical lines are shown vs.~pump detuning from the red sideband $\delta = \omega_\mathrm{p} - \left(\omega_0 - \Omega_0 \right)$ and we limit our considerations to detunings between $\delta_\mathrm{min} = 2\pi\times\qty{-225}{\mega\hertz}$ and $\delta_\mathrm{max} = 2\pi\times\qty{225}{\mega\hertz}$ (equivalent to $-\Omega_0/2$ and $+\Omega_0/2$), since this is the most relevant for our experiment.
For the on-chip pump power, we use $P_\mathrm{p} = \qty{13.5}{\nano\watt}$ or $\qty{-48.7}{\dBm}$, which was chosen such that in the dispersive-only case and with the pump exactly on the red sideband $\Gamma_\mathrm{pp}(\delta = 0) \approx \Gamma_0$, which is equivalent to $n_\mathrm{c}(\delta = 0) \approx 13750$ and $g_\omega(\delta = 0) = 2\pi \times \qty{3.52}{\mega\hertz}$.
The plots in Supplementary Fig.~\ref{fig:FigS9} reveal, that adding the dissipative interaction has considerable effects on the dynamical backaction.
Without the dissipative part (gray dashed lines), the backaction terms describe essentially a complex Lorentzian with a detuning-dependent scaling factor ${\propto} n_\mathrm{c}(\delta)/n_\mathrm{c}(0)$, which causes the asymmetry between positive and negative $\delta$.
With $g_\kappa \neq 0$ some characteristic features of coupling interference can be identified.
Most obviously there is a Fano-like distortion of the complex Lorentzian, which grows with increasing $|g_\kappa/g_\omega|$.
In other words, the presence of $g_\kappa$ changes the backaction magnitude as a function of frequency; for $\delta\lesssim 0$ the damping increases, while for $\delta \gtrsim 0$ the dissipative contribution counteracts the dispersive one.
This even implies that in the regime $\delta > 0$, but still around the red HF cavity sideband, there will be a parametric instability $\Gamma_\mathrm{eff} = \Gamma_0 + \Gamma_\mathrm{pp} < 0$, which we also found in experiments, cf.~main paper Fig.~5.
Another consequence of this frequency-dependent sign change is that the maximum of $\Gamma_\mathrm{pp}$ is both considerably red-detuned from $\omega_0 - \Omega_0$ and larger than the maximum for $g_\kappa = 0$.
For the PP frequency-shift $\delta\Omega_\mathrm{pp}$, the effect of $g_\kappa$ seems somewhat less complex, it mainly moves $\delta\Omega_\mathrm{pp}$ towards larger values and makes the overall shape of the frequency-dependence more peak-like.
The position of the maximum, however, also shifts as a function of $g_\kappa/g_\omega$.
The latter is no coincidence -- for very large ratios $|g_\kappa/g_\omega| \gtrsim 100$, the shapes of $\Gamma_\mathrm{pp}$ and $\delta\Omega_\mathrm{pp}$ would be identical to the dispersive-only case, but swapped (the relative scaling would not be swapped though).
It is useful to look at this swap on a more formal level, which will illustrate at the same time the meaning of the two cooperativities discussed in the main paper.
For simplicity, we do this in the sideband-resolved approximation, which describes our device sufficiently well.
In this limit, the backaction terms are given by
\begin{eqnarray}
	\Gamma_\mathrm{pp} & = & 2\mathrm{Re}\left[g_\omega g\chi_\mathrm{c0}\right] \\
	\delta\Omega_\mathrm{pp} & = & -\mathrm{Im}\left[g_\omega g\chi_\mathrm{c0}\right]
\end{eqnarray}
which can be explicitly written as
\begin{eqnarray}
	\Gamma_\mathrm{pp} & = & g_\omega^2 \kappa_0|\chi_\mathrm{c0}|^2 + g_\omega g_\kappa \delta |\chi_\mathrm{c0}|^2 \\
	\delta\Omega_\mathrm{pp} & = & g_\omega^2 \delta |\chi_\mathrm{c0}|^2 - \frac{g_\omega g_\kappa}{4}\kappa_0|\chi_\mathrm{c0}|^2
\end{eqnarray}
using
\begin{eqnarray}
	\chi_\mathrm{c0} & = & \frac{\frac{\kappa_0}{2}}{\frac{\kappa_0^2}{4} + \delta^2} - \i\frac{\delta}{\frac{\kappa_0^2}{4} + \delta^2} \\
	& = & \frac{\kappa_0}{2}|\chi_\mathrm{c0}|^2 - \i\delta |\chi_\mathrm{c0}|^2.
\end{eqnarray}
In the dispersive-only case, the relations would read
\begin{eqnarray}
	\Gamma_\mathrm{pp}^\omega & = & g_\omega^2 \kappa_0|\chi_\mathrm{c0}|^2\\
	\delta\Omega_\mathrm{pp}^\omega & = & g_\omega^2 \delta |\chi_\mathrm{c0}|^2,
\end{eqnarray}
where $\Gamma_\mathrm{pp}^\omega$ has its maximum for frequency-independent $g_\omega, g_\kappa$ at the maximum of $|\chi_\mathrm{c0}|^2$, i.e., at $\delta = 0$, and $\delta\Omega_\mathrm{pp}^\omega$ has its maximum at the maximum of $\delta |\chi_\mathrm{c0}|^2$, which is at $\delta = \kappa_0/2$.
So we can express
\begin{eqnarray}
	\Gamma_\mathrm{pp}^{\omega, \mathrm{max}} & = & \frac{4g_\omega^2}{\kappa_0} \\
	& = & \mathcal{C_\omega}\Gamma_0 \\
	\delta\Omega_\mathrm{pp}^{\omega, \mathrm{max}} & = & \frac{1}{4}\mathcal{C}_\omega\Gamma_0
\end{eqnarray}
with the dispersive cooperativity
\begin{equation}
	\mathcal{C}_\omega = \frac{4 g_\omega^2}{\kappa_0\Gamma_0}
\end{equation}
as a measure for the maximal dynamical backaction.
The new terms in the backaction arising from $g_\kappa \neq 0$ have the frequency-dependence and the position of the maxima swapped.
The second term in the photon-pressure damping is
\begin{eqnarray}
	\Gamma_\mathrm{pp}^\kappa & = & g_\omega g_\kappa \delta |\chi_\mathrm{c0}|^2
\end{eqnarray}
and so for frequency-independent $g_\omega, g_\kappa$ it has the maximum at $\delta = -\kappa_0/2$ for $g_\kappa < 0$.
Note that this new damping term is equivalent to the dispersive-only frequency-shift term, except for one $g_\omega$ being replaced by $g_\kappa$ (and the corresponding sign change, which would not be there for $g_\kappa > 0$).
Hence, we can write
\begin{eqnarray}
	\Gamma_\mathrm{pp}^{\kappa, \mathrm{max}} & = & -\frac{1}{2}C_{\omega\kappa}\Gamma_0
\end{eqnarray}
with the cross-cooperativity
\begin{equation}
	\mathcal{C}_{\omega\kappa} = \frac{2g_\omega g_\kappa}{\kappa_0\Gamma_0}.
\end{equation}
Similarly, we can write the cross-cooperativity induced frequency-shift
\begin{eqnarray}
	\delta\Omega_\mathrm{pp}^\kappa & = & - \frac{g_\omega g_\kappa}{4}\kappa_0|\chi_\mathrm{c0}|^2
\end{eqnarray}
at its maximum at $\delta = 0$ as
\begin{eqnarray}
	\delta\Omega_\mathrm{pp}^{\kappa, \mathrm{max}} & = & -\frac{1}{2}C_{\omega\kappa}\Gamma_0.
\end{eqnarray}
Note that choosing the prefactor 2 instead of 4 in the cross-cooperativity is somewhat arbitrary, but doing so reflects the factor $1/2$ entering the total complex coupling
\begin{equation}
	g = g_\omega + \i\frac{g_\kappa}{2}.
\end{equation}
It furthermore leads to the symmetric situation, that for $\mathcal{C}_\omega = \mathcal{C}_{\omega\kappa}$ the term ${\propto} \delta$ in each of the two dynamical backaction terms is half as large at $\delta = \kappa_0/2$ as the one ${\propto} \kappa_0$ at $\delta = 0$.
Note the different scalings though; while the cross-terms lead to identical maximum values of $\Gamma_\mathrm{pp}^\kappa$ and $\delta\Omega_\mathrm{pp}^\kappa$, the dispersive-only terms differ by a factor $4$ at their respective maxima.

\begin{figure*}
	\centerline{\includegraphics{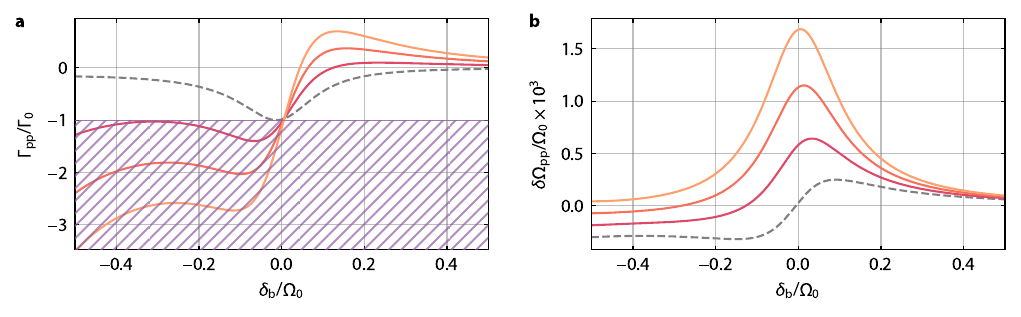}}
	\titlecaption{Blue-sideband dynamical backaction for increasing contribution of internal-dissipative photon-pressure}{
		\sublabel{a} Calculated photon-pressure damping $\Gamma_\mathrm{pp}$ and \sublabel{b} calculated photon-pressure frequency shift $\delta\Omega_\mathrm{pp}$ vs.~pump detuning $\delta_\mathrm{b} = \omega_\mathrm{p} - \left(\omega_0 + \Omega_0 \right)$ from the blue sideband for four different ratios of $g_\kappa/g_\omega = (0, -2, -4, -6)$ and $g_\omega(\delta_\mathrm{b} = 0) = 2\pi\times \qty{3.52}{\mega\hertz}$. Dashed gray line is the purely dispersive case $g_\kappa = 0$, and lines in red, dark orange, and bright orange correspond to $g_\kappa/g_\omega = -2$, $g_\kappa/g_\omega = -4$, and $g_\kappa/g_\omega = -6$, respectively. Quantities are normalized to the bare LF mode characteristics $\Gamma_0$ and $\Omega_0$, parameters of the calculation are $\omega_0 = 2\pi \times\qty{8.5}{\giga\hertz}$, $\kappa_0 = 2\pi\times\qty{100}{\mega\hertz}$, $\Omega_0 = 2\pi \times\qty{450}{\mega\hertz}$, $\Gamma_0 = 2\pi \times\qty{500}{\kilo\hertz}$, $g_{0\omega} = 2\pi\times\qty{30}{\kilo\hertz}$. Pump power $P_\mathrm{p} = \qty{15}{\nano\watt}$ is constant, which means $n_\mathrm{c}$ decreases with the pump moving to higher frequencies, cf.~Eq.~(\ref{eqn:nc_DBA}), which leads to e.g.~the curves for $g_\kappa = 0$ being clearly asymmetric with respect to $\delta_\mathrm{b} = 0$. With increasing $|g_\kappa/g_\omega|$, both the absolute values of the dynamical backaction and the Fano-like distortion grow. Note that $\Gamma_\mathrm{pp}$ is enhanced for $\delta_\mathrm{b} \lesssim 0$ by $g_\kappa$ and pushed to positive values for $\delta_\mathrm{b} \gtrsim 0$, which corresponds to blue-sideband cooling. For $\Gamma_\mathrm{pp}/\Gamma_0 \le -1$, the system will experience parametric instability (indicated by the hatched area in \sublabel{a}).
	}
	\label{fig:FigS10}
\end{figure*}

Although we did not measure the dynamical backaction with a blue-detuned pump tone, it is useful to look at the corresponding theoretical result using Eqs.~(\ref{eqn:DBA_noNL_Gpp}, \ref{eqn:DBA_noNL_dWpp}).
In Supplementary~Fig.~\ref{fig:FigS10} this is shown as a function of pump detuning from the blue sideband $\delta_\mathrm{b} = \omega_\mathrm{p} - \left(\omega_0 + \Omega_0 \right)$, and in complete correspondence to Supplementary Fig.~\ref{fig:FigS9}.
Again, for $g_\kappa = 0$ (gray dashed line) we find the familiar complex Lorentzian with negative photon-pressure damping (i.e.~amplification) and a photon-pressure frequency-shift that resembles the one on the red sideband.
All parameters are identical to the red-sideband case in Supplementary~Fig.~\ref{fig:FigS9}, except for the pump power $P_\mathrm{p} = \qty{15}{\nano\watt}$, which we slightly increased to again have $n_\mathrm{c}(\delta_\mathrm{b} = 0) \approx 13750$, $g_\omega(\delta_\mathrm{b} = 0) = 2\pi\times \qty{3.52}{\mega\hertz}$ and $\Gamma_\mathrm{pp}(\delta_\mathrm{b} = 0) = -\Gamma_0$.
Since we still calculate the pump photon number $n_\mathrm{c}$ via Eq.~(\ref{eqn:nc_DBA}), now the photon number increases with negative detunings, which causes the asymmetry between positive and negative $\delta_\mathrm{b}$ to show the opposite trend of the red-sideband case.
Aside from this, the frequency shift $\delta\Omega_\mathrm{pp}$ is modified by $g_\kappa \neq 0$ essentially the same way as on the red sideband; with increasing $|g_\kappa/g_\omega|$ it turns more and more into a Lorentzian-like peak and the maximum is approaching $\delta_\mathrm{b} \approx 0$.
The photon-pressure damping on the other hand looks not mirrored at all when comparing it to the red-sideband case, but there is nevertheless a certain symmetry.
For $\delta_\mathrm{b} > 0$, the dispersive backaction damping $\Gamma_\mathrm{pp} < 0$ is overcompensated by the dissipative contribution to $\Gamma_\mathrm{pp} > 0$, which in fact corresponds to (counterintuitive) cooling on the blue sideband.
For $\delta_\mathrm{b} < 0$ the negative backaction damping is strongly enhanced by the dissipative contribution, leading to parametric instability over the complete range of detunings for the parameters used here.
To conclude this section, we show in Supplementary Fig.~\ref{fig:FigS11} the dynamical backaction over the complete pump frequency range from $\omega_\mathrm{p} = \omega_0 - 1.5\Omega_0$ to $\omega_\mathrm{p} = \omega_0 + 1.5\Omega_0$.
In contrast to Supplementary Figs.~\ref{fig:FigS9} and \ref{fig:FigS10} we do it this time for constant pump photon number $n_\mathrm{c} = 13750$ as a function of pump detuning instead of constant pump power for two reasons.
First, for $g_\kappa = 0$, it reveals both the point-symmetry around $\Delta_\mathrm{p} = 0$ and the (nearly) symmetric Lorentzian around the two sidebands, which was masked above by the detuning-dependent $g_\omega$.
It also reveals the lack of that symmetry for $g_\kappa \neq 0$.
The remaining asymmetries in $\delta\Omega_\mathrm{pp}$ around the two sidebands originate from imperfect sideband-resolution.
And secondly, since otherwise the backaction on resonance $\Delta_\mathrm{p} = 0$ for $g_\kappa \neq 0$ is so large that the shape of the contributions on the sidebands are not easily recognizable anymore.
Most of the relevant features, that can be identified in the dynamical backaction, are already known at this point and discussed in the context of Supplementary Figs.~\ref{fig:FigS9} and \ref{fig:FigS10}.
What is new here is the backaction on and around resonance, i.e., for $\Delta_\mathrm{p} = 0$.
While for $g_\kappa = 0$ the dynamical backaction identically vanishes for $\omega_\mathrm{p} = \omega_0$, it is finite for $g_\kappa \neq 0$.
Due to the opposite signs of $g_\omega$ and $g_\kappa$ it leads to negative backaction damping and hence amplification on resonance, while for identical signs it would lead to cooling on resonance.
Such identical signs could be achieved in the future by e.g.~a properly designed external-dissipative coupling $g_\mathrm{\kappa_\mathrm{ext}}$ or by a modified circuit layout that contains a flux-dependent path through an additional on-chip resistor in series with the SQUID.
\clearpage

\begin{figure*}
	\centerline{\includegraphics{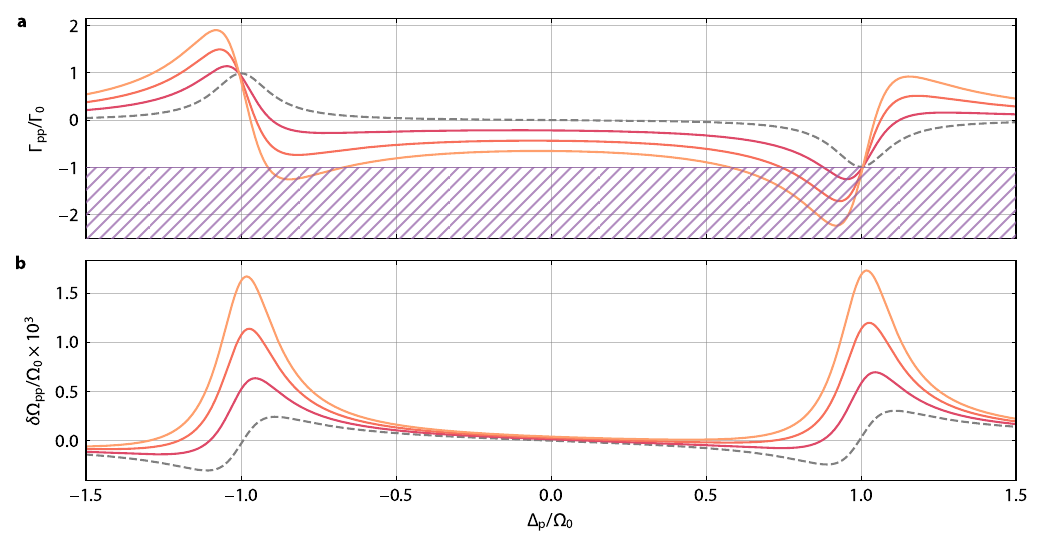}}
	\titlecaption{Broadband dynamical backaction for increasing contribution of internal-dissipative photon-pressure}{
		\sublabel{a} Calculated photon-pressure damping $\Gamma_\mathrm{pp}$ and \sublabel{b} calculated photon-pressure frequency shift $\delta\Omega_\mathrm{pp}$ vs.~pump detuning from the cavity resonance $\Delta_\mathrm{p} = \omega_\mathrm{p} - \omega_0$ for four different ratios of $g_\kappa/g_\omega = (0, -2, -4, -6)$ and $g_\omega = 2\pi\times \qty{3.52}{\mega\hertz}$ for all pump frequencies. Dashed gray line is the purely dispersive case $g_\kappa = 0$, and lines in red, dark orange, and bright orange correspond to $g_\kappa/g_\omega = -2$, $g_\kappa/g_\omega = -4$, and $g_\kappa/g_\omega = -6$, respectively. Quantities are normalized to the bare LF mode characteristics $\Gamma_0$ and $\Omega_0$, parameters of the calculation are $\omega_0 = 2\pi \times\qty{8.5}{\giga\hertz}$, $\kappa_0 = 2\pi\times\qty{100}{\mega\hertz}$, $\Omega_0 = 2\pi \times\qty{450}{\mega\hertz}$, $\Gamma_0 = 2\pi \times\qty{500}{\kilo\hertz}$, $g_{0\omega} = 2\pi\times\qty{30}{\kilo\hertz}$, and intracavity pump photon number $n_\mathrm{c} = 13750$ independent of $\Delta_\mathrm{p}$. With increasing $|g_\kappa/g_\omega|$, both the absolute values of the dynamical backaction and the Fano-like distortion grow. Interestingly, the dynamical backaction does not vanish on resonance, when $g_\kappa \neq 0$, which here and due to $g_\kappa < 0$ leads to LF amplification on resonance, and for $g_\kappa > 0$ would lead to cooling on resonance. For $\Gamma_\mathrm{pp}/\Gamma_0 \le -1$, the system will experience parametric instability (indicated by the hatched area in \sublabel{a}).
	}
	\label{fig:FigS11}
\end{figure*}
\subsection{Photon-pressure induced transparency}
\label{sec:Note6B}
Next, we plot the calculated HF reflection for the same parameters as the dynamical backaction in the previous section, but for a single pump frequency $\omega_\mathrm{p} = \omega_0 - \Omega_\mathrm{eff}$ in all four cases.
So again, the coupling ratios are $g_\kappa/g_\omega = (0, -2, -4, -6)$ and we plug everything into Eq.~(\ref{eqn:S11_PPIT})
\begin{equation}
	S_{11} = 1 - \kappa_\mathrm{ext}\chi_\mathrm{c}\left(1 - g_\omega g \chi_\mathrm{c}\chi_0^\mathrm{eff}\right).
	\label{eqn:PPIT_th_plot}
\end{equation}
The result is shown in Supplementary Fig.~\ref{fig:FigS12}, and illustrates the essential characteristics of the dissipative coupling contribution to PPIT.
\begin{figure*}
	\centerline{\includegraphics{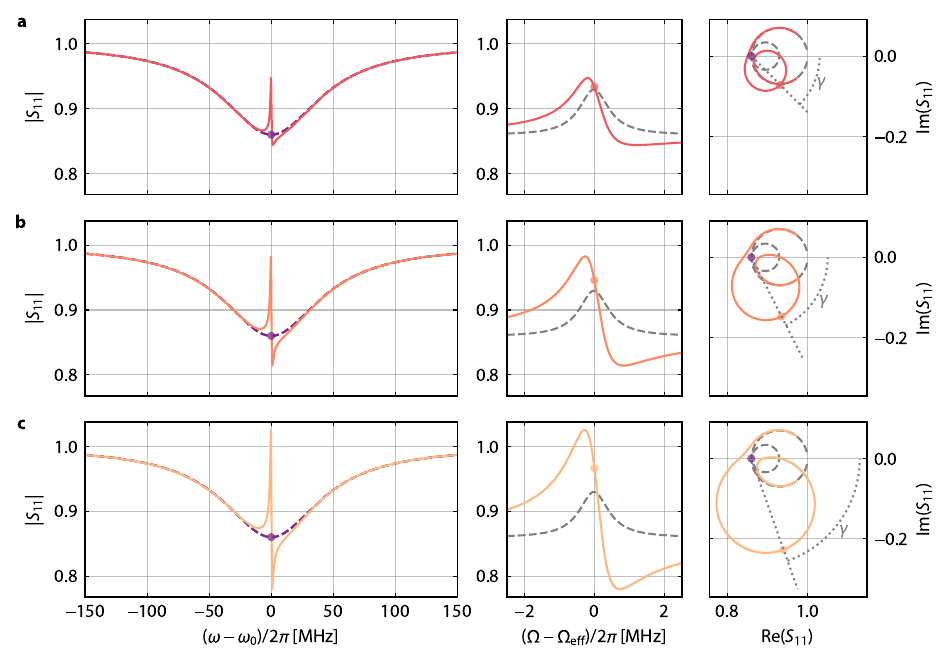}}
	\titlecaption{Photon-pressure-induced transparency for increasing contribution of internal-dissipative coupling}{
		All subpanels show the calculated HF cavity reflection $S_{11}$ using Eq.~(\ref{eqn:PPIT_th_plot}) in a resonant PPIT setting, i.e., with a pump tone at $\omega_\mathrm{p} = \omega_0 - \Omega_\mathrm{eff}$. Parameters for HF mode, LF mode and the coupling rates are given in subsection~\ref{sec:Note6A}. From top column to bottom column, the dissipative coupling increases, in \sublabel{a} $g_\kappa/g_\omega = -2$, in \sublabel{b} $g_\kappa/g_\omega = -4$, and in \sublabel{c} $g_\kappa/g_\omega = -6$. Left image shows magnitude of $S_{11}$ in a frequency span $3\kappa_0$ around $\omega_0$, center image shows a magnification of the transparency window with a frequency span $10\Gamma_0$, and right image shows the circle plot of the PPIT resonance and cavity resonance in the complex plane. Purple disks in left and right images show the point of $S_{11}$ on HF cavity resonance without the pump tone, orange disks in center and right panels show the point of PPIT resonance. Dotted straight lines connecting the two disks in the circle plots define the angle $\gamma$ between that line and the real axis. For reference, each zoom and complex plot (center and right, respectively) also shows the case $g_\kappa = 0$ as gray dashed line; the left images contain the cavity resonance for $g_\kappa = g_\omega = 0$ as purple dashed line. With increasing $g_\kappa/g_\omega$, the PPIT circle tilts more and more away from the real axis, i.e., the angle $\gamma = \arctan{g_\kappa/2g_\omega}$ increases, leading to an increasing Fano lineshape of the transparency resonance in $|S_{11}|$. For large $g_\kappa/g_\omega$, the PPIT circle gets larger than the cavity circle, despite being in the sideband-resolved regime, even leading to output gain $|S_{11}|>1$ in \sublabel{c}.
	}
	\label{fig:FigS12}
\end{figure*}
For $g_\kappa = 0$, the PPIT resonance is a symmetric peak in $|S_{11}|$, sitting at the bottom of the HF absorption dip; it becomes increasingly asymmetric and Fano-like with growing $g_\kappa/g_\omega$.
The origin of this asymmetry is the interference of the sidebands with amplitude ${\propto} g_\omega$ and ${\propto} \i g_\kappa/2$, which leads to a rotation of the PPIT resonance circle in the complex plane.
The PPIT resonance peak-height and circle diameter, respectively, also grow with increasing $g_\kappa$ due to the increased total coupling rate.
As a measure for $g_\kappa/g_\omega$ we can consider the tilt angle~$\gamma$ in the complex plane, which is given by
\begin{equation}
	\gamma = \arctan\frac{g_\kappa}{2g_\omega}
\end{equation}
and which could also be explicitly introduced into Eq.~(\ref{eqn:PPIT_th_plot}) by
\begin{eqnarray}
	g & = & g_\omega + \i\frac{g_\kappa}{2} \\
	& = & |g|\e^{\i\gamma}
\end{eqnarray}
and so
\begin{equation}
	S_{11} = 1 - \kappa_\mathrm{ext}\chi_\mathrm{c}\left(1 - g_\omega |g|\e^{\i\gamma} \chi_\mathrm{c}\chi_0^\mathrm{eff}\right).
	\label{eqn:PPIT_th_gamma}
\end{equation}
This form very much illustrates that the resonance circle of the PPIT, described by the second term in parenthesis, is rotated by $\gamma$ around its anchor point.
\subsection{The strong-coupling regime}
\label{sec:Note6C}
Similar to the dispersive-only case, we can calculate the normal modes of the system in the presence of dissipative coupling (even though we cannot yet reach the strong-coupling regime in the experiment).
We do this for simplicity again for the sideband-resolved regime $4\Omega_0^2 \gg \kappa_0^2$.
The effective susceptibility of the HF mode in this case is
\begin{equation}
	\chi_\mathrm{c}^\mathrm{eff} = \frac{\chi_\mathrm{c}}{1 + g_\omega g \chi_\mathrm{c}\chi_\mathrm{0}}.
\end{equation}
\begin{figure*}[b!]
	\includegraphics{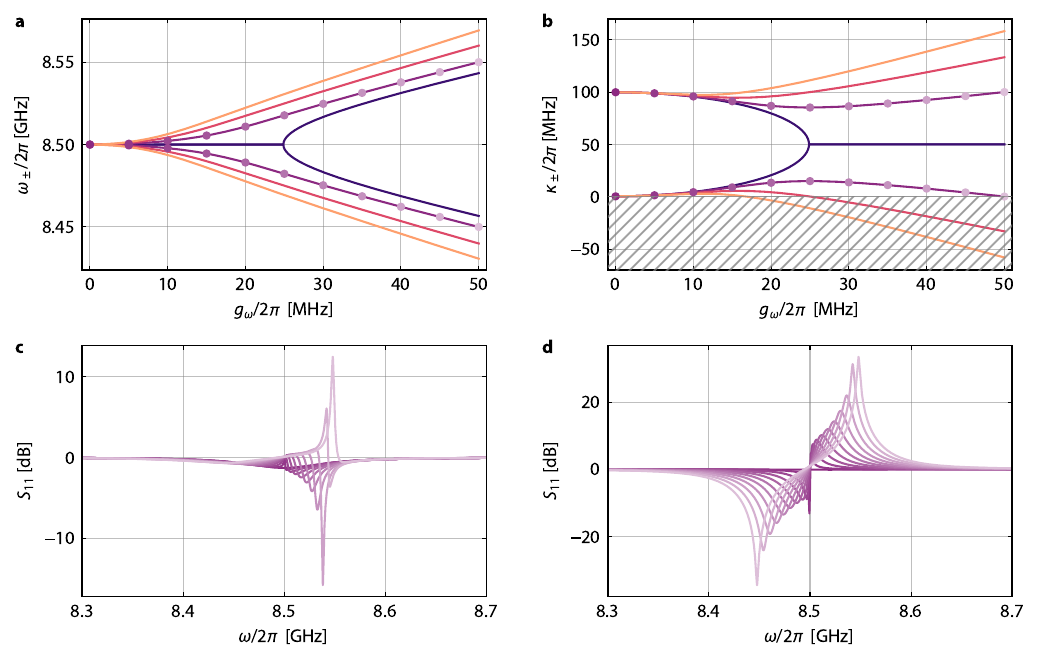}
	\titlecaption{Normal modes and amplification in the dispersive-plus-dissipative strong-coupling regime}{%
		\sublabel{a}~Normal-mode resonance frequencies $\omega_\pm = \mathrm{Re}(\tilde{\omega}_\pm)$ and \sublabel{b}~normal-mode linewidths $\kappa_\pm = 2\mathrm{Im}(\tilde{\omega}_\pm)$ of a photon-pressure system with dispersive-plus-dissipative coupling for the four ratios $g_\kappa/g_\omega = (0, -2, -4, -6)$ according to Eq.~(\ref{eqn:NMS}); from innermost to outermost curve $g_\kappa/g_\omega$ increases. System parameters are chosen close to those of our device, cf.~sections \ref{sec:Note6A} and \ref{sec:Note6B}, pump frequency is $\omega_\mathrm{p} = \omega_0 - \Omega_0$. For vanishing $g_\kappa$ (dark blue curves) the usual normal-mode splitting is observed, where for the coupling strength $g_\omega < (\kappa_0 - \Gamma_0)/4$ the resonance frequencies of the two modes coincide and the two linewidths differ, while for $g_\omega > (\kappa_0 - \Gamma_0)/4$ it is vice versa with both modes having the linewidth $(\kappa_0 + \Gamma_0)/2$. For finite $g_\kappa > 0$ these degeneracies are lifted and neither the frequencies nor the linewidths coincide for any $g_\omega$. Instead, both quantities show a linear splitting for large $g_\omega$, which naturally leads to a critical $g_\omega$, above which $\kappa_- < 0$, i.e., to an instability. When this point is approached from below, one of the two normal modes becomes a parametric amplifier, as can also be seen in panels \sublabel{c} and \sublabel{d}, where $S_{11}$ of the HF mode is plotted for eleven different $g_\omega$ between $2\pi\times\qty{50}{\hertz}$ and $2\pi\times\qty{50}{\mega\hertz}$ for the case $g_\kappa/g_\omega = -2$. The corresponding points are also marked in \sublabel{a} and \sublabel{b} with circles. In \sublabel{c} we show the case of $\kappa_\mathrm{ext} = 0.07\kappa_0$, which corresponds approximately to our current device, and in \sublabel{d} we use $\kappa_\mathrm{ext} = 0.99\kappa_0$, as one would choose for a dedicated parametric amplifier; $\kappa_0 = 2\pi\times\qty{100}{\mega\hertz}$.
	}
	\label{fig:FigS13}
\end{figure*}
The resonances can be found by the condition
\begin{equation}
	1 + g_\omega g \chi_\mathrm{c}\chi_0 = 0
\end{equation}
or
\begin{equation}
	\frac{1}{\chi_\mathrm{c}\chi_0} + g_\omega g  = 0.
\end{equation}
The latter is equivalent to a quadratic equation in $\omega$ and we can find the two complex-valued solutions $\tilde{\omega}_\pm$ from
\begin{eqnarray}
	0  & = & \left[\frac{\Gamma_0}{2} + \i\left(\tilde{\Omega} - \Omega_0 \right)\right]\left[\frac{\kappa_0}{2} + \i\left(\tilde{\omega} - \omega_0\right)\right] + g_\omega^2 + \i\frac{g_\omega g_\kappa}{2} \\
	& = & \left[\frac{\Gamma_0}{2} + \i\left(\tilde{\omega} - \omega_0 - \delta \right)\right]\left[\frac{\kappa_0}{2} + \i\left(\tilde{\omega} - \omega_0\right)\right] + g_\omega^2 + \i\frac{g_\omega g_\kappa}{2} \\
	& = & -\tilde{\omega}^2 + \tilde{\omega}\left[2\omega_0 + \delta + \i\frac{\kappa_0 + \Gamma_0}{2}\right] + \frac{\kappa_0\Gamma_0}{4} - \i\frac{\kappa_0 + \Gamma_0}{2}\omega_0 - \i\frac{\kappa_0}{2}\delta - \omega_0^2 - \omega_0\delta + g_\omega^2 + \i\frac{g_\omega g_\kappa}{2}
\end{eqnarray}
where $\delta = \omega_\mathrm{p} - \left( \omega_0 - \Omega_0 \right)$.
The two solutions are
\begin{equation}
	\tilde{\omega}_\pm = \omega_0 + \frac{\delta}{2} + \i\frac{\kappa_0 + \Gamma_0}{4} \pm \sqrt{g_\omega^2 + \i\frac{g_\omega g_\kappa}{2} - \left(\frac{\kappa_0 - \Gamma_0 + 2\i\delta}{4}\right)^2}
	\label{eqn:NMS}
\end{equation}
which differ from the dispersive normal-mode-splitting only via the term $\i g_\omega g_\kappa/2$ in the square-root.
The actual resonance frequencies of the two normal modes are then given by $\omega_\pm = \mathrm{Re}(\tilde{\omega}_\pm)$ and the linewidths by $\kappa_\pm = 2\mathrm{Im}(\tilde{\omega}_\pm)$.
The new term in the square root has considerable consequences, and we will discuss them for the most common case $\delta = 0$.
The additional term leads to the two normal modes splitting in resonance frequency for any $g_\omega > 0$ and never snapping to the same linewidth as would be the case for $g_\kappa = 0$.
Furthermore, the linewidth of one normal mode will decrease again after an initial increase until it becomes negative for sufficiently large $g_\omega$.
The two solutions of Eq.~(\ref{eqn:NMS}) in the case $\delta = 0$ are shown in Supplementary Fig.~\ref{fig:FigS13} for the usual four different $g_\kappa/g_\omega = (0, -2, -4, -6)$, showing the just described signatures.
Note that it is always possible to find a detuning $\delta = 2g_\omega g_\kappa/(\kappa_0 - \Gamma_0)$ for any $g_\omega g_\kappa/2$, such that the imaginary parts in the square root cancel and $\kappa_+ = \kappa_-$.
This however, is still different from the dispersive-only case for $\delta = 0$ and the two normal-modes will not be equal in e.g.~$|S_{11}|$ as would be the case for $g_\omega = 0$; they will furthermore not split linearly in frequency for large $g_\omega$.
We also plot $S_{11}$ of the HF mode for various pump strengths in the case $g_\kappa/g_\omega = -2$.
For comparison, we show the reflection for two different cases $\kappa_\mathrm{ext} = 2\pi\times \qty{7}{\mega\hertz}$, i.e., our strongly undercoupled experimental device, but also for $\kappa_\mathrm{ext} = 2\pi\times\qty{99}{\mega\hertz}$, i.e., a hypothetical overcoupled device.
All other parameters are identical to the subsections~\ref{sec:Note6A} and \ref{sec:Note6B}.
For both cases, one of the two normal modes becomes an amplifier, when the coupling is increased, a direct consequence of the decreasing decay rate $\kappa_-$ approaching an instability point.
It will be interesting to test this prediction in future devices and to investigate the noise properties of such an amplifier, especially since it could be implemented with mechanical oscillators, i.e., Josephson-free and therefore with a large dynamic range, and with a larger bandwidth than many other optomechanical amplifiers.
\subsection{Impact of an additional external-dissipative coupling}
\label{sec:Note6D}
As a last subsection here, we briefly discuss the expected modifications in case the external-dissipative coupling were not negligible.
For simplicity, we restrict ourselves again to the sideband-resolved regime with a highly undercoupled HF mode, since this is the relevant case for our experiment.
Also, we will only consider the case of a pump tone around the red HF mode sideband and omit all HF mode nonlinearities.
The derivation of the equations of motion with both $g_{0\kappa_\mathrm{int}}$ and  $g_{0\kappa_\mathrm{ext}}$ would be completely analogous to the one presented in Supplementary Note~\ref{sec:Note5}.
The only difference in Eqs.~(\ref{eqn:EOM_NL_a}) and (\ref{eqn:EOM_NL_b}) would be that each term containing $g_{0\kappa}$ will show up twice, once with $g_{0\kappa_\mathrm{int}}$ and once with $g_{0\kappa_\mathrm{ext}}$.
Correspondingly, the input terms for the $g_{0\kappa_\mathrm{ext}}$ terms will be $\hat{a}_\mathrm{ext, in}$ instead of $\hat{a}_\mathrm{int, in}$.
The main difference is, that these new input terms cannot be dropped in the linearization, since $\hat{a}_\mathrm{ext, in}$ contains the high-amplitude coherent pump tone and not just noise.
Hence, after linearization and dropping irrelevant terms, we get in the sideband-resolved limit
\begin{align}
	\dot{\hat{c}} & = \left(-\i\Delta_\mathrm{p} - \frac{\kappa_0}{2}\right)\hat{c} - \i\left(g_\omega + \i\frac{g_\kappa}{2}\right)\hat{d} - \frac{\i g_{0\kappa_\mathrm{ext}}\alpha_\mathrm{in}\e^{\i\phi_\mathrm{p}}}{2\sqrt{\kappa_\mathrm{ext}}}\hat{d} + \i\sqrt{\kappa_\mathrm{ext}}\hat{c}_\mathrm{in} \label{eqn:EOM_c_inclgext}\\
	\dot{\hat{d}} & = \left(\i\Omega_0 - \frac{\Gamma_0}{2} \right)\hat{d} - \i g_\omega\hat{c} - \frac{\i g_{0\kappa_\mathrm{ext}}\alpha_\mathrm{in}\e^{-\i\phi_\mathrm{p}}}{2\sqrt{\kappa_\mathrm{ext}}}\hat{c} + \i\sqrt{\Gamma_\mathrm{ext}}\hat{d}_\mathrm{in} - \i\frac{g_{\kappa_\mathrm{ext}}}{2\sqrt{\kappa_\mathrm{ext}}}\hat{c}_\mathrm{in} \label{eqn:EOM_d_inclgext}
\end{align}
with the total dissipative multi-photon coupling rate
\begin{equation}
	g_\kappa = g_{\kappa_\mathrm{int}} + g_{\kappa_\mathrm{ext}}
\end{equation}
The two new terms with $\alpha_\mathrm{in}$ originate from a direct coupling between the pump tone on the HF mode feedline and the LF circuit.
The third new term, the last one in the EOM for the LF mode, describes a direct excitation of the LF mode by the beating between the intracavity pump field and the probe tone on the feedline.
We can obtain the relation between $\alpha_0$ and $\alpha_\mathrm{in}$ in good approximation for the bare cavity as
\begin{equation}
	\alpha_0 = \frac{\i\sqrt{\kappa_\mathrm{ext}}}{\frac{\kappa_0}{2} + \i\Delta_\mathrm{p}}\alpha_\mathrm{in}\e^{\i\phi_\mathrm{p}}
\end{equation}
or in the sideband-resolved limit and with red-sideband pumping $|\Delta_\mathrm{p}| \gg \kappa_0/2$
\begin{equation}
	\alpha_\mathrm{in} \approx \frac{\Delta_\mathrm{p}}{\sqrt{\kappa_\mathrm{ext}}}\alpha_0, ~~~~~\phi_\mathrm{p} \approx 0.
\end{equation}
Plugging this result and its complex conjugate into the equations of motion leads to
\begin{align}
	\dot{\hat{c}} & = \left(-\i\Delta_\mathrm{p} - \frac{\kappa_0}{2}\right)\hat{c} - \i\left(g_\omega + \i\frac{g_\kappa}{2}\right)\hat{d} - \i g_{\kappa_\mathrm{ext}}\frac{\Delta_\mathrm{p}}{2\kappa_\mathrm{ext}}\hat{d} + \i\sqrt{\kappa_\mathrm{ext}}\hat{c}_\mathrm{in} \\
	\dot{\hat{d}} & = \left(\i\Omega_0 - \frac{\Gamma_0}{2} \right)\hat{d} - \i g_\omega\hat{c} - \i g_{\kappa_\mathrm{ext}}\frac{\Delta_\mathrm{p}}{2\kappa_\mathrm{ext}}\hat{c} + \i\sqrt{\Gamma_\mathrm{ext}}\hat{d}_\mathrm{in} - \i\frac{g_{\kappa_\mathrm{ext}}}{2\sqrt{\kappa_\mathrm{ext}}}\hat{c}_\mathrm{in}
\end{align}
and after defining the effective dispersive coupling rate
\begin{equation}
	\overline{g}_\omega = g_\omega + g_{\kappa_\mathrm{ext}}\frac{\Delta_\mathrm{p}}{2\kappa_\mathrm{ext}}
\end{equation}
the EOMS read
\begin{align}
	\dot{\hat{c}} & = \left(-\i\Delta_\mathrm{p} - \frac{\kappa_0}{2}\right)\hat{c} - \i\left(\overline{g}_\omega + \i\frac{g_\kappa}{2}\right)\hat{d} + \i\sqrt{\kappa_\mathrm{ext}}\hat{c}_\mathrm{in} \label{eqn:EOM_c_inclgext_eff}\\
	\dot{\hat{d}} & = \left(\i\Omega_0 - \frac{\Gamma_0}{2} \right)\hat{d} - \i \overline{g}_\omega\hat{c} + \i\sqrt{\Gamma_\mathrm{ext}}\hat{d}_\mathrm{in} - \i\frac{g_{\kappa_\mathrm{ext}}}{2\sqrt{\kappa_\mathrm{ext}}}\hat{c}_\mathrm{in}. \label{eqn:EOM_d_inclgext_eff}
\end{align}
Hence, aside from the additional input term in the LF EOM ${\propto}\hat{c}_\mathrm{in}$, we recover the equations of motion without external-dissipative coupling, just with a re-scaled and explicitly detuning-dependent dispersive coupling rate $\overline{g}_\omega$.
If the signs of $\Delta_\mathrm{p}$ and $g_{0\kappa_\mathrm{ext}}$ are identical, the effective dispersive coupling rate $\overline{g}_\omega$ is enhanced compared to $g_\omega$, if the signs are opposite $\overline{g}_\omega$ is reduced.
Note that this result is in agreement with Ref.~\cite{Primo2023x}, and also explains why the external-dissipative coupling there did not lead to significant Fano interference in the transparency data, since there $g_{\kappa_\mathrm{int}} = 0$ and $g_{\kappa_\mathrm{ext}}/g_\omega \sim 0.01$, i.e., $g_\kappa$ in Eq.~(\ref{eqn:EOM_c_inclgext_eff}) can be neglected.
In other words, the only relevant coupling term was the detuning-enhanced $\overline{g}_\omega$.
For the calculation of dynamical backaction we can set $\hat{c}_\mathrm{in} = 0$, and we get an identical result to vanishing external coupling, just with $\overline{g}_\omega$ instead of $g_\omega$.
For PPIT the situation is slightly more complicated, since have not only the additional probe-tone term in the LF EOM, but also a modified input-output relation at the probe tone frequency
\begin{equation}
	\hat{c}_\mathrm{out} = \hat{c}_\mathrm{in} + \i\sqrt{\kappa_\mathrm{ext}}\hat{c} - \i \frac{g_{\kappa_\mathrm{ext}}}{2\sqrt{\kappa_\mathrm{ext}}}\hat{d}.
\end{equation}
Solving the EOMs by Fourier transformation, using $\overline{g} = \overline{g}_\omega + \i g_\kappa/2$ as well as $\hat{d}_\mathrm{in} = 0$, and combining the equations leads to
\begin{align}
	\hat{c} & = \i\sqrt{\kappa_\mathrm{ext}}\chi_\mathrm{c}^\mathrm{eff}\left[1 + \i \frac{\overline{g}g_{\kappa_\mathrm{ext}}}{2\kappa_\mathrm{ext}}\chi_0\right]\hat{c}_\mathrm{in} \\
	\hat{d} & = \sqrt{\kappa_\mathrm{ext}}\chi_0^\mathrm{eff}\left[\overline{g}_\omega \chi_\mathrm{c} - \i\frac{g_{\kappa_\mathrm{ext}}}{2\kappa_\mathrm{ext}}\right]\hat{c}_\mathrm{in}
\end{align}
with
\begin{equation}
	\chi_\mathrm{c}^\mathrm{eff} = \frac{\chi_\mathrm{c}}{1 + \overline{g}_\omega \overline{g}\chi_0\chi_\mathrm{c}}, ~~~~~ \chi_0^\mathrm{eff} = \frac{\chi_0}{1 + \overline{g}_\omega \overline{g}\chi_0\chi_\mathrm{c}}.
\end{equation}
Finally, we can combine everything to find
\begin{align}
	S_{11} & = \frac{\langle \hat{c}_\mathrm{out} \rangle}{\langle \hat{c}_\mathrm{in} \rangle} \\
	& = 1 - \frac{g_{\kappa_\mathrm{ext}}^2}{2\kappa_\mathrm{ext}}\chi_0^\mathrm{eff} - \kappa_\mathrm{ext}\chi_\mathrm{c}^\mathrm{eff}\left[ 1 + \i\frac{\left(\overline{g}_\omega  + \overline{g}\right)g_{\kappa_\mathrm{ext}}}{2\kappa_\mathrm{ext}}\chi_0\right].
\end{align}
For parameters like in our experiment, we can approximate this further, if we reasonably assume that $|g_{0\kappa_\mathrm{ext}}|/2\pi < \qty{0.55}{\kilo\hertz}$, cf.~Supplementary Note~\ref{sec:Note3}.
This would correspond to $|g_{\kappa_\mathrm{ext}}|/2\pi < \qty{30}{\kilo\hertz}$ at the highest experimental pump photon numbers of ${\sim}2500$, and for the second term in the reflection equation we would find $g_{\kappa_\mathrm{ext}}^2|\chi_0^\mathrm{eff}|/2\kappa_\mathrm{ext} < 10^{-3}$.
At least for the lower pump photon number range, also the second term in the square brackets can be neglected, which is ${\lesssim}0.05 + 0.04\i$ for $n_\mathrm{c} \sim 2500$ (calculated at $\Omega = \Omega_0$) and is linear in $n_\mathrm{c}$.
Hence, in this approximation, we get
\begin{equation}
	S_{11} = 1 - \kappa_\mathrm{ext}\chi_\mathrm{c}^\mathrm{eff}
\end{equation}
which is formally identical to the case $g_{\kappa_\mathrm{ext}} = 0$.
It is not the same though as for $g_{\kappa_\mathrm{ext}} = 0$, the coupling rate $\overline{g}_\omega$ inside of $\chi_\mathrm{c}^\mathrm{eff}$ scales differently as a function of pump detuning compared to $g_\omega$, and so would the dynamical backaction, which even has a contribution ${\propto}\overline{g}_\omega^2$, cf. Eq.~(\ref{eqn:Gpp_lin}).
The difference would be expected to be maximal for the largest detunings between pump and HF mode and the dispersive coupling rates $g_\omega$ (we omit the primes here to avoid confusion and remain consistent within this section) we extracted from the PPIT data would not fall on top of a single line as in main paper Fig.~4\textbf{c}, but spread out due to the detuning-dependent scaling.
Within each power (color in main paper Fig.~4), the point of the lowest photon number would be pushed to higher $g_\omega$, while the points for the largest photon numbers would be pulled to smaller $g_\omega$ compared to the fit line through all of them, in case $g_{0\kappa_\mathrm{ext}}<0$.
For $g_{\kappa_\mathrm{ext}} > 0$ the trends would be opposite.
Furthermore, for the lowest photon numbers and largest detunings, i.e., the leftmost points in main paper Fig.~4\textbf{c}, $\overline{g}_\omega$ would be more than 1.5 times as large as $g_\omega$ if $g_\mathrm{0\kappa_\mathrm{ext}} /2\pi \sim \qty{-0.55}{\kilo\hertz}$ as suggested by the fit line in Supplementary Fig.~\ref{fig:FigS6}.
The dissipative coupling rate $g_\kappa$ would be essentially unmodified though, and hence the experimentally determined ratio $g_\kappa/g_\omega$ would not approach the value obtained from the flux arcs.
Since we observe none of these consequences, we believe that our conclusions -- the flux tuning of $\kappa_\mathrm{ext}$ is an analysis artifact and the external-dissipative coupling rate is truly negligible -- are well justified.
It will be very interesting to design and investigate such an external-dissipative coupling in the future though, also beyond the sideband-resolved limit and with adjustable sign, since there seems rich physics waiting to be explored.
\section{Supplementary Note VII: Data analysis and fitting routines}
\label{sec:Note7}
\subsection{Background correction and basic fitting routine}
\label{sec:Note7A}
\begin{figure*}
	\includegraphics{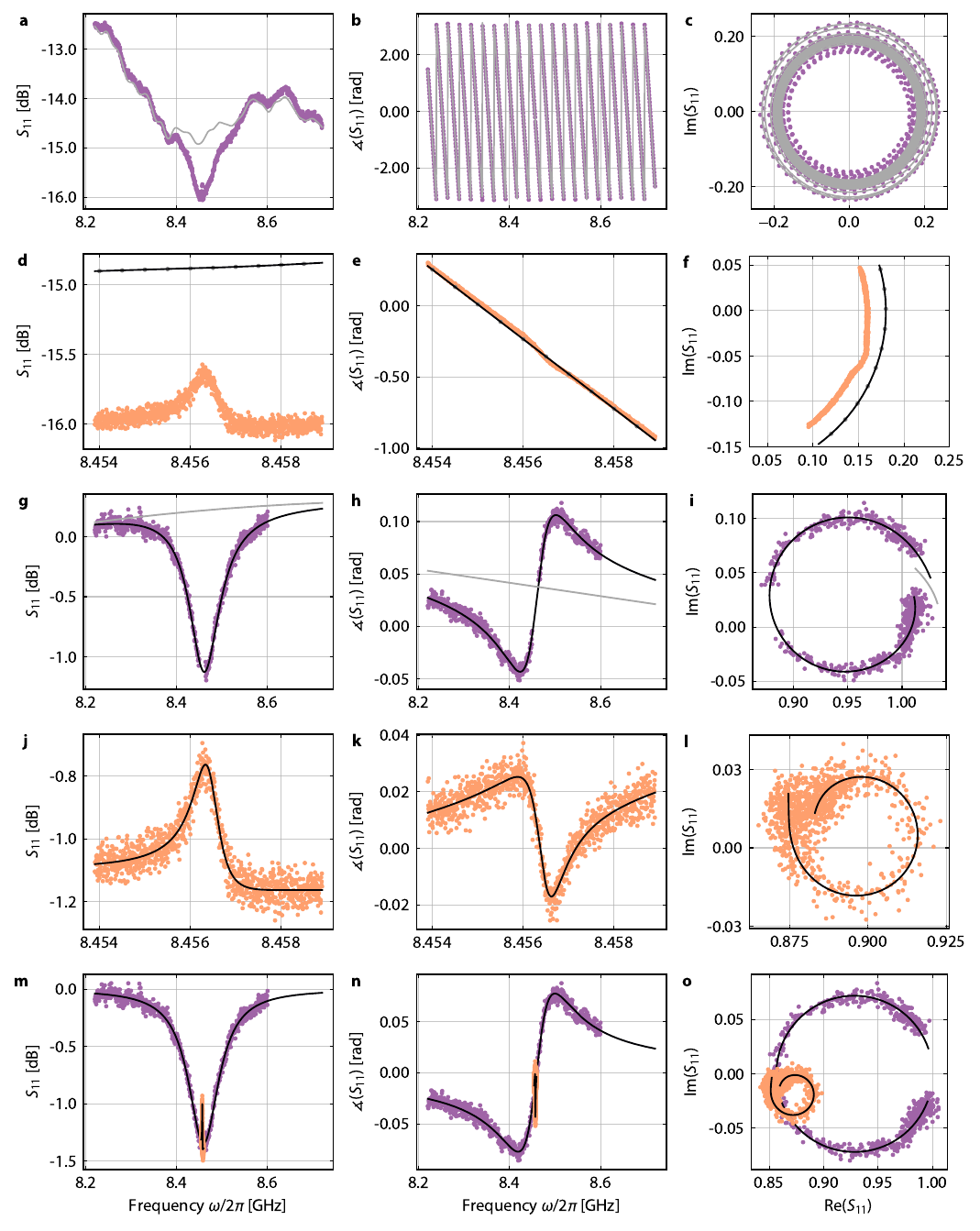}
	\titlecaption{Multi-step background correction procedure}{%
		Details are given in the text.
	}
	\label{fig:FigS14}
\end{figure*}
In an ideal world, the reflection data of LF mode, HF cavity and PPIT experiment would be described by the response functions Eqs.~(\ref{eqn:S11_LF}, \ref{eqn:S11_HF}, \ref{eqn:S11_PPIT}), respectively.
However, the actual signal is considerably modified by various accumulating effects of the experimental setup and multiple system components.
The cables have a finite length and a frequency-dependent attenuation, we have attenuators, directional couplers, amplifiers, dc blocks, wirebonds, a PCB and lots of SMA connectors mounted into the signal path, which all combined lead to attenuation, cable resonances, impedance mismatches and parasitic reflections or imperfect isolation, cf.~e.g.~Refs.~\cite{Wenner2011x, Rieger2023x}.
As a consequence, the actual reflection is approximately described by
\begin{equation}
	S_{11}^\mathrm{real} = A(\omega)\left[1 - f(\omega)\e^{\i\theta(\omega)}\right]\e^{\i\alpha(\omega)},
	\label{eqn:S11_real}
\end{equation}
where the ideal-world response would be
\begin{equation}
	S_{11}^\mathrm{ideal} = 1 - f(\omega)
\end{equation}
and $f(\omega)$ is defined by the experiment performed.
Here $A(\omega)$ is a real-valued and frequency-dependent attenuation/gain of the signal, $\alpha(\omega)$ is a frequency-dependent and real-valued phase accumulated by the signal on the path and $\theta(\omega)$ is a potentially frequency-dependent Fano-angle originating from impedance mismatches and/or interference with parasitic signal paths.
If the cavity has a high $Q$, i.e., its resonance is spread over a small frequency range ${\sim} 2\kappa_0$, it is usually sufficient to approximate $A(\omega)$ as a second-order polynomial, and $\alpha$ as a linear function of $\omega$.
Here, this is true for the direct measurement of the LF mode, but not for the HF cavity, and so we need a more elaborate approach to handle HF data.
The HF cavity has a linewidth of $\kappa_0 \sim 2\pi\times \qty{100}{\mega\hertz}$ and the background over such a large span can have a much more complicated shape than a second-order polynomial, cf.~Supplementary Fig.~\ref{fig:FigS14}\sublabel{a}.
To compensate for this, we perform a multi-step background correction for the HF cavity and the PPIT zoom-ins, which is illustrated step-by-step in Supplementary Fig.~\ref{fig:FigS14}.
To facilitate our approach, we take a high-power VNA trace $S_{11}^\mathrm{bg, exp}$ in addition to each low-power VNA dataset $S_{11}^\mathrm{exp}$ during the experiments, and the two are completely identical with respect to frequency points.
We choose the VNA power of the high-power trace sufficiently large, that the cavity absorption dip has vanished due to its nonlinear damping and its Kerr frequency shift (we verify the vanished cavity by adding even more power and check if the transmission is still changing).
In other words, $f(\omega) = 0$ and we measure the background reflection
\begin{equation}
	S_{11}^\mathrm{bg, exp} = \tilde{A}(\omega)\e^{\i\tilde{\alpha}(\omega)}.
\end{equation}
We label the (new) functions $\tilde{A}, \tilde{\alpha}$ with a tilde to indicate that they might be slightly different from the ones in the low-power regime, in our case for instance due to a slight cryo-HEMT compression.
The two traces, low-power and high-power, respectively, are shown for one of our datasets in Supplementary Fig.~\ref{fig:FigS14}\sublabel{a}-\sublabel{c} in amplitude, phase and complex representation.
The gray line is the high-power trace and the purple symbols are the low-power trace with the cavity resonance.
For the PPIT zooms, we go a slightly different way to have fully consistent backgrounds for wideband and narrowband data.
We take the high-power dataset of the HF cavity and select from it the corresponding narrow frequency range measured in each PPIT zoom.
Since this narrow frequency window of the HF dataset has a much smaller frequency resolution than the PPIT dataset, we fit the HF background in this span with a polynomial for both the amplitude and the phase.
This works very well, since the background does not vary much in such small frequency windows.
With the obtained fit parameters, we can calculate the background for all the experimental PPIT frequency points.
One example for a measured PPIT zoom with both its corresponding HF-snipped experimental background and its fit are shown in panels \sublabel{d}-\sublabel{f}.
Circle symbols are HF background frequency points, gray line is the fit for the HF background, and orange symbols are the low-power PPIT data.
Using the two datasets for each VNA scan (measured data and measured or fitted background, respectively), we perform as a next step a complex division and get as result
\begin{eqnarray}
	S_{11}^\mathrm{cor} & = & \frac{S_{11}^\mathrm{exp}}{S_{11}^\mathrm{bg, exp}} \\
	& = & \frac{A(\omega)}{\tilde{A}(\omega)}\left[1 - f(\omega)\e^{\i\theta(\omega)}\right]\e^{\i\left[\alpha(\omega) - \tilde{\alpha}(\omega)\right]} \\
	& = & B(\omega)\left[1 - f(\omega)\e^{\i\theta(\omega)}\right]\e^{\i\phi(\omega)}
\end{eqnarray}
where we usually get a very weak frequency dependence of the remaining parameters $B(\omega) \approx 1$ and $\phi(\omega) \approx 0$ for the cavity, cf.~panels \sublabel{g}-\sublabel{i}, which show $S_{11}^\mathrm{cor}$ in amplitude, phase and complex representation of the original data in \sublabel{a}-\sublabel{c}.
The corresponding first-step corrected PPIT data are shown in panels \sublabel{j}-\sublabel{l}.
In principle, we can now fit the cavity dataset to obtain $B(\omega)$, $\phi(\omega)$ and $\theta(\omega)$, but in practice we need to set $\theta(\omega) = \theta$ to be frequency-independent for our fit routine to run properly.
Furthermore, we remove the data points, for which $\omega > 2\pi\times\qty{8.6}{\giga\hertz}$, since there seems to be another small resonance, which modifies the cavity fits.
For the first cavity fit, we remove the PPIT window from the cavity dataset, i.e., we remove all data points within a span of $15\Gamma_\mathrm{eff}$ around $\Omega_\mathrm{eff}$ from the cavity scan.
Then, we fit the remaining datapoints with
\begin{equation}
	S_{11}^\mathrm{cor, fit} = B(\omega)\left[1 - f(\omega)\e^{\i\theta}\right]\e^{\i\phi(\omega)}
	\label{eqn:S11_corr_fitA},
\end{equation}
where
\begin{eqnarray}
	B(\omega) & = & b_0 + b_1\omega + b_2\omega^2 \label{eqn:S11_corr_fitB}\\
	\phi(\omega) & = & \phi_0 + \phi_1 \omega \label{eqn:S11_corr_fitC}\\
	f(\omega) & = & \frac{\kappa_\mathrm{ext}}{\frac{\kappa_0}{2} + \i(\omega - \omega_0)}
	\label{eqn:S11_corr_fitD}
\end{eqnarray}
and get $b_0, b_1, b_2, \phi_0, \phi_1$ and $\theta$ as fit parameters.
Simultaneously, we get first values for the cavity parameters $\omega_0, \kappa_\mathrm{ext}$ and $\kappa_0$ (or $\omega_0'$, $\kappa_0'$, and $\kappa_\mathrm{ext}'$, but we do not explicitly mention the primed quantities here).
We perform this procedure for all sideband pump powers and detunings.
During this procedure, we find a small and nearly constant $\theta = 0.08$ for most datasets, only for the highest pump powers and largest PPIT signatures, the values deviate considerably from that, which we attribute to the large PPIT signature modifying the shape of the cavity even beyond the cutout window of $15\Gamma_\mathrm{eff}$.
To deal with this, we re-run the fitting routine with Eqs.~(\ref{eqn:S11_corr_fitA})-(\ref{eqn:S11_corr_fitD}), but keep $\theta = \theta_0 = 0.08$ fixed for all fits this time.
As a result, we have the updated background $S_{11}^\mathrm{bg, fit} = (b_0 + b_1\omega + b_2\omega^2)\e^{\i(\phi_0 + \phi_1\omega)}$ as well as the updated cavity parameters $\omega_0, \kappa_\mathrm{ext}$ and $\kappa_0$ from this second fit round.
Then, we divide all experimentally corrected datasets by this fit background and rotate off the $\theta_0$ angle
\begin{equation}
	S_{11}^\mathrm{final} = \left(\frac{S_{11}^\mathrm{cor}}{S_{11}^\mathrm{bg, fit}} - 1\right)\e^{-\i\theta_0} + 1.
\end{equation}
We do this for all HF datasets equally, wideband or narrowband alike, and the result of this multi-step correction $S_{11} = S_{11}^\mathrm{final}$ is what is plotted as HF resonance or PPIT resonance in all figures of the main paper and Supplementary Material, except for the resonance before cJJ cutting in Supplementary Fig.~\ref{fig:FigS3}.
The fully adjusted and rotated data and fits are also included for the exemplary dataset of Supplementary Fig.~\ref{fig:FigS14} as panels \sublabel{m}-\sublabel{o}, wideband and narrowband data are overlaid in the graph to also show that they form one consistent dataset.
\vspace{-2mm}
\subsection{PPIT fitting routine}
\label{sec:Note7B}
\vspace{-2mm}
As described in the previous section, we first remove the PPIT signature from the cavity resonance data, and then fit the background-corrected cavity twice with Eqs.~(\ref{eqn:S11_corr_fitA})-(\ref{eqn:S11_corr_fitD}), in the first step with $\theta$ as fit parameter, in the second step with fixed $\theta_0$ obtained from the $\theta$ analysis.
From this routine, we obtain $\omega_0$, $\kappa_0$ and $\kappa_\mathrm{ext}$ in the last step, and we keep those parameters fixed for all subsequent analysis steps.
Note that we just generically use the unprimed variables in this and the previous subsection, in reality of course we obtain e.g.~$\omega_0'$, $\kappa_0'$ and $\kappa_\mathrm{ext}'$ in case of a sideband pump.
Similarly, we fit the PPIT zoom-ins with Eqs.~(\ref{eqn:S11_corr_fitA})-(\ref{eqn:S11_corr_fitD}) once and with $\theta$ being a fit parameter, and obtain from that fit $\Gamma_\mathrm{eff}$ and $\Omega_\mathrm{eff}$.
All further fit parameters like the background contributions, the Fano-angle and $\Gamma_\mathrm{ext}$ are meaningless for the zoom fit and will be dropped, they just describe the small part of the cavity absorption dip, in which the PPIT signature is located for each pump frequency and power.
This cavity-resonance-background we of course want to keep and thus we do not remove it from the PPIT datasets.
Then, we generate a combined single datafile from the completely background-corrected wideband and narrowband datasets and fit this combined dataset with the PPIT reflection Eq.~(\ref{eqn:PPIT_th_plot}).
The individual resonance parameters $\omega_0$, $\kappa_0$, $\kappa_\mathrm{ext}$, $\Gamma_\mathrm{eff}$ and $\Omega_\mathrm{eff}$ are fixed from the previous fits and the remaining two unknowns $g_\omega$ and $g_\kappa$ are obtained as fit parameters.
The result of combining the data and the corresponding $S_{11}$ PPIT-fits are used for main paper Fig.~3 and in Sec.~\ref{sec:Note8A}.
Now, we have the complete multi-photon parameter set to describe the PPIT data $\omega_0, \kappa_0, \kappa_\mathrm{ext}, \Omega_\mathrm{eff}, \Gamma_\mathrm{eff}, g_\omega$ and $g_\kappa$.
We limit our PPIT analysis to the pump detuning range $\delta\in [-\kappa_0'/4, +\kappa_0'/4]$, because for $\delta < \kappa_0'/4$ the transparency resonance gets very small and for $\delta > \kappa_0'/4$ we approach the dissipative red-sideband instability at high pump powers, cf.~Sec.~\ref{sec:Note6A}.

\vspace{-2mm}
\subsection{LF reflection fitting routine for dynamical backaction}
\label{sec:Note7C}
\vspace{-2mm}
The LF reflection datasets taken during the experiment for the characterization of dynamical backaction, cf.~main paper Fig.~5\textbf{b} for one example, are first fitted using Eqs.~(\ref{eqn:S11_corr_fitA})-(\ref{eqn:S11_corr_fitD}), i.e., with a linear circuit response.
Of course, we use different variables and quantities here, i.e. $\kappa_0 \rightarrow \Gamma_\mathrm{eff}$, $\kappa_\mathrm{ext} \rightarrow \Gamma_\mathrm{ext}$, $\omega \rightarrow \Omega$ and $\omega_0 \rightarrow \Omega_\mathrm{eff}$.
In a next step, the background is divided off and a possible $\theta$ is rotated off; the background-corrected datasets are stored.
Although the data were taken at a sufficiently low LF probe power that without dynamical backaction there are no signs of the intrinsic LF Kerr nonlineaerity in the response resonances, we found that close to the parametric instability, the resonances slightly and asymmetrically tilt to the left as typical for Duffing/Kerr nonlinearities.
Hence, we re-fit the background-corrected resonances using Eqs.~(\ref{eqn:charac_poly_beta})-(\ref{eqn:phi_beta}) with the replacements $\Gamma_0 \rightarrow \Gamma_\mathrm{eff}$, $\Omega_0 \rightarrow \Omega_\mathrm{eff}$ and $P_\beta = \qty{-97}{\dBm}$, which is our best estimate for the LF on-chip probe power.
The fit parameters in this fit are $\Gamma_\mathrm{ext}$, $\Gamma_\mathrm{eff}$, $\Omega_\mathrm{eff}$ and $\mathcal{K}_\mathrm{LF}$, of which $\Gamma_\mathrm{eff}$ and $\Omega_\mathrm{eff}$ are used for plotting and further analyses, cf.~main paper Fig.~5 and Supplementary Fig.~\ref{fig:FigS17}.
The data points, that are actually impacted by the slight LF self-Kerr tilt are roughly the last $10$ points of each pump power before instability sets in; note also that mainly $\Omega_\mathrm{eff}$ is modified by this, but not $\Gamma_\mathrm{eff}$.
\vspace{-2mm}
\subsection{Photon number calibration}
\vspace{-2mm}
For much of our analysis, it is important to know the pump photon number $n_\mathrm{c}$, in particular for the analysis of the nonlinearity-enhanced coupling rates and the dynamical backaction evaluation.
In principle, we could calculate it via Eqs.~(\ref{eqn:charac_poly}) numerically, if we knew the on-chip pump power, i.e., the attenuation between the signal generator providing the pump tone and the chip, as well as the nonlinear coefficients $\mathcal{K}, \kappa_1$, $\kappa_2$ and $\kappa_3$.
Unfortunately, we do not.
This is obvious for the nonlinear coefficients, especially for the $\kappa_1, \kappa_2, \kappa_3$, since there is no theory available that could be used to estimate their values, and to extract their values from data, we would need to know $n_\mathrm{c}$ in the first place.
Regarding the attenuation, it can in principle be calibrated with e.g.~a temperature-controlled white noise source at the input of the HEMT, but since we do not have space for a microwave switch in the setup and measure directly in liquid helium (making accurate $T$-control impractical) this was also not a reasonable option.
Furthermore, the attenuation is considerably frequency-dependent over the more than $\qty{300}{\mega\hertz}$ wide range of relevant $\omega_\mathrm{p}$ for main paper Fig.~5, which is supported by the frequency-dependence of the VNA probe tone reflection, cf.~Supplementary Fig.~\ref{fig:FigS14}\sublabel{a}.
Therefore, we use a different approach to get $n_\mathrm{c}$, which only requires the value for $g_{0\omega}$, that we get from the HF flux arc derivative and the LF zero-point-fluctuations, and that we assume the only dispersive nonlinearity of the HF mode to be a Kerr anharmonicity.
From the experiment with a sideband pump tone, we know that the increase of $\kappa_0'$ with respect to $\kappa_0$ is sufficiently small compared to the other frequencies that we can apply the approximations of Sec.~\ref{sec:Note4F} and use Eq.~(\ref{eqn:nc_Kerr}) to calculate
\begin{equation}
	\mathcal{K}n_\mathrm{c} = \frac{ \delta\omega_0\left(2 - \delta\omega_0/\Delta_\mathrm{p} \right)}{4}
	\label{eqn:Kerr_shift_nc}
\end{equation}
where $\delta\omega_0 = \omega_0' - \omega_0$ is the pump-induced HF frequency shift, $\Delta_\mathrm{p} = \omega_\mathrm{p} - \omega_0$ is the detuning between pump tone and unpumped HF cavity and $\mathcal{K}$ is the Kerr anharmonicity.
As a next step, we normalize the total dispersive multiphoton coupling rate data $g_\omega'$ (obtained from the PPIT experiments) as
\begin{equation}
	\frac{g_\omega'}{\sqrt{\mathcal{K}n_\mathrm{c}}} = g_{0\omega}\alpha_\mathcal{K} + g_{0\mathcal{K}}\mathcal{K}n_\mathrm{c}\alpha_\mathcal{K}^3
\end{equation}
where $\alpha_\mathcal{K} = 1/\sqrt{\mathcal{K}}$, and then fit the normalized coupling rates as a function of $\mathcal{K}n_\mathrm{c}$ with $\alpha_\mathcal{K}$ and $g_{0\mathcal{K}}$ as the fit parameters and $g_{0\omega} = 2\pi\times \qty{27.1}{\kilo\hertz}$.
From this approach, we obtain $\mathcal{K} = 2\pi\times\qty{-5.4}{\kilo\hertz}$ and subsequently $n_\mathrm{c}$ for all datasets.
Note, that equivalently we could plug Eq.~(\ref{eqn:Kerr_shift_nc}) into the equation for $g_\omega'$ and then fit $g_\omega'$ as a function of $\delta\omega_0$ with $\mathcal{K}$ (and $g_{0\mathcal{K}}$) as fit parameter.
This approach would give slightly different weights to the data points, but the result for $\mathcal{K}$ and $n_\mathrm{c}$ only differs by~${<}5\%$.
\begin{figure*}
	\includegraphics{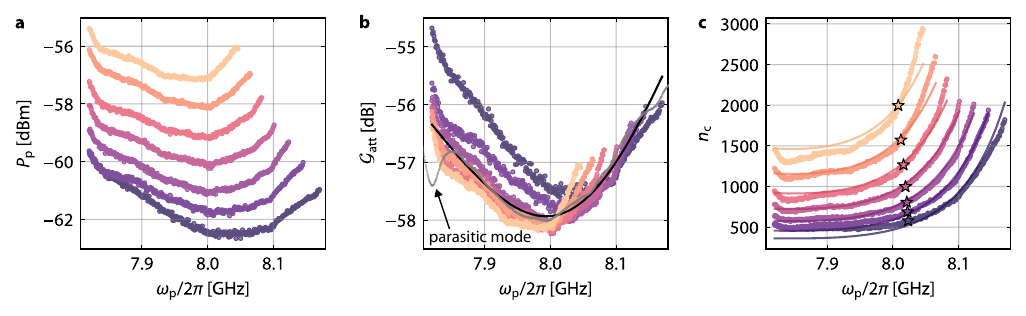}
	\titlecaption{On-chip pump power, pump-line attenuation and intracavity photon number as a function of frequency}{%
		\sublabel{a}~Inferred on-chip pump power as a function of pump frequency calculated with Eq.~(\ref{eqn:P_onchip}), using the experimental HF mode parameters $\omega_0'$, $\kappa_0'$ as well as their fits as a function of $n_\mathrm{c}$. Different colors correspond to different signal generator powers between $P_\mathrm{sg}^\mathrm{min} = \qty{-5}{\dBm}$ and $P_\mathrm{sg}^\mathrm{max} = +\qty{1}{\dBm}$ in steps of $\qty{1}{\decibel}$. Lowest $P_\mathrm{p}$ corresponds to lowest $P_\mathrm{sg}$, highest $P_\mathrm{p}$ corresponds to highest $P_\mathrm{sg}$.
		\sublabel{b}~Pump-line attenuation $\mathcal{G}_\mathrm{att} = P_\mathrm{p}/P_\mathrm{sg}$ as a function of pump frequency $\omega_\mathrm{p}$ and for seven different $P_\mathrm{sg}$, color-code is analogous to \sublabel{a}. Symbols are data, black line is a simultaneous fit of the six highest $P_\mathrm{p}$ datasets with a fourth-order polynomial, which we use as average pump attenuation $\overline{\mathcal{G}}_\mathrm{att}$ for further processing. Small gray symbols show the VNA probe tone attenuation in that frequency window, shifted manually by $\qty{-43.4}{\decibel}$ to get to the same level of magnitude; despite the much longer path of the probe tone, that includes also the way back from the sample to the VNA, the qualitative agreement in shape is considerable. It also shows a parasitic on-chip mode at ${\sim} \qty{7.82}{\giga\hertz}$, which might have an impact to the power-dependence of $\mathcal{G}_\mathrm{att}$.
		\sublabel{c}~Experimentally obtained (symbols) and numerically calculated (lines) pump photon $n_\mathrm{c}$ vs.~pump frequency $\omega_\mathrm{p}$. For the calculated numbers, we use Eq.~(\ref{eqn:charac_poly}) and the mean attenuation $\overline{\mathcal{G}}_\mathrm{att}$ shown in \sublabel{b} as fit line.
		Star symbols correspond to an effective pump-sideband detuning $\delta' = 0$, cf.~main paper Fig.~5 and Supplementary Fig.~\ref{fig:FigS17}.
	}
	\label{fig:FigS15}
\end{figure*}
To model the dynamical backaction in main paper Fig.~5, which is measured and presented as a function of pump detuning $\Delta_\mathrm{p}'$, we furthermore need a function for $n_\mathrm{c}(\Delta_\mathrm{p}')$.
Combined with the fit/theory functions for $g_\omega'(n_\mathrm{c})$, $g_\kappa'(n_\mathrm{c})$, $\omega_0'(n_\mathrm{c})$, and $\kappa_0'(n_\mathrm{c})$, which can be performed/calculated now that $n_\mathrm{c}$ is calibrated, we have everything we need to model $\Gamma_\mathrm{eff}(\Delta_\mathrm{p}')$ and $\Omega_\mathrm{eff}(\Delta_\mathrm{p}')$.
The fit values for the nonlinear coefficients are $\kappa_1 = 2\pi\times\qty{3.7}{\kilo\hertz}$, $\kappa_2 \approx 0$, $\kappa_3 = 2\pi\times\qty{0.37}{\milli\hertz}$, $g_\mathrm{nl1} = 2\pi\times\qty{-22.1}{\hertz}$, $g_\mathrm{nl2} \approx 0$, $g_\mathrm{nl3} = 2\pi\times\qty{-3.5}{\micro\hertz}$, and $g_{0\mathcal{K}} = 2\pi\times\qty{10.0}{\hertz}$.
The procedure to obtain $n_\mathrm{c}(\Delta_\mathrm{p}')$ will be discussed in the next subsection.
\vspace{-2mm}
\subsection{On-chip pump power calibration}
\vspace{-2mm}
To calibrate the on-chip pump power, we need to find an estimate for the pump attenuation between signal generator and device.
For this, we use Eq.~(\ref{eqn:nc_new}) and solve for the pump power
\begin{equation}
	P_\mathrm{p} = \frac{n_\mathrm{c}\hbar\omega_\mathrm{p}}{4}\frac{\kappa_\mathrm{eff}^2 + 4\tilde{\Delta}_\mathrm{p}^2}{\kappa_\mathrm{ext}}.
	\label{eqn:P_onchip}
\end{equation}
For each datapoint $(\omega_\mathrm{p}, P_\mathrm{p})$, we take now the experimental values for $n_\mathrm{c}$, $\omega_\mathrm{p}$ and $\kappa_\mathrm{ext}$, and we calculate
\begin{eqnarray}
	\kappa_\mathrm{eff} & = & \kappa_0 + \kappa_1 n_\mathrm{c} + \kappa_2 n_\mathrm{c}^2 + \kappa_3 n_\mathrm{c}^3 \\
	\tilde\Delta_\mathrm{p} & = & \Delta_\mathrm{p} - \mathcal{K}n_\mathrm{c}
\end{eqnarray}
using the experimental value for $\Delta_\mathrm{p} = \omega_\mathrm{p} - \omega_0$ and the fit values for $\mathcal{K}, \kappa_1, \kappa_2$ and $\kappa_3$.
The frequency-dependent attenuation $\mathcal{G}_\mathrm{att}$ can then be calculated for each datapoint individually by
\begin{equation}
	\mathcal{G}_\mathrm{att} = \frac{P_\mathrm{p}}{P_\mathrm{sg}}
\end{equation}
with the signal generator output power $P_\mathrm{sg}$.
The result of this calculation is shown for all data points in Supplementary Fig.~\ref{fig:FigS15}.
Interestingly, we find not only a frequency-dependent attenuation, but it seems also to be a pump-power dependent attenuation.
We believe this is not just an artifact, but at least partly the result of a second chip mode located at around $\qty{7.82}{\giga\hertz}$.
This second mode is much less nonlinear compared to the target HF mode, but since the pump is much closer to it, it will still absorb photons from the pump tone and if it shifts as a function of pump power, this absorption itself is nonlinear.
Of course, other factors might also contribute to a seemingly power-dependent attenuation, such as a nonlinear response of the HF feedline or the fingers in the interdigitated coupling capacitor, a power-dependent input impedance of the cryogenic HEMT amplifier, a power-dependent $\kappa_\mathrm{ext}$, higher-order dispersive nonlinearities in the HF mode, or deviations of the actual cavity parameters $\kappa_0'$ and $\omega_0'$ from their fit values due to uncertainties in the background correction routine.
For the purpose of modeling the dynamical backaction, however, and to demonstrate that theory and experiment are consistent with each other, we believe it is most convincing to work with a single averaged attenuation.
To have a general smooth function for $\mathcal{G}_\mathrm{att}$, we fit all the obtained data points combined with a fourth order polynomial
\begin{equation}
	\overline{\mathcal{G}}_\mathrm{att}(\omega_\mathrm{p}) = \mathcal{G}_0 + \mathcal{G}_1\omega_\mathrm{p} + \mathcal{G}_2\omega_\mathrm{p}^2 + \mathcal{G}_3\omega_\mathrm{p}^3 + \mathcal{G}_4\omega_\mathrm{p}^4,
\end{equation}
the result is overlaid to the data points in Supplementary Fig.~\ref{fig:FigS15} and shows acceptable agreement.
We neglect the lowest-power dataset for this fit, since it deviates strongly from the others for $\omega_\mathrm{p}\lesssim 2\pi\times\qty{8}{\giga\hertz}$.
Whenever we give on-chip pump power values in the main manuscript, we calculate this value using the generator output power and $\overline{\mathcal{G}}_\mathrm{att}(\omega = 2\pi\times\qty{8}{\giga\hertz})$, and we round the power values to integers.
To finally calculate $n_\mathrm{c}(\omega_\mathrm{p})$, we numerically solve the polynomial Eq.~(\ref{eqn:charac_poly}) with all the parameters as obtained from the fit of $\kappa_0'$ and the analysis of $\omega_0'$, while using as input photon flux
\begin{equation}
	n_\mathrm{in} = \overline{\mathcal{G}}_\mathrm{att}(\omega_\mathrm{p})\frac{P_\mathrm{sg}}{\hbar\omega_\mathrm{p}}.
\end{equation}
The result of this calculation in direct comparison with the values obtained from the experiment and $g_{0\omega}$ is shown in Supplementary Fig.~\ref{fig:FigS15}\sublabel{c}.
In particular for the lower generator powers, the agreement is very good over the most important range of $\omega_\mathrm{p}$ between $\qty{7.9}{\giga\hertz}$ and $\qty{8.1}{\giga\hertz}$.
The remaining power-dependent deviations of up to ${\sim} 15\%$ originate mainly from the deviation between the individual $\mathcal{G}_\mathrm{att}$ and $\overline{\mathcal{G}}_\mathrm{att}$.
\section{Supplementary Note VIII: Additional data}
\label{sec:Note8}
\vspace{-2mm}
\subsection{Resonant PPIT data for all pump powers}
\label{sec:Note8A}
\vspace{-2mm}
\begin{figure*}[h!]
	\includegraphics{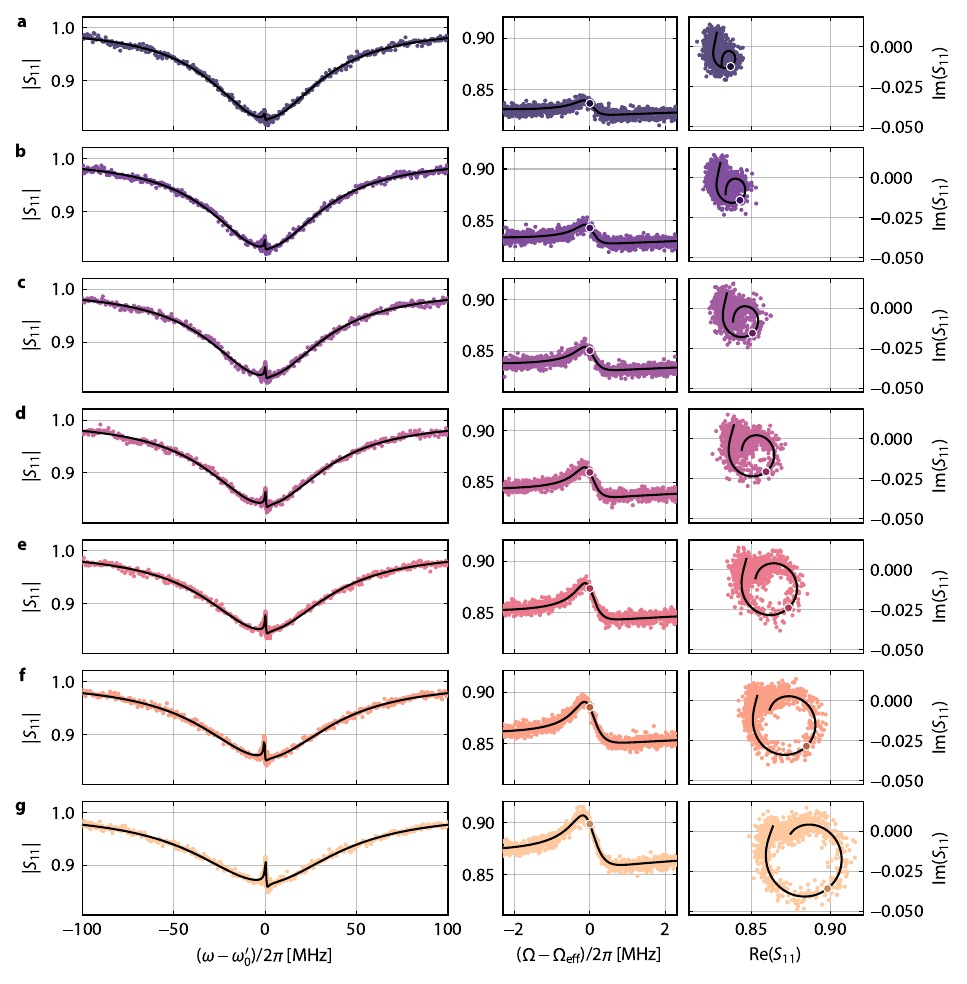}
	\titlecaption{Near-resonant PPIT datasets for all investigated pump powers}{%
		All panels show $S_{11}$ of a photon-pressure-induced transparency experiment, from \sublabel{a} to \sublabel{g} the on-chip pump power increases from $P_\mathrm{p} = \qty{-63}{\dBm}$ to  $P_\mathrm{p} = \qty{-57}{\dBm}$ in steps of $\qty{1}{\decibel}$. Left column shows $|S_{11}|$ with a wide frequency span of $\qty{200}{\mega\hertz}$, middle column shows corresponding zooms to the PPIT resonance with a narrow frequency span of $\qty{4.5}{\mega\hertz}$, right column shows the zoom of $S_{11}$ in the complex plane. Colored dots are experimental data, lines are fit curves, and colored disks with white boundaries mark the points of the PPIT resonances. With increasing $P_\mathrm{p}$, the transparency window grows and becomes more asymmetric, indicating increasing coupling rates $g_\omega'$, $g_\kappa'$ and a slightly increasing coupling rate ratio $g_\kappa'/g_\omega'$. Since $\delta_\mathrm{eff} \approx 0$, but still $\neq 0$ for most of the datasets, the angle $\gamma$ is omitted here.
	}
	\label{fig:FigS16}
\end{figure*}
In Fig.~3 of the main paper, we present the PPIT data for $\delta_\mathrm{eff} \approx 0$ and an on-chip pump power of $P_\mathrm{p} = \qty{-58}{\dBm}$.
In Supplementary Fig.~\ref{fig:FigS16} we show for completeness the datasets and fits for all pump powers between $P_\mathrm{p} = \qty{-63}{\dBm}$ and $P_\mathrm{p} = \qty{-57}{\dBm}$.
First, we notice that with increasing $P_\mathrm{p}$ the (internal) cavity linewidth increases as also discussed in main paper Fig.~4.
Furthermore, we observe that the PPIT signature both in reflection magnitude and in the complex plane grows with $P_\mathrm{p}$ due to the increased coupling rates $g_\omega'$ and $g_\kappa'$.
For all pump powers, the PPIT circle is considerably rotated away from the real axis.
However, a graphical and quantitative $\gamma$-analysis analogous to main paper Fig.~3 is difficult here, especially for the low pump-powers, due to the small but still finite $\delta_\mathrm{eff}$ of all datasets (pump frequency step width was $\qty{2}{\mega\hertz}$).
A small offset between PPIT and cavity resonance is particularly noticeable for small PPIT circles, i.e., low pump powers, and using the actual reference point of the cavity circle ($-\delta_\mathrm{eff}$ away from resonance) is not very insightful, since then none of the resonances is on the real axis and the definition of $\gamma$ requires two angled lines instead of one and the real axis.
Therefore, we omit $\gamma$ in the plot.
\vspace{-2mm}
\subsection{Dynamical backaction for all pump powers}
\label{sec:Note8B}
\vspace{-2mm}
Finally, we show the effective LF linewidth $\Gamma_\mathrm{eff}$ and effective LF resonance frequency $\Omega_\mathrm{eff}$, which include the dynamical backaction $\Gamma_\mathrm{pp}$ and $\delta\Omega_\mathrm{pp}$, for all pump powers in Supplementary Fig.~\ref{fig:FigS17}.
As described in the main paper in the context of Fig.~5, we obtain the experimental values for $\Gamma_\mathrm{eff}$ and $\Omega_\mathrm{eff}$ by probing directly the LF resonance with a VNA, while stepping the pump tone in steps of $\qty{2}{\mega\hertz}$ through the red sideband of the HF mode.
The on-chip LF probe-tone power we estimate to be $P_\beta = \qty{-97}{\dBm}$, which at $\Omega = \Omega_0 = 2\pi\times \qty{446}{\mega\hertz}$, $\Gamma_0/2\pi = \qty{600}{\kilo\hertz}$ and $\Gamma_\mathrm{ext}/2\pi = \qty{37}{\kilo\hertz}$ corresponds to a intracircuit probe-photon number $n_\mathrm{pr} \sim 44000$.
This is sufficiently low to stay below the onset of a LF self-Kerr nonlinearity or a modification of the dynamical backaction, except for the ${\sim}10$ datasets closest to the parametric instability, where slight Duffing-like deformations of the LF resonances appear, cf. Supplementary Note~\ref{sec:Note7C} for the corresponding fitting routine.
Then, we model the data using
\begin{eqnarray}
	\Gamma_\mathrm{eff} & = & \Gamma_0 + \kappa_\mathrm{c}n_\mathrm{c} + \Gamma_\mathrm{pp} \\
	\Omega_\mathrm{eff} & = & \Omega_0 + \mathcal{K}_\mathrm{c}n_\mathrm{c} + \delta\Omega_\mathrm{pp}
\end{eqnarray}
with the cross-Kerr constant $\mathcal{K}_\mathrm{c}$, the cross-nonlinear-damping $\kappa_\mathrm{c}$ and $\Omega_0$ and $\Gamma_0$ as fit parameters.
The photon-pressure dynamical backaction contributions $\Gamma_\mathrm{pp}$ and $\delta\Omega_\mathrm{pp}$ we calculate using the independently obtained functions for $g_\omega'(n_\mathrm{c})$, $g_\kappa'(n_\mathrm{c})$, $\omega_0'(n_\mathrm{c})$ and $\kappa_0'(n_\mathrm{c})$ in combination with $n_\mathrm{c}(\omega_\mathrm{p})$ as described in the main paper Fig.~4 and Sec.~\ref{sec:Note7} of the Supplementary Material.
The fit is performed with a single function for $\Gamma_\mathrm{eff}$ and $\delta\Omega_\mathrm{eff}$ and simultaneously for all powers, and the result agrees well with the experimental data.
The remaining deviations originate from uncertainties in the pump attenuation and the intracavity pump photon number, respectively, and especially for high $n_\mathrm{c}$ from possible higher-order nonlinearities and the onset of parametric instability.
\clearpage
\begin{figure*}
	\includegraphics[trim={0 2mm 0 2mm},clip = true]{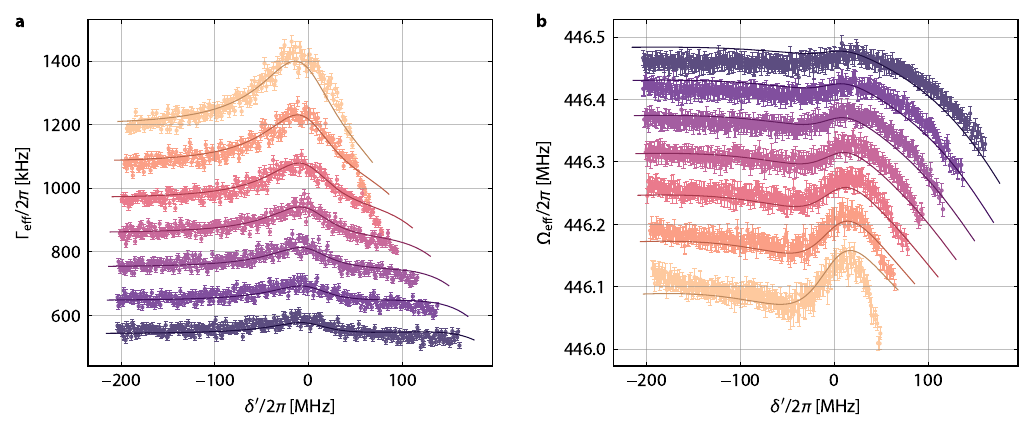}
	\titlecaption{Dynamical backaction for all investigated pump powers}{%
		\sublabel{a}~Effective LF mode linewidth $\Gamma_\mathrm{eff}$ and \sublabel{b}~effective LF mode resonance frequency $\Omega_\mathrm{eff}$ for all investigated pump powers, plotted vs.~pump detuning from the red sideband $\delta' = \omega_\mathrm{p} - \left(\omega_0' - \Omega_0'\right)$. On-chip pump powers are between $\qty{-63}{\dBm}$ (dark blue) and $\qty{-57}{\dBm}$ (yellow) in steps of $\qty{1}{\decibel}$ (power values at $\qty{8}{\giga\hertz}$). Data acquisition was identical to main paper Fig.~5, error bars show the standard error from the fit routine. Symbols are data points, solid lines are simultaneous fit curves of all $\Gamma_\mathrm{eff}$ and $\Omega_\mathrm{eff}$ with the free parameters $\Omega_0 = 2\pi\times\qty{446.53}{\mega\hertz}$, $\Gamma_0 = 2\pi\times\qty{527.0}{\kilo\hertz}$, $\mathcal{K}_\mathrm{c} = 2\pi\times\qty{-130.8}{\hertz}$, and $\kappa_\mathrm{c} = 2\pi\times\qty{38.4}{\hertz}$. Adjacent curves are manually shifted by $+\qty{100}{\kilo\hertz}$ each for $\Gamma_\mathrm{eff}$ and by $-\qty{40}{\kilo\hertz}$ each for $\Omega_\mathrm{eff}$ for higher visibility.\\
	}
	\vspace{-8mm}
	\label{fig:FigS17}
\end{figure*}

\renewcommand\refname{Supplementary References}


\begin{thebibliography}{45}

	\expandafter\ifx\csname natexlab\endcsname\relax\def\natexlab#1{#1}\fi
	\expandafter\ifx\csname bibnamefont\endcsname\relax
	\def\bibnamefont#1{#1}\fi
	\expandafter\ifx\csname bibfnamefont\endcsname\relax
	\def\bibfnamefont#1{#1}\fi
	\expandafter\ifx\csname citenamefont\endcsname\relax
	\def\citenamefont#1{#1}\fi
	\expandafter\ifx\csname url\endcsname\relax
	\def\url#1{\texttt{#1}}\fi
	\expandafter\ifx\csname urlprefix\endcsname\relax\def\urlprefix{URL }\fi
	\providecommand{\bibinfo}[2]{#2}
	\providecommand{\eprint}[2][]{\url{#2}}


	\bibitem[{\citenamefont{Krantz et~al.}(2019)\citenamefont{Krantz, Kjaergaard, Yan, Orlando, Gustavsson, Oliver}}]{Krantz2019}
	\bibinfo{author}{\bibnamefont{Krantz},~\bibfnamefont{P.}},
	\bibinfo{author}{\bibnamefont{Kjaergaard},~\bibfnamefont{M.}},
	\bibinfo{author}{\bibnamefont{Yan},~\bibfnamefont{F.}},
	\bibinfo{author}{\bibnamefont{Orlando},~\bibfnamefont{T.~P.}},
	\bibinfo{author}{\bibnamefont{Gustavsson},~\bibfnamefont{S.}},
	\bibnamefont{and}
	\bibinfo{author}{\bibnamefont{Oliver},~\bibfnamefont{W.~D.}},
	\hyperref{https://pubs.aip.org/aip/apr/article/6/2/021318/570326/A-quantum-engineer-s-guide-to-superconducting}
	\bibinfo{title}{\bibfnamefont{A quantum engineer's guide to superconducting qubits}}.
	\bibinfo{journal}{Applied Physics Reviews} \textbf{\bibinfo{volume}{6}},
	\bibinfo{pages}{021318} (\bibinfo{year}{2019})

	\bibitem[{\citenamefont{Blais et~al.}(2021)\citenamefont{Blais, Grimsmo, Girvin, Wallraff}}]{Blais2021}
	\bibinfo{author}{\bibnamefont{Blais},~\bibfnamefont{A.}},
	\bibinfo{author}{\bibnamefont{Grimsmo},~\bibfnamefont{A.~L.}},
	\bibinfo{author}{\bibnamefont{Girvin},~\bibfnamefont{S.~M.}},
	\bibnamefont{and}
	\bibinfo{author}{\bibnamefont{Wallraff},~\bibfnamefont{A.}},
	\hyperref{https://journals.aps.org/rmp/abstract/10.1103/RevModPhys.93.025005}
	\bibinfo{title}{\bibfnamefont{Circuit quantum electrodynamics}}.
	\bibinfo{journal}{Review of Modern Physics} \textbf{\bibinfo{volume}{93}},
	\bibinfo{pages}{025005} (\bibinfo{year}{2021})

	\bibitem[{\citenamefont{Arute et~al.}(2019)\citenamefont{Arute, Arya, Babbush, Bacon, Bardin, Barends, Biswas, Boixo, Brandao, Buell, Burkett, Chen, Chen, Chiaro, Collins, Courtney, Dunsworth, Farhi, Foxen, Fowler, Gidney, Giustina, Graff, Guerin, Habegger, Harrigan, Hartmann, Ho, Hoffmann, Huang, Humble, Isakov, Jeffrey, Jiang, Kafri, Kechedzhi, Kelly, Klimov, Knysh, Korotkov, Kostritsa, Landhuis, Lindmark, Lucero, Lyakh, Mandr\'a, McClean, McEwen, Megrant, Mi, Michielsen, Mohseni, Mutus, Naaman, Neeley, Neill, Niu, Ostby, Petukhov, Platt, Quintana, Rieffel, Roushan, Rubin, Sank, Satzinger, Smelyianskiy, Sung, Trevithick, Vainsencher, Villalonga, White, Yao, Yeh, Zalcman, Neven, Martinis}}]{Arute2019}
	\bibinfo{author}{\bibnamefont{Arute},~\bibfnamefont{F.}},
	\bibinfo{author}{\bibnamefont{Arya},~\bibfnamefont{K.}},
	\bibinfo{author}{\bibnamefont{Babbush},~\bibfnamefont{R.}},
	\bibinfo{author}{\bibnamefont{Bacon},~\bibfnamefont{D.}},
	\bibinfo{author}{\bibnamefont{Bardin},~\bibfnamefont{J.~C.}},
	\bibinfo{author}{\bibnamefont{Barends},~\bibfnamefont{R.}},
	\bibinfo{author}{\bibnamefont{Biswas},~\bibfnamefont{R.}},
	\bibinfo{author}{\bibnamefont{Boixo},~\bibfnamefont{S.}},
	\bibinfo{author}{\bibnamefont{Brandao},~\bibfnamefont{F.~G.~S.~L.}},
	\bibinfo{author}{\bibnamefont{Buell},~\bibfnamefont{D.~A.}},
	\bibinfo{author}{\bibnamefont{Burkett},~\bibfnamefont{B.}},
	\bibinfo{author}{\bibnamefont{Chen},~\bibfnamefont{Y.}},
	\bibinfo{author}{\bibnamefont{Chen},~\bibfnamefont{Z.}},
	\bibinfo{author}{\bibnamefont{Chiaro},~\bibfnamefont{B.}},
	\bibinfo{author}{\bibnamefont{Collins},~\bibfnamefont{R.}},
	\bibinfo{author}{\bibnamefont{Courtney},~\bibfnamefont{W.}},
	\bibinfo{author}{\bibnamefont{Dunsworth},~\bibfnamefont{A.}},
	\bibinfo{author}{\bibnamefont{Farhi},~\bibfnamefont{E.}},
	\bibinfo{author}{\bibnamefont{Brooks},~\bibfnamefont{F.}},
	\bibinfo{author}{\bibnamefont{Fowler},~\bibfnamefont{A.}},
	\bibinfo{author}{\bibnamefont{Gidney},~\bibfnamefont{C.}},
	\bibinfo{author}{\bibnamefont{Giustina},~\bibfnamefont{M.}},
	\bibinfo{author}{\bibnamefont{Graff},~\bibfnamefont{R.}},
	\bibinfo{author}{\bibnamefont{Guerin},~\bibfnamefont{K.}},
	\bibinfo{author}{\bibnamefont{Habegger},~\bibfnamefont{S.}},
	\bibinfo{author}{\bibnamefont{Harrigan},~\bibfnamefont{M.~P.}},
	\bibinfo{author}{\bibnamefont{Hartmann},~\bibfnamefont{M.~J.}},
	\bibinfo{author}{\bibnamefont{Ho},~\bibfnamefont{A.}},
	\bibinfo{author}{\bibnamefont{Hoffmann},~\bibfnamefont{M.}},
	\bibinfo{author}{\bibnamefont{Huang},~\bibfnamefont{T.}},
	\bibinfo{author}{\bibnamefont{Humble},~\bibfnamefont{T.~S.}},
	\bibinfo{author}{\bibnamefont{Isakov},~\bibfnamefont{S.~V.}},
	\bibinfo{author}{\bibnamefont{Jeffrey},~\bibfnamefont{E.}},
	\bibinfo{author}{\bibnamefont{Jiang},~\bibfnamefont{Z.}},
	\bibinfo{author}{\bibnamefont{Kafri},~\bibfnamefont{D.}},
	\bibinfo{author}{\bibnamefont{Kechedzhi},~\bibfnamefont{K.}},
	\bibinfo{author}{\bibnamefont{Kelly},~\bibfnamefont{J.}},
	\bibinfo{author}{\bibnamefont{Klimov},~\bibfnamefont{P.~V.}},
	\bibinfo{author}{\bibnamefont{Knysh},~\bibfnamefont{S.}},
	\bibinfo{author}{\bibnamefont{Korotkov},~\bibfnamefont{A.}},
	\bibinfo{author}{\bibnamefont{Kostritsa},~\bibfnamefont{F.}},
	\bibinfo{author}{\bibnamefont{Landhuis},~\bibfnamefont{D.}},
	\bibinfo{author}{\bibnamefont{Lindmark},~\bibfnamefont{M.}},
	\bibinfo{author}{\bibnamefont{Lucero},~\bibfnamefont{E.}},
	\bibinfo{author}{\bibnamefont{Lyakh},~\bibfnamefont{D.}},
	\bibinfo{author}{\bibnamefont{Mandr\'a},~\bibfnamefont{S.}},
	\bibinfo{author}{\bibnamefont{McClean},~\bibfnamefont{J.~R.}},
	\bibinfo{author}{\bibnamefont{McEwen},~\bibfnamefont{M.}},
	\bibinfo{author}{\bibnamefont{Megrant},~\bibfnamefont{A.}},
	\bibinfo{author}{\bibnamefont{Mi},~\bibfnamefont{X.}},
	\bibinfo{author}{\bibnamefont{Michielsen},~\bibfnamefont{K.}},
	\bibinfo{author}{\bibnamefont{Mohseni},~\bibfnamefont{M.}},
	\bibinfo{author}{\bibnamefont{Mutus},~\bibfnamefont{J.}},
	\bibinfo{author}{\bibnamefont{Naaman},~\bibfnamefont{O.}},
	\bibinfo{author}{\bibnamefont{Neeley},~\bibfnamefont{M.}},
	\bibinfo{author}{\bibnamefont{Neill},~\bibfnamefont{C.}},
	\bibinfo{author}{\bibnamefont{Niu},~\bibfnamefont{M.~Y.}},
	\bibinfo{author}{\bibnamefont{Ostby},~\bibfnamefont{E.}},
	\bibinfo{author}{\bibnamefont{Pethukov},~\bibfnamefont{A.}},
	\bibinfo{author}{\bibnamefont{Platt},~\bibfnamefont{J.~C.}},
	\bibinfo{author}{\bibnamefont{Quintana},~\bibfnamefont{C.}},
	\bibinfo{author}{\bibnamefont{Rieffel},~\bibfnamefont{E.~G.}},
	\bibinfo{author}{\bibnamefont{Roushan},~\bibfnamefont{P.}},
	\bibinfo{author}{\bibnamefont{Rubin},~\bibfnamefont{N.~C.}},
	\bibinfo{author}{\bibnamefont{Sank},~\bibfnamefont{D.}},
	\bibinfo{author}{\bibnamefont{Satzinger},~\bibfnamefont{K.~J.}},
	\bibinfo{author}{\bibnamefont{Smelyanskiy},~\bibfnamefont{V.}},
	\bibinfo{author}{\bibnamefont{Sung},~\bibfnamefont{K.~J.}},
	\bibinfo{author}{\bibnamefont{Trevithick},~\bibfnamefont{M.~D.}},
	\bibinfo{author}{\bibnamefont{Vainsencher},~\bibfnamefont{A.}},
	\bibinfo{author}{\bibnamefont{Villalonga},~\bibfnamefont{B.}},
	\bibinfo{author}{\bibnamefont{White},~\bibfnamefont{T.}},
	\bibinfo{author}{\bibnamefont{Yao},~\bibfnamefont{Z.~J.}},
	\bibinfo{author}{\bibnamefont{Yeh},~\bibfnamefont{P.}},
	\bibinfo{author}{\bibnamefont{Zalcman},~\bibfnamefont{A.}},
	\bibinfo{author}{\bibnamefont{Neven},~\bibfnamefont{H.}},
	\bibnamefont{and}
	\bibinfo{author}{\bibnamefont{Martinis},~\bibfnamefont{J.~M.}},
	\hyperref{https://www.nature.com/articles/s41586-019-1666-5}
	\bibinfo{title}{\bibfnamefont{Quantum supremacy using a programmable superconducting processor}}.
	\bibinfo{journal}{Nature} \textbf{\bibinfo{volume}{574}},
	\bibinfo{pages}{505-510} (\bibinfo{year}{2019})

	\bibitem[{\citenamefont{Kim et~al.}(2023)\citenamefont{Kim, Eddins, Anand, Wei, van~den~Berg, Rosenblatt, Nayfeh, Wu, Zaletel, Temme, Kandala}}]{Kim2023}
	\bibinfo{author}{\bibnamefont{Kim},~\bibfnamefont{Y.}},
	\bibinfo{author}{\bibnamefont{Eddins},~\bibfnamefont{A.}},
	\bibinfo{author}{\bibnamefont{Anand},~\bibfnamefont{S.}},
	\bibinfo{author}{\bibnamefont{Wei},~\bibfnamefont{K.~X.}},
	\bibinfo{author}{\bibnamefont{van~den~Berg},~\bibfnamefont{E.}},
	\bibinfo{author}{\bibnamefont{Rosenblatt},~\bibfnamefont{S.}},
	\bibinfo{author}{\bibnamefont{Nayfeh},~\bibfnamefont{H.}},
	\bibinfo{author}{\bibnamefont{Wu},~\bibfnamefont{Y.}},
	\bibinfo{author}{\bibnamefont{Zaletel},~\bibfnamefont{M.}},
	\bibinfo{author}{\bibnamefont{Temme},~\bibfnamefont{K.}},
	\bibnamefont{and}
	\bibinfo{author}{\bibnamefont{Kandala},~\bibfnamefont{A.}},
	\hyperref{https://www.nature.com/articles/s41586-023-06096-3}
	\bibinfo{title}{\bibfnamefont{Evidence for the utility of quantum computing before fault tolerance}}.
	\bibinfo{journal}{Nature} \textbf{\bibinfo{volume}{618}},
	\bibinfo{pages}{500-505} (\bibinfo{year}{2023})

	\bibitem[{\citenamefont{Houck et~al.}(2012)\citenamefont{Houck, Türeci, Koch}}]{Houck2012}
	\bibinfo{author}{\bibnamefont{Houck},~\bibfnamefont{A.~A.}},
	\bibinfo{author}{\bibnamefont{Türeci},~\bibfnamefont{H.~E.}},
	\bibnamefont{and}
	\bibinfo{author}{\bibnamefont{Koch},~\bibfnamefont{J.}},
	\hyperref{https://www.nature.com/articles/nphys2251}
	\bibinfo{title}{\bibfnamefont{On-chip quantum simulation with superconducting circuits}}.
	\bibinfo{journal}{Nature Physics} \textbf{\bibinfo{volume}{8}},
	\bibinfo{pages}{292-299} (\bibinfo{year}{2012})

	\bibitem[{\citenamefont{Carusotto et~al.}(2020)\citenamefont{Carusotto, Houck, Kollar, Roushan, Schuster, Simon}}]{Carusotto2020}
	\bibinfo{author}{\bibnamefont{Carusotto},~\bibfnamefont{I.}},
	\bibinfo{author}{\bibnamefont{Houck},~\bibfnamefont{A.~A.}},
	\bibinfo{author}{\bibnamefont{Kollar},~\bibfnamefont{A.~J.}},
	\bibinfo{author}{\bibnamefont{Roushan},~\bibfnamefont{P.}},
	\bibinfo{author}{\bibnamefont{Schuster},~\bibfnamefont{D.~I.}},
	\bibnamefont{and}
	\bibinfo{author}{\bibnamefont{Simon},~\bibfnamefont{J.}},
	\hyperref{https://www.nature.com/articles/s41567-020-0815-y}
	\bibinfo{title}{\bibfnamefont{Photonic materials in circuit quantum electrodynamics}}.
	\bibinfo{journal}{Nature Physics} \textbf{\bibinfo{volume}{16}},
	\bibinfo{pages}{268-279} (\bibinfo{year}{2020})

	\bibitem[{\citenamefont{Andrews et~al.}(2014)\citenamefont{Andrews, Peterson, Purdy, Cicak, Simmonds, Regal, Lehnert}}]{Andrews2014}
	\bibinfo{author}{\bibnamefont{Andrews},~\bibfnamefont{R.~W.}},
	\bibinfo{author}{\bibnamefont{Peterson},~\bibfnamefont{R.~W.}},
	\bibinfo{author}{\bibnamefont{Purdy},~\bibfnamefont{T.~P.}},
	\bibinfo{author}{\bibnamefont{Cicak},~\bibfnamefont{K.}},
	\bibinfo{author}{\bibnamefont{Simmonds},~\bibfnamefont{R.~W.}},
	\bibinfo{author}{\bibnamefont{Regal},~\bibfnamefont{C.~A.}},
	\bibnamefont{and}
	\bibinfo{author}{\bibnamefont{Lehnert},~\bibfnamefont{K.~W.}},
	\hyperref{https://www.nature.com/articles/nphys2911}
	\bibinfo{title}{\bibfnamefont{Bidirectional and efficient conversion between microwave and optical light}}.
	\bibinfo{journal}{Nature Physics} \textbf{\bibinfo{volume}{10}},
	\bibinfo{pages}{321-326} (\bibinfo{year}{2014})

	\bibitem[{\citenamefont{Forsch et~al.}(2020)\citenamefont{Forsch, Stockill, Wallucks, Marinkovic, Gärtner, Norte, van~Otten, Fiore, Srinivasan, Gröblacher}}]{Forsch2020}
	\bibinfo{author}{\bibnamefont{Forsch},~\bibfnamefont{M.}},
	\bibinfo{author}{\bibnamefont{Stockill},~\bibfnamefont{R.}},
	\bibinfo{author}{\bibnamefont{Wallucks},~\bibfnamefont{A.}},
	\bibinfo{author}{\bibnamefont{Marinkovic},~\bibfnamefont{I.}},
	\bibinfo{author}{\bibnamefont{Gärtner},~\bibfnamefont{C.}},
	\bibinfo{author}{\bibnamefont{Norte},~\bibfnamefont{R.~A.}},
	\bibinfo{author}{\bibnamefont{van~Otten},~\bibfnamefont{F.}},
	\bibinfo{author}{\bibnamefont{Fiore},~\bibfnamefont{A.}},
	\bibinfo{author}{\bibnamefont{Srinivasan},~\bibfnamefont{K.}},
	\bibnamefont{and}
	\bibinfo{author}{\bibnamefont{Gröblacher},~\bibfnamefont{S.}},
	\hyperref{https://www.nature.com/articles/s41567-019-0673-7}
	\bibinfo{title}{\bibfnamefont{Microwave-to-optics conversion using a mechanical oscillator in its quantum ground state}}.
	\bibinfo{journal}{Nature Physics} \textbf{\bibinfo{volume}{16}},
	\bibinfo{pages}{69-74} (\bibinfo{year}{2020})

	\bibitem[{\citenamefont{Bienfait et~al.}(2016)\citenamefont{Bienfait, Pla, Kubo, Stern, Zhou, Lo, Weis, Schenkel, Thewalt, Vion, Esteve, Julsgaard, Molmer, Morton, Bertet}}]{Bienfait2016}
	\bibinfo{author}{\bibnamefont{Bienfait},~\bibfnamefont{A.}},
	\bibinfo{author}{\bibnamefont{Pla},~\bibfnamefont{J.~J.}},
	\bibinfo{author}{\bibnamefont{Kubo},~\bibfnamefont{Y.}},
	\bibinfo{author}{\bibnamefont{Stern},~\bibfnamefont{M.}},
	\bibinfo{author}{\bibnamefont{Zhou},~\bibfnamefont{X.}},
	\bibinfo{author}{\bibnamefont{Lo},~\bibfnamefont{C.~C.}},
	\bibinfo{author}{\bibnamefont{Weis},~\bibfnamefont{C.~D.}},
	\bibinfo{author}{\bibnamefont{Schenkel},~\bibfnamefont{T.}},
	\bibinfo{author}{\bibnamefont{Thewalt},~\bibfnamefont{M.~L.~W.}},
	\bibinfo{author}{\bibnamefont{Vion},~\bibfnamefont{D.}},
	\bibinfo{author}{\bibnamefont{Esteve},~\bibfnamefont{D.}},
	\bibinfo{author}{\bibnamefont{Julsgaard},~\bibfnamefont{B.}},
	\bibinfo{author}{\bibnamefont{Molmer},~\bibfnamefont{K.}},
	\bibinfo{author}{\bibnamefont{Morton},~\bibfnamefont{J.~J.~L.}},
	\bibnamefont{and}
	\bibinfo{author}{\bibnamefont{Bertet},~\bibfnamefont{P.}},
	\hyperref{https://www.nature.com/articles/nnano.2015.282}
	\bibinfo{title}{\bibfnamefont{Reaching the quantum limit of sensitivity in electron spin resonance}}.
	\bibinfo{journal}{Nature Nanotechnology} \textbf{\bibinfo{volume}{11}},
	\bibinfo{pages}{253-257} (\bibinfo{year}{2016})

	\bibitem[{\citenamefont{Wang et~al.}(2023)\citenamefont{Wang, Balembois, Rancic, Billaud, Le~Dantec, Ferrier, Goldner, Bertaina, Chaneliere, Esteve, Dion, Bertet, Flurin}}]{Wang2023}
	\bibinfo{author}{\bibnamefont{Wang},~\bibfnamefont{Z.}},
	\bibinfo{author}{\bibnamefont{Balembois},~\bibfnamefont{L.}},
	\bibinfo{author}{\bibnamefont{Rancic},~\bibfnamefont{M.}},
	\bibinfo{author}{\bibnamefont{Billaud},~\bibfnamefont{E.}},
	\bibinfo{author}{\bibnamefont{Le~Dantec},~\bibfnamefont{M.}},
	\bibinfo{author}{\bibnamefont{Ferrier},~\bibfnamefont{A.}},
	\bibinfo{author}{\bibnamefont{Goldner},~\bibfnamefont{P.}},
	\bibinfo{author}{\bibnamefont{Bertaina},~\bibfnamefont{S.}},
	\bibinfo{author}{\bibnamefont{Chaneliere},~\bibfnamefont{T.}},
	\bibinfo{author}{\bibnamefont{Esteve},~\bibfnamefont{D.}},
	\bibinfo{author}{\bibnamefont{Vion},~\bibfnamefont{D.}},
	\bibinfo{author}{\bibnamefont{Bertet},~\bibfnamefont{P.}},
	\bibnamefont{and}
	\bibinfo{author}{\bibnamefont{Flurin},~\bibfnamefont{E.}},
	\hyperref{https://www.nature.com/articles/s41586-023-06097-2}
	\bibinfo{title}{\bibfnamefont{Single-electron spin resonance detection by microwave photon counting}}.
	\bibinfo{journal}{Nature} \textbf{\bibinfo{volume}{619}},
	\bibinfo{pages}{276-281} (\bibinfo{year}{2023})

	\bibitem[{\citenamefont{Castellanos-Beltran et~al.}(2008)\citenamefont{Castellanos-Beltran, Irwin, Hilton, Vale, Lehnert}}]{CastellanosBeltran2008}
	\bibinfo{author}{\bibnamefont{Castellanos-Beltran},~\bibfnamefont{M.~A.}},
	\bibinfo{author}{\bibnamefont{Irwin},~\bibfnamefont{K.~D.}},
	\bibinfo{author}{\bibnamefont{Hilton},~\bibfnamefont{G.~C.}},
	\bibinfo{author}{\bibnamefont{Vale},~\bibfnamefont{L.~R.}},
	\bibnamefont{and}
	\bibinfo{author}{\bibnamefont{Lehnert},~\bibfnamefont{K.~W.}},
	\hyperref{https://www.nature.com/articles/nphys1090}
	\bibinfo{title}{\bibfnamefont{Amplification and squeezing of quantum noise with a tunable Josephson metamaterial}}.
	\bibinfo{journal}{Nature Physics} \textbf{\bibinfo{volume}{4}},
	\bibinfo{pages}{929-931} (\bibinfo{year}{2008})

	\bibitem[{\citenamefont{Bergeal et~al.}(2010)\citenamefont{Bergeal, Schackert, Metcalfe, Vijay, Manucharyan, Frunzio, Prober, Schoelkopf, Girvin, Devoret}}]{Bergeal2010}
	\bibinfo{author}{\bibnamefont{Bergeal},~\bibfnamefont{N.}},
	\bibinfo{author}{\bibnamefont{Schackert},~\bibfnamefont{F.}},
	\bibinfo{author}{\bibnamefont{Metcalfe},~\bibfnamefont{M.}},
	\bibinfo{author}{\bibnamefont{Vijay},~\bibfnamefont{R.}},
	\bibinfo{author}{\bibnamefont{Manucharyan},~\bibfnamefont{V.~E.}},
	\bibinfo{author}{\bibnamefont{Frunzio},~\bibfnamefont{L.}},
	\bibinfo{author}{\bibnamefont{Prober},~\bibfnamefont{D.~E.}},
	\bibinfo{author}{\bibnamefont{Schoelkopf},~\bibfnamefont{R.~J.}},
	\bibinfo{author}{\bibnamefont{Girvin},~\bibfnamefont{S.~M.}},
	\bibnamefont{and}
	\bibinfo{author}{\bibnamefont{Devoret},~\bibfnamefont{M.~H.}},
	\hyperref{https://www.nature.com/articles/nature09035}
	\bibinfo{title}{\bibfnamefont{Phase-preserving amplification near the quantum limit with a Josephson ring modulator}}.
	\bibinfo{journal}{Nature} \textbf{\bibinfo{volume}{465}},
	\bibinfo{pages}{64-68} (\bibinfo{year}{2010})

	\bibitem[{\citenamefont{Macklin et~al.}(2015)\citenamefont{Macklin, O'Brien, Hover, Schwartz, Bolkhovsky, Zhang, Oliver, Siddiqi}}]{Macklin2015}
	\bibinfo{author}{\bibnamefont{Macklin},~\bibfnamefont{C.}},
	\bibinfo{author}{\bibnamefont{O'Brien},~\bibfnamefont{K.}},
	\bibinfo{author}{\bibnamefont{Hover},~\bibfnamefont{D.}},
	\bibinfo{author}{\bibnamefont{Schwartz},~\bibfnamefont{M.~E.}},
	\bibinfo{author}{\bibnamefont{Bolkhovsky},~\bibfnamefont{V.}},
	\bibinfo{author}{\bibnamefont{Zhang},~\bibfnamefont{X.}},
	\bibinfo{author}{\bibnamefont{Oliver},~\bibfnamefont{W.~D.}},
	\bibnamefont{and}
	\bibinfo{author}{\bibnamefont{Siddiqi},~\bibfnamefont{I.}},
	\hyperref{https://www.science.org/doi/10.1126/science.aaa8525}
	\bibinfo{title}{\bibfnamefont{A near-quantum-limited Josephson traveling-wave parametric amplifier}}.
	\bibinfo{journal}{Science} \textbf{\bibinfo{volume}{350}},
	\bibinfo{pages}{307-310} (\bibinfo{year}{2015})

	\bibitem[{\citenamefont{Johansson et~al.}(2014)\citenamefont{Johansson, Johansson, Nori}}]{Johansson2014}
	\bibinfo{author}{\bibnamefont{Johansson},~\bibfnamefont{J.~R.}},
	\bibinfo{author}{\bibnamefont{Johansson},~\bibfnamefont{G.}},
	\bibnamefont{and}
	\bibinfo{author}{\bibnamefont{Nori},~\bibfnamefont{F.}},
	\hyperref{https://journals.aps.org/pra/abstract/10.1103/PhysRevA.90.053833}
	\bibinfo{title}{\bibfnamefont{Optomechanical-like coupling between superconducting resonators}}.
	\bibinfo{journal}{Physical Review A} \textbf{\bibinfo{volume}{90}},
	\bibinfo{pages}{053833} (\bibinfo{year}{2014})

	\bibitem[{\citenamefont{Eichler and Petta}(2018)\citenamefont{Eichler, Petta}}]{Eichler2018}
	\bibinfo{author}{\bibnamefont{Eichler},~\bibfnamefont{C.}},
	\bibnamefont{and}
	\bibinfo{author}{\bibnamefont{Petta},~\bibfnamefont{J.~R.}},
	\hyperref{https://journals.aps.org/prl/abstract/10.1103/PhysRevLett.120.227702}
	\bibinfo{title}{\bibfnamefont{Realizing a Circuit Analog of an Optomechanical System with Longitudinally Coupled Superconducting Resonators}}.
	\bibinfo{journal}{Physical Review Letters} \textbf{\bibinfo{volume}{120}},
	\bibinfo{pages}{227702} (\bibinfo{year}{2018})

	\bibitem[{\citenamefont{Bothner et~al.}(2021)\citenamefont{Bothner, Rodrigues, Steele}}]{Bothner2021}
	\bibinfo{author}{\bibnamefont{Bothner},~\bibfnamefont{D.}},
	\bibinfo{author}{\bibnamefont{Rodrigues},~\bibfnamefont{I.~C.}},
	\bibnamefont{and}
	\bibinfo{author}{\bibnamefont{Steele},~\bibfnamefont{G.~A.}},
	\hyperref{https://www.nature.com/articles/s41567-020-0987-5}
	\bibinfo{title}{\bibfnamefont{Photon-pressure strong coupling between two superconducting circuits}}.
	\bibinfo{journal}{Nature Physics} \textbf{\bibinfo{volume}{17}},
	\bibinfo{pages}{85-91} (\bibinfo{year}{2021})

	\bibitem[{\citenamefont{Aspelmeyer et~al.}(2014)\citenamefont{Aspelmeyer, Kippenberg, Marquardt}}]{Aspelmeyer2014}
	\bibinfo{author}{\bibnamefont{Aspelmeyer},~\bibfnamefont{M.}},
	\bibinfo{author}{\bibnamefont{Kippenberg},~\bibfnamefont{T.~J.}},
	\bibnamefont{and}
	\bibinfo{author}{\bibnamefont{Marquardt},~\bibfnamefont{F.}},
	\hyperref{https://journals.aps.org/rmp/abstract/10.1103/RevModPhys.86.1391}
	\bibinfo{title}{\bibfnamefont{Cavity optomechanics}}.
	\bibinfo{journal}{Reviews of Modern Physics} \textbf{\bibinfo{volume}{86}},
	\bibinfo{pages}{1391} (\bibinfo{year}{2014})

	\bibitem[{\citenamefont{Rodrigues et~al.}(2021)\citenamefont{Rodrigues, Bothner, Steele}}]{Rodrigues2021}
	\bibinfo{author}{\bibnamefont{Rodrigues},~\bibfnamefont{I.~C.}},
	\bibinfo{author}{\bibnamefont{Bothner},~\bibfnamefont{D.}},
	\bibnamefont{and}
	\bibinfo{author}{\bibnamefont{Steele},~\bibfnamefont{G.~A.}},
	\hyperref{https://www.science.org/doi/10.1126/sciadv.abg6653}
	\bibinfo{title}{\bibfnamefont{Cooling photon-pressure circuits into the quantum regime}}.
	\bibinfo{journal}{Science Advances}
	\textbf{\bibinfo{volume}{7}},
	\bibinfo{pages}{eabg6653} (\bibinfo{year}{2021})

	\bibitem[{\citenamefont{Potts et~al.}(2025)\citenamefont{Potts, Dekker, Deve, Strijbis, Steele}}]{Potts2025}
	\bibinfo{author}{\bibnamefont{Potts},~\bibfnamefont{C.~A.}},
	\bibinfo{author}{\bibnamefont{Dekker},~\bibfnamefont{R.~C.}},
	\bibinfo{author}{\bibnamefont{Deve},~\bibfnamefont{S.}},
	\bibinfo{author}{\bibnamefont{Strijbis},~\bibfnamefont{E.~W.}},
	\bibnamefont{and}
	\bibinfo{author}{\bibnamefont{Steele},~\bibfnamefont{G.~A.}},
	\hyperref{https://journals.aps.org/prl/abstract/10.1103/PhysRevLett.134.153603}
	\bibinfo{title}{\bibfnamefont{Strong Intrinsic Longitudinal Coupling in Circuit Quantum Electrodynamics}}.
	\bibinfo{journal}{Physical Review Letters} \textbf{\bibinfo{volume}{134}},
	\bibinfo{pages}{153603} (\bibinfo{year}{2025})

	\bibitem[{\citenamefont{Rodrigues et~al.}(2022)\citenamefont{Rodrigues, Steele, Bothner}}]{Rodrigues2022}
	\bibinfo{author}{\bibnamefont{Rodrigues},~\bibfnamefont{I.~C.}},
	\bibinfo{author}{\bibnamefont{Steele},~\bibfnamefont{G.~A.}},
	\bibnamefont{and}
	\bibinfo{author}{\bibnamefont{Bothner},~\bibfnamefont{D.}},
	\hyperref{https://www.science.org/doi/10.1126/sciadv.abq1690}
	\bibinfo{title}{\bibfnamefont{Parametrically enhanced interactions and nonreciprocal bath dynamics in a photon-pressure Kerr amplifier}}.
	\bibinfo{journal}{Science Advances} \textbf{\bibinfo{volume}{8}},
	\bibinfo{pages}{eabq1690} (\bibinfo{year}{2022})

	\bibitem[{\citenamefont{Rodrigues et~al.}(2024)\citenamefont{Rodrigues, Steele, Bothner}}]{Rodrigues2024}
	\bibinfo{author}{\bibnamefont{Rodrigues},~\bibfnamefont{I.~C.}},
	\bibinfo{author}{\bibnamefont{Steele},~\bibfnamefont{G.~A.}},
	\bibnamefont{and}
	\bibinfo{author}{\bibnamefont{Bothner},~\bibfnamefont{D.}},
	\hyperref{https://journals.aps.org/prl/abstract/10.1103/PhysRevLett.132.203603}
	\bibinfo{title}{\bibfnamefont{Photon Pressure with an Effective Negative Mass Microwave Mode}}.
	\bibinfo{journal}{Physical Review Letters} \textbf{\bibinfo{volume}{132}},
	\bibinfo{pages}{203603} (\bibinfo{year}{2024})

	\bibitem[{\citenamefont{Elste et~al.}(2009)\citenamefont{Elste, Girvin, Clerk}}]{Elste2009}
	\bibinfo{author}{\bibnamefont{Elste},~\bibfnamefont{F.}},
	\bibinfo{author}{\bibnamefont{Girvin},~\bibfnamefont{S.~M.}},
	\bibnamefont{and}
	\bibinfo{author}{\bibnamefont{Clerk},~\bibfnamefont{A.~A.}},
	\hyperref{https://journals.aps.org/prl/abstract/10.1103/PhysRevLett.102.207209}
	\bibinfo{title}{\bibfnamefont{Quantum Noise Interference and Backaction Cooling in Cavity Nanomechanics}}.
	\bibinfo{journal}{Physical Review Letters} \textbf{\bibinfo{volume}{102}},
	\bibinfo{pages}{207209} (\bibinfo{year}{2009})

	\bibitem[{\citenamefont{Xuereb et~al.}(2011)\citenamefont{Xuereb, Schnabel, Hammerer}}]{Xuereb2011}
	\bibinfo{author}{\bibnamefont{Xuereb},~\bibfnamefont{A.}},
	\bibinfo{author}{\bibnamefont{Schnabel},~\bibfnamefont{R.}},
	\bibnamefont{and}
	\bibinfo{author}{\bibnamefont{Hammerer},~\bibfnamefont{K.}},
	\hyperref{https://journals.aps.org/prl/abstract/10.1103/PhysRevLett.107.213604}
	\bibinfo{title}{\bibfnamefont{Dissipative Optomechanics in a Michelson-Sagnac Interferometer}}.
	\bibinfo{journal}{Physical Review Letters} \textbf{\bibinfo{volume}{107}},
	\bibinfo{pages}{239901} (\bibinfo{year}{2011})

	\bibitem[{\citenamefont{Weiss and Nunnenkamp}(2013)\citenamefont{Weiss, Nunnenkamp}}]{Weiss2013}
	\bibinfo{author}{\bibnamefont{Weiss},~\bibfnamefont{T.}}
	\bibnamefont{and}
	\bibinfo{author}{\bibnamefont{Nunnenkamp},~\bibfnamefont{A.}},
	\hyperref{https://journals.aps.org/pra/abstract/10.1103/PhysRevA.88.023850}
	\bibinfo{title}{\bibfnamefont{Quantum limit of laser cooling in dispersively and dissipatively coupled optomechanical systems}}.
	\bibinfo{journal}{Physical Review A} \textbf{\bibinfo{volume}{88}},
	\bibinfo{pages}{023850} (\bibinfo{year}{2013})

	\bibitem[{\citenamefont{Kilda and Nunnenkamp}(2015)\citenamefont{Kilda, Nunnenkamp}}]{Kilda2015}
	\bibinfo{author}{\bibnamefont{Kilda},~\bibfnamefont{D.}}
	\bibnamefont{and}
	\bibinfo{author}{\bibnamefont{Nunnenkamp},~\bibfnamefont{A.}},
	\hyperref{https://iopscience.iop.org/article/10.1088/2040-8978/18/1/014007}
	\bibinfo{title}{\bibfnamefont{Squeezed light and correlated photons from dissipatively coupled optomechanical systems}}.
	\bibinfo{journal}{Journal of Optics} \textbf{\bibinfo{volume}{18}},
	\bibinfo{pages}{014007} (\bibinfo{year}{2015})

	\bibitem[{\citenamefont{Yanay et~al.}(2016)\citenamefont{Yanay, Sankey, Clerk}}]{Yanay2016}
	\bibinfo{author}{\bibnamefont{Yanay},~\bibfnamefont{Y.}},
	\bibinfo{author}{\bibnamefont{Sankey},~\bibfnamefont{J.~C.}},
	\bibnamefont{and}
	\bibinfo{author}{\bibnamefont{Clerk},~\bibfnamefont{A.~A.}},
	\hyperref{https://journals.aps.org/pra/abstract/10.1103/PhysRevA.93.063809}
	\bibinfo{title}{\bibfnamefont{Quantum backaction and noise interference in asymmetric two-cavity optomechanical systems}}.
	\bibinfo{journal}{Physical Review A} \textbf{\bibinfo{volume}{93}},
	\bibinfo{pages}{063809} (\bibinfo{year}{2016})

	\bibitem[{\citenamefont{Tagantsev and Fedorov}(2019)\citenamefont{Tagantsev, Fedorov}}]{Tagantsev2019}
	\bibinfo{author}{\bibnamefont{Tagantsev},~\bibfnamefont{A.~K.}}
	\bibnamefont{and}
	\bibinfo{author}{\bibnamefont{Fedorov},~\bibfnamefont{S.~A.}},
	\hyperref{https://journals.aps.org/prl/abstract/10.1103/PhysRevLett.123.043602}
	\bibinfo{title}{\bibfnamefont{Quantum-Limited Measurements Using an Optical Cavity with Modulated Intrinsic Loss}}.
	\bibinfo{journal}{Physical Review Letters} \textbf{\bibinfo{volume}{123}},
	\bibinfo{pages}{043602} (\bibinfo{year}{2019})

	\bibitem[{\citenamefont{Baraillon et~al.}(2020)\citenamefont{Baraillon, Taurel, Labeye, Duraffourg}}]{Baraillon2020}
	\bibinfo{author}{\bibnamefont{Baraillon},~\bibfnamefont{J.}},
	\bibinfo{author}{\bibnamefont{Taurel},~\bibfnamefont{B.}},
	\bibinfo{author}{\bibnamefont{Labeye},~\bibfnamefont{P.}},
	\bibnamefont{and}
	\bibinfo{author}{\bibnamefont{Duraffourg},~\bibfnamefont{L.}},
	\hyperref{https://journals.aps.org/pra/abstract/10.1103/PhysRevA.102.033509}
	\bibinfo{title}{\bibfnamefont{Linear analytical approach to dispersive, external dissipative and intrinsic dissipative couplings in optomechanical systems}}.
	\bibinfo{journal}{Physical Review A} \textbf{\bibinfo{volume}{102}},
	\bibinfo{pages}{033509} (\bibinfo{year}{2020})

	\bibitem[{\citenamefont{Tagantsev and Polzik}(2021)\citenamefont{Tagantsev, Polzik}}]{Tagantsev2021}
	\bibinfo{author}{\bibnamefont{Tagantsev},~\bibfnamefont{A.~K.}}
	\bibnamefont{and}
	\bibinfo{author}{\bibnamefont{Polzik},~\bibfnamefont{E.~S.}},
	\hyperref{https://journals.aps.org/pra/abstract/10.1103/PhysRevA.103.063503}
	\bibinfo{title}{\bibfnamefont{Dissipative optomechanical coupling with a membrane outside of an optical cavity}}.
	\bibinfo{journal}{Physical Review A} \textbf{\bibinfo{volume}{103}},
	\bibinfo{pages}{063503} (\bibinfo{year}{2021})

	\bibitem[{\citenamefont{Monsel et~al.}(2024)\citenamefont{Monsel, Ciers, Manjeshwar, Wieczorek, Splettstoesser}}]{Monsel2024}
	\bibinfo{author}{\bibnamefont{Monsel},~\bibfnamefont{J.}},
	\bibinfo{author}{\bibnamefont{Ciers},~\bibfnamefont{A.}},
	\bibinfo{author}{\bibnamefont{Manjeshwar},~\bibfnamefont{S.~K.}},
	\bibinfo{author}{\bibnamefont{Wieczorek},~\bibfnamefont{W.}},
	\bibnamefont{and}
	\bibinfo{author}{\bibnamefont{Splettstoesser},~\bibfnamefont{J.}},
	\hyperref{https://journals.aps.org/pra/abstract/10.1103/PhysRevA.109.043532}
	\bibinfo{title}{\bibfnamefont{Dissipative and dispersive cavity optomechanics with a frequency-dependent mirror}}.
	\bibinfo{journal}{Physical Review A} \textbf{\bibinfo{volume}{109}},
	\bibinfo{pages}{043532} (\bibinfo{year}{2024})

	\bibitem[{\citenamefont{Li et~al.}(2009)\citenamefont{Li, Pernice, Tang}}]{Li2009}
	\bibinfo{author}{\bibnamefont{Li},~\bibfnamefont{M.}},
	\bibinfo{author}{\bibnamefont{Pernice},~\bibfnamefont{W.~H.~P.}},
	\bibnamefont{and}
	\bibinfo{author}{\bibnamefont{Tang},~\bibfnamefont{H.~X.}},
	\hyperref{https://journals.aps.org/prl/abstract/10.1103/PhysRevLett.103.223901}
	\bibinfo{title}{\bibfnamefont{Reactive Cavity Optical Force on Mikrodisk-Coupled Nanomechanical Beam Waveguides}}.
	\bibinfo{journal}{Physical Review Letters} \textbf{\bibinfo{volume}{103}},
	\bibinfo{pages}{223901} (\bibinfo{year}{2009})

	\bibitem[{\citenamefont{Wu et~al.}(2014)\citenamefont{Wu, Hryciw, Healey, Lake, Jayakumar, Freeman, Davis, Barclay}}]{Wu2014}
	\bibinfo{author}{\bibnamefont{Wu},~\bibfnamefont{M.}},
	\bibinfo{author}{\bibnamefont{Hryciw},~\bibfnamefont{A.~C.}},
	\bibinfo{author}{\bibnamefont{Healey},~\bibfnamefont{C.}},
	\bibinfo{author}{\bibnamefont{Lake},~\bibfnamefont{D.~P.}},
	\bibinfo{author}{\bibnamefont{Jayakumar},~\bibfnamefont{H.}},
	\bibinfo{author}{\bibnamefont{Freeman},~\bibfnamefont{M.~R.}},
	\bibinfo{author}{\bibnamefont{Davis},~\bibfnamefont{J.~P.}},
	\bibnamefont{and}
	\bibinfo{author}{\bibnamefont{Barclay},~\bibfnamefont{P.~E.}},
	\hyperref{https://journals.aps.org/prx/abstract/10.1103/PhysRevX.4.021052}
	\bibinfo{title}{\bibfnamefont{Dissipative and Dispersive Optomechanics in a Nanocavity Torque Sensor}}.
	\bibinfo{journal}{Physical Review X} \textbf{\bibinfo{volume}{4}},
	\bibinfo{pages}{021052} (\bibinfo{year}{2014})

	\bibitem[{\citenamefont{Sawadsky et~al.}(2015)\citenamefont{Sawadsky, Kaufer, Moghadas~Nia, Tarabrin, Khalili, Hammerer, Schnabel}}]{Sawadsky2015}
	\bibinfo{author}{\bibnamefont{Sawadsky},~\bibfnamefont{A.}},
	\bibinfo{author}{\bibnamefont{Kaufer},~\bibfnamefont{H.}},
	\bibinfo{author}{\bibnamefont{Moghadas~Nia},~\bibfnamefont{R.}},
	\bibinfo{author}{\bibnamefont{Tarabrin},~\bibfnamefont{S.~P.}},
	\bibinfo{author}{\bibnamefont{Khalili},~\bibfnamefont{F.~Ya..}},
	\bibinfo{author}{\bibnamefont{Hammerer},~\bibfnamefont{K.}},
	\bibnamefont{and}
	\bibinfo{author}{\bibnamefont{Schnabel},~\bibfnamefont{R.}},
	\hyperref{https://journals.aps.org/prl/abstract/10.1103/PhysRevLett.114.043601}
	\bibinfo{title}{\bibfnamefont{Observation of Generalized Optomechanical Coupling and Cooling on Cavity Resonance}}.
	\bibinfo{journal}{Physical Review Letters} \textbf{\bibinfo{volume}{114}},
	\bibinfo{pages}{043601} (\bibinfo{year}{2015})

	\bibitem[{\citenamefont{Primo et~al.}(2023)\citenamefont{Primo, Pinho, Benevides, Gröblacher, Wiederhecker, Mayer~Alegre}}]{Primo2023}
	\bibinfo{author}{\bibnamefont{Primo},~\bibfnamefont{A.~G.}},
	\bibinfo{author}{\bibnamefont{Pinho},~\bibfnamefont{P.~V.}},
	\bibinfo{author}{\bibnamefont{Benevides},~\bibfnamefont{R.}},
	\bibinfo{author}{\bibnamefont{Gröblacher},~\bibfnamefont{S.}},
	\bibinfo{author}{\bibnamefont{Wiederhecker},~\bibfnamefont{G.~S.}},
	\bibnamefont{and}
	\bibinfo{author}{\bibnamefont{Mayer~Alegre},~\bibfnamefont{T.~P.}},
	\hyperref{https://www.nature.com/articles/s41467-023-41127-7}
	\bibinfo{title}{\bibfnamefont{Dissipative optomechanics in high-frequency nanomechanical resonators}}.
	\bibinfo{journal}{Nature Communications} \textbf{\bibinfo{volume}{14}},
	\bibinfo{pages}{5793} (\bibinfo{year}{2023})

	\bibitem[{\citenamefont{Chaudhuri}(2019)\citenamefont{Chaudhuri}}]{Chaudhuri2019}
	\bibinfo{author}{\bibnamefont{Chaudhuri},~\bibfnamefont{S.}},
	\hyperref{https://purl.stanford.edu/qm978sp7183}
	\bibinfo{title}{\bibfnamefont{The dark matter radio: a quantum-enhanced search for QCD axion dark matter}}.
	\bibinfo{journal}{PhD thesis},
	\bibinfo{pages}{Stanford University} (\bibinfo{year}{2019})

	\bibitem[{\citenamefont{Brouwer et~al.}(2022)\citenamefont{Brouwer, Chaudhuri, Cho, Corbin, Dawson, Droster, Foster, Fry, Graham, Henning, Irwin, Kadribasic, Kahn, Keller, Kolevatov, Kuenstner, Leder, Li, Ouellet, Pappas, Phipps, Rapidis, Safdi, Salemi, Simanovskaia, Singh, van~Assendelft, van~Bibber, Wells, Winslow, Wisniewski, Young}}]{Brouwer2022}
	\bibinfo{author}{\bibnamefont{Brouwer},~\bibfnamefont{L.}},
	\bibinfo{author}{\bibnamefont{Chaudhuri},~\bibfnamefont{S.}},
	\bibinfo{author}{\bibnamefont{Cho},~\bibfnamefont{H.-M.}},
	\bibinfo{author}{\bibnamefont{Corbin},~\bibfnamefont{J.}},
	\bibinfo{author}{\bibnamefont{Dawson},~\bibfnamefont{C.~S.}},
	\bibinfo{author}{\bibnamefont{Droster},~\bibfnamefont{A.}},
	\bibinfo{author}{\bibnamefont{Foster},~\bibfnamefont{J.~W.}},
	\bibinfo{author}{\bibnamefont{Fry},~\bibfnamefont{J.~T.}},
	\bibinfo{author}{\bibnamefont{Graham},~\bibfnamefont{P.~W.}},
	\bibinfo{author}{\bibnamefont{Henning},~\bibfnamefont{R.}},
	\bibinfo{author}{\bibnamefont{Irwin},~\bibfnamefont{K.~D.}},
	\bibinfo{author}{\bibnamefont{Kadribasic},~\bibfnamefont{F.}},
	\bibinfo{author}{\bibnamefont{Kahn},~\bibfnamefont{Y.}},
	\bibinfo{author}{\bibnamefont{Keller},~\bibfnamefont{A.}},
	\bibinfo{author}{\bibnamefont{Kolevatov},~\bibfnamefont{R.}},
	\bibinfo{author}{\bibnamefont{Kuenstner},~\bibfnamefont{S.}},
	\bibinfo{author}{\bibnamefont{Leder},~\bibfnamefont{A.~F.}},
	\bibinfo{author}{\bibnamefont{Li},~\bibfnamefont{D.}},
	\bibinfo{author}{\bibnamefont{Ouellet},~\bibfnamefont{J.~L.}},
	\bibinfo{author}{\bibnamefont{Pappas},~\bibfnamefont{K.~M.~W.}},
	\bibinfo{author}{\bibnamefont{Phipps},~\bibfnamefont{A.}},
	\bibinfo{author}{\bibnamefont{Rapidis},~\bibfnamefont{N.~M.}},
	\bibinfo{author}{\bibnamefont{Safdi},~\bibfnamefont{B.~R.}},
	\bibinfo{author}{\bibnamefont{Salemi},~\bibfnamefont{C.~P.}},
	\bibinfo{author}{\bibnamefont{Simanovskaia},~\bibfnamefont{M.}},
	\bibinfo{author}{\bibnamefont{Singh},~\bibfnamefont{J.}},
	\bibinfo{author}{\bibnamefont{van~Assendelft},~\bibfnamefont{E.~C.}},
	\bibinfo{author}{\bibnamefont{van~Bibber},~\bibfnamefont{K.}},
	\bibinfo{author}{\bibnamefont{Wells},~\bibfnamefont{K.}},
	\bibinfo{author}{\bibnamefont{Winslow},~\bibfnamefont{L.}},
	\bibinfo{author}{\bibnamefont{Wisniewski},~\bibfnamefont{W.~J.}},
	\bibnamefont{and}
	\bibinfo{author}{\bibnamefont{Young},~\bibfnamefont{B.~A.}},
	\hyperref{https://journals.aps.org/prd/abstract/10.1103/PhysRevD.106.112003}
	\bibinfo{title}{\bibfnamefont{Proposal for a definitive search for GUT-scale QCD axions}}.
	\bibinfo{journal}{Physical Review D} \textbf{\bibinfo{volume}{106}},
	\bibinfo{pages}{112003} (\bibinfo{year}{2022})

	\bibitem[{\citenamefont{Kuenstner et~al.}(2025)\citenamefont{Kuenstner, van~Assendelft, Chaudhuri, Cho, Corbin, Henderson, Kadribasic, Li, Phipps, Rapidis, Simanovskaia, Singh, Yu, Irwin}}]{Kuenstner2025}
	\bibinfo{author}{\bibnamefont{Kuenstner},~\bibfnamefont{S.}},
	\bibinfo{author}{\bibnamefont{van~Assendelft},~\bibfnamefont{E.~C.}},
	\bibinfo{author}{\bibnamefont{Chaudhuri},~\bibfnamefont{S.}},
	\bibinfo{author}{\bibnamefont{Cho},~\bibfnamefont{H.-M.}},
	\bibinfo{author}{\bibnamefont{Corbin},~\bibfnamefont{J.}},
	\bibinfo{author}{\bibnamefont{Henderson},~\bibfnamefont{S.~W.}},
	\bibinfo{author}{\bibnamefont{Kadribasic},~\bibfnamefont{F.}},
	\bibinfo{author}{\bibnamefont{Li},~\bibfnamefont{D.}},
	\bibinfo{author}{\bibnamefont{Phipps},~\bibfnamefont{A.}},
	\bibinfo{author}{\bibnamefont{Rapidis},~\bibfnamefont{N.~M.}},
	\bibinfo{author}{\bibnamefont{Simanovskaia},~\bibfnamefont{M.}},
	\bibinfo{author}{\bibnamefont{Singh},~\bibfnamefont{J.}},
	\bibinfo{author}{\bibnamefont{Yu},~\bibfnamefont{C.}},
	\bibnamefont{and}
	\bibinfo{author}{\bibnamefont{Irwin},~\bibfnamefont{K.~D.}},
	\hyperref{https://journals.aps.org/prresearch/abstract/10.1103/PhysRevResearch.7.013281}
	\bibinfo{title}{\bibfnamefont{Quantum metrology of low-frequency electromagnetic modes with frequency upconverters}}.
	\bibinfo{journal}{Physical Review Research} \textbf{\bibinfo{volume}{7}},
	\bibinfo{pages}{013281} (\bibinfo{year}{2025})

	\bibitem[{\citenamefont{Mikkelsen et~al.}(2017)\citenamefont{Mikkelsen, Fogarty, Twamley, Busch}}]{Mikkelsen2017}
	\bibinfo{author}{\bibnamefont{Mikkelsen},~\bibfnamefont{M.}},
	\bibinfo{author}{\bibnamefont{Fogarty},~\bibfnamefont{T.}},
	\bibinfo{author}{\bibnamefont{Twamley},~\bibfnamefont{J.}},
	\bibnamefont{and}
	\bibinfo{author}{\bibnamefont{Busch},~\bibfnamefont{Th.}},
	\hyperref{https://journals.aps.org/pra/abstract/10.1103/PhysRevA.96.043832}
	\bibinfo{title}{\bibfnamefont{Optomechanics with a position-modulated Kerr-type nonlinear coupling}}.
	\bibinfo{journal}{Physical Review A} \textbf{\bibinfo{volume}{96}},
	\bibinfo{pages}{043832} (\bibinfo{year}{2017})

	\bibitem[{\citenamefont{Djorwe et~al.}(2019)\citenamefont{Djorwe, Pennec, Djafari-Rouhani}}]{Djorwe2019}
	\bibinfo{author}{\bibnamefont{Djorwe},~\bibfnamefont{P.}},
	\bibinfo{author}{\bibnamefont{Pennec},~\bibfnamefont{Y.}},
	\bibnamefont{and}
	\bibinfo{author}{\bibnamefont{Djafari-Rouhani},~\bibfnamefont{B.}},
	\hyperref{https://www.nature.com/articles/s41598-019-38578-8}
	\bibinfo{title}{\bibfnamefont{Low-power phonon lasing through position-modulated Kerr-type nonlinearity}}.
	\bibinfo{journal}{Scientific Reports} \textbf{\bibinfo{volume}{9}},
	\bibinfo{pages}{1684} (\bibinfo{year}{2019})

	\bibitem[{\citenamefont{Rodrigues et~al.}(2019)\citenamefont{Rodrigues, Bothner, Steele}}]{Rodrigues2019}
	\bibinfo{author}{\bibnamefont{Rodrigues},~\bibfnamefont{I.~C.}},
	\bibinfo{author}{\bibnamefont{Bothner},~\bibfnamefont{D.}},
	\bibnamefont{and}
	\bibinfo{author}{\bibnamefont{Steele},~\bibfnamefont{G.~A.}},
	\hyperref{https://www.nature.com/articles/s41467-019-12964-2}
	\bibinfo{title}{\bibfnamefont{Coupling microwave photons to a mechanical resonator using quantum interference}}.
	\bibinfo{journal}{Nature Communications} \textbf{\bibinfo{volume}{10}},
	\bibinfo{pages}{5359} (\bibinfo{year}{2019})

	\bibitem[{\citenamefont{Zoepfl et~al.}(2020)\citenamefont{Zoepfl, Juan, Schneider, Kirchmair}}]{Zoepfl2020}
	\bibinfo{author}{\bibnamefont{Zoepfl},~\bibfnamefont{D.}},
	\bibinfo{author}{\bibnamefont{Juan},~\bibfnamefont{M.~L.}},
	\bibinfo{author}{\bibnamefont{Schneider},~\bibfnamefont{C.~M.~F.}},
	\bibnamefont{and}
	\bibinfo{author}{\bibnamefont{Kirchmair},~\bibfnamefont{G.}},
	\hyperref{https://journals.aps.org/prl/abstract/10.1103/PhysRevLett.125.023601}
	\bibinfo{title}{\bibfnamefont{Single-Photon Cooling in Microwave Magnetomechanics}}.
	\bibinfo{journal}{Physical Review Letters} \textbf{\bibinfo{volume}{125}},
	\bibinfo{pages}{023601} (\bibinfo{year}{2020})

	\bibitem[{\citenamefont{Schmidt et~al.}(2020)\citenamefont{Schmidt, Amawi, Pogorzalek, Deppe, Marx, Gross, Huebl}}]{Schmidt2020}
	\bibinfo{author}{\bibnamefont{Schmidt},~\bibfnamefont{P.}},
	\bibinfo{author}{\bibnamefont{Amawi},~\bibfnamefont{M.~T.}},
	\bibinfo{author}{\bibnamefont{Pogorzalek},~\bibfnamefont{S.}},
	\bibinfo{author}{\bibnamefont{Deppe},~\bibfnamefont{F.}},
	\bibinfo{author}{\bibnamefont{Marx},~\bibfnamefont{A.}},
	\bibinfo{author}{\bibnamefont{Gross},~\bibfnamefont{R.}},
	\bibnamefont{and}
	\bibinfo{author}{\bibnamefont{Huebl},~\bibfnamefont{H.}},
	\hyperref{https://www.nature.com/articles/s42005-020-00501-3}
	\bibinfo{title}{\bibfnamefont{Sideband-resolved resonator electromechanics based on a nonlinear Josephson inductance probed on the single-photon level}}.
	\bibinfo{journal}{Communications Physics} \textbf{\bibinfo{volume}{3}},
	\bibinfo{pages}{233} (\bibinfo{year}{2020})

	\bibitem[{\citenamefont{Bera et~al.}(2021)\citenamefont{Bera, Majumder, Sahu, Singh}}]{Bera2021}
	\bibinfo{author}{\bibnamefont{Bera},~\bibfnamefont{T.}},
	\bibinfo{author}{\bibnamefont{Majumder},~\bibfnamefont{S.}},
	\bibinfo{author}{\bibnamefont{Sahu},~\bibfnamefont{S.~K.}},
	\bibnamefont{and}
	\bibinfo{author}{\bibnamefont{Singh},~\bibfnamefont{V.}},
	\hyperref{https://www.nature.com/articles/s42005-020-00514-y}
	\bibinfo{title}{\bibfnamefont{Large flux-mediated coupling in hybrid electromechanical system with a transmon qubit}}.
	\bibinfo{journal}{Communications Physics} \textbf{\bibinfo{volume}{4}},
	\bibinfo{pages}{12} (\bibinfo{year}{2021})

	\bibitem[{\citenamefont{Bothner et~al.}(2022)\citenamefont{Bothner, Rodrigues, Steele}}]{Bothner2022}
	\bibinfo{author}{\bibnamefont{Bothner},~\bibfnamefont{D.}},
	\bibinfo{author}{\bibnamefont{Rodrigues},~\bibfnamefont{I.~C.}},
	\bibnamefont{and}
	\bibinfo{author}{\bibnamefont{Steele},~\bibfnamefont{G.~A.}},
	\hyperref{https://www.nature.com/articles/s42005-022-00808-3}
	\bibinfo{title}{\bibfnamefont{Four-wave-cooling to the single phonon level in Kerr optomechanics}}.
	\bibinfo{journal}{Communications Physics} \textbf{\bibinfo{volume}{5}},
	\bibinfo{pages}{33} (\bibinfo{year}{2022})

	\bibitem[{\citenamefont{Luschmann et~al.}(2022)\citenamefont{Luschmann, Schmidt, Deppe, Marx, Sanchez, Gross, Huebl}}]{Luschmann2022}
	\bibinfo{author}{\bibnamefont{Luschmann},~\bibfnamefont{T.}},
	\bibinfo{author}{\bibnamefont{Schmidt},~\bibfnamefont{P.}},
	\bibinfo{author}{\bibnamefont{Deppe},~\bibfnamefont{F.}},
	\bibinfo{author}{\bibnamefont{Marx},~\bibfnamefont{A.}},
	\bibinfo{author}{\bibnamefont{Sanchez},~\bibfnamefont{A.}},
	\bibinfo{author}{\bibnamefont{Gross},~\bibfnamefont{R.}},
	\bibnamefont{and}
	\bibinfo{author}{\bibnamefont{Huebl},~\bibfnamefont{H.}},
	\hyperref{https://www.nature.com/articles/s41598-022-05438-x}
	\bibinfo{title}{\bibfnamefont{Mechanical frequency control in inductively coupled electromechanical systems}}.
	\bibinfo{journal}{Scientific reports} \textbf{\bibinfo{volume}{12}},
	\bibinfo{pages}{1608} (\bibinfo{year}{2022})

	\bibitem[{\citenamefont{Zoepfl et~al.}(2023)\citenamefont{Zoepfl, Juan, Diaz-Naufal, Schneider, Deeg, Sharafiev, Metelmann, Kirchmair}}]{Zoepfl2023}
	\bibinfo{author}{\bibnamefont{Zoepfl},~\bibfnamefont{D.}},
	\bibinfo{author}{\bibnamefont{Juan},~\bibfnamefont{M.~L.}},
	\bibinfo{author}{\bibnamefont{Diaz-Naufal},~\bibfnamefont{N.}},
	\bibinfo{author}{\bibnamefont{Schneider},~\bibfnamefont{C.~M.~F.}},
	\bibinfo{author}{\bibnamefont{Deeg},~\bibfnamefont{L.~F.}},
	\bibinfo{author}{\bibnamefont{Sharafiev},~\bibfnamefont{A.}},
	\bibinfo{author}{\bibnamefont{Metelmann},~\bibfnamefont{A.}},
	\bibnamefont{and}
	\bibinfo{author}{\bibnamefont{Kirchmair},~\bibfnamefont{G.}},
	\hyperref{https://journals.aps.org/prl/abstract/10.1103/PhysRevLett.130.033601}
	\bibinfo{title}{\bibfnamefont{Kerr Enhanced Backaction Cooling in Magneomechanics}}.
	\bibinfo{journal}{Physical Review Letters} \textbf{\bibinfo{volume}{130}},
	\bibinfo{pages}{033601} (\bibinfo{year}{2023})

	\bibitem[{\citenamefont{Schmidt et~al.}(2024)\citenamefont{Schmidt, Claessen, Higgins, Hofer, Hansen, Asenbaum, Zemlicka, Uhl, Kleiner, Gross, Huebl, Trupke, Aspelmeyer}}]{Schmidt2024}
	\bibinfo{author}{\bibnamefont{Schmidt},~\bibfnamefont{P.}},
	\bibinfo{author}{\bibnamefont{Claessen},~\bibfnamefont{R.}},
	\bibinfo{author}{\bibnamefont{Higgins},~\bibfnamefont{G.}},
	\bibinfo{author}{\bibnamefont{Hofer},~\bibfnamefont{J.}},
	\bibinfo{author}{\bibnamefont{Hansen},~\bibfnamefont{J.~J.}},
	\bibinfo{author}{\bibnamefont{Asenbaum},~\bibfnamefont{P.}},
	\bibinfo{author}{\bibnamefont{Zemlicka},~\bibfnamefont{M.}},
	\bibinfo{author}{\bibnamefont{Uhl},~\bibfnamefont{K.}},
	\bibinfo{author}{\bibnamefont{Kleiner},~\bibfnamefont{R.}},
	\bibinfo{author}{\bibnamefont{Gross},~\bibfnamefont{R.}},
	\bibinfo{author}{\bibnamefont{Huebl},~\bibfnamefont{H.}},
	\bibinfo{author}{\bibnamefont{Trupke},~\bibfnamefont{M.}},
	\bibnamefont{and}
	\bibinfo{author}{\bibnamefont{Aspelmeyer},~\bibfnamefont{M.}},
	\hyperref{https://journals.aps.org/prapplied/abstract/10.1103/PhysRevApplied.22.014078}
	\bibinfo{title}{\bibfnamefont{Remote sensing of a levitated superconductor with a flux-tunable microwave cavity}}.
	\bibinfo{journal}{Physical Review Applied} \textbf{\bibinfo{volume}{22}},
	\bibinfo{pages}{014078} (\bibinfo{year}{2024})

	\bibitem[{\citenamefont{Uhl et~al.}(2024)\citenamefont{Uhl, Hackenbeck, Peter, Kleiner, Koelle, Bothner}}]{Uhl2024}
	\bibinfo{author}{\bibnamefont{Uhl},~\bibfnamefont{K.}},
	\bibinfo{author}{\bibnamefont{Hackenbeck},~\bibfnamefont{D.}},
	\bibinfo{author}{\bibnamefont{Peter},~\bibfnamefont{J.}},
	\bibinfo{author}{\bibnamefont{Kleiner},~\bibfnamefont{R.}},
	\bibinfo{author}{\bibnamefont{Koelle},~\bibfnamefont{D.}},
	\bibnamefont{and}
	\bibinfo{author}{\bibnamefont{Bothner},~\bibfnamefont{D.}},
	\hyperref{https://journals.aps.org/prapplied/abstract/10.1103/PhysRevApplied.21.024051}
	\bibinfo{title}{\bibfnamefont{Niobium quantum interference microwave circuits with monolithic three-dimensional nanobridge junctions}}.
	\bibinfo{journal}{Physical Review Applied} \textbf{\bibinfo{volume}{21}},
	\bibinfo{pages}{024051} (\bibinfo{year}{2024})

	\bibitem[{\citenamefont{Uhl et~al.}(2024)\citenamefont{Uhl, Hackenbeck, Koelle, Kleiner, Bothner}}]{Uhl2024a}
	\bibinfo{author}{\bibnamefont{Uhl},~\bibfnamefont{K.}},
	\bibinfo{author}{\bibnamefont{Hackenbeck},~\bibfnamefont{D.}},
	\bibinfo{author}{\bibnamefont{Koelle},~\bibfnamefont{D.}},
	\bibinfo{author}{\bibnamefont{Kleiner},~\bibfnamefont{R.}},
	\bibnamefont{and}
	\bibinfo{author}{\bibnamefont{Bothner},~\bibfnamefont{D.}},
	\hyperref{https://journals.aps.org/prapplied/abstract/10.1103/PhysRevApplied.22.064052}
	\bibinfo{title}{\bibfnamefont{Extracting the current-phase relation of a monolithic three-dimensional nanoconstriction using a dc-current-tunable superconducting microwave cavity}}.
	\bibinfo{journal}{Physical Review Applied} \textbf{\bibinfo{volume}{22}},
	\bibinfo{pages}{064052} (\bibinfo{year}{2024})

	\bibitem[{\citenamefont{Fleischhauer et~al.}(2005)\citenamefont{Fleischhauer, Imamoglu, Marangos}}]{Fleischhauer2005}
	\bibinfo{author}{\bibnamefont{Fleischhauer},~\bibfnamefont{M.}},
	\bibinfo{author}{\bibnamefont{Imamoglu},~\bibfnamefont{A.}},
	\bibnamefont{and}
	\bibinfo{author}{\bibnamefont{Marangos},~\bibfnamefont{J.~P.}},
	\hyperref{https://journals.aps.org/rmp/abstract/10.1103/RevModPhys.77.633}
	\bibinfo{title}{\bibfnamefont{Electromagnetically induced transparency: Optics in coherent media}}.
	\bibinfo{journal}{Reviews of Modern Physics} \textbf{\bibinfo{volume}{77}},
	\bibinfo{pages}{633} (\bibinfo{year}{2005})

	\bibitem[{\citenamefont{Weis et~al.}(2010)\citenamefont{Weis, Riviere, Deleglise, Gavartin, Arcizet, Schliesser, Kippenberg}}]{Weis2010}
	\bibinfo{author}{\bibnamefont{Weis},~\bibfnamefont{S.}},
	\bibinfo{author}{\bibnamefont{Riviere},~\bibfnamefont{R.}},
	\bibinfo{author}{\bibnamefont{Deleglise},~\bibfnamefont{S.}},
	\bibinfo{author}{\bibnamefont{Gavartin},~\bibfnamefont{E.}},
	\bibinfo{author}{\bibnamefont{Arcizet},~\bibfnamefont{O.}},
	\bibinfo{author}{\bibnamefont{Schliesser},~\bibfnamefont{A.}},
	\bibnamefont{and}
	\bibinfo{author}{\bibnamefont{Kippenberg},~\bibfnamefont{T.~J.}},
	\hyperref{https://www.science.org/doi/10.1126/science.1195596}
	\bibinfo{title}{\bibfnamefont{Optomechanically Induced Transparency}}.
	\bibinfo{journal}{Science} \textbf{\bibinfo{volume}{330}},
	\bibinfo{pages}{1520-1523} (\bibinfo{year}{2010})

	\bibitem[{\citenamefont{Frattini et~al.}(2018)\citenamefont{Frattini, Sivak, Lingenfelter, Shankar, Devoret}}]{Frattini2018}
	\bibinfo{author}{\bibnamefont{Frattini},~\bibfnamefont{N.~E.}},
	\bibinfo{author}{\bibnamefont{Sivak},~\bibfnamefont{V.~V.}},
	\bibinfo{author}{\bibnamefont{Lingenfelter},~\bibfnamefont{A.}},
	\bibinfo{author}{\bibnamefont{Shankar},~\bibfnamefont{S.}},
	\bibnamefont{and}
	\bibinfo{author}{\bibnamefont{Devoret},~\bibfnamefont{M.~H.}},
	\hyperref{https://journals.aps.org/prapplied/abstract/10.1103/PhysRevApplied.10.054020}
	\bibinfo{title}{\bibfnamefont{Optimizing the Nonlinearity and Dissipation of a SNAIL Parametric Amplifier for Dynamic Range}}.
	\bibinfo{journal}{Physical Review Applied} \textbf{\bibinfo{volume}{10}},
	\bibinfo{pages}{054020} (\bibinfo{year}{2018})

	\bibitem[{\citenamefont{Teufel et~al.}(2008)\citenamefont{Teufel, Harlow, Regal, Lehnert}}]{Teufel2008}
	\bibinfo{author}{\bibnamefont{Teufel},~\bibfnamefont{J.~D.}},
	\bibinfo{author}{\bibnamefont{Harlow},~\bibfnamefont{J.~W.}},
	\bibinfo{author}{\bibnamefont{Regal},~\bibfnamefont{C.~A.}},
	\bibnamefont{and}
	\bibinfo{author}{\bibnamefont{Lehnert},~\bibfnamefont{K.~W.}},
	\hyperref{https://journals.aps.org/prl/abstract/10.1103/PhysRevLett.101.197203}
	\bibinfo{title}{\bibfnamefont{Dynamical Backaction of Microwave Fields on a Nanomechanical Oscillator}}.
	\bibinfo{journal}{Physical Review Letters} \textbf{\bibinfo{volume}{101}},
	\bibinfo{pages}{197203} (\bibinfo{year}{2008})

	\bibitem[{\citenamefont{Igreja and Dias}(2004)\citenamefont{Igreja, Dias}}]{Igreja2004}
	\bibinfo{author}{\bibnamefont{Igreja},~\bibfnamefont{R.}} \bibnamefont{and}
	\bibinfo{author}{\bibnamefont{Dias},~\bibfnamefont{C.~J.}},
	\hyperref{https://www.sciencedirect.com/science/article/pii/S0924424704000779?via\%3Dihub}
	\bibinfo{title}{\bibfnamefont{Analytical evaluation of the interdigital electrodes capacitance for a multi-layered structure}}.
	\bibinfo{journal}{Sensors and Actuators A} \textbf{\bibinfo{volume}{112}},
	\bibinfo{pages}{291-301} (\bibinfo{year}{2004})

	\bibitem[{\citenamefont{Wenner et~al.}(2011)\citenamefont{Wenner, Neeley, Bialczak, Lenander, Lucero, O'Connell, Sank, Wang, Weides, Cleland, Martinis}}]{Wenner2011}
	\bibinfo{author}{\bibnamefont{Wenner},~\bibfnamefont{J.}},
	\bibinfo{author}{\bibnamefont{Neeley},~\bibfnamefont{M.}},
	\bibinfo{author}{\bibnamefont{Bialczak},~\bibfnamefont{Radoslav~C.}},
	\bibinfo{author}{\bibnamefont{Lenander},~\bibfnamefont{M.}},
	\bibinfo{author}{\bibnamefont{Lucero},~\bibfnamefont{E.}},
	\bibinfo{author}{\bibnamefont{O'Connell},~\bibfnamefont{A.~D.}},
	\bibinfo{author}{\bibnamefont{Sank},~\bibfnamefont{D.}},
	\bibinfo{author}{\bibnamefont{Wang},~\bibfnamefont{H.}},
	\bibinfo{author}{\bibnamefont{Weides},~\bibfnamefont{M.}},
	\bibinfo{author}{\bibnamefont{Cleland},~\bibfnamefont{A.~N.}},
	\bibnamefont{and}
	\bibinfo{author}{\bibnamefont{Martinis},~\bibfnamefont{J.~M.}},
	\hyperref{https://iopscience.iop.org/article/10.1088/0953-2048/24/6/065001}
	\bibinfo{title}{\bibfnamefont{Wirebond crosstalk and cavity modes in large chip mounts for superconducting qubits}}.
	\bibinfo{journal}{Superconductor Science and Technology} \textbf{\bibinfo{volume}{24}},
	\bibinfo{pages}{065001} (\bibinfo{year}{2011})

	\bibitem[{\citenamefont{Rieger et~al.}(2023)\citenamefont{Rieger, Günzler, Spiecker, Nambisan, Wernsdorfer, Pop}}]{Rieger2023}
	\bibinfo{author}{\bibnamefont{Rieger},~\bibfnamefont{D.}},
	\bibinfo{author}{\bibnamefont{Günzler},~\bibfnamefont{S.}},
	\bibinfo{author}{\bibnamefont{Spiecker},~\bibfnamefont{M.}},
	\bibinfo{author}{\bibnamefont{Nambisan},~\bibfnamefont{A.}},
	\bibinfo{author}{\bibnamefont{Wernsdorfer},~\bibfnamefont{W.}},
	\bibnamefont{and}
	\bibinfo{author}{\bibnamefont{Pop},~\bibfnamefont{I.~M.}},
	\hyperref{https://journals.aps.org/prapplied/abstract/10.1103/PhysRevApplied.20.014059}
	\bibinfo{title}{\bibfnamefont{Fano Interference in Microwave Resonator Measurements}}.
	\bibinfo{journal}{Physical Review Applied} \textbf{\bibinfo{volume}{20}},
	\bibinfo{pages}{014059} (\bibinfo{year}{2023})

	\bibitem[{\citenamefont{Khapaev et~al.}(2001)\citenamefont{Khapaev, Kidiyarova-Shevchenko, Magnelind, Kupriyanov}}]{Khapaev2001}
	\bibinfo{author}{\bibnamefont{Khapaev},~\bibfnamefont{M.~M.}},
	\bibinfo{author}{\bibnamefont{Kidiyarova-Shevchenko},~\bibfnamefont{A.~Yu.}},
	\bibinfo{author}{\bibnamefont{Magnelind},~\bibfnamefont{P.}},
	\bibnamefont{and}
	\bibinfo{author}{\bibnamefont{Kupriyanov},~\bibfnamefont{M.~Yu.}},
	\hyperref{https://ieeexplore.ieee.org/document/919537}
	\bibinfo{title}{\bibfnamefont{3D-MLSI: software package for inductance calculation in multilayer superconducting integrated circuits}}.
	\bibinfo{journal}{IEEE Transactions on Applied Superconductivity} \textbf{\bibinfo{volume}{11}},
	\bibinfo{pages}{1090-1093} (\bibinfo{year}{2001})

	\bibitem[{\citenamefont{Uhl et~al.}(2023)\citenamefont{Uhl, Hackenbeck, Füger, Kleiner, Koelle, Bothner}}]{Uhl2023}
	\bibinfo{author}{\bibnamefont{Uhl},~\bibfnamefont{K.}},
	\bibinfo{author}{\bibnamefont{Hackenbeck},~\bibfnamefont{D.}},
	\bibinfo{author}{\bibnamefont{Füger},~\bibfnamefont{C.}},
	\bibinfo{author}{\bibnamefont{Kleiner},~\bibfnamefont{R.}},
	\bibinfo{author}{\bibnamefont{Koelle},~\bibfnamefont{D.}},
	\bibnamefont{and}
	\bibinfo{author}{\bibnamefont{Bothner},~\bibfnamefont{D.}},
	\hyperref{https://pubs.aip.org/aip/apl/article/122/18/182603/2887638/A-flux-tunable-YBa2Cu3O7-quantum-interference}
	\bibinfo{title}{\bibfnamefont{A flux-tunable YBa$_2$Cu$_3$O$_{7}$ quantum interference microwave circuit}}.
	\bibinfo{journal}{Applied Physics Letters} \textbf{\bibinfo{volume}{122}},
	\bibinfo{pages}{182603} (\bibinfo{year}{2023})

	\bibitem[{\citenamefont{Gely et~al.}(2024)\citenamefont{Gely, Sanz~Mora, Yanai, van~der~Spek, Bothner, Steele}}]{Gely2024}
	\bibinfo{author}{\bibnamefont{Gely},~\bibfnamefont{M.~F.}},
	\bibinfo{author}{\bibnamefont{Sanz~Mora},~\bibfnamefont{A.}},
	\bibinfo{author}{\bibnamefont{Yanai},~\bibfnamefont{S.}},
	\bibinfo{author}{\bibnamefont{van~der~Spek},~\bibfnamefont{R.}},
	\bibinfo{author}{\bibnamefont{Bothner},~\bibfnamefont{D.}},
	\bibnamefont{and}
	\bibinfo{author}{\bibnamefont{Steele},~\bibfnamefont{G.~A.}},
	\hyperref{https://www.nature.com/articles/s41467-023-43128-y}
	\bibinfo{title}{\bibfnamefont{Apparent nonlinear damping triggered by quantum fluctuations}}.
	\bibinfo{journal}{Nature Communications} \textbf{\bibinfo{volume}{14}},
	\bibinfo{pages}{7566} (\bibinfo{year}{2024})

	\bibitem[{\citenamefont{Gardiner and Collett}(1985)\citenamefont{Gardiner, Collett}}]{Gardiner1985}
	\bibinfo{author}{\bibnamefont{Gardiner},~\bibfnamefont{C.~W.}}
	\bibnamefont{and}
	\bibinfo{author}{\bibnamefont{Collett},~\bibfnamefont{M.~J.}},
	\hyperref{https://journals.aps.org/pra/abstract/10.1103/PhysRevA.31.3761}
	\bibinfo{title}{\bibfnamefont{Input and output in damped quantum systems: Quantum stochastic differential equations and the master equation}}.
	\bibinfo{journal}{Physical Review A} \textbf{\bibinfo{volume}{31}},
	\bibinfo{pages}{3761} (\bibinfo{year}{1985})


\end{thebibliography}

\begin{thebibliography}{26}
	\expandafter\ifx\csname natexlab\endcsname\relax\def\natexlab#1{#1}\fi
	\expandafter\ifx\csname bibnamefont\endcsname\relax
	\def\bibnamefont#1{#1}\fi
	\expandafter\ifx\csname bibfnamefont\endcsname\relax
	\def\bibfnamefont#1{#1}\fi
	\expandafter\ifx\csname citenamefont\endcsname\relax
	\def\citenamefont#1{#1}\fi
	\expandafter\ifx\csname url\endcsname\relax
	\def\url#1{\texttt{#1}}\fi
	\expandafter\ifx\csname urlprefix\endcsname\relax\def\urlprefix{URL }\fi
	\providecommand{\bibinfo}[2]{#2}
	\providecommand{\eprint}[2][]{\url{#2}}

	\bibitem[{\citenamefont{Igreja and Dias}(2004)\citenamefont{Igreja, Dias}}]{Igreja2004x}
	\bibinfo{author}{\bibnamefont{Igreja},~\bibfnamefont{R.}} \bibnamefont{and}
	\bibinfo{author}{\bibnamefont{Dias},~\bibfnamefont{C.~J.}},
	\hyperref{https://www.sciencedirect.com/science/article/pii/S0924424704000779?via\%3Dihub}
	\bibinfo{title}{\bibfnamefont{Analytical evaluation of the interdigital electrodes capacitance for a multi-layered structure}}.
	\bibinfo{journal}{Sensors and Actuators A} \textbf{\bibinfo{volume}{112}},
	\bibinfo{pages}{291-301} (\bibinfo{year}{2004})

	\bibitem[{\citenamefont{Wenner et~al.}(2011)\citenamefont{Wenner, Neeley, Bialczak, Lenander, Lucero, O'Connell, Sank, Wang, Weides, Cleland, Martinis}}]{Wenner2011x}
	\bibinfo{author}{\bibnamefont{Wenner},~\bibfnamefont{J.}},
	\bibinfo{author}{\bibnamefont{Neeley},~\bibfnamefont{M.}},
	\bibinfo{author}{\bibnamefont{Bialczak},~\bibfnamefont{Radoslav~C.}},
	\bibinfo{author}{\bibnamefont{Lenander},~\bibfnamefont{M.}},
	\bibinfo{author}{\bibnamefont{Lucero},~\bibfnamefont{E.}},
	\bibinfo{author}{\bibnamefont{O'Connell},~\bibfnamefont{A.~D.}},
	\bibinfo{author}{\bibnamefont{Sank},~\bibfnamefont{D.}},
	\bibinfo{author}{\bibnamefont{Wang},~\bibfnamefont{H.}},
	\bibinfo{author}{\bibnamefont{Weides},~\bibfnamefont{M.}},
	\bibinfo{author}{\bibnamefont{Cleland},~\bibfnamefont{A.~N.}},
	\bibnamefont{and}
	\bibinfo{author}{\bibnamefont{Martinis},~\bibfnamefont{J.~M.}},
	\hyperref{https://iopscience.iop.org/article/10.1088/0953-2048/24/6/065001}
	\bibinfo{title}{\bibfnamefont{Wirebond crosstalk and cavity modes in large chip mounts for superconducting qubits}}.
	\bibinfo{journal}{Superconductor Science and Technology} \textbf{\bibinfo{volume}{24}},
	\bibinfo{pages}{065001} (\bibinfo{year}{2011})

	\bibitem[{\citenamefont{Rieger et~al.}(2023)\citenamefont{Rieger, Günzler, Spiecker, Nambisan, Wernsdorfer, Pop}}]{Rieger2023x}
	\bibinfo{author}{\bibnamefont{Rieger},~\bibfnamefont{D.}},
	\bibinfo{author}{\bibnamefont{Günzler},~\bibfnamefont{S.}},
	\bibinfo{author}{\bibnamefont{Spiecker},~\bibfnamefont{M.}},
	\bibinfo{author}{\bibnamefont{Nambisan},~\bibfnamefont{A.}},
	\bibinfo{author}{\bibnamefont{Wernsdorfer},~\bibfnamefont{W.}},
	\bibnamefont{and}
	\bibinfo{author}{\bibnamefont{Pop},~\bibfnamefont{I.~M.}},
	\hyperref{https://journals.aps.org/prapplied/abstract/10.1103/PhysRevApplied.20.014059}
	\bibinfo{title}{\bibfnamefont{Fano Interference in Microwave Resonator Measurements}}.
	\bibinfo{journal}{Physical Review Applied} \textbf{\bibinfo{volume}{20}},
	\bibinfo{pages}{014059} (\bibinfo{year}{2023})

	\bibitem[{\citenamefont{Khapaev et~al.}(2001)\citenamefont{Khapaev, Kidiyarova-Shevchenko, Magnelind, Kupriyanov}}]{Khapaev2001x}
	\bibinfo{author}{\bibnamefont{Khapaev},~\bibfnamefont{M.~M.}},
	\bibinfo{author}{\bibnamefont{Kidiyarova-Shevchenko},~\bibfnamefont{A.~Yu.}},
	\bibinfo{author}{\bibnamefont{Magnelind},~\bibfnamefont{P.}},
	\bibnamefont{and}
	\bibinfo{author}{\bibnamefont{Kupriyanov},~\bibfnamefont{M.~Yu.}},
	\hyperref{https://ieeexplore.ieee.org/document/919537}
	\bibinfo{title}{\bibfnamefont{3D-MLSI: software package for inductance calculation in multilayer superconducting integrated circuits}}.
	\bibinfo{journal}{IEEE Transactions on Applied Superconductivity} \textbf{\bibinfo{volume}{11}},
	\bibinfo{pages}{1090-1093} (\bibinfo{year}{2001})

	\bibitem[{\citenamefont{Rodrigues et~al.}(2021)\citenamefont{Rodrigues, Bothner, Steele}}]{Rodrigues2021x}
	\bibinfo{author}{\bibnamefont{Rodrigues},~\bibfnamefont{I.~C.}},
	\bibinfo{author}{\bibnamefont{Bothner},~\bibfnamefont{D.}},
	\bibnamefont{and}
	\bibinfo{author}{\bibnamefont{Steele},~\bibfnamefont{G.~A.}},
	\hyperref{https://www.science.org/doi/10.1126/sciadv.abg6653}
	\bibinfo{title}{\bibfnamefont{Cooling photon-pressure circuits into the quantum regime}}.
	\bibinfo{journal}{Science Advances} \textbf{\bibinfo{volume}{7}},
	\bibinfo{pages}{eabg6653} (\bibinfo{year}{2021})

	\bibitem[{\citenamefont{Uhl et~al.}(2023)\citenamefont{Uhl, Hackenbeck, Füger, Kleiner, Koelle, Bothner}}]{Uhl2023x}
	\bibinfo{author}{\bibnamefont{Uhl},~\bibfnamefont{K.}},
	\bibinfo{author}{\bibnamefont{Hackenbeck},~\bibfnamefont{D.}},
	\bibinfo{author}{\bibnamefont{Füger},~\bibfnamefont{C.}},
	\bibinfo{author}{\bibnamefont{Kleiner},~\bibfnamefont{R.}},
	\bibinfo{author}{\bibnamefont{Koelle},~\bibfnamefont{D.}},
	\bibnamefont{and}
	\bibinfo{author}{\bibnamefont{Bothner},~\bibfnamefont{D.}},
	\hyperref{https://pubs.aip.org/aip/apl/article/122/18/182603/2887638/A-flux-tunable-YBa2Cu3O7-quantum-interference}
	\bibinfo{title}{\bibfnamefont{A flux-tunable YBa$_2$Cu$_3$O$_{7}$ quantum interference microwave circuit}}.
	\bibinfo{journal}{Applied Physics Letters} \textbf{\bibinfo{volume}{122}},
	\bibinfo{pages}{182603} (\bibinfo{year}{2023})

	\bibitem[{\citenamefont{Uhl et~al.}(2024)\citenamefont{Uhl, Hackenbeck, Peter, Kleiner, Koelle, Bothner}}]{Uhl2024x}
	\bibinfo{author}{\bibnamefont{Uhl},~\bibfnamefont{K.}},
	\bibinfo{author}{\bibnamefont{Hackenbeck},~\bibfnamefont{D.}},
	\bibinfo{author}{\bibnamefont{Peter},~\bibfnamefont{J.}},
	\bibinfo{author}{\bibnamefont{Kleiner},~\bibfnamefont{R.}},
	\bibinfo{author}{\bibnamefont{Koelle},~\bibfnamefont{D.}},
	\bibnamefont{and}
	\bibinfo{author}{\bibnamefont{Bothner},~\bibfnamefont{D.}},
	\hyperref{https://journals.aps.org/prapplied/abstract/10.1103/PhysRevApplied.21.024051}
	\bibinfo{title}{\bibfnamefont{Niobium quantum interference microwave circuits with monolithic three-dimensional nanobridge junctions}}.
	\bibinfo{journal}{Physical Review Applied} \textbf{\bibinfo{volume}{21}},
	\bibinfo{pages}{024051} (\bibinfo{year}{2024})

	\bibitem[{\citenamefont{Uhl et~al.}(2024)\citenamefont{Uhl, Hackenbeck, Koelle, Kleiner, Bothner}}]{Uhl2024ax}
	\bibinfo{author}{\bibnamefont{Uhl},~\bibfnamefont{K.}},
	\bibinfo{author}{\bibnamefont{Hackenbeck},~\bibfnamefont{D.}},
	\bibinfo{author}{\bibnamefont{Koelle},~\bibfnamefont{D.}},
	\bibinfo{author}{\bibnamefont{Kleiner},~\bibfnamefont{R.}},
	\bibnamefont{and}
	\bibinfo{author}{\bibnamefont{Bothner},~\bibfnamefont{D.}},
	\hyperref{https://journals.aps.org/prapplied/abstract/10.1103/PhysRevApplied.22.064052}
	\bibinfo{title}{\bibfnamefont{Extracting the current-phase relation of a monolithic three-dimensional nanoconstriction using a dc-current-tunable superconducting microwave cavity}}.
	\bibinfo{journal}{Physical Review Applied} \textbf{\bibinfo{volume}{22}},
	\bibinfo{pages}{064052} (\bibinfo{year}{2024})

	\bibitem[{\citenamefont{Primo et~al.}(2023)\citenamefont{Primo, Pinho, Benevides, Gröblacher, Wiederhecker, Mayer~Alegre}}]{Primo2023x}
	\bibinfo{author}{\bibnamefont{Primo},~\bibfnamefont{A.~G.}},
	\bibinfo{author}{\bibnamefont{Pinho},~\bibfnamefont{P.~V.}},
	\bibinfo{author}{\bibnamefont{Benevides},~\bibfnamefont{R.}},
	\bibinfo{author}{\bibnamefont{Gröblacher},~\bibfnamefont{S.}},
	\bibinfo{author}{\bibnamefont{Wiederhecker},~\bibfnamefont{G.~S.}},
	\bibnamefont{and}
	\bibinfo{author}{\bibnamefont{Mayer~Alegre},~\bibfnamefont{T.~P.}},
	\hyperref{https://www.nature.com/articles/s41467-023-41127-7}
	\bibinfo{title}{\bibfnamefont{Dissipative optomechanics in high-frequency nanomechanical resonators}}.
	\bibinfo{journal}{Nature Communications} \textbf{\bibinfo{volume}{14}},
	\bibinfo{pages}{5793} (\bibinfo{year}{2023})

	\bibitem[{\citenamefont{Gely et~al.}(2023)\citenamefont{Gely, Sanz~Mora, Yanai, van~der~Spek, Bothner, Steele}}]{Gely2023x}
	\bibinfo{author}{\bibnamefont{Gely},~\bibfnamefont{M.~F.}},
	\bibinfo{author}{\bibnamefont{Sanz~Mora},~\bibfnamefont{A.}},
	\bibinfo{author}{\bibnamefont{Yanai},~\bibfnamefont{S.}},
	\bibinfo{author}{\bibnamefont{van~der~Spek},~\bibfnamefont{R.}},
	\bibinfo{author}{\bibnamefont{Bothner},~\bibfnamefont{D.}},
	\bibnamefont{and}
	\bibinfo{author}{\bibnamefont{Steele},~\bibfnamefont{G.~A.}},
	\hyperref{https://www.nature.com/articles/s41467-023-43128-y}
	\bibinfo{title}{\bibfnamefont{Apparent nonlinear damping triggered by quantum fluctuations}}.
	\bibinfo{journal}{Nature Communications} \textbf{\bibinfo{volume}{14}},
	\bibinfo{pages}{7566} (\bibinfo{year}{2023})

	\bibitem[{\citenamefont{Rodrigues et~al.}(2022)\citenamefont{Rodrigues, Steele, Bothner}}]{Rodrigues2022x}
	\bibinfo{author}{\bibnamefont{Rodrigues},~\bibfnamefont{I.~C.}},
	\bibinfo{author}{\bibnamefont{Steele},~\bibfnamefont{G.~A.}},
	\bibnamefont{and}
	\bibinfo{author}{\bibnamefont{Bothner},~\bibfnamefont{D.}},
	\hyperref{https://www.science.org/doi/10.1126/sciadv.abq1690}
	\bibinfo{title}{\bibfnamefont{Parametrically enhanced interactions and nonreciprocal bath dynamics in a photon-pressure Kerr amplifier}}.
	\bibinfo{journal}{Science Advances} \textbf{\bibinfo{volume}{8}},
	\bibinfo{pages}{eabq1690} (\bibinfo{year}{2022})

	\bibitem[{\citenamefont{Rodrigues et~al.}(2024)\citenamefont{Rodrigues, Steele, Bothner}}]{Rodrigues2024x}
	\bibinfo{author}{\bibnamefont{Rodrigues},~\bibfnamefont{I.~C.}},
	\bibinfo{author}{\bibnamefont{Steele},~\bibfnamefont{G.~A.}},
	\bibnamefont{and}
	\bibinfo{author}{\bibnamefont{Bothner},~\bibfnamefont{D.}},
	\hyperref{https://journals.aps.org/prl/abstract/10.1103/PhysRevLett.132.203603}
	\bibinfo{title}{\bibfnamefont{Photon Pressure with an Effective Negative Mass Microwave Mode}}.
	\bibinfo{journal}{Physical Review Letters} \textbf{\bibinfo{volume}{132}},
	\bibinfo{pages}{203603} (\bibinfo{year}{2024})

	\bibitem[{\citenamefont{Gardiner and Collett}(1985)\citenamefont{Gardiner, Collett}}]{Gardiner1985x}
	\bibinfo{author}{\bibnamefont{Gardiner},~\bibfnamefont{C.~W.}}
	\bibnamefont{and}
	\bibinfo{author}{\bibnamefont{Collett},~\bibfnamefont{M.~J.}},
	\hyperref{https://journals.aps.org/pra/abstract/10.1103/PhysRevA.31.3761}
	\bibinfo{title}{\bibfnamefont{Input and output in damped quantum systems: Quantum stochastic differential equations and the master equation}}.
	\bibinfo{journal}{Physical Review A} \textbf{\bibinfo{volume}{31}},
	\bibinfo{pages}{3761} (\bibinfo{year}{1985})

	\bibitem[{\citenamefont{Elste et~al.}(2009)\citenamefont{Elste, Girvin, Clerk}}]{Elste2009x}
	\bibinfo{author}{\bibnamefont{Elste},~\bibfnamefont{F.}},
	\bibinfo{author}{\bibnamefont{Girvin},~\bibfnamefont{S.~M.}},
	\bibnamefont{and}
	\bibinfo{author}{\bibnamefont{Clerk},~\bibfnamefont{A.~A.}},
	\hyperref{https://journals.aps.org/prl/abstract/10.1103/PhysRevLett.102.207209}
	\bibinfo{title}{\bibfnamefont{Quantum Noise Interference and Backaction Cooling in Cavity Nanomechanics}}.
	\bibinfo{journal}{Physical Review Letters} \textbf{\bibinfo{volume}{102}},
	\bibinfo{pages}{207209} (\bibinfo{year}{2009})

	\bibitem[{\citenamefont{Xuereb et~al.}(2011)\citenamefont{Xuereb, Schnabel, Hammerer}}]{Xuereb2011x}
	\bibinfo{author}{\bibnamefont{Xuereb},~\bibfnamefont{A.}},
	\bibinfo{author}{\bibnamefont{Schnabel},~\bibfnamefont{R.}},
	\bibnamefont{and}
	\bibinfo{author}{\bibnamefont{Hammerer},~\bibfnamefont{K.}},
	\hyperref{https://journals.aps.org/prl/abstract/10.1103/PhysRevLett.107.213604}
	\bibinfo{title}{\bibfnamefont{Dissipative Optomechanics in a Michelson-Sagnac Interferometer}}.
	\bibinfo{journal}{Physical Review Letters} \textbf{\bibinfo{volume}{107}},
	\bibinfo{pages}{213604} (\bibinfo{year}{2011})

	\bibitem[{\citenamefont{Weiss and Nunnenkamp}(2013)\citenamefont{Weiss, Nunnenkamp}}]{Weiss2013x}
	\bibinfo{author}{\bibnamefont{Weiss},~\bibfnamefont{T.}}
	\bibnamefont{and}
	\bibinfo{author}{\bibnamefont{Nunnenkamp},~\bibfnamefont{A.}},
	\hyperref{https://journals.aps.org/pra/abstract/10.1103/PhysRevA.88.023850}
	\bibinfo{title}{\bibfnamefont{Quantum limit of laser cooling in dispersively and dissipatively coupled optomechanical systems}}.
	\bibinfo{journal}{Physical Review A} \textbf{\bibinfo{volume}{88}},
	\bibinfo{pages}{023850} (\bibinfo{year}{2013})

	\bibitem[{\citenamefont{Bothner et~al.}(2022)\citenamefont{Bothner, Rodrigues, Steele}}]{Bothner2022x}
	\bibinfo{author}{\bibnamefont{Bothner},~\bibfnamefont{D.}},
	\bibinfo{author}{\bibnamefont{Rodrigues},~\bibfnamefont{I.~C.}},
	\bibnamefont{and}
	\bibinfo{author}{\bibnamefont{Steele},~\bibfnamefont{G.~A.}},
	\hyperref{https://www.nature.com/articles/s42005-022-00808-3}
	\bibinfo{title}{\bibfnamefont{Four-wave-cooling to the single phonon level in Kerr optomechanics}}.
	\bibinfo{journal}{Communications Physics} \textbf{\bibinfo{volume}{5}},
	\bibinfo{pages}{33} (\bibinfo{year}{2022})

\end{thebibliography}
\end{document}